\documentclass[12pt,a4paper]{book}
\usepackage[T1]{fontenc}
\usepackage{graphicx}
\usepackage{indentfirst}
\usepackage{amsfonts}
\newcommand{\field}[1]{\mathbb{#1}}
\newcommand{\RR}{\field{R}}
\newcommand{\CC}{\field{C}}
\newcommand{\ID}{\field{I}}
\newcommand{\LL}{\field{L}}
\newcommand{\bi}[1]{\mbox{\boldmath$#1$}}
\newcommand{\balpha}{\mbox{\boldmath$\alpha$}}
\newcommand{\bbeta}{\mbox{\boldmath$\beta$}}
\newcommand{\bphi}{\mbox{\boldmath$\phi$}}
\newcommand{\bomega}{\mbox{\boldmath$\omega$}}
\newcommand{\bmu}{\mbox{\boldmath$\mu$}}
\newcommand{\bnu}{\mbox{\boldmath$\nu$}}
\newcommand{\brho}{\mbox{\boldmath$\rho$}}
\newcommand{\bsigma}{\mbox{\boldmath$\sigma$}}
\newcommand{\bgamma}{\mbox{\boldmath$\gamma$}}
\newcommand{\cfigl}[3]{\begin{figure}[!hbtp]\centering
 \includegraphics[width=.5\textwidth]{#2}\caption{\small{\bf{#3}}}\label{#1}\end{figure}}
\newcommand{\dfigsin}[2]{\begin{figure}[!hbtp]\centering%
 \includegraphics[width=.35\textwidth]{#1}\hspace{0.5cm}\includegraphics[width=.35\textwidth]{#2}%
\end{figure}}

\begin{document} 
\newpage
\frontmatter

\begin{center}

\dfigsin{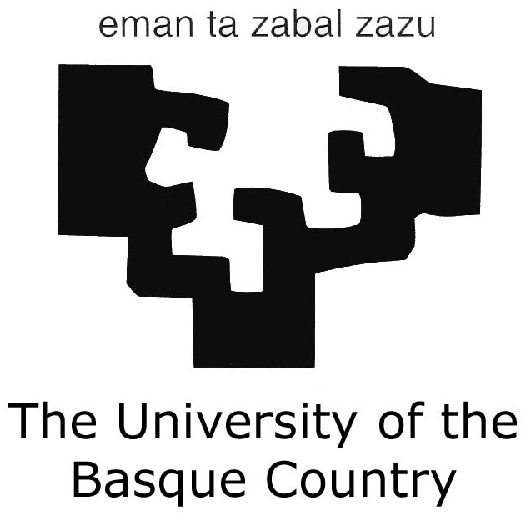}{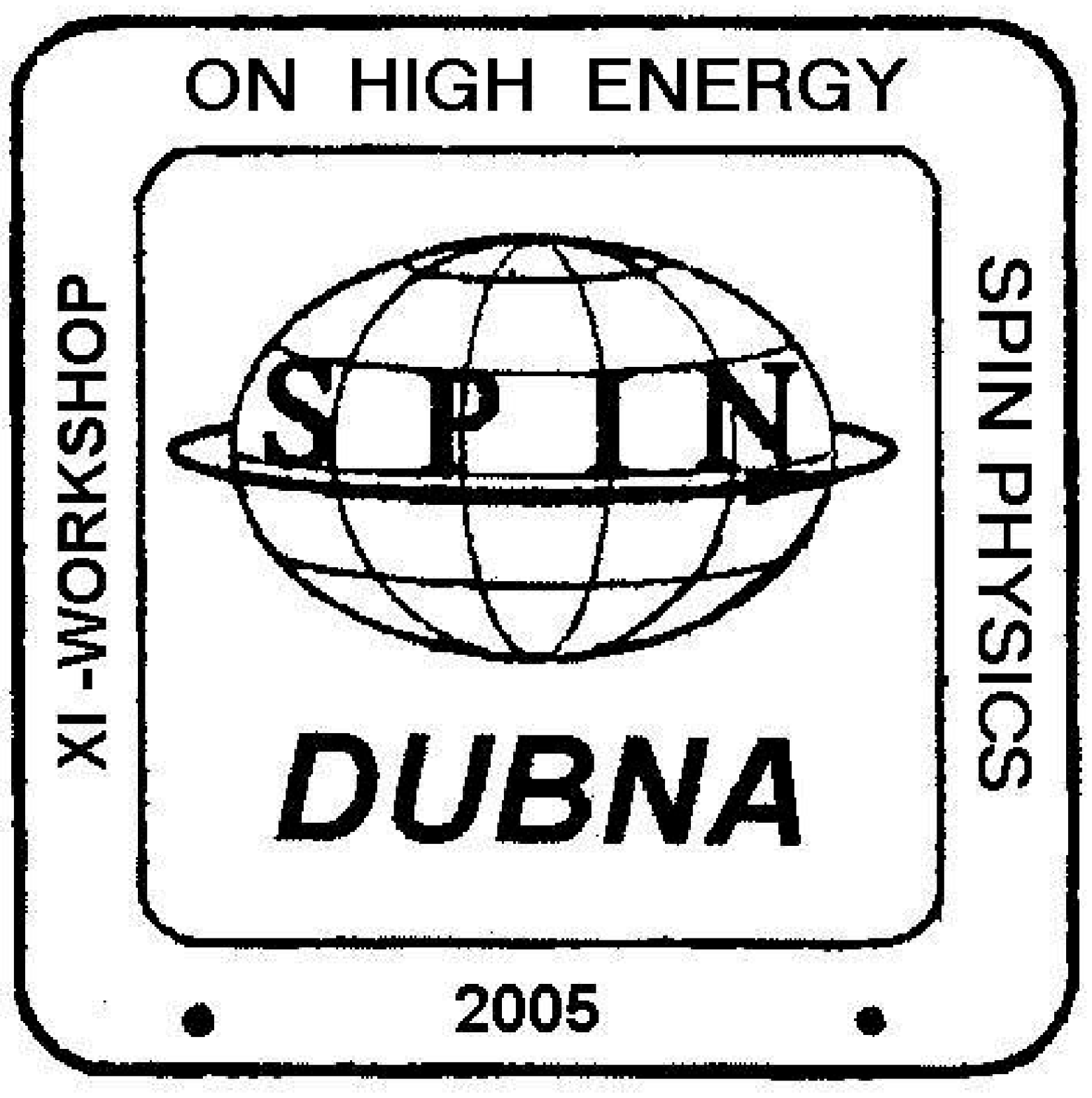}

\vspace{1cm}
\bf{\Huge{Kinematical formalism of \\ elementary spinning particles}}\\
\vspace{1cm}
(\large{Pre-Workshop Lecture Course})

\vspace{2cm}
{\Large{Mart\'{\i}n Rivas}\\
Theoretical Physics Department\\University of the Basque Country\\ 
Apdo.~644, 48080 Bilbao, Spain\\
\vspace{0.5cm}
{e-mail: martin.rivas@ehu.es}}

\vspace{4cm}
\huge{JINR, Dubna, 19-23 September 2005}
\end{center}

\strut\thispagestyle{empty}
\vfill
\pagebreak
\setcounter{page}{1}

\tableofcontents	

\newpage
\vspace{5cm}
\begin{itemize}

\item{If I can't picture it, I can't understand it.\\
You know, it would be sufficient to really understand the electron.
{\begin{flushright} A. Einstein \\
\end{flushright}}}
\vspace{0.5cm}
\item{If a spinning particle is not quite a point particle, nor a solid three dimensional top, what can it be?
What is the structure which can appear under probing with electromagnetic fields as a point charge,
yet as far as spin and wave properties are concerned exhibits a size of the order of the Compton wavelength?
{\begin{flushright} A.O. Barut \\
\end{flushright}}}
\end{itemize}
\newpage
\mainmatter
\chapter*{Preface}

The present notes contain some basic materials, physical and mathematical, of the general formalism
for analyzing elementary particles, which under the general name of {\it Kinematical Formalism of Elementary
Spinning Particles}, I have been working during the last years. 
The name {\it kinematical} makes reference
to its close relationship with the kinematical group of space-time transformations associated to
the Restricted Relativity Principle which a theoretical framework must necessarily satisfy.

In a certain sense it is a revision of the basic fundamentals of the Lagrangian formalism which leads
to Euler-Lagrange equations, Noether's theorem, etc., but looking for solutions which go through
the postulated initial and final states of the variational formalism. 
This produces a classical formalism
which is going to be expressed in terms of the end point variables of the dynamical evolution. This formalism 
is, therefore, closer to the quantum mechanical dynamical theory and it is through Feynman's path integral approach
that we can find the bridge between them. 

These end point variables, which I propose to call them {\it kinematical variables}, in the case
of elementary particles will necessarilly span a homogeneous space of the kinematical group. 
In this way, the kinematical group not only reflects the symmetries of the system. It also supplies
the necessary variables to describe elementary matter. It is crucial for the description of matter
to improve in our knowledge of this kinematical group. In the present notes we shall deal mainly with the
Galilei and Poincar\'e groups, but the formalism is so general that it can accomodate any further
group we consider as the basic symmetry group of matter. 

The notes pretend to be selfcontained and in this way we have included at the end of the first three chapters
some mathematical appendices which contain not very well spread materials. 
The lecture notes are organised as follows. We begin with an enumeration of some pros and cons the formalism
has. Probably the right place of this section will be at the end of the notes as a kind of 
conclusions and comments. I like to put it first because it suggests many of the features we can find
when analyzing the classical models and may create some provocation in the reader.
The remaining materials are collected into five chapters which more or less will cover one lecture each.

We begin with the general Lagrangian formalism just to enhance the role of the kinematical variables
in defining a concept of elementary particle. Lecture two will be devoted to the analysis of several
relativistic and nonrelativistic models, to show how the standard methods of analyzing symmetries leads to the
definition of the relevant observables. In particular, we shall pay attention to the definition
of the spin. The spin, as any other observable, will be defined in the classical case
in terms of the degrees of freedom and their derivatives, and we shall analyze its mathematical structure.

The next two lectures will cover the quantization of the formalism and the analysis of some relativistic and
nonrelativistic examples.
The separate fourth chapter is devoted to the model which satisfies Dirac's equation. Special attention is paid
to the analysis of Dirac's algebra and its relationship with the classical observables and to show
a geometrical interpretation of the difference in chirality between matter and antimatter.

Finally, some physical features which are related to the spin of the elementary particles, are described.
In some places, the lectures will be complemented with numerical simulations whenever the theoretical solution
is not available or very difficult to interpret because of the mathematical complexity.

I am very grateful to professor Oleg Teryaev for his kind invitation to the Joint Institute for Nuclear
Research at Dubna.

\vspace{3cm}

\begin{center}
Mart\'{\i}n Rivas\\
\vspace{1cm}
Bilbao-Dubna, September 2005.
\end{center}

\chapter*{Pros and cons of the kinematical formalism}

I will present in what follows a kind of general introduction to the kinematical formalism, 
in the form of some pros and cons the formalism has, 
from my point of view, with some additional comments.
There will be probably many more cons than the ones quoted, and the pros I consider
could be nonsense for other readers. You will probably
put some of the pros in the cons part. Do it, please. Science is a collective job and probably
the greatest endeavour of mankind.
Any new proposal needs a thorough analysis to find its contradictions. 
It is the only way to improve knowledge.
I am strongly convinced that the kinematical formalism I propose
has more advantages than previous approaches and it is simpler. It opens new perspectives
to establish a deeper formalism to deal with spinning particles from the very begining. 

Things like the prediction
of formation of electromagnetic bound pairs of electrons, the chiral difference between particles and antiparticles 
or the justification of $g=2$ by pure
kinematical arguments are sufficient, from my point of view, to deserve
some time to it, even to analyze its contradictions. One feature of the formalism is that it is not finished yet. 
Another is that it shows that the space-time symmetry group of 
the described elementary particles is larger than the
group we start with. We have to accomplish, and finish, this task.

We accept a {\bf variational formalism} for describing the dynamics of elementary particles.
To be consistent with the variational statements we have to look for solutions
of the corresponding differential equations, passing through the fixed end points. This implies
that the variables which define the end points of the formalism, which will be called from now on {\bf kinematical
variables}, will play in the classical case the same role as the wave function in the quantum formalism,
to characterising the states of the system. Our first task will be to rewrite the variational
formalism in terms of these kinematical variables. It is for this reason we have included a first
lecture about a generalized Lagrangian formalism.

\begin{center}
\large\bf{PROS}
\end{center}

\begin{itemize}

\item{An elementary particle is by definition the simplest mechanical system. 
An elementary particle can be annihilated but it can never be deformed.
Its intrinsic attributes are not modified by any interaction.
It has no excited states
and therefore all its possible states are just kinematical modifications of any one of them.
If the state of the particle changes it is always possible to find a new inertial observer who describes 
the particle in the same state as before.
In the quantum case this leads to the conclusion that the Hilbert space which describes its pure states
carries an {\bf irreducible representation} of the kinematical group.
In the classical case the kinematical Lagrangian space is a {\bf homogeneous space} of the kinematical group.}

\item{It is the {\bf Restricted Relativity Principle} which characterizes not only the space-time symmetry
group of the theory, usually called the {\bf kinematical group}, but also which supplies 
the classical variables to describe an elementary spinning particle.}

\item{The formalism is independent of the kinematical group. The classical variables which characterize
the kinematical state of the system are not postulated. They are related to 
the variables which characterize the parameterization of the kinematical group, and the manifold they span is necessarily a homogeneous
space of the kinematical group.}

\item{There is no need to use Grassmann or spinor variables to characterize the states
of a classical elementary spinning particle. If we restrict ourselves to the Galilei or Poincar\'e
groups, the only and most general variables to characterize the states of a classical elementary particle are
the 10 variables $t,{\bi r},{\bi v},\balpha$, interpreted respectively as the time, 
position of a point where the charge is located, velocity of this point
and finally the orientation of the system around this point.
These variables are in fact, the variables which define a 
parametrization of any of the mentioned groups.
}

\item{The point particle, either relativistic or non-relativistic, is an elementary particle
according to this formalism. It is not postulated, it is a consequence of the formalism. 
Its initial and final states are just characterized by the kinematical variables $t$ and ${\bi r}$.
It is the simplest localized system the formalism allows to describe but it 
represents an spinless object. It seems that there are no
spinless elementary particles in nature and therefore the use of spinless point particles
for the description of physical phenomena is an approximate one. If the whole formalism
of particle physics starts by dealing at first with spinless point particles and afterwards this system
is endowed with spin and other internal properties in some ad hoc manner, this will produce
a different framework than starting with spinning elementary particles from the very beginning.
There is a quotation by Albert Einstein that:
"Things should be made as simple as possible, but not simpler."}

\item{The spin is an angular momentum and therefore it is always defined with respect to some
fixed point. This point has to be clearly identified.
}

\item{If we understand as the spin of the electron as the angular momentum of the electron with 
respect to its center of mass, then the spin of the electron is twofold. One part is related to the rotation
of the electron ($\balpha$ variables and its derivative the angular velocity $\bomega$) and the other to the separation between the position of the charge 
and the center of mass, which is a different point, and its relative orbital motion. This comes from the use
of the ${\bi v}$ variables and its derivative, the acceleration ${\bi a}$.
The rotation does not produce magnetic moment.
The magnetic moment is the result of the motion of the charge around the center of mass, so that
when the magnetic moment is expressed in terms of the total spin this produces a 
clear interpretation of the gyromagnetic ratio.}

\item{The size and shape of any elementary particle are approximate observables. They are geometrical
aspects which can be defined for macroscopic objects but they loose their geometrical meaning
when talking about elementary particles.}

\item{The center of mass of the spinning particle, which is a very well defined
point from the classical point of view, can be related to the Newton-Wigner position operator
or to the position vector defined by the Foldy-Wouthuysen transformation, in the quantum case. 
Dirac spin operator is not the angular momentum of the electron with respect to its center of mass.
It is the angular momentum with respect to the center of charge.}

\item{The photon and the charge of the electron move at the speed of light. This motion
is not altered by any external interaction. Only the center of mass motion is affected.
This is not contradictory with special relativity because the center of mass or center of energy
moves always at a velocity $v<c$.}

\item{For any inertial observer the charge of the electron is never at rest and 
therefore it always measures 
magnetic moment and electric dipole moment with respect to the center of mass.}

\item{In a certain sense electromagnetism means speed of light for the carriers
of the electromagnetic interaction and also for the sources of the field.
}

\item{The charge of the electron is located at a single point and therefore we have no problems
associated to the charge distribution of the extended models. But at the same time it moves
and oscillates in
a region of radius half Compton's wavelength, where at this scale quantum phenomena appear.}

\item{In the relativistic case the most invariant way to define the internal structure of the electron
is precisely to assume that the charge is moving at the speed of light. This velocity cannot be altered
by any interaction, and therefore this corroborates the idea that an elementary particle 
cannot be deformed.}

\item{Classical particles which move in straight lines at the speed of light are massless particles which 
rotate around the direction of motion with the spin ${\bi S}$ either
parallel or antiparallel to the velocity, but in the same direction as the angular velocity ${\bomega}$. 
Spin is invariant under Poincar\'e transformations
but the angular velocity transforms to produce Doppler effect.
The spin, when quantized, is not restricted. It can take all greater than zero values 
$1/2$, $1$, $3/2$, $\ldots$. The spinless case is not predicted. The energy of the spin 1 object is just ${\bi S}\cdot{\bomega}=h\nu$, 
being $\nu$ the frequency of the rotation of the particle.}

\item{Particles whose charge moves in circles at the speed of light, in a plane orthogonal to the spin, 
although the center of mass is moving below $c$, satisfy when quantized, Dirac's equation.
All of them represent massive charged spin $1/2$ particles. Mass and charge are unrestricted in this formalism.
They are the only systems, predicted by this formalism, which satisfy Dirac equation when quantized.}

\item{Particles whose charge moves faster than light, although the center of mass is moving below $c$,
all of them represent, when quantized, massive charged spin 1 particles. Mass and charge are also unrestricted.
Is it a plausible description of classical massive $W^{\pm}$ bosons?}

\item{Elementary spinning particles show a clear chirality. 
Once the spin direction is fixed, the position of the charge of the
particle has a definite direction of motion, while the antiparticle has the opposite one.
They are mirror images of each other.}

\item{The electron is just a moving charge which can also rotate. The electric and magnetic moment 
are not intrinsic properties. They are derived observables produced by the separation between the center of mass
and the position of the charge and the motion of the charge, respectively. 
This justifies that quantum electrodynamics is completely determined by the minimal coupling 
between the electron current and the external potentials, 
and without any anomalous electric or magnetic coupling.}

\item{Dirac algebra of $4\times 4$ matrices is completely generated by complex linear combination
of 16 linearly independent hermitian matrices which represent very precise observables. 
Their real linear combinations will produce all translation invariant observables of the electron.
One of them is the unit matrix and the other 
are 15 traceless hermitian matrices which represent the following observables: 3 are the spin components
in the laboratory frame, other 3 are the spin components in the body frame and the remaining 9 are
the 9 components of the three unit vectors of the body frame. We must remember that the elementary
particle has as degrees of freedom the position of a point and its orientation in space which 
can be characterized by the description of the body frame. Once the translational degrees of freedom
are supressed, by analyzing for instance the electron in the center of mass frame,
the orientation observables completely characterize its translation invariant structure.
From the algebraic point of view Dirac algebra 
can be generated by products and linear combinations of any 4 of these 9 
components of its body frame.
}
\item{When analized the interaction between two spinning electrons different processes can be described
according to the energy involved and the separation between particles. In low energy we can describe 
elastic scattering and also the possibility of formation of a spin 1 bound state of 
two electrons, provided the energy is below a certain value, the spins are parallel 
and their respective center of
masses are separated by a distance below Compton's wavelength.
In high energy, for relative velocities of the center of masses of the particles greater than $0.1 c$, 
deep inelastic scattering processes appear, where by deep we mean that the two charges approach each other
below Compton's wavelength.
}
\item{The bound state of two equal charged particles is metastable. The mass of the system
is greater than the sum of the two masses of the elementary particles.}

\item{The formalism is complete in the sense that all Lagrangian systems whose kinematical space
is a homogeneous space of the kinematical group can be quantized and the corresponding Hilbert space
carries a projective unitary irreducible representation of the kinematical group. All known
one-particle wave equations are produced by quantization of the corresponding Lagrangian models.}

\item{To quantize these systems it is not necessary to use any constrained Hamiltonian formalism
like the one proposed by Dirac, because for every system we have a very well 
defined non-singular Lagrangian, as suggested by Feynman, and the quantization is performed through
Feynman's path integral approach.}

\item{The formalism predicts, if $PCT$ invariance is assumed, 
that the magnetic moment and spin of an elementary particle and its antiparticle 
must necessarily have the same relative orientation, either parallel or antiparallel. 
It is postulated that electrons and positrons
have opposite relative orientation.
To my knowledge no clear experimental evidence
of this relative measurement for free electrons and positrons have ever been performed. 
The same thing happens for $\mu^+$ and $\mu^-$. All very accurate 
measurements of $g-2$ are precession experiments 
which do not discriminate whether spin and magnetic moment are parallel or antiparallel.
}

\item{The description of the electron can be done in terms of dimensionless variables, so that the symmetry group
of the model is larger than the Poincar\'e group. Space-time dilations are among the new symmetries. Because
the description of the orientation has to be independent of how we choose the local inertial body frame, we have at least,
in the quantum formulation an additional $SU(2)\times U(1)$ kinematical group of space-time transformations.
What is the physical meaning of the generators of these groups? 
Although we started the formalism by assuming Poincar\'e invariance we find that the elementary spinning 
objects it describes have a larger kinematical group of symmetries. 
We have to start again, so that the new group will
give us new classical variables to describe more internal structure. But this additional 
'internal' structure is related to space-time symmetries. Poincar\'e invariance
only describes mass and spin as intrinsic (observer independent) properties. It is through the analysis
of the Casimir operators of the new enlarged group that new intrinsic properties could be defined. 
This new group does not commute with the Poincar\'e group so that the enlargement is 
not simply a direct product of the two groups. This has not been completed yet.}

\end{itemize}

\vspace{1cm}

\begin{center}
\large\bf{CONS}
\end{center}

\begin{itemize}
\item{We consider massive particles whose charge is moving at the speed of light. 
The charge of a massive elementary spinning particle is located at a point. The classical system
that when quantized satisfies Dirac equation is that one whose charge moves at the speed of light.
This is not contradictory with special relativity because no information is traveling faster than light
for distances greater than Compton's wavelength, which is the spatial domain for the quantum phenomena. 
Energy moves with velocity below $c$.
This possibility of motion at the  speed of light
is also contained in Dirac's analysis of the electron.}

\item{The charge of a free electron moves in circles at the speed of light in the center of mass frame.
It is therefore accelerated and classical mechanics predicts radiation. But classical mechanics
predicts that this internal motion of the charge is stationary, and therefore quantum 
mechanics allows nonradiating stationary states. The classical theory of radiation has to be revisited
and radiation has to be associated with the acceleration of the center of mass.}

\item{The electric field created by the electron is neither static nor Coulomb like. It behaves
like $1/r$ from the retarded charge position. Nevertheless, the time average value 
of the electric field over one complete 
turn of the charge, is static and Coulomb like from around a distance of five 
Compton's wavelength from the center of mass
up to infinity and in any direction. 
It does not diverge at the origin where it goes to zero. The instantaneous electromagnetic 
energy density in the surroundings of the charge goes like $1/r^2$ and is 
greater than that for the strict Coulomb field
which goes like $1/r^4$. 
}

\item{The only divergences of the electromagnetic field appear in the zitterbewegung plane but all they go
like $1/r$ when $r\to 0$ and therefore the electromagnetic energy density has no divergence there.
Although they are promissing features I have not been able for the moment,
to renormalize these classical models.}

\item{The position of the charge of the electron satisfies a fourth order differential equation.
This fourth order differential equation is most difficult to analyze. But it can be separated
in a system of coupled second order differential equations for the center of mass and center of charge.
One possibility is that we have Newton-like differential equations for the center of mass in terms of the external force
and a harmonic oscillator like equations for the motion of the charge around the center of mass, and which
is independent of the external force, thus confirming the idea that the internal structure is unaffected
by the interaction.
Nevertheless a fourth order differential equation is the most general differential equation a point
can satisfy in three-dimensional space, as derived from Frenet-Serret equations.
External interactions on a spinning particle will produce forces and torques. It will therefore
change the curvature and torsion of the trajectory, thus justifying the need of a fourth order differential equation.}

\item{It is usually said that a point cannot rotate. It is clear that all matter 
that surround us moves and rotates. In the usual classical approach,
it seems that the only exception are the elementary particles. 
It is out of logic that the basic constituents of matter are excluded from rotation. Why?
This implies that the consideration of the point particle (and therefore spinless) as the starting
object to build all material systems as is done in Newtonian mechanics is only an approximate
formalism.
In our formalism we have to attach orientation
to the point to describe its possible rotation. And this rotation contributes to the spin of the system,
although there is another contribution coming from the orbital motion of the charge around the center of
mass.
}

\end{itemize}

\chapter{Lagrangian formalism}

\section{Generalized Lagrangian formalism} 
\index{Lagrangian formalism}
\index{Lagrangian formalism!generalized}\index{generalized!Lagrangian formalism}
\label{sec:formalism}

The Lagrangian formalism of generalized 
systems depending on higher order derivatives
was already worked out by Ostrogradsky.~\footnote{\hspace{0.1cm}M. Ostrogradsky, {\sl 
M\'emoire sur les \'equations diff\'erentielles
relatives au probl\`eme des isop\'erim\`etres,}{ Mem. Acad. St. 
Petersburg}, {\bf 6}(4), 385-517 (1850).} 
We shall 
outline it briefly here, mainly to analyze the generalized Lagrangians 
 \index{Lagrangian!generalized}
not only in terms of the independent degrees 
of freedom but also as functions of what we shall call the {\bf kinematical variables} of the 
system, {\sl i.e.}, of the end point variables of the variational formulation. 

Let us consider a mechanical system of $n$ degrees of freedom, 
characterized by a Lagrangian that depends on time $t$ and on the $n$ 
essential coordinates $q_i(t)$, that represent the $n$ 
independent degrees of freedom, and their derivatives up 
to a finite order $k$. Because we can have
time derivatives of arbitrary order we use a superindex enclosed in brackets to represent
the corresponding $k$-th derivative, {\sl i.e.}, $q_i^{(k)}(t)=d^kq_i(t)/dt^k$. The {\bf action functional}
 \index{action functional} 
is defined by: 
 \begin{equation} 
 {\cal A}[q] =\int_{t_1}^{t_2}L(t,q_i(t),q_i^{(1)}(t),\ldots,q_i^{(k)}(t))dt,
  \label{eq:accion1}
 \end{equation} 
where $i=1,\ldots,n$. Using a more compact notation we define $q_i^{(0)}\equiv q_i$, and therefore
we shall write 
\[
L(t,q_i(t),q_i^{(1)}(t),\ldots,q_i^{(k)}(t))\equiv L(t,q_i^{(s)}(t)),
\]
for $s=0,\ldots,k$.

The trajectory followed by 
the mechanical system is that path which passing through the fixed end-points at initial
and final times $t_1$ and $t_2$,
$q_i^{(s)}(t_1)$ and $q_i^{(s)}(t_2),\ i=1,...,n,\ s= 0,1,...,k-1$, makes 
extremal the action functional (\ref{eq:accion1}). Note that we need to fix as
boundary values of the variational principle some particular 
values of time $t$, the $n$ degrees of freedom $q_i$
and their derivatives up to order $k-1$, {\sl i.e.}, one order less than the highest derivative of
each variable $q_i$ in the Lagrangian, at both end points of the problem. In other words we can say 
that the Lagrangian of any arbitrary generalized system is in general an explicit function of the 
variables we keep fixed as end points of the variational formulation and also of their
next order time derivative.

\begin{quote}
\footnotesize{Once the action functional (\ref{eq:accion1}) is defined for some particular path $q_i(t)$, 
to analyze its variation let us produce an infinitesimal modification of the functions 
$q_i(t)$, $q_i(t)\to q_i(t)+\delta q_i(t)$ while leaving fixed the 
end-points of the variational problem, {\sl i.e.}, such that at $t_1$ and
$t_2$ the modification of the generalized coordinates and their derivatives
up to order $k-1$ vanish, and thus $\delta q_i^{(s)}(t_1)=\delta q_i^{(s)}(t_2)=0$, for
$i= 1,\ldots,n$ and $s=0,1,\ldots,k-1$. Then, the variation of the 
derivatives of the $q_i(t)$ is given  by $q_i^{(s)}(t)\to q_i^{(s)}(t) + \delta 
q_i^{(s)}(t)=q_i^{(s)}(t) + d^s\delta q_i(t)/dt^s$, since the modification
of the $s$-th derivative function is just the $s$-th derivative of the modification of the
corresponding function. This produces a 
variation in the action functional $\delta{\cal A}={\cal A}[q+\delta q]-{\cal A}[q]$, given by:
  \[ \delta {\cal A} 
=\int_{t_1}^{t_2}L(t,q_i^{(s)}(t)+\delta q_i^{(s)}(t))dt-
\int_{t_1}^{t_2}L(t,q_i^{(s)}(t))dt
 \]  
 \begin{equation}
 =\int_{t_1}^{t_2}dt\sum_{i=1}^n\left[\frac{\partial L}{\partial q_i}\delta 
q_i+\frac{\partial L}{\partial q_i^{(1)}}\delta q_i^{(1)}+\cdots+\frac{\partial 
L}{\partial q_i^{(k)}}\delta q_i^{(k)}\right],
\label{eq:sumaccion}
 \end{equation}  
after expanding to lowest order the first integral. The term
 \[ 
 \frac{\partial 
 L}{\partial q_i^{(1)}}\,\delta q_i^{(1)}=\frac{\partial 
 L}{\partial q_i^{(1)}}\,\frac{d}{ dt}\delta q_i=\frac{d}{ 
 dt}\left(\frac{\partial L}{\partial q_i^{(1)}}\delta q_i\right)-\frac{d}{ 
 dt}\left(\frac{\partial L}{\partial q_i^{(1)}}\right)\delta q_i,
 \]  and by partial integration of this expression between $t_1$ and $t_2$, 
it gives: 
 \[ 
 \int_{t_1}^{t_2}\frac{\partial L}{\partial 
q_i^{(1)}}\delta q_i^{(1)}dt=\frac{\partial L}{\partial q_i^{(1)}}\delta 
q_i(t_2)-\frac{\partial L}{\partial q_i^{(1)}}\delta q_i(t_1)-
\int_{t_1}^{t_2}\frac{d}{ dt}\left(\frac{\partial L}{\partial 
q_i^{(1)}}\right)\delta q_idt
 \]
 \[ 
=-\int_{t_1}^{t_2}\frac{d}{ 
dt}\left(\frac{\partial L}{\partial q_i^{(1)}}\right)\delta q_i\,dt, 
 \]  
because the variations $\delta q_i(t_1)$ and $\delta q_i(t_2)$, vanish. 
Similarly for the next term: 
 \[ 
\frac{\partial 
L}{\partial q_i^{(2)}}\delta q_i^{(2)}=\frac{\partial L}{\partial 
q_i^{(2)}}\frac{d}{ dt}\delta q_i^{(1)}=\frac{d}{ dt}\left(\frac{\partial 
L}{\partial q_i^{(2)}}\delta q_i^{(1)}\right)-\frac{d}{ dt}\left(\frac{\partial 
L}{\partial q_i^{(2)}}\right)\delta q_i^{(1)},
 \] 
 \[ 
\int_{t_1}^{t_2}\frac{\partial 
L}{\partial q_i^{(2)}}\delta q_i^{(2)}dt=-\int_{t_1}^{t_2}\frac{d}{ 
dt}\left(\frac{\partial L}{\partial q_i^{(2)}}\right)\delta 
q_i^{(1)}dt=\int_{t_1}^{t_2}\frac{d^2}{ dt^2}\left(\frac{\partial L}{\partial 
q_i^{(2)}}\right)\delta q_i\,dt, 
 \]
because $\delta q_i$ and $\delta q_i^{(1)}$ vanish at $t_1$ and $t_2$, 
and finally for the last term
 \[ 
\int_{t_1}^{t_2}\frac{\partial L}{\partial q_i^{(k)}}\delta q_i^{(k)}dt=
(-1)^k\int_{t_1}^{t_2}\frac{d^k}{ dt^k}\left(\frac{\partial 
L}{\partial q_i^{(k)}}\right)\delta q_i\,dt,
 \]  
so that each term of (\ref{eq:sumaccion}) is written only in terms of the variations of the
degrees of freedom $\delta q_i$ and not of their higher order derivatives. 
Remark that to reach these final expressions, it has been necessary to assume the vanishing
of all $\delta q_i^{(s)}$, for $s=0,\ldots,k-1$, at times $t_1$ and $t_2$. By collecting all terms we get
 \[ 
 \delta {\cal A} =\int_{t_1}^{t_2}dt\sum_{i=1}^n\left[\frac{\partial L}{\partial q_i}-
\frac{d}{ dt}\left(\frac{\partial L}{\partial q_i^{(1)}}\right)+
\cdots+(-1)^k\frac{d^k}{ 
dt^k}\left(\frac{\partial L}{\partial q_i^{(k)}}\right)\right]\delta 
q_i.
 \]
                                    
If the action functional is extremal along the path $q_i(t)$,
its variation must vanish, $\delta {\cal A}=0$. The variations $\delta q_i$ are arbitrary
and therefore all
terms between squared brackets cancel out. We obtain a system of $n$ differential equations,
    \begin{equation} 
\frac{\partial L}{\partial q_i}-
\frac{d}{ dt}\left(\frac{\partial L}{\partial q_i^{(1)}}\right)+
\cdots+(-1)^k\frac{d^k}{ 
dt^k}\left(\frac{\partial L}{\partial q_i^{(k)}}\right)=0,\quad 
i=1,\ldots,n,
\label{eq:Euler1}
 \end{equation}}
\end{quote} 
the Euler-Lagrange equations, \index{Euler-Lagrange equations}\index{generalized!Euler-Lagrange equations}\index{Euler-Lagrange equations!generalized}
which can be written in condensed form as:
 \begin{equation} 
\sum_{s=0}^k(-1)^s\frac{d^s}{ dt^s}\left(\frac{\partial L}{\partial 
q_i^{(s)}}\right)=0,\qquad i=1,\ldots,n.\label{eq:Euler}
 \end{equation}  

\section{Kinematical variables} 
 \index{kinematical variables} 
 \index{kinematical!variables} 
\label{sec:kinematvar}

In general, the system (\ref{eq:Euler}) is a system of $n$ 
ordinary differential equations of
order $2k$, and thus existence and uniqueness theorems guarantee only 
the existence of a solution of this system for the $2kn$ boundary conditions 
$q_i^{(s)}(t_1)$, $i=1,\ldots,n$ and $s=0,1,\ldots,2k-1$, at the 
initial instant $t_1$. 
However the variational problem has been stated by the requirement 
that the solution goes through the two fixed endpoints, a 
condition that does not guarantee neither the existence nor the 
uniqueness of the solution. Nevertheless, let us assume that with the 
fixed endpoint conditions of the variational problem, $q_i^{(s)}(t_1)$ and $q_i^{(s)}(t_2)$, $i= 
1,\ldots,n$ and $s=0,1,\ldots,k-1$, at times $t_1$ and $t_2$, respectively, there exists a solution of
(\ref{eq:Euler}) perhaps non-unique. This implies that the $2kn$ boundary 
conditions at time $t_1$ required by the existence and uniqueness theorems, can be 
expressed perhaps in a non-uniform 
way, as functions of the $kn$ conditions at each of the two endpoints. 
From now on, we shall consider 
systems in which this condition is satisfied. It turns out that a 
particular solution  passing through these points will be expressed as
a function of time with some explicit dependence of the end point values
\begin{equation}
\widetilde{q}_i(t)\equiv q_i(t;\, q_j^{(r)}(t_1),q_l^{(r)}(t_2)), 
\label{eq:gensoluci}
\end{equation}
$i,j,l=1,\ldots,n,\ r=0,1,\ldots k-1$, in 
terms of these boundary end point conditions.

\begin{quotation}
{\normalsize{
\noindent{\bf Definition:} 
The {\bf Action Function} \index{action function}\footnote{Please remark that we use the same
letter $A(\;)$ for the action function, followed by normal brackets containing the variables
of which it depends, and for the action functional $A[\;]$ which is followed by squared brackets to enhance
that it is not a function but rather a functional.}
of the system along a classical path is
the value of the action functional (\ref{eq:accion1}) when we introduce 
in the integrand a particular solution (\ref{eq:gensoluci}) passing 
through those endpoints: 
 \begin{equation} 
 \int_{t_1}^{t_2}L\left(t,\widetilde{q}_i(t)\right)dt = 
A\left(t_1,q_i^{(r)}(t_1);t_2,q_i^{(r)}(t_2)\right).
 \label{eq:funcionaccion}
 \end{equation}}}\\
\end{quotation}

Once the time integration is performed, we see that it will be an explicit function 
of the $kn+1$ variables at the initial instant, $q_j^{(r)}(t_1)$, $r=0,\ldots,k-1$ 
including the time $t_1$, and of the corresponding $kn+1$ variables at final time $t_2$.
We write it as 
\[
A\left(t_1,q_i^{(r)}(t_1);t_2,q_i^{(r)}(t_2)\right)\equiv A(x_1,x_2).
\]
We thus arrive at the following\\

\begin{quotation}
{\normalsize{
\noindent{\bf Definition:} 
The {\bf kinematical variables}\index{kinematical!variables}
of the system are the time $t$ and the $n$ 
degrees of freedom $q_i$ and their time derivatives up to order $k-1$. The 
manifold $X$ they span is the {\bf kinematical space}
 \index{kinematical!space} 
of the system.}}\\
\end{quotation}

The kinematical space for ordinary Lagrangians is just the configuration space spanned by variables $q_i$
enlarged with the time variable $t$. It is usually called the {\bf enlarged configuration space}.\index{enlarged configuration space} 
But for generalized Lagrangians it also 
includes higher order derivatives up to one order less than the highest derivative. 
Thus, {\bf the action function of a system becomes a function of the values the 
kinematical variables take at the end points of the trajectory}, $x_1$ 
and $x_2$. From now on we shall consider systems for which the action 
function is defined and is a continuous and differentiable function of 
the kinematical variables at the end points of its possible evolution. 
This function clearly has the property $A(x,x)=0$. 

\subsection{Replacement of time as evolution parameter}

The constancy of speed of light in special relativity brings space and time variables
on the same footing. So, the next step is to remove the time observable as the evolution parameter of the 
variational formalism and express the evolution as a function of some arbitrary parameter to be chosen properly. 
Then, let us assume that the trajectory of the system can be expressed in parametric form, 
in terms of some arbitrary evolution parameter $\tau$, 
$\{t(\tau),q_i(\tau)\}$. The functional 
(\ref{eq:accion1}) can be rewritten in terms of the kinematical variables
and their derivatives and becomes: 
 \[ 
{\cal A}[t,q] =\int_{\tau_1}^{\tau_2}L\left(t(\tau),q_i(\tau),\frac{\dot 
q_i(\tau)}{\dot t(\tau)},\ldots,\frac{\dot q_i^{(k-1)}(\tau)}{\dot 
t(\tau)}\right)\dot t(\tau)d\tau
 \]  
 \begin{equation} 
= \int_{\tau_1}^{\tau_2}\widehat L\left(x(\tau),\dot 
x(\tau)\right)d\tau,
 \label{eq:lkinemat}
 \end{equation} 
where the dot means derivative with respect to the evolution variable $\tau$ 
that without loss of
generality can be taken dimensionless.
 \index{evolution parameter} 
Therefore $\widehat L\equiv L(t(\tau),{\dot q_i^{(s)}}/{\dot t}(\tau))\,\dot t(\tau)$
has dimensions of action.

It seems that (\ref{eq:lkinemat}) represents the  
variational problem of a Lagrangian system depending only 
on first order derivatives and of $kn+1$ 
degrees of freedom. However the kinematical variables, considered as 
generalized coordinates, are not all independent. There exist among 
them the following $(k-1)n$ differential constraints 
\begin{equation}
q_i^{(s)}(\tau)=\dot q_i^{(s-1)}(\tau)/\dot t(\tau),\quad i=1,\ldots,n,\quad s 
= 1,\ldots,k-1. 
\label{eq:ligues}
\end{equation}

We can also see that the integrand $\widehat L$ is 
a homogeneous function of first degree as a function
of the derivatives of the kinematical variables. In fact, each 
time derivative function $q_i^{(s)}(t)$ has been replaced by the quotient
$ \dot q_i^{(s-1)}(\tau)/\dot t(\tau)$ of two derivatives with respect 
to $\tau$. Even the highest order $k$-th derivative function
$q^{(k)}_i={\dot{q}}^{(k-1)}_i/\dot{t}$ is expressed in terms of the
derivatives of the kinematical variables ${q}^{(k-1)}_i$ and ${t}$.
Thus the original function $L$ is a homogeneous function of zero 
degree in the derivatives of the kinematical variables. Finally, the 
last term $\dot t(\tau)$, gives to the new defined $\widehat L$ the character of a homogeneous
function of first degree. Then, Euler's theorem on homogeneous 
functions gives rise to the additional relation:
 \index{homogeneity condition}
 \begin{equation} 
 \widehat L(x(\tau),\dot x(\tau)) =\sum_j \frac{\partial\widehat L}
{\partial\dot x^j}\dot{x}^j=\sum_j F_j(x,\dot x)\dot{x}^j.
\label{eq:homogen}
 \end{equation}  
With the above $(k-1)n$ differentiable constraints among the 
kinematical variables (\ref{eq:ligues}) and condition (\ref{eq:homogen}), 
it reduces to $n$ the number of essential degrees of freedom of the 
system (\ref{eq:lkinemat}).

This possibility of expressing the Lagrangian as a homogeneous function of first degree of the
derivatives was already considered in 1933 by Dirac~\footnote{\hspace{0.1cm}P.A.M. Dirac, {\sl Proc. Cam. Phil. Soc.}
 {\bf 29}, 389 (1933):
``a greater elegance is obtained'', ``a symmetrical treatment suitable for relativity.''} on aesthetical grounds. 
It is this homogeneity of first degree in terms of the derivatives which will allow us
later to transform the variational formalism into a geodesic problem on the kinematical space $X$, but where
the metric $g_{ij}(x,\dot{x})$ will be direction dependent, and thus the particle trajectory
is a geodesic, not in a Riemannian manifold but rather in a Finsler space.\footnote{ G.S. Asanov, {\sl Finsler geometry, Relativity and Gauge theories}, Reidel 
Pub. Co, Dordrecht (1985).}

Function $\widehat L$ is not an explicit function of the evolution parameter 
 \index{evolution parameter}
$\tau$ and thus we can see that the variational problem (\ref{eq:lkinemat}), is invariant with 
respect to any arbitrary change of evolution parameter $\tau$.~{\footnote{\hspace{0.1cm}R. Courant, D. Hilbert, 
{\sl  Methods of Mathematical Physics}, Vol. 1, Interscience, N.Y. 
(1970); I.M. Gelfand, S.V. Fomin, {\sl  Calculus of Variations} 
Prentice Hall, Englewood Cliffs, N.J. (1963).}} 

\begin{quotation}\noindent\footnotesize{
In fact, if we change the evolution parameter $\tau=\tau(\theta)$, 
then the derivative $\dot t(\tau)= (dt/d\theta)(d\theta/d\tau)$ and $\dot 
q_i^{(s)}(\tau)=(dq_i^{(s)}(\theta)/d\theta)(d\theta/d\tau)$ such that 
the quotients 
\[
\frac{{\dot{q}}_i^{(s)}(\tau)}{\dot{t}(\tau)}=\frac{(dq_i^{(s)}(\theta)/d\theta)\,\dot{\theta}(\tau)}{(dt(\theta)/d\theta)\,\dot{\theta}(\tau)}
\equiv\frac{{\dot{q}}_i^{(s)}(\theta)}{\dot{t}(\theta)},
\]
where once again this last dot means derivation with respect to $\theta$.
It turns 
out that (\ref{eq:lkinemat}) can be written as:
 \[ 
A[t,q]=\int_{\tau_1}^{\tau_2} L(t(\theta),q_i(\theta),\ldots,\dot q_i^{(k-
1)}(\theta)/\dot t(\theta))\frac{dt(\theta)}{ d\theta}{d\theta} 
 \]
 \begin{equation} 
=\int_{\theta_1}^{\theta_2}\widehat L(x(\theta),\dot 
x(\theta))d\theta.
 \label{eq:cambiopar}
 \end{equation} }
\end{quotation}

\subsection{Recovering the Lagrangian from the Action function}

The formalism thus stated has the advantage that it is independent of the 
evolution parameter, and if we want to come back to a time evolution 
description, we just use our time as the evolution parameter and make the 
replacement $\tau=t$, and therefore $\dot t=1$. From now on we shall 
consider those systems for which the evolution can be described in a
parametric form, and we shall delete the symbol $ \;\widehat{\;\;}$ over the 
Lagrangian, which is understood as written in terms of the 
kinematical variables and their first order derivatives.

If what we know is the action function of any system $A(x_1,x_2)$, as a 
function of the kinematical variables at the end points we can proceed 
conversely and recover the Lagrangian $L(x,\dot x)$ by the 
limiting process:
 \begin{equation} 
L(x,\dot x) = \lim_{y\to x}\frac{\partial A(x,y)}{\partial 
y^j}\dot x^j,
\label{eq:limit}
 \end{equation} 
where the usual addition convention on repeated or dummy index $j$, 
extended to the whole set of kinematical variables, has been assumed.
 
\begin{quotation}\footnotesize{
If in (\ref{eq:lkinemat}) we consider two very close points
$x_1\equiv x$ and $x_2\equiv x+dx$,
we have that the action function $A(x,x+dx)=A(x,x+\dot{x}d\tau)=L(x,\dot x)d\tau$ 
and making a Taylor expansion of the 
function $A$ with the condition $A(x,x)=0$ we get (\ref{eq:limit}).}
\end{quotation}

\section{Generalized Noether's theorem}
 \index{Noether's theorem} \index{Noether's theorem!generalized} \index{generalized!Noether's theorem}
\label{sec:Noether}

Noether's analysis for generalized Lagrangian systems also states the following

\begin{quotation}\normalsize{
\noindent{\bf Theorem: } 
To every one-parameter group of continuous transformations that 
transform the action function of the system, leaving dynamical equations invariant, in the form 
\[
A(\delta gx_1,\delta gx_2)=
A(x_1,x_2)+B(x_2)\delta g-B(x_1)\delta g,
\] 
and where $B(x)$ is a function defined
on the kinematical space, there is associated a classical 
observable $N$, which is a constant of the motion. \\}
\end{quotation}

Let us assume the existence of a $r$-parameter continuous group of 
transformations $G$, of the enlarged configuration space $(t,q_i)$, that 
 \index{group prolongation}
can be extended as a transformation group to the whole kinematical 
space $X$. Let $\delta g$ be an infinitesimal element of $G$ with 
coordinates $\delta g^\alpha,\ \alpha=1,\ldots,r$ and its action on these 
variables be given by: 
 \begin{eqnarray} 
t\to t'&=&t+\delta t=t+M_\alpha(t,q)\delta 
g^\alpha,\label{eq:timepar}\\
q_i(t)\to q_i'(t')&=&q_i(t)+\delta q_i(t)=q_i(t)+M_{i\alpha}^{(0)}(t,q)\delta 
g^\alpha,\label{eq:tpar}
 \end{eqnarray} 
and its extension on the remaining kinematical variables by
 \begin{eqnarray} 
{q'}_i^{(1)}(t')&=&q_i^{(1)}(t)+\delta 
q_i^{(1)}(t)=q_i^{(1)}(t)+M_{i\alpha}^{(1)}(t,q,q^{(1)})\delta 
g^\alpha,\label{eq:q1par}
 \end{eqnarray} 
and in general
 \begin{equation} 
{q'}_i^{(s)}(t')=q_i^{(s)}(t)+\delta 
q_i^{(s)}(t)=q_i^{(s)}(t)+M_{i\alpha}^{(s)}(t,q,\ldots,q^{(s)})\delta 
q^\alpha,\qquad s=0,1,\ldots,k-1,
\label{eq:qspar}
 \end{equation} 
where $M_\alpha$ and $M_{i\alpha}^{(0)}$ are 
functions only of $q_i$ and $t$ while the functions $M_{i\alpha}^{(s)}$ 
with $s\ge 1$, obtained in terms of the derivatives of the previous ones,
will be functions of the time $t$ and of the variables 
$q_i$ and their time derivatives up to order $s$. 

\begin{quotation}\noindent\footnotesize{
For instance, \[
{q'}_i^{(1)}(t')\equiv\frac{d{q'}_i(t')}{dt'}=\frac{d({q}_i(t)+M_{i\alpha}^{(0)}\delta g^\alpha)}{dt}\,\frac{dt}{dt'},
\]
but up to first order in $\delta g$
\[
\frac{dt}{dt'}=1-M_\alpha(t,q)\delta g^\alpha,
\]
and thus
\[
{q'}_i^{(1)}(t')=q_i^{(1)}(t)+\left(\frac{dM_{i\alpha}^{(0)}(t,q)}{dt}-q_i^{(1)}M_\alpha(t,q)\right)\delta g^\alpha,
\]
and comparing with (\ref{eq:q1par}) we get
\[
M_{i\alpha}^{(1)}(t,q,q^{(1)})=\frac{dM_{i\alpha}^{(0)}(t,q)}{dt}-q_i^{(1)}M_\alpha(t,q),
\]
where the total time derivative
\[
\frac{dM_{i\alpha}^{(0)}(t,q)}{dt}=\frac{\partial M_{i\alpha}^{(0)}(t,q)}{\partial t}+
\sum_j\frac{\partial M_{i\alpha}^{(0)}(t,q)}{\partial q_j}\,q_j^{(1)}.
\]
The remaining $M_{i\alpha}^{(s)}$ for $s>1$, are obtained in the same way from the previous
$M_{i\alpha}^{(s-1)}$.\\}

\end{quotation}


Under $\delta g$ the change of the action functional of the system is:
 \begin{eqnarray*}
\delta{\cal A}[\,q\,]&=&\int_{t'_1}^{t'_2}L(t',{q'}_i^{(s)}(t'))dt'-
\int_{t_1}^{t_2}L(t,q_i^{(s)}(t))dt\\
&=&\int_{t'_1}^{t'_2}L(t+\delta t,q_i^{(s)}(t)+\delta q_i^{(s)}(t))dt'-
\int_{t_1}^{t_2}L(t,q_i^{(s)}(t))dt.
 \end{eqnarray*}
By replacing in the first integral the integration range 
$(t_1',t_2')$ by $(t_1,t_2)$ having in mind the Jacobian of
$t'$ in terms of $t$, this implies that the differential $dt'=(1+d(\delta 
t)/dt)dt$, and thus: 
 \begin{eqnarray*}
\delta{\cal A}[\,q\,]&=&\int_{t_1}^{t_2}L(t+\delta t,q_i^{(s)}+\delta q_i^{(s)})\left(1+
\frac{d(\delta t)}{dt}\right)dt-\int_{t_1}^{t_2}L(t,q_i^{(s)})dt\\
&=&\int_{t_1}^{t_2} \left(L\frac{d(\delta t)}{ dt}+\frac{\partial L}{\partial t}\delta 
t+\frac{\partial L}{\partial q_i^{(s)}}\delta q_i^{(s)}(t)\right)dt,
 \end{eqnarray*}
keeping only for the Lagrangian
$L(t+\delta t, q^{(s)}+\delta q^{(s)})$, first order terms in its Taylor expansion.

Now, in the total variation of $\delta q_i^{(s)}(t)={q'}_i^{(s)}(t')-
q_i^{(s)}(t)$ is contained a variation in the form of the function 
$q_i^{(s)}(t)$ and a variation in its argument $t$, that is also affected by the
transformation of the group, {\sl i.e.},
 \begin{eqnarray*}
\delta q_i^{(s)}= {q'}_i^{(s)}(t+\delta t)-q_i^{(s)}(t)&=&{q'}_i^{(s)}(t)-
q_i^{(s)}(t)+(dq_i^{(s)}(t)/dt)\delta t\\
&=&\bar\delta q_i^{(s)}(t)+q_i^{(s+1)}(t)\delta t,
 \end{eqnarray*} 
where $\bar\delta q_i^{(s)}(t)$ is the variation in form of the function $q_i^{(s)}(t)$ 
at the instant of time $t$. 
Taking into account that for the variation in form
 \[
\bar\delta q_i^{(s)}(t)=d^s(\bar\delta 
q_i(t))/dt^s=d(\bar\delta q_i^{(s-1)}(t))/dt,
 \] 
it follows that 
 \[
\delta{\cal A}[\,q\,]=\int_{t_1}^{t_2} \left(L\frac{d(\delta t)}{ dt}+\frac{\partial L}{\partial 
t}\delta t+\frac{\partial L}{\partial q_i^{(s)}}\bar\delta 
q_i^{(s)}(t)+\frac{\partial L}{\partial q_i^{(s)}}\frac{dq_i^{(s)}}{ dt}\delta 
t\right)dt
 \] 
 \begin{equation} 
=\int_{t_1}^{t_2} \left(\frac{d(L\delta t)}{ dt}+ \frac{\partial 
L}{\partial q_i^{(s)}}\bar\delta q_i^{(s)}(t)\right)dt.
\label{eq:actifunc}
 \end{equation} 
Making the replacements
 \begin{eqnarray*}
\frac{\partial L}{\partial q_i}\bar\delta q_i&=&\frac{\partial 
L}{\partial q_i}\bar\delta q_i,\\  \frac{\partial L}{\partial 
q_i^{(1)}}\bar\delta q_i^{(1)}&=&\frac{\partial L}{\partial 
q_i^{(1)}}\frac{d(\bar\delta q_i)}{dt}=\frac{d}{dt}\left(\frac{\partial L}{\partial 
q_i^{(1)}}\bar\delta q_i\right)-\frac{d}{dt}\left(\frac{\partial L}{\partial 
q_i^{(1)}}\right)\bar\delta q_i,\\    \frac{\partial L}{\partial 
q_i^{(2)}}\bar\delta q_i^{(2)}&=&\frac{d}{dt}\left(\frac{\partial L}{\partial 
q_i^{(2)}}\bar\delta q_i^{(1)}\right)-\frac{d}{dt}\left(\frac{\partial 
L}{\partial q_i^{(2)}}\right)\bar\delta q_i^{(1)}  
  \end{eqnarray*} 
\[
=\frac{d}{dt}
\left(\frac{\partial L}{\partial q_i^{(2)}}\bar\delta q_i^{(1)}\right)-
\frac{d}{ dt}\left(\frac{d}{ dt}\left(\frac{\partial L}{\partial 
q_i^{(2)}}\right)\bar\delta q_i\right)+\frac{d^2}{ dt^2}\left(\frac{\partial 
L}{\partial q_i^{(2)}}\right)\bar\delta q_i,
\]
 \[
\frac{\partial L}{\partial 
q_i^{(k)}}\bar\delta q_i^{(k)}=\frac{d}{ dt}\left(\frac{\partial L}{\partial 
q_i^{(k)}}\bar\delta q_i^{(k-1)}\right)-\frac{d}{ dt}\left(\frac{d}{ 
dt}\left(\frac{\partial L}{\partial q_i^{(k)}}\right)\bar\delta q_i^{(k-
2)}\right)+\cdots,
  \]
and collecting terms we get 
\[
\delta{\cal A}[\,q\,]
= \int_{t_1}^{t_2}dt\left\{\frac{d(L\delta t)}{dt}\right.\hbox{\hspace{7cm}}
\]\vspace{-0.4cm}
\begin{eqnarray*}
&\,&+ \bar{\delta} 
q_i\left[\frac{\partial L}{\partial q_i}-\frac{d}{ dt}\left(\frac{\partial 
L}{\partial q_i^{(1)}}\right)+\cdots+(-1)^k\frac{d^k}{ dt^k}\left(\frac{\partial L}{\partial 
q_i^{(k)}}\right)\right]
\\
&\,&+\frac{d}{dt}\left(\bar\delta q_i\left[\frac{\partial 
L}{\partial q_i^{(1)}}-\frac{d}{ dt}\left(\frac{\partial L}{\partial 
q_i^{(2)}}\right)+\cdots+(-1)^{k-1}\frac{d^{k-1}}{ dt^{k-1}}\left(\frac{\partial 
L}{\partial q_i^{(k)}}\right)\right]\right)
\\
&\,&+\frac{d}{dt}\left(\bar\delta 
q_i^{(1)}\left[\frac{\partial L}{\partial q_i^{(2)}}-\frac{d}{ dt}\left(\frac{\partial 
L}{\partial q_i^{(3)}}\right)+\cdots+(-1)^{k-2}\frac{d^{k-2}}{ dt^{k-
2}}\left(\frac{\partial L}{\partial q_i^{(k)}}\right)\right]\right)
\\
&\,&+\left.\cdots+\frac{d}{dt}\left(\bar\delta q_i^{(k-1)}\left[\frac{\partial 
L}{\partial q_i^{(k)}}\right]\right)\right\}.
\end{eqnarray*}

\noindent The terms between squared brackets are precisely the conjugate 
momenta of order $s$, $p^i_{(s)}$, except the first one, which is the left-hand side 
of (\ref{eq:Euler1}) and vanishes identically if 
the functions $q_i$ satisfy the dynamical equations. 

\begin{quote}\footnotesize{In ordinary Lagrangian systems that depend only on first order
derivatives of the independent degrees of freedom, the canonical
approach associates to every generalized coordinate $q_i$ a dynamical
variable $p_i$, called its canonical conjugate momentum and defined by
\[
p_i=\frac{\partial L}{\partial{\dot q_i}}.
\]

As a generalization of this, in Lagrangian systems with higher order derivatives, 
a generalized canonical formalism can be obtained by defining various 
canonical conjugate momenta (up to a total of $k$ of them)
 \index{conjugate momenta!canonical} 
 \index{conjugate momenta!generalized} 
associated to each of the independent degrees of freedom $q_i$:~\footnote{\hspace{0.1cm}E.T.Whittaker, 
{\sl  Analytical Dynamics}, Cambridge 
University Press, Cambridge (1927), p.~265.}
 \begin{equation} 
p_{(s)}^i=\sum_{r=0}^{k-s}(-1)^r\frac{d^r}{ 
dt^r}\left(\frac{\partial L}{\partial q_i^{(r+s)}}\right),\quad 
s=1,\ldots,k,\quad i=1,\ldots,n, \label{eq:momenta1}
 \end{equation} 
which are precisely the above terms between the squared barckets. 
It is said that $p_{(s)}^i$ is the conjugate momentum of order $s$ of the variable $q_i$. }
\end{quote}

Now if we introduce in the integrand 
the variables $q_i$ that satisfy Euler-Lagrange equations, the variation of the action functional (\ref{eq:actifunc})
is transformed into the variation of the action function along the classical trajectory, and therefore,
the variation of the action function can be written as, 
 \begin{equation} 
\delta A(x_1,x_2)=\int_{t_1}^{t_2}\frac{d}{ 
dt}\left\{L\delta t+ \left(\bar\delta q_i p_{(1)}^i+\bar\delta 
q_i^{(1)}p_{(2)}^i+\cdots+\bar\delta q_i^{(k-1)}p_{(k)}^i\right)\right\}dt, 
 \label{eq:accion2}
 \end{equation}  
with  $p^i_{(s)}$ given in (\ref{eq:momenta1}). If we replace in (\ref{eq:accion2}) the form 
variation $\bar\delta q_i^{(s)}= \delta q_i^{(s)}- q_i^{(s+1)}$, then
 \begin{equation} 
\delta A(x_1,x_2)=\int_{t_1}^{t_2}\frac{d}{ 
dt}\left\{L\delta t+ \delta q_i^{(s)} p_{(s+1)}^i-q_i^{(s+1)}p_{(s+1)}^i\delta 
t\right\} dt
 \end{equation} 
with the usual addition convention. By substitution
of the variations $\delta t$ and $\delta q^{(s)}_i$ 
in terms of the infinitesimal element of the group $\delta g^\alpha$,  
(\ref{eq:tpar}-\ref{eq:qspar}), we get:
 \begin{equation} 
\delta A(x_1,x_2)=\int_{t_1}^{t_2}\frac{d}{ dt}\left\{\left(L-p_{(s)}^i q_i^{(s)}\right)M_\alpha 
+p_{(u+1)}^i M_{i\alpha}^{(u)}\right\}\delta g^\alpha dt,
 \label{eq:accion3}
 \end{equation}  
with the following range for repeated indexes for the addition 
convention, $i=1,\ldots,n$, $s=1,\ldots,k$, $u=0,1,\ldots,k-1$ and
$\alpha=1,\ldots,r$.

In the above integral we are using the solution of 
the dynamical equations, and therefore the variation of the action function is
\[
\delta A(x_1,x_2)=A(\delta g x_1,\delta g x_2)-A(x_1,x_2).
\] 
If it happens to be of first order in the group parameters in the form
\begin{equation}
\delta A(x_1,x_2)=B_\alpha(x_2)\delta g^\alpha-B_\alpha(x_1)\delta g^\alpha,
\label{eq:deltaAB}
\end{equation}
then equating to (\ref{eq:accion3}) we can perform the trivial time integral on the right hand side.
By considering that the group parameters $\delta g^\alpha$ are arbitrary, rearranging terms depending 
on $t_1$ and $t_2$ on the left- and 
right-hand side, respectively,
we get several observables that take the same values at the two 
arbitrary times $t_1$ and $t_2$.
They are thus constants of the motion and represent the time conserved
physical quantities, 
 \index{constants of the motion}
 \begin{equation} 
N_\alpha=B_\alpha(x)-\left(L-p_{(s)}^i q_i^{(s)}\right)M_\alpha-
p_{(s+1)}^iM_{i\alpha}^{(s)},\qquad \alpha=1,\ldots,r,
\label{eq:constant} 
 \end{equation} 
where the term within brackets $H=p_{(s)}^i q_i^{(s)}-L$ is the generalized Hamiltonian.

These are the $r$ Noether constants of the motion related to 
the infinitesimal transformations (\ref{eq:deltaAB}) of the action function under
the corresponding $r$-parameter Lie group.

To express the different magnitudes in terms of the kinematical variables, let us define the 
variables $x^j$ according to the 
rule: $x^0=t$, $x^i=q_i$, $x^{n+i}=q_i^{(1)},\ldots,x^{(k-1)n+i}=q_i^{(k-
1)}$. Since $L=\widehat L/\dot x^0$, and $q_i^{(s)}=\dot q_i^{(s-1)}/\dot 
x^0$, the derivatives in the definition of the canonical momenta can 
be written as:
 \begin{equation} 
\frac{\partial L}{\partial q_i^{(s)}}=\frac{\partial(\widehat L/\dot 
x^0)}{\partial\left(\dot x^{(s-1)n+i}/\dot x^0\right)}=\frac{\partial\widehat 
L}{\partial\dot x^{(s-1)n+i}}= F_{(s-1)n+i},
 \end{equation} 
in terms of the functions $F_i$ of the expansion (\ref{eq:homogen}) of the
Lagrangian.
The different conjugate momenta appear in the form:
 \begin{equation} 
p_{(s)}^i= \sum_{r=0}^{k-s}(-1)^r\frac{d^r}{ dt^r}F_{(r+s-1)n+i},
\label{eq:momenta2}
 \end{equation} 
in terms of the functions $F_i$ and their time derivatives.
Therefore the Noether constants of the motion are written as
 \index{Noether!constants of the motion}
 \begin{equation} 
N_\alpha=B_\alpha(x)-\left(F_j\frac{\dot x^j}{\dot x^0}-p_{(s)}^i\; \frac{\dot x^{(s-
1)n+i}}{\dot x^0}\right)M_\alpha-p_{(s+1)}^iM_{i\alpha}^{(s)}.
 \label{eq:constants}
 \end{equation}  

We see that the Noether constants of the motion $N_\alpha$ are finally expressed in terms of
the functions $F_i$ and their time derivatives, of the functions $M_{i\alpha}^{(s)}$
which represent the way the different kinematical variables transform under infinitesimal
transformations, and of the functions $B_\alpha$ which, as we shall see below, are related to the 
exponents of the group $G$. Functions $F_i$ and their time derivatives
are homogeneous functions of zero degree in 
terms of the derivatives of the kinematical variables $\dot x^i$. Functions
$B_\alpha(x)$ and $M_{i\alpha}^{(s)}(x)$ depend only on the kinematical variables.
Consequently, Noether constants of the 
motion are also homogeneous functions of zero degree in terms of the derivatives of kinematical variables and thus invariant under
arbitrary changes of evolution parameter. 

\section{Lagrangian gauge functions}
 \index{gauge function} \index{Lagrangian!gauge function}
\label{sec:Fungauge}

In the variational formulation of classical mechanics
 \begin{equation} 
{\cal A}[\,q\,]=\int_{t_1}^{t_2}L(t,q_i^{(s)}(t))dt\equiv 
\int_{\tau_1}^{\tau_2}L(x,\dot x)d\tau, 
\label{eq:2.5.1}
 \end{equation} 
${\cal A}[\,q\,]$ is a path functional, {\sl i.e.}, it takes in general 
different values for the different paths joining the fixed end points 
$x_1$ and $x_2$. Then it is necessary that $Ld\tau$ be a non-exact 
differential. Otherwise, if $Ldt=d\lambda$, then ${\cal A}[\,q\,]=\lambda_2-
\lambda_1$ and the functional does not distinguish between the 
different paths and the action function of the system from $x_1$ to 
$x_2$, $A(x_1,x_2)=\lambda(x_2)-\lambda(x_1)$, is expressed in terms 
of the potential function $\lambda(x)$, and is thus, path independent. 

If $\lambda(x)$ is a real function defined on the kinematical space 
$X$ of a Lagrangian system with action function $A(x_1,x_2)$, then 
the function $A'(x_1,x_2)=A(x_1,x_2)+\lambda(x_2)-
\lambda(x_1)$ is another action function equivalent to $A(x_1,x_2)$.
In fact it gives rise by (\ref{eq:limit}) to the Lagrangian $L'$ that differs 
from $L$ in a total $\tau$-derivative.~\footnote{\hspace{0.1cm}J.M. Levy-Leblond, {\sl Comm. Math. Phys.} {\bf 12}, 64 
(1969).}

Using (\ref{eq:limit}), we have
 \begin{equation} 
L'(x,\dot x)=L(x,\dot x)+\frac{d\lambda}{d\tau},
\label{eq:2.5.2}
 \end{equation}  
and therefore $L$ and $L'$ produce the same dynamical 
equations and $A(x_1,x_2)$ and $A'(x_1,x_2)$ are termed as equivalent action
functions.

Let $G$ be a transformation group of the enlarged configuration space 
$(t,q_i)$, that can be extended to a transformation group of the 
kinematical space $X$. Let $g\in G$ be an arbitrary element of $G$ and 
$x'=gx$, the transform of $x$. Consider a mechanical system 
characterized by the action function $A(x_1,x_2)$ that under the 
transformation $g$ is changed into $A(x'_1,x'_2)$. If $G$ is a symmetry 
group of the system, {\sl i.e.}, the dynamical equations in terms of the 
variables $x'$ are the same as those in terms of the variables $x$, 
this implies that $A(x'_1,x'_2)$ and $A(x_1,x_2)$ are necessarily equivalent 
action functions, and thus they will be related by: 
 \begin{equation} 
A(gx_1,gx_2)=A(x_1,x_2)+\alpha(g;x_2)-\alpha(g;x_1). 
\label{eq:2.5.3}
 \end{equation} 
The function $\alpha$ \index{gauge function} will be in general a 
continuous function of $g$ and $x$. This real function $\alpha(g;x)$ 
defined on $G\times X$ is called a {\bf gauge function} of the group $G$ for 
the kinematical space $X$. Because of the continuity of the group 
it satisfies $\alpha(e;x)=0$, $e$ being the neutral element of $G$. If 
the transformation $g$ is infinitesimal, let us represent it by 
the coordinates $\delta g^\sigma$, then 
$\alpha(\delta g;x)=\delta g^\sigma B_\sigma(x)$ to first order in the group parameters.
The transformation of the action function takes the form 
 \[ 
A(\delta gx_1,\delta gx_2)=A(x_1,x_2)+\delta g^\sigma
B_\sigma(x_2)-\delta g^\sigma B_\sigma(x_1),
 \] 
{\sl i.e.}, in the form required by Noether's theorem to obtain the 
corresponding conserved quantities. In general, $B_\sigma$ functions for gauge-variant
Lagrangians are obtained by
\begin{equation}
B_\sigma(x)=\left.\frac{\partial\alpha(g;x)}{\partial g^\sigma}\right|_{g=0}.
\label{eq:1B}
\end{equation}

Because of the associative property of the group law, 
any gauge function satisfies the identity
 \begin{equation}
 \alpha(g';gx)+\alpha(g;x)-\alpha(g'g;x)=\xi(g',g),
\label{eq:2.6.2}
 \end{equation} 
where the function $\xi$, defined on $G\times G$, is independent of 
$x$ and is an exponent of the group $G$. 
\begin{quotation}\footnotesize{
This can be seen by the mentioned
 associative property of the group law. From (\ref{eq:2.5.3}) we get:
 \begin{equation}
A(g'gx_1,g'gx_2) =A(x_1,x_2)+\alpha(g'g;x_2)-\alpha(g'g;x_1),
\label{eq:alfa1}
 \end{equation}
and also
\[
A(g'gx_1,g'gx_2)= A(gx_1,gx_2)+\alpha(g';gx_2)-\alpha(g';gx_1)\]
\[= A(x_1,x_2)+\alpha(g;x_2)-\alpha(g;x_1)+\alpha(g';gx_2)-\alpha(g';gx_1),
 \]
and therefore by identification of this with the above (\ref{eq:alfa1}), when
collecting terms with the same $x$ argument we get
 \[ 
\alpha(g';gx_2)+\alpha(g;x_2)-\alpha(g'g;x_2)=\alpha(g';gx_1)+\alpha(g;x_1)-
\alpha(g'g;x_1),
 \]  
and since $x_1$ and $x_2$ are two arbitrary points of $X$,
this expression is (\ref{eq:2.6.2}) and defines a function $\xi(g',g)$, independent of $x$.}
\end{quotation}

If we substitute this function $\xi(g',g)$ into (\ref{eq:2.2.2}) we see 
that it is satisfied identically. For $g'=g=e$, it reduces to
$\xi(e,e)=\alpha(e;x)=0$, and thus $\xi$ is an exponent of $G$. 

It is shown by Levy-Leblond in the previous reference that if $X$ is a homogeneous space of $G$, {\sl i.e.}, if there
exists a subgroup $H$ of $G$ such that
$X=G/H$, then, the exponent $\xi$ is equivalent to zero on the 
subgroup $H$, and gauge functions for homogeneous spaces become:
 \begin{equation} 
\alpha(g;x)=\xi(g,h_x),
\label{eq:2.6.4}
 \end{equation} 
where $h_x$ is any group element of the coset space represented by $x\in G/H$.

For the Poincar\'e group ${\cal P}$ all its exponents are equivalent to zero and 
thus the gauge functions when $X$ is a homogeneous space of ${\cal P}$ 
are identically zero. Lagrangians of relativistic systems whose 
kinematical spaces are homogeneous spaces of ${\cal P}$ can be taken 
strictly invariant.

However, the Galilei group ${\cal G}$ has nontrivial exponents, that 
are characterized by a parameter $m$ that is interpreted as the total 
mass of the system, and thus Galilei Lagrangians for massive systems 
are not in general invariant under ${\cal G}$. In the quantum 
formalism, the Hilbert space of states of a massive nonrelativistic 
system carries a projective unitary representation of the Galilei 
group instead of a true unitary representation.~\footnote{\hspace{0.1cm}see ref.7 and also
J.M. Levy-Leblond, {\sl Galilei Group and 
Galilean Invariance}, in E.M. Loebl, {\sl Group Theory and its 
applications}, Acad. Press, NY (1971), vol. 2, p.~221.}

\section{Elementary systems}
 \index{elementary systems}
 \index{elementary particle}
\label{sec:elemental}

In Newtonian mechanics the simplest geometrical object is a point of 
mass $m$. Starting with massive points we can construct arbitrary 
systems of any mass and shape, and thus any distribution of matter. The massive point
can be considered as the elementary particle of Newtonian mechanics. In the modern view of 
particle physics it corresponds to a spinless particle. We 
know that there exist spinning objects like electrons, muons, photons, 
neutrinos, quarks and perhaps many others, that can be considered as 
elementary particles in the sense that they cannot be considered as compound systems of 
other objects. Even more, we do not find in Nature any spinless elementary particles.
It is clear that the Newtonian point does 
not give account of the spin structure of particles and the 
existence of spin is a fundamental intrinsic attribute of an elementary 
particle, which is lacking in Newtonian mechanics, but it has to be accounted for.

In quantum mechanics, Wigner's work~\footnote{\hspace{0.1cm}see ref.1.} on the representations
of the inhomogeneous 
Lorentz group provides a very precise mathematical 
definition of the concept of elementary particle. An {\bf elementary 
particle} \index{elementary particle!quantum} is a quantum mechanical system whose Hilbert space 
of pure states is the representation space of a projective unitary 
irreducible representation of the Poincar\'e group. Irreducible 
representations of the Poincar\'e group are characterized by two 
invariant parameters $m$ and $S$, the mass and the spin of the system, 
respectively. By finding the different irreducible representations, we 
can obtain the quantum description of massless and massive particles of any 
spin. 

The very important expression of the above 
mathematical definition, with physical consequences, 
lies in the term {\bf irreducible}. 
 \index{irreducible}
Mathematically it means that the Hilbert space is an invariant vector 
space under the group action and that it has no other invariant 
subspaces. But it also means that there are no other states for a 
single particle than those that can be obtained by just taking any 
arbitrary vector state, form all its possible images in the different 
inertial frames and finally produce the closure of all finite linear combinations of these 
vectors. 

We see that starting from a single state and by a simple change of 
inertial observer, we obtain the state of the particle described in 
this new frame. Take the orthogonal part of this vector to 
the previous one and normalize it. Repeat this operation with another 
kinematical transformation acting on the same first state, followed by 
the corresponding orthonormalization procedure, as many times as 
necessary to finally obtain a complete orthonormal basis of the whole 
Hilbert space of states. All states in this basis are characterized by 
the physical parameters that define the first state and a countable 
collection of group transformations of the kinematical group $G$. And 
this can be done starting from any arbitrary state. 

This idea allows us to define a concept of physical equivalence among 
 \index{physical equivalence}
states of any arbitrary quantum mechanical system in the following 
way: Two states are said to be physically equivalent if they can 
produce by the above method an orthonormal basis of the same Hilbert 
subspace, or in an equivalent way, if they belong to the same 
invariant subspace under the group action. It is easy to see that this 
is an equivalence relation. But if the representation is irreducible, 
all states are equivalent as basic pieces of physical information for 
describing the elementary system. There is one and only one single 
piece of basic physical information to describe an elementary object. 
That is what the term elementary might mean. 

But this definition of elementary particle is a pure group theoretical 
one. The only quantum mechanical ingredient is that the group 
operates on a Hilbert space. Then one question arises. Can we 
translate this quantum mechanical definition into the classical domain 
and obtain an equivalent group theoretical definition for a 
classical elementary particle? 

Following with the above idea, in classical mechanics we have no 
vector space structure to describe the states of a system. What we 
have are manifolds of points where each point represents either the 
configuration state, the kinematical state or the phase state of the 
system depending on which manifold we work. But the idea that any 
point that represents the state of an elementary particle is 
physically equivalent to any other, is in fact the very mathematical 
concept of homogeneity of the manifold under the corresponding group 
action. In this way, the irreducibility assumption of the quantum 
mechanical definition is translated into the realm of classical 
mechanics in the concept of homogeneity of the corresponding manifold 
under the Poincar\'e group or any other kinematical group we consider 
as the symmetry group of the theory. But, what manifold? Configuration 
space? Phase space? The answer as has been shown in previous 
works,~\footnote{\hspace{0.1cm}M. Rivas, {\sl J. Phys.} A {\bf 18}, 1971 (1985);
{\sl J. Math. Phys.} {\bf 30}, 318 (1989); {\sl J. Math. 
Phys.} {\bf 35}, 3380 (1994).} is that the 
appropriate manifold is the {\bf kinematical space}. 

In the Lagrangian approach of classical mechanics, the kinematical 
space $X$ is the manifold where the dynamics is developed as an input-output formalism. When 
quantizing the system we will obtain the natural link between the classical 
and quantum formalisms through Feynman's path integral approach, 
as will be shown later. This manifold is the natural space on which to 
define the Hilbert space structure of the quantized system. In a 
formal way we can say that each point $x\in X$ that represents the 
kinematical state of a system is spread out and is transformed through Feynman's 
quantization into the particle wave 
function $\psi(x)$ defined around $x$. 
This wave function is a squared integrable complex function
defined on $X$.

We can also analyze the elementarity condition from a different point 
of view. Let us consider an inertial observer $O$ that is measuring a 
certain observable $A(\tau)$ of an arbitrary system at an instant $\tau$. 
This observable takes the value $A'(\tau)$ for a different inertial 
observer $O'$. It can be expressed in terms of $A(\tau)$ in the 
form $A'(\tau)=f(A(\tau),g)$, where $g$ is the kinematical 
transformation between both observers. At instant $\tau+d\tau$, the 
corresponding measured values of that observable will have changed but $A'(\tau+d\tau)=f(A(\tau+d\tau),g)$ with 
the same $g$ as before, and assuming that the evolution parameter $\tau$ is group invariant. 

But if the system is elementary, we take as an assumption that
the modifications of the observables
produced by the dynamics can always be compensated by a change of inertial reference frame.
Then, given an observer $O$, 
it is always possible to find at instant $\tau+d\tau$ another inertial 
observer $O'$ who measures the value of an essential observable 
$A'(\tau+d\tau)$ with the same value as $O$ does at instant $\tau$, {\sl i.e.}, 
$A'(\tau+d\tau)\equiv A(\tau)$. If the system is not 
elementary, this will not be possible in general because the external 
interaction might change its internal structure, and thus it will not 
be possible to compensate the modification of the observable by a simple 
change of inertial observer. Think
about a non-relativistic description of an atom that goes into some excited state. 
The new internal energy, which is Galilei invariant, cannot be transformed into the 
old one by a simple change of reference frame.

But the essential observables are the kinematical variables. From the 
dynamical point of view we can take as initial and final points any 
$x_1$ and $x_2\in X$, compatible with the causality requirements. This 
means that any $x$ can be considered as the initial point of the 
variational formalism. In this way, at any instant $\tau$ if the system is elementary, we can find 
an infinitesimal kinematical transformation $\delta g(\tau)$ such that
 \[ 
x'(\tau+d\tau)=f(x(\tau+d\tau),\delta g(\tau))\equiv x(\tau),
 \] 
or by taking the inverse of this transformation,
 \[
x(\tau+d\tau)=f^{-1}(x(\tau),\delta g(\tau)).
 \] 
This equation represents the dynamical evolution equation in $X$ space. Knowledge of the initial state 
$x_1$ and the function $\delta g(\tau)$ completely determines the 
evolution of the system. In general, $\delta g(\tau)$ will depend on 
the instant $\tau$, because the change of the observables depends 
on the external interaction. But if the system is elementary and the 
motion is free, all $\delta g(\tau)$ have necessarily to be the same, 
and thus $\tau$ independent. We cannot distinguish in a free motion 
one instant from any other. Then, starting from $x_1$ we shall arrive 
at $x_2$ by the continuous action of the same infinitesimal group element 
$\delta g$, and the free particle motion is the action of the one-parameter 
group generated by $\delta g$ on the initial state. Therefore, 
there should exist a finite group element $g\in G$ such that $x_2=gx_1$. 
 \index{free motion}
If the evolution is not free, the composition of all infinitesimal group elements $\delta g(\tau)$
for all intermediate values of $\tau\in[\tau_1,\tau_2]$, will also produce a finite group element $g$,
and thus, $x_2=gx_1$.
We thus arrive at the:

\begin{quotation}\normalsize{\noindent{\bf Definition: } A {\bf classical elementary particle} \index{elementary particle!classical}
is a Lagrangian system whose kinematical space $X$ is a {\bf homogeneous 
space} of the kinematical group $G$.\\}
\end{quotation}

Usually the Lagrangian of any classical Newtonian system is 
restricted to depend only on the first order derivative of each of the 
coordinates $q_i$ that represent the independent 
degrees of freedom, or equivalently, that the $q_i$ satisfy second 
order differential equations. But at this stage, if we do not know 
what are the basic variables we need to describe our elementary system,
how can we state that they necessarily satisfy second order 
differential equations? If some of the degrees of freedom, say $q_1$, 
$q_2$ and $q_3$, represent the center of mass position of the system, 
Newtonian mechanics implies that in this particular case $L$ will 
depend on the first order derivatives of these three variables. But 
what about other degrees of freedom? It is this condition 
on the kinematical space to be considered as a homogeneous space of $G$, as the mathematical statement of 
elementarity, that will restrict the dependence of the Lagrangian on 
these higher order derivatives. It is this definition of elementary 
particle with the proper election of the kinematical group, which 
will supply information about the structure of the Lagrangian. 

The Galilei and Poincar\'e groups are ten-parameter Lie groups and therefore
the largest homogeneous space we can find for these groups is a ten-dimensional manifold.
The variables that define the different homogeneous spaces will share the 
same domains and dimensions as the 
corresponding variables we use to parameterize the group. Both groups, as we shall see later, are 
parameterized in terms of the following variables $(b,{\bi a},{\bi 
v},{\balpha})$ with domains and dimensions respectively like 
$b\in\RR$ 
that represents the time parameter of the time translation and ${\bi a}\in 
\RR^3$, the three spatial coordinates for the space translation. Parameter ${\bi 
v}\in\RR^3$ are the three components of the relative velocity between the 
inertial observers, restricted to $v<c$ in the Poincar\'e
case. Finally ${\balpha}\in SO(3)$ are three dimensionless variables 
which characterize the relative orientation of the corresponding 
Cartesian frames and whose compact domain is expressed in terms of a 
suitable parametrization of the rotation group. 

In this way the maximum number of kinematical variables, for a classical elementary particle, 
is also ten.
We represent them by $x\equiv(t,{\bi r},{\bi u},{\balpha})$ with 
the same domains and dimensions as above and interpret them 
respectively as the {\bf time, position, velocity and orientation} of the 
particle. 
 \index{orientation}

Because the Lagrangian must also depend on the next order 
derivatives of the kinematical variables, we arrive at the conclusion 
that $L$ must also depend on the acceleration and angular velocity of 
the particle. The particle is a system of six degrees of freedom, 
three ${\bi r}$, represent the position of a point and other three 
${\balpha}$, its orientation in space. We can visualize this by 
assuming a system of three orthogonal unit vectors linked to point 
${\bi r}$ as a body frame. But the Lagrangian will depend up to the 
second time derivative of ${\bi r}$, or acceleration of that point, 
and on the first derivative of ${\balpha}$, {\sl i.e.}, on the angular 
velocity. The Galilei and Poincar\'e groups lead to generalized
Lagrangians depending up to second order derivatives of the position.

By this definition it is the kinematical group $G$ that 
implements the special Relativity Principle that completely determines 
the structure of the kinematical space where the Lagrangians that represent 
classical elementary particles have to be defined.~\footnote{\hspace{0.1cm}see ref.14.}
Point particles are particular 
cases of the above definition and their kinematical spaces are just 
the quotient structures between the group $G$ and 
subgroup of rotations and boosts, and thus their 
kinematical variables reduce only to time and position $(t,{\bi r})$. 
Therefore, the larger the kinematical group of space-time transformations,
the greater the number of allowed classical variables to describe 
elementary objects with a more detailed and complex structure. In this way, the 
proposed formalism can be accommodated to any symmetry group. It is the proper definition
of this group which contains the physical information of the elementary particles.

\subsection{Elementary Lagrangian systems}
 \index{elementary!Lagrangian systems}
\label{sec:elementary}

An elementary Lagrangian system will be characterized by the 
Lagrangian function 
$L(x,\dot x)$ where the variables $x\in X$ lie in a homogeneous space 
$X$ of $G$. $L$ is a homogeneous function of first degree of the 
derivatives of the 
kinematical variables, and this allows us to write
 \begin{equation} 
L(x,\dot{x})=F_i(x,\dot{x})\,\dot{x}^i.
 \end{equation} 
Functions $F_i(x,\dot{x})$ are therefore homogeneous functions of zero 
degree in the variables $\dot{x}^i$ and summation convention on repeated indexes as usual is assumed.

Under $G$, $x$ transforms as $x'=gx$ or more explicitly its coordinates by
${x}^{'i}=f^i(g,x)$, and their derivative variables
 \begin{equation} 
{\dot{x}}^{'i}=\frac{\partial{x}^{'i}}{\partial x^j}\dot{x}^j,
\label{eq:(1.7.1)}
 \end{equation} 
transform like the components of a contravariant vector.

The Lagrangian transforms under $G$, 
 \begin{equation} 
L(x'(x),\dot{x}'(x,\dot{x}))=L(x,\dot{x})+\frac{d\alpha(g;x)}{ d\tau},
 \end{equation} 
{\sl i.e.},
 \begin{equation} 
F_i(x',\dot{x}'){\dot{x}}^{'i}=F_j(x,\dot x)\dot x^j+\frac{\partial\alpha(g;x)}{\partial x^j}\dot x^j.
\label{eq:(1.7.2)}
 \end{equation} 
Taking into account the way the different variables transform, we 
thus arrive at:
 \begin{equation} 
F_i(x',\dot{x}')=\frac{\partial x^j}{ \partial{x}^{'i}}\,\left 
[F_j(x,\dot{x}) +\frac{\partial\alpha(g;x)}{\partial x^j}\right].
\label{eq:(1.7.3)}
 \end{equation}  
In the case when $\alpha(g;x)=0$, they transform like the components of a covariant vector over
the kinematical space $X$. 
But in general this will not be the case and $\alpha(g;x)$ contains basic physical information
about the system.

We thus find that for a fixed kinematical space $X$, the knowledge of 
the group action of $G$ on $X$, and the gauge function $\alpha(g;x)$, 
will give us information about the possible structure of the functions 
$F_i(x,\dot x)$, and therefore about the structure of the Lagrangian. 

In  practice, if we restrict ourselves to the Galilei ${\cal 
G}$ and Poincar\'e ${\cal P}$ groups, we see that ${\cal P}$ has 
gauge functions 
equivalent to zero and thus Poincar\'e Lagrangians that describe 
elementary particles can be taken strictly invariant. In the case of the 
Galilei group, it has only one class of gauge functions that define the 
mass of the system, and thus nonrelativistic Lagrangians will be in general not invariant.
In the particular case of Galilei invariant Lagrangians, they 
will describe massless systems.

\section{Appendix: Lie groups of transformations}
 \index{Lie group}
 \index{Lie group!of transformations}\label{sec:Liegroups}

Let us introduce the notation and general features
of the action of Lie groups on continuous manifolds to analyze the transformation 
properties of the different magnitudes we can work with
in either classical or quantum mechanics. We shall use these features all 
throughout this book.

Let us consider the transformation of an $n$-dimensional manifold $X$, $x'=gx$ given by $n$ continuous 
and differentiable functions depending on a set $g\in G$ of $r$ continuous parameters of the form
\[
x'^i=f^i(x^j;g^\sigma),\quad\forall x\in X,\quad\forall 
g\in G,\quad i,j=1,\ldots,n,\quad \sigma=1,\ldots,r.
\]
This transformation is said to be the action of a Lie group of transformations if it
fulfils the two conditions:

\noindent{\bf(i)} $G$ is a Lie group, {\sl i.e.}, there
exists a group composition law $c=\phi(a,b)\in G$, $\forall a,b\in G$, in terms of $r$ continuous
and differentiable functions $\phi^\sigma$. 

\noindent{\bf(ii)} The transformation equations satisfy
\[
x''=f(x';b)=f(f(x;a);b)=f(x;c)=f(x;\phi(a,b)).
\]

The group parametrization can be chosen such that the coordinates that characterize the neutral 
element $e$ of the group are $e\equiv(0,\ldots,0)$, so that an infinitesimal element
of the group is the one with infinitesimal coordinates $\delta g^\sigma, \sigma=1,\ldots,r$.

Under the action of an infinitesimal element $\delta g$ of the group $G$, the change in the coordinates $x^i$
of a point $x\in X$ is given by
\[
x^i+dx^i=f^i(x;\delta g)=x^i+\left.\frac{\partial f^i(x;g)}{\partial g^\sigma}
\right|_{g=e}\delta g^\sigma,
\]
after a Taylor expansion up to first order in the group parameters and with $x^i=f^i(x;0)$. 
There are $nr$ auxiliary functions of the group that are defined as
 \begin{equation}
u^i_\sigma(x)=\left.\frac{\partial f^i(x;g)}{\partial g^\sigma}\right|_{g=e},
\label{eq:auxilifunc}
 \end{equation}
and therefore to first order in the group parameters, $dx^i=u^i_\sigma(x)\delta g^\sigma$.

The group action on the manifold $X$ can be extended to the action on the set ${\cal F}(X)$ 
of continuous and differentiable functions defined on $X$ by means of:
\begin{equation}
g:h(x)\to h'(x)\equiv h(gx).
\label{eq:transforh}
\end{equation}
If the group element is infinitesimal, then
 \index{Lie group!infinitesimal element}
\[
h'(x)=h(x^i+dx^i)=h(x^i+u^i_\sigma(x)\delta g^\sigma)=h(x)+\frac{\partial h(x)}{\partial x^i}\;
u^i_\sigma(x)\delta g^\sigma,
\]
after a Taylor expansion to first order in the infinitesimal group parameters. 
The infinitesimal transformation on 
${\cal F}(X)$ can be represented by the action of a differential operator in the form
 \index{Lie group!differential operator}
\[
h'(x)=\left(\ID+\delta g^\sigma\, u^i_\sigma(x)\frac{\partial}{\partial x^i}\right)h(x)=
\left(\ID+\delta g^\sigma X_\sigma\right)h(x)=U(\delta g)h(x),
\]
where $\ID$ is the identity operator and the linear differential operators 
\begin{equation}
X_\sigma= u^i_\sigma(x)\frac{\partial}{\partial x^i}.
\label{eq:generat}
\end{equation}
In particular, when acting with the operator $U(\delta g)\equiv\left(\ID+\delta g^\sigma X_\sigma\right)$ 
on the coordinate $x^j$ we get $x^j+dx^j=x^j+u^j_\sigma(x)\delta g^\sigma$.

The operators $X_\sigma$ are called the {\bf generators}\index{generators}\index{Lie group!generators}
 \index{Lie group!generator} 
of the infinitesimal transformations.
 \index{Lie group!infinitesimal transformations} 
They are $r$ linearly independent operators that span an $r$-dimensional real vector space 
such that its commutator $[X_\sigma,X_\lambda]$
also belongs to the same vector space, {\sl i.e.},
\begin{equation}
[X_\sigma,X_\lambda]=c_{\sigma\lambda}^{\alpha}\;X_\alpha, \quad \alpha,\sigma,\lambda=1,\ldots,r.
\label{eq:Liealge}
\end{equation}
The coefficients $c_{\sigma\lambda}^{\alpha}$ are a set of real constant numbers, 
called the {\bf structure constants}
 \index{Lie group!structure constants} 
of the group, and the vector space spanned by the generators
is named the {\bf Lie algebra}
 \index{Lie algebra} \index{Lie group!Lie algebra of a}
${\cal L}(G)$, associated to the Lie group $G$. The structure
constants are antisymmetric in their lower indexes 
$c_{\sigma\lambda}^{\alpha}=-c_{\lambda\sigma}^{\alpha}$, and satisfy Jacobi's indentitites:
\[
c_{\sigma\lambda}^{\alpha}c_{\mu\alpha}^{\beta}+c_{\lambda\mu}^{\alpha}c_{\sigma\alpha}^{\beta}+
c_{\mu\sigma}^{\alpha}c_{\lambda\alpha}^{\beta}=0,\qquad \forall \sigma,\lambda,\mu,\beta=1,\ldots,r.
\]
Equations (\ref{eq:Liealge}) are the commutation relations that characterize the structure
of the Lie algebra of the group. 

If a finite group transformation of parameters $g^\sigma$ can be 
done in $n$ smaller steps of parameters $g^\sigma/n$, with $n$ sufficiently large, then a
finite transformation $U(g)h(x)$ can be obtained as
\[
U(g)h(x)\equiv\lim_{n\to\infty}\left(\ID+\frac{g^\sigma}{n} X_\sigma\right)^n\;h(x)=
\exp(g^\sigma X_\sigma)\,h(x).
\]
This defines the exponential mapping and in this case the group parameters $g^\sigma$ 
are called {\bf normal} or {\bf canonical} parameters. 
 \index{Lie group!normal parameters}
 \index{Lie group!canonical parameters}\index{canonical!parameters}\index{normal!parameters}
In the normal parameterization the composition law of one-parameter subgroups
reduces to the addition of the corresponding parameters of the involved
group elements.

Consider ${\cal F}(X)$ a Hilbert space
 \index{Hilbert space}
of states of a quantum system; (\ref{eq:transforh}) can be interpreted as the transformed wave
function under the group element $g$. Then if the operator $U(g)$ is unitary it is usually 
written in the explicit form
\[
U(g)=\exp\left(\frac{i}{\hbar}\;g^\sigma\widetilde X_\sigma\right),
\]
in terms of the imaginary unit $i$ and Planck's constant $\hbar$,
such that in this case the new $\widetilde X_\sigma$ above are self-adjoint operators and 
therefore represent 
certain observables of the system. The physical dimensions of these observables depend
on the dimensions of the group parameters $g^\sigma$, since the argument of the exponential 
function is dimensionless and because of the introduction of Planck's constant this implies that
$g^\sigma\widetilde X_\sigma$ has dimensions of action.
These observables, taking into account (\ref{eq:generat}), are represented in a unitary representation
by the differential operators
\begin{equation}
\quad\widetilde X_\sigma=\frac{\hbar}{i}\,u^i_\sigma(x)\frac{\partial}{\partial x^i}.
\label{eq:generquant}
\end{equation}
However, (\ref{eq:transforh}) is not the most general form of transformation of the wave
function of a quantum system, as we shall see in Chapter 3, but once we
know the way it transforms we shall be able to obtain the explicit expression of the 
group generators by a similar procedure as the one developed so far. In general the wave
function transforms under continuous groups with what is called
a projective unitary representation of the group, which involves in general some additional
phase factors.

\subsection{Casimir operators}
\label{sec:casimir}
\index{Casimir!operators}

When we have a representation of a Lie group either by linear operators or by matrices
acting on a linear space, we can define there what are 
called the Casimir operators.
\index{Casimir!operators}
They are operators $C$ that can be expressed as functions 
of the generators $X_\sigma$ of the Lie algebra with the property that they commute 
with all of them, {\sl i.e.}, they satisfy $[C,X_\sigma]=0,\quad\forall\sigma=1,\ldots,r$. 
In general they are not expressed as real linear combinations of the $X_\sigma$ 
and therefore they do not belong to the Lie 
algebra of the group. They belong to what is called the
{\bf group algebra}, {\sl i.e.}, the associative, but in general non-commutative algebra,
spanned by the real or complex linear combinations of products of the $X_\sigma$, in the
corresponding group representation. 

In those representations where the $X_\sigma$ are represented 
by self-adjoint 
operators as in a quantum formalism, the Casimir operators may be also self-adjoint and will
represent those observables
that remain invariant under the group transformations. In particular, when we consider 
later the kinematical
groups that relate the space-time measurements between inertial observers, the Casimir
operators of these groups will represent the intrinsic properties of the
system. They are those properties of the physical system
whose measured values are independent of the inertial observers. 

For semisimple groups, {\sl i.e.}, for groups that do not 
have Abelian invariant subgroups like the rotation group $SO(3)$, the unitary groups $SU(n)$
and many others, it is shown that the Casimir operators are real homogeneous polynomials of the 
generators $X_\sigma$, but this is no longer the case for general Lie groups. Nevertheless,
for most of the interesting Lie groups in physics, like Galilei, Poincar\'e, De Sitter, 
$SL(4,\RR)$, the inhomogeneous $ISL(4,\RR)$ and Conformal $SU(2,2)$ groups, the Casimir operators 
can be taken as real polynomial functions of the generators.

\subsection{Exponents of a group}
 \index{Lie group!exponent}
 \index{exponent}
 \label{sec:Exponents}

The concept of exponent of a continuous group $G$ was developed by  
Bargmann in his work on the projective unitary representations of continuous 
groups.~\footnote{\hspace{0.1cm}V.Bargmann, {\sl Ann. Math.} {\bf 59}, 1 (1954).} 
 \index{Lie group!projective representation}
 \index{projective representation}
 
Wigner's theorem \index{Wigner's theorem}
about the symmetries of 
a physical system is well known in  Quantum Mechanics.~\footnote{\hspace{0.1cm}E.P. Wigner, 
{\sl Group theory and its application to the 
quantum mechanics of atomic spectra}, Acad. Press, NY (1959); V. Bargmann, 
{\sl J. Math. Phys.} {\bf 5}, 862 (1964).} 

It states that if ${\cal H}$ is a Hilbert space 
that characterizes the pure quantum states of a system, and the 
system has a symmetry $S$, then there exists a unitary or antiunitary 
operator $U(S)$, defined up to a phase, that implements that symmetry 
on ${\cal H}$, {\sl i.e.}, if $\phi$ and $\psi\in {\cal H}$ are two  
possible vector states of the system and  $|<\phi|\psi>|^2$ is the transition 
probability between them and $U(S)\phi$ and $U(S)\psi$ represent the 
transformed states under the operation $S$, then
 \index{transition probability}
 \[
|<U(S)\phi|U(S)\psi>|^2=|<\phi|\psi>|^2.
 \]

If the system has a whole group of symmetry operations $G$, then
to each  element $g\in G$ there is associated an operator $U(g)$ unitary or 
antiunitary, but if $G$ is a continuous group, in that case $U(g)$ is 
necessarily unitary. This can be seen by the fact that the product of 
two antiunitary operators is a unitary one. 

Because there is an ambiguity in the election of the phase of the 
unitary operator $U(g)$, it implies that in general $U(g_1)U(g_2)\neq 
U(g_1g_2)$ and therefore the transformation of the wave function is not 
given by an expression of the form 
(\ref{eq:transforh}), but it also involves in general a phase factor. 
However in the case of continuous groups we can properly 
choose the corresponding phases of all elements in such a way that
 \index{Lie group!exponent}
 \begin{equation} U(g_1)U(g_2)=\omega(g_1,g_2)U(g_1g_2), \label{eq:2.1.1} \end{equation} 
where $\omega(g_1,g_2)=\exp\{i\xi(g_1,g_2)\}$ is a phase that is a 
continuous function of its arguments. The real continuous function on 
$G\times G$, $\xi(g_1,g_2)$ is called an {\bf exponent} of $G$. The 
operators $U(g)$ do not reproduce the composition law of the group $G$ 
and (\ref{eq:2.1.1}) represents what Bargmann calls a {\bf projective 
representation} of the group. 
 \index{Lie group!projective representation}

If we use the associative property of the group law, we get
 \begin{eqnarray*} 
(U(g_1)U(g_2))\,U(g_3)&=&\omega(g_1,g_2)U(g_1g_2)U(g_3) \\
&=&\omega(g_1,g_2)\omega(g_1g_2,g_3)U(g_1g_2g_3),
 \end{eqnarray*} 
and also
 \begin{eqnarray*} 
U(g_1)\,(U(g_2)U(g_3))&=&U(g_1)\omega(g_2,g_3)U(g_2g_3) \\
&=&\omega(g_1,g_2g_3)\omega(g_2,g_3)U(g_1g_2g_3).
 \end{eqnarray*} 
Therefore
 \begin{equation} 
\omega(g_1,g_2)\omega(g_1g_2,g_3)=\omega(g_1,g_2g_3)\omega(g_2,g_3),
\label{eq:2.2.1}
 \end{equation}  
which in terms of the exponents becomes: 
 \begin{equation} 
\xi(g_1,g_2)+\xi(g_1g_2,g_3)=\xi(g_1,g_2g_3)+\xi(g_2,g_3).
\label{eq:2.2.2}
 \end{equation} 
Because of the continuity of the exponents,
 \begin{equation} 
\xi(g,e)=\xi(e,g)=0,\quad \forall g\in G,
\label{eq:2.2.3}
 \end{equation} 
where $e$ is the neutral element of the group.

Any continuous function on $G$, $\phi(g)$, with the condition $\phi(e)=0$, 
can generate a trivial exponent by
 \index{exponent!trivial}
 \[ 
\xi(g,g')=\phi(gg')-\phi(g)-\phi(g'),
 \] 
that satisfies 
(\ref{eq:2.2.2}) and (\ref{eq:2.2.3}). All trivial exponents are 
equivalent to zero exponents, and in a unitary representation 
(\ref{eq:2.1.1}) can be compensated into the phases of the factors, 
thus transforming the projective representation (\ref{eq:2.1.1}) into 
a true unitary one.

Given a continuous group, the existence or not of non-trivial exponents is an intrinsic
group property related to the existence or not of central extensions of the 
group.~\footnote{\hspace{0.1cm}see ref.7 and J.M. Levy-Leblond, {\sl Comm. Math. Phys.}, {\bf 12}, 64 (1969); 
A.A. Kirillov, {\sl \'Elements de la theorie des
repr\'esentations}, Mir, Moscow (1974).}

\subsection{Homogeneous space of a group}
 \index{homogeneous space}
\label{sec:Espahomo}

A manifold $X$ is called a homogeneous space of a group $G$, if 
$\forall\, x_1,\,x_2\in X$ there exists at least one element $g\in G$ 
such that $x_2=gx_1$. In that case it is said that $G$ acts on $X$ in 
a transitive way. The term homogeneous reminds us that the local
properties of the manifold at a point $x$ are translated to any other
point of the manifold by means of the group action, and therefore all
points of $X$ share the same local properties.

The {\bf orbit} of a point $x$ is the set of points of the form $gx$, $\forall g\in G$,
such that if $X$ is a homogeneous space of $G$, then the whole $X$ is the orbit of any of its points.
 \index{orbit} \index{Lie group!orbit}

Given a point $x_0\in X$, the {\bf stabilizer group} (little group) of $x_0$ 
is the subgroup $H_{x_0}$ of $G$, that leaves invariant the 
point $x_0$, {\sl i.e.}, $\forall h\in H_{x_0},\,  hx_0=x_0$. 
 \index{stabilizer group}
 \index{little group}

If $H$ is a subgroup of $G$, then every element $g\in G$ can be 
written as $g=g'h$, where $h\in H$, and $g'$ is an element of 
$G/H$, the set of left cosets generated by the subgroup $H$. 
If $X$ is a homogeneous space of $G$, it can be generated by the 
action of $G$ on an arbitrary point $x_0\in X$. Then $\forall x\in X$,
$x=gx_0=g'hx_0=g'x_0$, and thus the homogeneous space $X$ is 
isomorphic to the manifold $G/H_{x_0}$. 

The homogeneous spaces of a group can be constructed as quotient 
manifolds of the group by all its possible continuous subgroups. 
Conversely, it can also be shown that if $X$ a homogeneous space of a 
group $G$, then there exists a subgroup $H$ of $G$ such that 
$X$ is isomorphic to $G/H$. Therefore, the largest homogeneous space
of a group is the group itself.

\chapter{Soluble examples of spinning particles}
\section*{Nonrelativistic particles}

\section{Nonrelativistic point particle} 
 \index{point particle} \index{point particle!nonrelativistic}  
\label{sec:galipoint} 

See the Appendix about the Galilei group at the end of this chapter for the notation used
through this chapter.

Let us consider a mechanical system whose kinematical space is the 
manifold $X={\cal G}/{\cal H}$, where ${\cal H}$ is
the six-dimensional subgroup of the homogeneous Galilei transformations 
of elements of the form $(0,{\bi 0},{\bi v},\bmu)$. See the Appendix at the end of this chapter
for the notation related to the Galilei group.
Then $X$ is a four-dimensional manifold spanned by the variables 
$(t,{\bi r})\equiv x$, with domains $t\in\RR$, ${\bi r}\in \RR^3$, 
similar to the group parameters $b$ and ${\bi a}$ respectively.
We assume that they are functions of some evolution parameter $\tau$ and at any instant $\tau$ 
of the evolution two different inertial observers relate their measurements by:
 \begin{eqnarray} 
t'(\tau)&=&t(\tau)+b,\\ 
{\bi r}'(\tau)&=&R(\bmu){\bi r}(\tau)+{\bi v}t(\tau)+{\bi a}.
\label{nr.1}
 \end{eqnarray}  
Because of the way they transform, we can interpret them respectively as the 
time and position of the system. If we assume that the evolution 
parameter $\tau$ is group invariant, by taking the $\tau-$derivative
of both sides of the above expressions, it turns out that the derivatives of the 
kinematical variables at any instant $\tau$ transform as: 
 \begin{eqnarray} 
\dot t'(\tau)&=&\dot t(\tau),\label{nr.21}\\ 
\dot{\bi r}'(\tau)&=&R(\bmu)\dot{\bi r}(\tau)+{\bi v}\dot t(\tau).
\label{nr.22}
 \end{eqnarray} 
There are no constraints among these variables. It is only the 
homogeneity of the Lagrangian in terms of their derivatives (\ref{eq:homogen})
which reduces to three the number of independent degrees of freedom. This homogeneity 
leads to the general form:
 \begin{equation}  
L=T\dot t+{\bi R}\cdot\dot{\bi r},
\label{poinLag}
 \end{equation} 
where $T=\partial L/\partial\dot t$ and $R_i=\partial L/\partial\dot r_i$ 
are still some unknown functions of the kinematical variables and their derivatives, 
which are homogeneous of zero degree in terms of the derivatives.

Associated to this manifold $X$, the gauge function for this system is
 \begin{equation}  
\alpha(g;x)=\xi(g,x)=m\left(v^2t/2+{\bi v}\cdot R(\bmu){\bi r}\right),
\label{nrgauge}
 \end{equation}  
where the parameter $m$ is interpreted as the mass of the 
system and $\xi(g,g')$ is the exponent of ${\cal G}$, so that the 
transformation of the Lagrangian under the Galilei group is 
 \begin{equation}  
L(x',\dot x')=L(x,\dot x)+m\left(v^2\dot t/2+{\bi v}\cdot R(\bmu)\dot{\bi r}\right).
\label{nrL2}
 \end{equation} 
Then 
 \begin{equation}  
 T'=\frac{\partial L'}{\partial\dot t'}=\left(\frac{\partial L}{\partial\dot 
t}+\frac{1}{2}mv^2\right)\frac{\partial\dot t}{\partial\dot t'}+\left(\frac{\partial L}{\partial\dot 
r_i}+mv_jR(\bmu)_{ji}\right)\frac{\partial\dot r_i}{\partial\dot t'},
\label{eq:Ttra1}
 \end{equation} 
but from (\ref{nr.21}) and (\ref{nr.22}) we get ${\partial\dot t}/{\partial\dot t'}=1$ and
${\partial\dot r_i}/{\partial\dot t'}=-R^{-1}(\bmu)_{ik}v_k$, respectively, and thus
 \begin{equation}  
 T'=T-\frac{1}{2}\,mv^2-{\bi v}\cdot R(\bmu){\bi R}.
\label{eq:Ttr1}
 \end{equation} 
Similarly
 \begin{equation}  
 {\bi R}'=R(\bmu){\bi R}+m{\bi v}.
\label{eq:Rtr1}
 \end{equation} 
The conjugate momenta of the independent degrees of freedom $q_i=r_i$, 
are $p_i=\partial L/\partial\dot r_i$, and 
consequently Noether's theorem leads to the following constants of the 
motion:

\noindent {\bf a)} Under time translations the gauge function (\ref{nrgauge}) vanishes, 
$\delta t=\delta b$, $M=1$, while $\delta r_i=0$
and the constant reduces to the following expression ${\bi R}\cdot d{\bi r}/dt-L/\dot t=-T$.

\noindent {\bf b)} Under space translations also $\alpha(g;x)\equiv0$, $\delta t=0$, $M=0$, while $\delta r_i=\delta a_i$,
$M_{ij}=\delta_{ij}$ and the conserved observable is ${\bi R}$.

\noindent {\bf c)} Under pure Galilei transformations $\delta t=\delta b$ and $M=0$, while $\delta r_i=t\delta v_i$
and $M_{ij}=t\delta_{ij}$, but now the gauge function to first order in the velocity parameters
is $\alpha(\delta{\bi v};x)=m{\bi r}\cdot\delta{\bi v}$, 
and we get $m{\bi r}-{\bi R}t$.

\noindent {\bf d)} Under rotations $\alpha(g;x)\equiv0$, $\delta t=0$ and $M=0$, while $\delta r_i=-\varepsilon_{ijk}r_jn_k\delta\alpha$
and $M_{ik}=-\varepsilon_{ijk}r_j$ the conserved quantity is ${\bi r}\times{\bi R}$.

Collecting all terms we can give them the following names:
 \begin{eqnarray} 
\hbox{\rm Energy}\quad H&=&-T,\label{H1}\\
\hbox{\rm linear momentum}\quad {\bi P}&=&{\bi R}\;=\;{\bi p},\label{p1}\\
\hbox{\rm kinematical momentum}\quad {\bi K}&=&m{\bi r}-{\bi P}t,\label{k1}\\
\hbox{\rm angular momentum}\quad {\bi J}&=&{\bi r}\times{\bi P}.\label{J1}
 \end{eqnarray}\index{kinematical momentum}\index{kinematical!momentum}\index{total energy}\index{total linear momentum}\index{total angular momentum}

We reserve for these observables the same symbols as the corresponding group generators
which produce the space-time transformations that leave dynamical equations invariant. 
Even their names make reference to the corresponding group 
transformation parameter, except the energy which in
this context should be called the `temporal momentum'. For the kinematical momentum we can find in
the literature alternative names like `Galilei momentum' or `static momentum'. 
Being consistent with this notation,
we should call it `Poincar\'e or Lorentz momentum' in a relativistic approach. Nevertheless
we shall use the name of kinematical momentum for this 
observable ${\bi K}$ in either relativistic or non-relativistic formalism. None of these conserved quantities
associated to the symmetry under one-parameter subgroups are definite positive so that the observable $H$
can take any sign.

The linear momentum takes the general expression ${\bi P}=m\dot{\bi r}/\dot t=m{\bi u}$ 
because taking the $\tau$-derivative in (\ref{k1}) of the
kinematical momentum, $\dot{\bi K}=0$, implies ${\bi 
P}=m{\bi u}$, where ${\bi u}$ is the time 
derivative of the position of the system, {\sl i.e.}, the velocity of the 
particle.

The six conditions ${\bi P}=0$ and ${\bi K}=0$, imply ${\bi u}=0$ and 
${\bi r}=0$, so that the system is at rest and placed at the origin of 
the observer's reference frame. There is still an arbitrary rotation and a time 
translation to fix a unique inertial observer. Nevertheless we call
this class of observers, for which ${\bi P}=0$ and ${\bi K}=0$, the 
center of mass observer. These six conditions will be also used
as the definition of the center of mass observer for any other system
even in a relativistic approach.
 \index{center of mass!observer}

From (\ref{eq:Ttr1}) and (\ref{eq:Rtr1}) we see that the energy and linear momentum transform as:
 \begin{eqnarray} 
H'&=&H+{\bi v}\cdot R(\bmu){\bi P}+\frac{1}{2}mv^2,\label{eq:Htr1}\\
 {\bi P}'&=&R(\bmu){\bi P}+m{\bi v}.\label{eq:Ptr1}
 \end{eqnarray}
Then, if $H_0$  and ${\bi P}=0$ are the energy and linear momentum
measured by the center of mass observer, for any arbitrary observer who
sees the particle moving with velocity ${\bi u}$, it follows from (\ref{eq:Htr1}) and
(\ref{eq:Ptr1}) that
\[
H=H_0+\frac{1}{2}mu^2=H_0+{\bi P}^2/2m,\quad {\bi P}=m{\bi u}.
\]
The Lagrangian for the point particle is thus
 \begin{equation}
L=T\dot{t}+{\bi R}\cdot\dot{\bi r}=-H\dot{t}+{\bi P}\cdot\dot{\bi r}=-H_0\dot{t}+\frac{m}{2}\frac{{\dot{\bi r}}^2}{\dot{t}},
\label{GaliL}
 \end{equation}
with $H_0$ an arbitrary constant which plays no role in the dynamics and can be taken $H_0=0$.
It will be related to the $mc^2$ term of the relativistic point particle.

If we define the spin of the system, as in (\ref{spin2}), by 
 \begin{equation} 
{\bi S}\equiv{\bi J}-\frac{1}{m}{\bi K}\times{\bi P}={\bi J}-{\bi r}\times{\bi P}=0,
 \end{equation} 
it represents the angular momentum of the system with respect to the center of mass
${\bi r}$. It vanishes, so that the point particle is a spinless system.

\section{Galilei free spinning particle}
\label{sec:galispin}

The most general nonrelativistic particle~\footnote{\hspace{0.1cm}M. Rivas, {\sl 
J. Phys.} {\bf A 18}, 1971 (1985).} is the system
whose kinematical space $X$ is the whole Galilei group ${\cal G}$. Then the 
kinematical variables are the ten real variables 
$x(\tau)\equiv(t(\tau),{\bi r}(\tau),{\bi 
u}(\tau),\brho(\tau))$ with domains $t\in\RR$, ${\bi r}\in \RR^3$, ${\bi 
u}\in \RR^3$ and $\brho\in\RR_c^3$ similarly as the corresponding 
group parameters. The relationship between the values $x'(\tau)$ and $x(\tau)$ 
they take at any instant $\tau$
for two arbitrary inertial observers, is given by:
 \begin{eqnarray} 
t'(\tau)&=&t(\tau)+b,\label{eq:gt1}\\
{\bi r}'(\tau)&=&R(\bmu){\bi r}(\tau)+{\bi v}t(\tau)+{\bi a},\\ 
{\bi u}'(\tau) &=&R(\bmu){\bi u}(\tau)+{\bi v},\\ 
\brho'(\tau)&=&{\bmu+\brho(\tau)+\bmu\times\brho(\tau)\over 1-
\bmu\cdot\brho(\tau)}.
\label{eq:4.1.1}
 \end{eqnarray} 

Among these kinematical variables there exist the differential 
constraints ${\bi u}(\tau)=\dot{\bi r}(\tau)/\dot t(\tau)$, that 
together with the homogeneity condition of the Lagrangian $L$ in terms of 
the derivatives of the kinematical variables: 
 \begin{equation}  
L(x,\dot x)=(\partial L/\partial\dot x_i)\dot x_i, 
 \label{4.1.2}
 \end{equation}   
reduce from ten to six the essential degrees of freedom of the system.

These degrees of freedom 
are the position ${\bi r}(t)$ and the orientation $\brho(t)$.
The Lagrangian depends on the second derivative of ${\bi r}(t)$ 
and the first derivative of $\brho(t)$. Expression (\ref{4.1.2}) is explicitly given by:
 \begin{equation}  
L=T\dot t+{\bi R}\cdot\dot{\bi r}+{\bi U}\cdot\dot{\bi u}+{\bi V}\cdot\dot{\brho},
 \label{4.1.3}
 \end{equation}   
where the functions $T=\partial L/\partial\dot t$, $R_i=\partial L/\partial\dot r^i$, 
$U_i=\partial L/\partial\dot u^i$, $V_i=\partial L/\partial\dot{\rho}^i$ 
will be in general functions of the 
ten kinematical variables $(t,{\bi r},{\bi u},\brho)$ and homogeneous 
functions of zero degree in terms of the derivatives 
$(\dot t,\dot{\bi r},\dot{\bi u},\dot{\brho})$. By assuming that 
the evolution parameter $\tau$ is group invariant, these derivatives transform 
under ${\cal G}$:
 \begin{eqnarray} 
\dot t'(\tau)&=&\dot t(\tau),\\
 \dot{\bi r}'(\tau)&=&R(\bmu)\dot{\bi r}(\tau)+{\bi v}\dot t(\tau),\\
 \dot{\bi u}'(\tau)&=&R(\bmu)\dot{\bi u}(\tau),\\
 \dot{\brho}'(\tau)&=&\frac{(\dot{\brho}(\tau)+\bmu\times\dot{\brho}(\tau))(1-
\bmu\cdot\brho(\tau))}{(1-\bmu\cdot\brho(\tau))^2}\nonumber+\\ 
&\;& \frac{\bmu\cdot\dot{\brho}(\tau)(\bmu+\brho(\tau)+\bmu\times\brho(\tau))}
{(1-\bmu\cdot\brho(\tau))^2}.
 \label{4.1.4}
 \end{eqnarray} 

Instead of the derivative $\dot{\brho}(\tau)$ that transforms in a 
complicated way, we can define the angular velocity of the particle $\bomega$ as a function of it in 
the form
 \index{angular velocity}
 \begin{equation}  
\bomega=\frac{2}{1+\brho^2}(\dot{\brho}+\brho\times\dot{\brho}).
\label{eq:4.1.5}
 \end{equation}  
It is a linear function of $\dot{\brho}$, and transforms as:
 \begin{equation}  
\bomega'(\tau)=R(\bmu)\bomega(\tau).
 \label{eq:4.1.6}
 \end{equation}  
We interpret the rotation matrix $R(\brho)$ as the rotation that carries the initial frame
linked to the body at instant $\tau=0$ to the frame at instant $\tau$, as in a rigid body. Then,
the three columns of matrix $R(\brho)$ represent the Cartesian 
components of the three unit 
vectors linked to the body when chosen parallel to the laboratory frame at instant $\tau=0$. 

\begin{quotation}\noindent\footnotesize{
If ${\bi k}(\tau)$ is any internal vector of a rigid body with origin
at point ${\bi r}$, then its dynamics is contained in the expression
${\bi k}(\tau)=R(\brho(\tau)){\bi k}(0)$. The velocity of point ${\bi k}$
is 
\[
\dot{\bi k}(\tau)=\dot{R}(\brho(\tau)){\bi k}(0)=\dot{R}(\brho(\tau))R^{-1}(\brho(\tau)){\bi k}(\tau)=\Omega(\tau){\bi k}(\tau)
\]
where matrix $\Omega=\dot{R}R^{-1}=\dot{R}R^T$ is an antisymmetric matrix. 
At any instant $\tau$, $R(\brho(\tau))R^T(\brho(\tau))=\ID$, where superscript $T$ 
means the transposed matrix and $\ID$ is the $3\times3$ unit matrix. 
Taking the $\tau$-derivative of this expression, $\dot R R^T+R\dot 
R^T=\Omega+\Omega^T=0$, and thus the three essential components of the 
antisymmetric matrix $\Omega$ define a three-vector $\bomega$
\[
\Omega=\pmatrix{0&-\omega_z&\omega_y\cr \omega_z&0&-\omega_x\cr -\omega_y&\omega_x&0},
\]
such that we can also write $\dot{\bi k}(\tau)=\Omega(\tau){\bi k}(\tau)
\equiv\bomega(\tau)\times{\bi k}(\tau)$ and $\bomega$ is interpreted as the instantaneous angular velocity.
The different components of
$\bomega$, expressed as functions of the
variables $\brho$ and $\dot{\brho}$ are given in (\ref{eq:4.1.5}). }
\end{quotation}

Expression (\ref{eq:4.1.1}) corresponds to $R(\brho'(\tau))=R(\bmu)R(\brho(\tau))$. Therefore
\begin{eqnarray*}
\Omega'&=&\dot R(\brho'(\tau))R^T(\brho'(\tau))=R(\bmu)\dot R(\brho(\tau))
R^T(\brho(\tau))R^T(\bmu) \\
\,&=&R(\bmu)\Omega R^{-1}(\bmu),
 \end{eqnarray*}
and this leads to the equation (\ref{eq:4.1.6}) in terms of the essential 
components $\bomega$ of the antisymmetric matrix $\Omega$.

In this way the last part of the Lagrangian $(\partial L/\partial\dot{
\rho}^i)\dot{\rho}^i$ can be writen as
 \begin{equation}  
{\bi V}\cdot\dot{\brho}\equiv\frac{\partial L}{\partial\dot{\rho}^i}\,\dot{\rho}^i
=\frac{\partial L}{\partial \omega^j}\,\frac{\partial\omega^j}{\partial\dot{\rho}^i}
\,\dot{\rho}^i={\bi W}\cdot\bomega,
 \end{equation}
due to the linearity of $\bomega$ in terms of
$\dot{\brho}$ and where $W_i=\partial L/\partial\omega^i$. Thus the most 
general form of the Lagrangian of a nonrelativistic particle can also be written instead of
(\ref{4.1.3}) as:
 \begin{equation}
L=T\dot t+{\bi R}\cdot\dot{\bi r}+{\bi U}\cdot\dot{\bi u}+{\bi W}\cdot\bomega.
 \label{eq:4.1.7}
 \end{equation}   

Since $X$ is the whole Galilei group
${\cal G}$ the most general gauge function 
is just the group exponent:
 \index{Galilei!gauge function}
 \begin{equation}
\alpha(g;x)=\xi(g,h_x)=m({\bi v}^2t(\tau)/2+{\bi v}\cdot R(\bmu){\bi r}(\tau)),
 \label{eq:4.1.9}
 \end{equation}
similar to (\ref{nrgauge}), and this allows us to interpret the parameter $m$ as the mass of the 
system. Under the action of an arbitrary element of the Galilei group, 
the Lagrangian $L$ transforms according to: 
 \begin{equation}
L(gx(\tau),d(gx(\tau))/d\tau)=L(x(\tau),\dot x(\tau))+d\alpha(g;x(\tau))/d\tau.
 \label{4.1.10}
 \end{equation}
This leads through some straightforward calculations,
similar to the ones performed in (\ref{eq:Ttra1})-(\ref{eq:Rtr1}),
to the following form of transformation of the functions:
 \begin{eqnarray} 
T'(\tau)&=&T(\tau)-{\bi v}\cdot R(\bmu){\bi R}(\tau)-m{\bi v}^2/2,\\ 
{\bi R}'(\tau)&=&R(\bmu){\bi R}(\tau)+m{\bi v},\\
 {\bi U}'(\tau)&=&R(\bmu){\bi U}(\tau),\\
 {\bi W}'(\tau)&=&R(\bmu){\bi W}(\tau).
 \label{4.1.11}
 \end{eqnarray} 

\subsection{Noether constants of the motion}

Using the action of the Galilei group on the kinematical
space given by (\ref{eq:gt1})-(\ref{eq:4.1.1}), Noether's theorem defines the following constants of the motion:

\noindent{\bf a)} Under time translation the action function is
invariant and as usual we call the corresponding conserved quantity, 
the {\bf total energy} of the system $H$. Since $\delta t=\delta b$ and 
$\delta q_i^{(s)}=0$, $M=1$ and $M_i^{(s)}=0$, by applying 
(\ref{eq:constants}) 
we have:
 \index{total energy}
 \[ 
H=-(L-p_{(s)}^i q_i^{(s)})M=-(\hat L/\dot t-p_{(s)}^i q_i^{(s)})=-T-{\bi R}\cdot{\bi u}
-{\bi U}\cdot\dot{\bi u}/\dot t-{\bi W}\cdot\bomega/\dot t
 \] 
 \[ 
+({\bi R}-d{\bi U}/dt)\cdot{\bi u}+{\bi U}\cdot\dot{\bi u}/\dot t+
{\bi V}\cdot\dot{\brho}/\dot t,
 \]
and since ${\bi W}\cdot{\bomega}={\bi V}\cdot\dot{\brho}$, it turns out that
 \begin{equation}
H=-T-\frac{d{\bi U}}{ dt}\cdot{\bi u}.
 \label{eq:4.1.13}
 \end{equation}  

\noindent{\bf b)} Under spatial translation, $A(x_1,x_2)$ is
invariant and this defines the {\bf total linear momentum} of the system. We have now: 
 \index{total linear momentum}
\[ 
\delta t=0,\ M=0,\ \delta r_i=\delta a_i,\ M_{ij}^{(0)}=\delta_{ij},\ 
\delta u_i=0,\ M_{ij}^{(1)}=0,\
 \]
 \[
 \delta\rho_i=0,\ M_{ij}^{(\rho)}=0,
 \]  
and then
 \begin{equation}  
{\bi P}={\bi R}-\frac{d{\bi U}}{ dt}.
 \label{eq:4.1.14}
 \end{equation}   

\noindent{\bf c)} Under a pure Galilei transformation of velocity 
$\delta{\bi v}$, $A(x_1,x_2)$ is no longer invariant but taking into account (\ref{eq:2.5.3})
and the gauge function (\ref{eq:4.1.9}), it transforms 
as $\delta A=m{\bi r}_2\cdot\delta{\bi v}-m{\bi r}_1\cdot\delta{\bi 
v}$ and this defines the {\bf total kinematical momentum} ${\bi K}$, in the following way:
 \index{total kinematical momentum}
 \[
\delta t=0,\ M=0,\ \delta r_i=\delta v_it,\ M_{ij}^{(0)}=\delta_{ij}t,\ \delta u_i=
\delta v_i,\ M_{ij}^{(1)}=\delta_{ij},
 \]
  \[
 \delta\rho_i=0,\ M_{ij}^{(\rho)}=0,
 \]
and thus
 \begin{equation}
{\bi K}=m{\bi r}-{\bi P}\,t-{\bi U}. 
 \label{eq:4.1.15}
 \end{equation}  

From $\dot{\bi K}=0$, this leads 
to ${\bi P}=m{\bi u}-d{\bi U}/dt$, and thus by identification with 
(\ref{eq:4.1.14}), the function ${\bi R}=m{\bi u}$ irrespective of the particular Lagrangian. 
The total linear momentum 
does not lie along the velocity of point ${\bi r}$.

\noindent{\bf d)} Finally, under rotations $A(x_1,x_2)$ remains invariant and 
the corresponding constant of the motion, the {\bf total angular momentum} 
of the system, comes from the infinitesimal transformation of value 
$\delta\mu_i=\delta\alpha_i/2$, {\sl i.e.}, half of the rotated infinitesimal angle, and then
 \index{total angular momentum}
 \[ 
\delta t=0,\ M_i=0, \delta r_i=\epsilon_{ikj}\delta\alpha_jr_k,\ M_{ij}^{(0)}=
\epsilon_{ikj}r_k,
 \]
\[
\delta u_i=\epsilon_{ikj}\delta\alpha_ju_k,\quad
M_{ij}^{(1)}=\epsilon_{ikj}u_k,\ 
\]
\[
\delta\rho_i=\delta\alpha^j(\delta_{ij}+
\epsilon_{ikj}\rho^k+\rho_i\rho_j)/2,\quad
M_{ij}^{(\rho)}=(\delta_{ij}+\epsilon_{ikj}\rho^k+\rho_i\rho_j)/2,
 \]
which leads to
 \[ 
V_iM_{ij}^{(\rho)}=\frac{\partial L}{\partial\omega^k}\,\frac{\partial\omega^k}
{\partial\dot{\rho}_i}M_{ij}^{(\rho)}=W_j,
 \]
and therefore
 \begin{equation}
{\bi J}={\bi r}\times{\bi P}+{\bi u}\times{\bi U}+{\bi W}={\bi r}\times{\bi P}+{\bi Z}.
 \label{eq:4.1.16}
 \end{equation}  
We are tempted to consider ${\bi Z}$ as the spin of the system. Since $\dot{\bi J}=0$, this 
function ${\bi Z}$ satisfies $d{\bi Z}/dt={\bi P}\times{\bi u}$ and 
is not a constant of the motion for a free particle. It is the classical angular momentum
equivalent to Dirac's spin operator in the quantum case. Because ${\bi J}$ is the angular momentum of the
particle with respect to the origin of the observer frame, ${\bi Z}$ represents the angular momentum
of the particle with respect to the point ${\bi r}$.
We shall define
the spin for a free particle as the angular momentum with respect to its center of mass, 
once we accurately
identify the center of mass of the particle.

The center of mass observer is defined as that inertial observer for 
whom ${\bi P}=0$ and ${\bi K}=0$. These six conditions do not define 
uniquely an inertial observer but rather a class of them up to a   
rotation and an arbitrary time translation. In fact, the
condition ${\bi P}=0$ establishes the class of observers for which the 
center of mass is at rest, and ${\bi K}=0$ is the additional condition
to locate it at the origin of coordinates.
This comes from the 
analysis of (\ref{eq:4.1.15}), where ${\bi k}={\bi U}/m$ is an observable 
with dimensions of length, and taking the derivative with respect to 
$\tau$ of both sides, taking into account that $\dot{\bi P}=0$, we have:
 \begin{equation}
\dot{\bi K}=0=m\dot{\bi r}-{\bi P}\,\dot t-m\dot{\bi k}, 
 \qquad\hbox{\sl i.e.,}\qquad {\bi P}=m\frac{d({\bi r}-{\bi k})}{dt}.
 \end{equation}   
Then the point ${\bi q}={\bi r}-{\bi k}$ is moving at constant speed and we 
say that it represents the position of the center of mass of the 
system. Thus, the observable ${\bi k}={\bi r}-{\bi q}$ is just the 
relative  position of point ${\bi r}$ with respect to the center of 
mass. Therefore ${\bi P}=0$ and ${\bi K}=0$ give rise to $d{\bi 
q}/dt=0$, and ${\bi r}={\bi k}$, {\sl i.e.}, ${\bi q}=0$, as we pointed out. 
With this definition, the kinematical momentum can be written 
as ${\bi K}=m{\bi q}-{\bi P}t$, 
in terms of the center of mass position ${\bi q}$ and the total linear momentum ${\bi P}$.

The spin of the system is defined as the difference between the total
angular momentum ${\bi J}$ and the orbital angular momentum of the
center of mass motion ${\bi q}\times{\bi P}$, and thus
 \index{spin}
 \begin{equation} 
{\bi S}={\bi J}-{\bi q}\times{\bi P}={\bi J}-\frac{1}{m}{\bi K}\times{\bi 
P}={\bi Z}+{\bi k}\times{\bi P}=-m{\bi k}\times\frac{d{\bi k}}{dt}+{\bi W}.
 \label{spin1}
 \end{equation} 
The spin ${\bi S}$, expressed in terms of the constants of the motion ${\bi J}$, 
${\bi K}$ and ${\bi P}$, is also a constant of the motion.

It is the sum of two terms, one ${\bi Z}={\bi u}\times{\bi U}+{\bi W}$,
coming from the new degrees of freedom and another ${\bi k}\times{\bi
P}$, which is the angular momentum of the linear momentum located at point ${\bi r}$
with respect to the center of mass. Alternatively we can
describe the spin according to the last expression in which
the term $-{\bi k}\times md{\bi k}/dt$ suggests a contribution of (anti)orbital 
type coming from the motion around 
the center of mass. It is related to the zitterbewegung or more precisely 
 \index{zitterbewegung}
to the function ${\bi U}=m{\bi k}$ which reflects the dependence of the Lagrangian on the acceleration.
The other term ${\bi W}$ comes from the dependence on the other three 
degrees of freedom $\rho_i$, and thus on the angular velocity. This 
zitterbewegung is the motion of the center of charge around the center 
 \index{center of charge}
of mass. Point ${\bi r}$, as 
representing the position of the center of charge, has been also 
suggested in previous works for the relativistic electron.~\footnote{\hspace{0.1cm}A.O. Barut 
and A.J. Bracken, {\sl Phys. Rev. D} {\bf 23}, 
2454 (1981).} 

Because $\dot{\bi J}=0$, and that ${d{\bi W}}/{d\tau}=\bomega\times{\bi W}$ and the expression of
${\bi P}$, (\ref{eq:4.1.14}),
this implies the general relation for a free particle
 \begin{equation}
\dot{\bi r}\times{\bi R}+\dot{\bi u}\times{\bi U}+{\bomega}\times{\bi W}=0,
 \label{eq:ruw}
 \end{equation}
which reflects the fact that velocity, acceleration and angular 
velocity are not independent magnitudes, and taking into account that 
${\bi R}$ and $\dot{\bi r}$ have the same direction, it reduces to
 \begin{equation}  
\dot{\bi u}\times{\bi U}+{\bomega}\times{\bi W}=0.
\label{eq:UxW}
 \end{equation}  

\subsection{Galilei spinning particle of (anti)orbital spin}

To analyze the spin structure of the particle, and therefore
the different contributions to spin coming from these functions ${\bi U}$ and ${\bi W}$,
let us consider the following simpler example.

Consider a Galilei 
particle whose kinematical space is $X={\cal G}/SO(3)$, so that any 
point $x\in X$ can be characterized by the seven variables 
$x\equiv(t,{\bi r},{\bi u})$, ${\bi u}=d{\bi r}/dt$, which are 
interpreted as time, position and velocity of the particle 
respectively. In this example we have no orientation variables.
The Lagrangian will also depend 
on the next order derivatives, {\sl i.e.}, on the velocity which is already 
considered as a kinematical variable and
on the acceleration of the 
particle. Rotation and translation invariance implies that $L$ will be 
a function of only ${\bi u}^2$, $(d{\bi u}/dt)^2$ and ${\bi u}\cdot 
d{\bi u}/dt= d(u^2/2)/dt$, but this last term is a total time derivative 
and it will not be considered here. 

Since from condition (\ref{eq:UxW}) ${\bi U}\sim\dot{\bi u}$, let us assume that our elementary system is represented by the 
following Lagrangian, which when written in terms of the three degrees of freedom
and their derivatives is expressed as
 \begin{equation} 
L=\frac{m}{2}\left(\frac{d{\bi r}}{ dt}\right)^2-
\frac{m}{2\omega^2}\left(\frac{d^2{\bi r} }{ dt^2}\right)^2.
\label{eq:n1}
 \end{equation} 
Parameter $m$ is the mass of the particle because
the first term is gauge variant in terms of the gauge function (\ref{eq:4.1.9}) defined
by this constant $m$, while parameter $\omega$ of dimensions of 
time$^{-1}$ represents an internal frequency. It is the frequency of the 
internal zitterbewegung.\index{zitterbewegung} 

In terms of the kinematical 
variables and their derivatives, and in terms of some group invariant 
evolution parameter $\tau$, the Lagrangian can also be written as 
 \begin{equation} 
L=\frac{m}{2}\frac{\dot{\bi r}^2}{\dot t}-
\frac{m}{2\omega^2}\frac{\dot{\bi u}^2 }{\dot t},
 \label{eq:n2}
 \end{equation} 
where the dot means $\tau$-derivative. If we consider that the 
evolution parameter is dimensionless, all terms in the Lagrangian have 
dimensions of action. Because the Lagrangian is a homogeneous function 
of first degree in terms of the derivatives of the kinematical 
variables, $L$ can also be written as 
 \begin{equation} 
L=T\dot t+{\bi R}\cdot\dot{\bi r}+{\bi U}\cdot\dot{\bi u},
\label{eq:n3}
 \end{equation} 
where the functions accompanying the derivatives of the kinematical 
variables are defined and explicitly given by 
 \begin{eqnarray} 
T&=&\frac{\partial L}{\partial\dot t}=-\frac{m}{2}\left(\frac{d{\bi r}}{ dt}\right)^2+
\frac{m}{2\omega^2}\left(\frac{d^2{\bi r} }{ dt^2}\right)^2,\nonumber\\ 
{\bi R}&=&\frac{\partial L}{\partial\dot{\bi r}}=m\frac{d{\bi r}}{ 
dt},\label{eq:n4}\\ {\bi U}&=&\frac{\partial L}{\partial\dot{\bi u}}=-
\frac{m}{\omega^2}\frac{d^2{\bi r}}{ dt^2}.\label{eq:n4bis}
 \end{eqnarray} 
Dynamical equations obtained from Lagrangian (\ref{eq:n1}) are: 
 \begin{equation} 
\frac{1}{\omega^2}\frac{d^4{\bi r}}{ dt^4}+\frac{d^2{\bi r}}{ dt^2}=0,
\label{eq:n5}
 \end{equation} 
whose general solution is: 
 \begin{equation} 
{\bi r}(t)={\bi A}+{\bi B}t+{\bi C}\cos\omega t+{\bi D}\sin\omega t,
 \label{eq:n6} 
 \end{equation} 
in terms of the 12 integration constants ${\bi A}$, 
${\bi B}$, ${\bi C}$ and ${\bi D}$. 

When applying Noether's theorem to the invariance of dynamical 
equations under the Galilei group, the corresponding constants of the 
motion can be written in terms of the above functions in the form: 
 \begin{eqnarray} 
\hbox{\rm Energy}\quad H&=&-T-{\bi u}\cdot\frac{d{\bi U}}{ dt},\label{eq:n71}\\ 
\hbox{\rm linear momentum}\quad {\bi P}&=&{\bi R}-\frac{d{\bi U}}{ dt}=
m{\bi u}-\frac{d{\bi U}}{ dt},\label{eq:n72}\\ 
\hbox{\rm kinematical momentum}\quad {\bi K}&=&m{\bi r}-{\bi P}t-{\bi U},\label{eq:n73}\\ 
\hbox{\rm angular momentum}\quad {\bi J}&=&{\bi r}\times{\bi P}+{\bi u}\times{\bi U}.
\label{eq:n7} 
 \end{eqnarray} 
It is the presence of the ${\bi U}$ function that distinguishes the 
features of this system with respect to the point particle case. We 
find that the total linear momentum is not lying along the direction of 
the velocity ${\bi u}$, and the spin structure is directly related to
the dependence of the Lagrangian on the acceleration.

If we substitute the general solution (\ref{eq:n6}) in 
(\ref{eq:n71}-\ref{eq:n7}) we see in fact that 
the integration constants are related to the above conserved 
quantities 
\begin{eqnarray} 
H&=&\frac{m}{2}{\bi B}^2-\frac{m\omega^2}{2}({\bi C}^2+{\bi D}^2),\\ 
{\bi P}&=&m{\bi B},\\ 
{\bi K}&=&m{\bi A},\label{defK}\\ 
{\bi J}&=&{\bi A}\times m{\bi B}-m\omega{\bi C}\times{\bi D}.
 \end{eqnarray} 

We see that the kinematical momentum ${\bi K}$ in (\ref{eq:n73}) differs from the point particle 
case (\ref{k1}) in the term $-{\bi U}$, such that if we define the vector ${\bi 
k}={\bi U}/m$, with dimensions of length, then $\dot{\bi K}=0$ leads 
from (\ref{eq:n73}) to the equation: 
 \[
{\bi P}=m\frac{d({\bi r}-{\bi k})}{ dt},
 \] 
and ${\bi q}={\bi r}-{\bi k}$, defines the position of the center 
of mass of the particle that is a different point than ${\bi r}$ and 
using (\ref{eq:n4bis}) is given by 
 \begin{equation}
{\bi q}={\bi r}-\frac{1}{m}{\bi U}={\bi
r}+\frac{1}{\omega^2}\;\frac{d^2{\bi r}}{ dt^2}.
\label{eq:n10}
 \end{equation} 
In terms of it, dynamical equations (\ref{eq:n5}) can be 
separated into the form: 
\begin{eqnarray} 
 \frac{d^2{\bi q}}{ 
dt^2}&=&0,\label{eq:n110}\\ \frac{d^2{\bi r}}{ dt^2}&+&\omega^2({\bi
r}-{\bi q})=0, 
\label{eq:n111} 
 \end{eqnarray} 
where (\ref{eq:n110}) is just eq. (\ref{eq:n5}) after twice
differentiating (\ref{eq:n10}), and Equation (\ref{eq:n111}) is (\ref{eq:n10})
after collecting all terms on the left hand side.

From (\ref{eq:n110}) we see that point ${\bi q}$ moves in a straight 
trajectory at constant velocity 
while the motion of point ${\bi r}$, given in (\ref{eq:n111}), is an isotropic harmonic motion
of angular frequency $\omega$ around point ${\bi q}$. 

The spin of the system ${\bi S}$ is defined as
 \begin{equation} 
{\bi S}={\bi J}-{\bi q}\times{\bi P}={\bi J}-\frac{1}{m}{\bi K}\times{\bi P},
 \label{spin01}
 \end{equation} 
and since it is written in terms of constants of the motion 
it is clearly a constant of the motion, and its magnitude $S^2$ is also a Galilei invariant 
quantity that characterizes the system. In terms of the 
integration constants it is expressed as
 \begin{equation} 
{\bi S}=-m\omega\,{\bi C}\times{\bi D}.
 \end{equation} 
From its definition we get
 \begin{equation} 
{\bi S}={\bi u}\times{\bi U}+{\bi k}\times{\bi P}=-m({\bi r}-{\bi q})
\times\frac{d}{dt}\left({\bi r}-{\bi q}\right)=-{\bi k}\times m\frac{d{\bi k}}{dt},
 \label{eq:spin02}
 \end{equation} 
which appears as the (anti)orbital angular momentum of the relative 
motion of point ${\bi r}$ around the center of mass position ${\bi 
q}$ at rest, so that the total angular momentum can be written as
 \begin{equation} 
{\bi J}={\bi q}\times{\bi P}+{\bi S}={\bi L}+{\bi S}.
\label{angulJ}
 \end{equation} 
It is the sum of the orbital angular momentum ${\bi L}$ associated to the motion
of the center of mass and the spin part ${\bi S}$. For a free particle
both ${\bi L}$ and ${\bi S}$ are separately constants of the motion. We use the term (anti)orbital
to suggest that if vector ${\bi k}$ represents the position of a point mass $m$, the angular momentum
of this motion is in the opposite direction as the obtained spin observable. But as we shall see in a moment,
vector ${\bi k}$ does not represent the position of the mass $m$ but rather the position of the charge $e$
of the particle.

\subsection{Interacting with an external electromagnetic field} 
\label{sec:galispinmag}

But if ${\bi q}$ represents the center of mass position, then what 
position does point ${\bi r}$ represent? Point ${\bi r}$ represents 
the position of the charge of the particle. This can be seen by 
considering some interaction with an external field. The homogeneity 
condition of the Lagrangian in terms of the derivatives of the 
kinematical variables leads us to consider an interaction term of the 
form 
 \begin{equation} 
L_I=-e\phi(t,{\bi r})\dot t+e{\bi A}(t,{\bi r})\cdot\dot{\bi r},
 \label{eq:n13}
 \end{equation}
which is linear in the derivatives of the kinematical
variables $t$ and ${\bi r}$ and where the external potentials are only
functions of $t$ and ${\bi r}$. We can also consider more general 
interaction terms of the form ${\bi N}(t,{\bi r},{\bi u})\cdot\dot{\bi 
u}$, and also more general terms in which functions $\phi$ and ${\bi A}$ also depend 
on ${\bi u}$ and $\dot{\bi 
u}$. If the interaction Lagrangian depends on $\dot{\bi u}$ this implies that the interaction
modifies the definition of the observable ${\bi U}=m{\bi k}$ which defines the spin of the free system. 
But if the system is elementary the spin definition cannot be changed, so that (\ref{eq:n13})
is the most general interaction term.

Dynamical equations obtained from $L+L_I$ are 
 \begin{equation} 
 \frac{1}{\omega^2}\frac{d^4{\bi r}}{ dt^4}+\frac{d^2{\bi 
r}}{ dt^2}=\frac{e}{ m} \left({\bi E}(t,{\bi r})+{\bi u}\times{\bi 
B}(t,{\bi r})\right),
 \label{eq:n14} 
 \end{equation}
where the electric field ${\bi E}$ and magnetic field ${\bi B}$ are 
expressed in terms of the potentials in the usual form, ${\bi E}=-
\nabla\phi-\partial{\bi A}/\partial t$, ${\bi B}=\nabla\times{\bi A}$. 
Dynamical equations 
(\ref{eq:n14}) can again be separated into the form 
 \begin{eqnarray} 
 \frac{d^2{\bi q}}{ dt^2}&=&\frac{e}{ m}\left({\bi E}(t,{\bi r})+
{\bi u}\times{\bi B}(t,{\bi r})\right),\label{eq:n151}\\ 
\frac{d^2{\bi r}}{ dt^2}&+&\omega^2({\bi r}-{\bi q})=0.\label{eq:n152} 
 \end{eqnarray} 
The center of mass ${\bi q}$ satisfies Newton's equations under 
the action of the total external Lorentz force, while point ${\bi r}$ 
still satisfies the isotropic harmonic motion of angular frequency 
$\omega$ around point ${\bi q}$. But the external force and the 
fields are defined at point ${\bi r}$ and not at point ${\bi q}$. It 
is the velocity ${\bi u}$ of point ${\bi r}$ that appears in the 
magnetic term of the Lorentz force. Point ${\bi r}$ clearly represents 
the position of the charge. In fact, this minimal coupling we have 
considered is the coupling of the electromagnetic potentials with the 
particle current, that in the relativistic case can be written as 
$j_\mu A^\mu$, but the current $j_\mu$ is associated to the motion of 
a charge $e$ at point ${\bi r}$. 

\cfigl{fig:zitter}{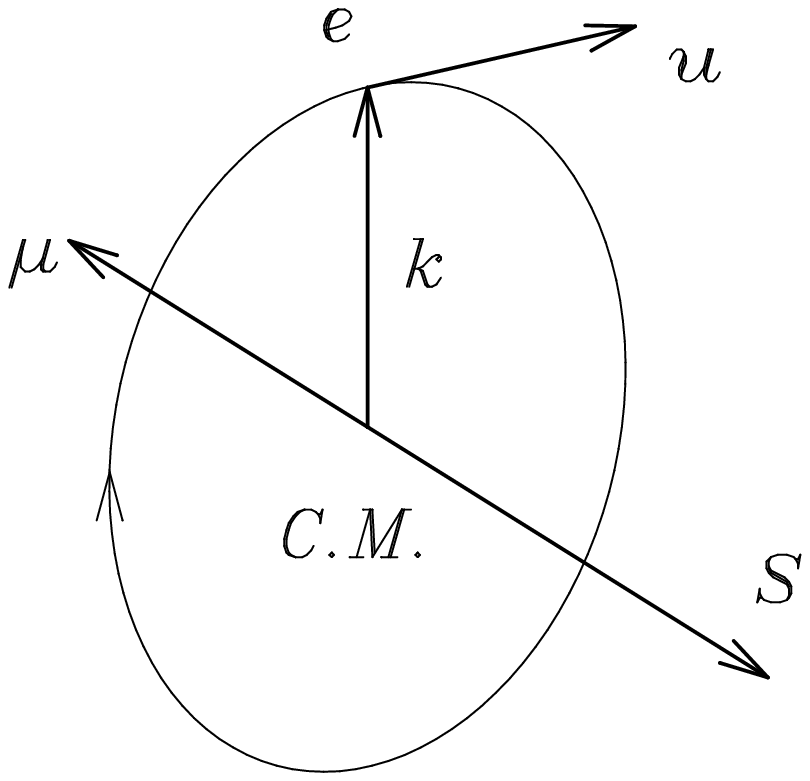}{Charge motion in the C.M. frame.}

This charge has an oscillatory motion of very high frequency $\omega$ 
that, in the case of the relativistic electron 
is $\omega=2mc^2/\hbar\simeq 
1.55\times10^{21}$s$^{-1}$. The average position of the charge is the 
center of mass, but it is this internal orbital motion, usually 
known as the zitterbewegung, that gives rise to the spin structure 
for this model and 
also to the magnetic properties of the particle, as we shall see later.

When analyzed in the center of mass frame (see 
Fig.~\ref{fig:zitter}), ${\bi q}=0$, ${\bi r}={\bi k}$, the system reduces to a point charge whose 
motion is in general an ellipse, but if we choose $C=D$, and ${\bi 
C}\cdot{\bi D}=0$, it reduces to a circle of radius $a=C=D$, 
orthogonal to the spin. Then if the 
particle has charge $e$, it has a magnetic moment that according to the
usual classical definition\index{magnetic moment} is:~\footnote{\hspace{0.1cm}J.D. Jackson, {\sl Classical Electrodynamics}, 
John Wiley \& Sons, NY (1998), 3rd. ed. p.186.}
 \begin{equation} 
{\bmu}=\frac{1}{2}\int{\bi r}\times{\bi j}\, d^3{\bi r}=\frac{e}{2}\,{\bi k}\times\frac{d{\bi k}}{dt}= 
-\frac{e}{2m}{\bi S},
\label{eq:magneticmoment}
 \end{equation}  
where ${\bi j}=e\delta^3({\bi r}-{\bi k})d{\bi k}/dt$ is the current
associated to the motion of a charge $e$ located at point ${\bi k}$.
The magnetic moment is orthogonal to the zitterbewegung plane\index{zitterbewegung}\index{magnetic moment}
and opposite to the spin if $e>0$. It also has
a non-vanishing oscillating electric dipole ${\bi d}=e{\bi k}$,
orthogonal to\index{electric dipole}\index{electric dipole!oscillating}
${\bmu}$ and therefore to ${\bi S}$ in the center of mass frame, such that its time average value 
vanishes for times larger than the natural period of this internal 
motion. 
\index{electric dipole moment}\index{moment!electric dipole}
Although this is a nonrelativistic example it is interesting to point out
and compare with Dirac's relativistic analysis of the electron,~\footnote{\hspace{0.1cm}P.A.M. Dirac, {\sl The Principles of Quantum 
mechanics}, Oxford Univ. Press, 4th ed. (1967).} in which
both momenta $\bmu$ and ${\bi d}$ appear, giving rise to two possible interacting terms
in Dirac's Hamiltonian. We shall come back to this analysis later when we study the 
elementary relativistic particles.

\subsection{Spinning Galilei particle with orientation}
\label{sec:galiorien}

Another simple example of spinning particles is the one in which the spin
is related only to the angular variables that describe orientation. 

Let us assume now a dynamical system whose kinematical space is 
$X={\cal G}/\RR^3_v$, where $\RR^3_v\equiv\{\RR^3,+\}$ is the 3-parameter Abelian subgroup of pure Galilei 
transformations. Then, the kinematical variables are $x\equiv(t,{\bi 
r},\brho)$, which are interpreted as the time, position and orientation 
respectively. 

The Lagrangian for this model takes the general form
\[
L=T\dot t+{\bi R}\cdot\dot{\bi r}+{\bi W}\cdot\bomega.
\]
Because of the structure of the exponent (\ref{exponent}),
the gauge function for this system can be taken the same as before.
The general relationship (\ref{eq:UxW}) leads to ${\bi W}\times\bomega=0$, 
because the Lagrangian is independent
of $\dot{\bi u}$, and therefore ${\bi W}$ and $\bomega$ must be collinear. According to
the transformation properties of the Lagrangian, the third
term ${\bi W}\cdot\bomega$ is Galilei invariant
and since ${\bi W}$ and $\bomega$ are collinear, we can take ${\bi W}\sim\bomega$ and one possible 
Lagrangian that describes this model is of the form:
 \begin{equation}  
L=\frac{m}{2}\frac{\dot{\bi r}^2}{\dot t}+\frac{I}{2}\frac{\bomega^2}{\dot t}.
 \label{4.5.1}
 \end{equation}  
    
The different Noether's constants are
\[
H=\frac{m}{2}\left(\frac{d{\bi r}}{dt}\right)^2+\frac{I}{2}{\bf\Omega}^2,\quad 
{\bi P}=m{\bi u},
\]
\[
{\bi K}=m{\bi r}-{\bi P}t,\quad {\bi J}={\bi r}\times{\bi P}+{\bi W},
\]
where ${\bi u}=d{\bi r}/dt$ is the velocity of point ${\bi r}$, 
and ${\bf\Omega}=\bomega/\dot t$ is the time evolution angular velocity.
Point ${\bi r}$ is moving at a constant speed and it also represents 
the position of the center of mass. The spin 
is just the observable ${\bi S}\equiv{\bi W}$ 
that satisfies the dynamical equation $d{\bi S}/dt=\bomega\times{\bi S}=0$, and
thus the frame linked to 
the body rotates with a constant angular velocity ${\bf\Omega}$.

The spin \index{spin} takes the constant value ${\bi S}=I{\bf\Omega}$, whose 
absolute value is independent of the inertial observer and also the 
angular velocity ${\bf\Omega}=\bomega/\dot t$ is constant. The 
parameter $I$ plays the role of a principal moment of inertia, 
suggesting a linear relationship between the spin and the angular 
velocity, which corresponds to a particle with spherical symmetry. 
The particle can also be considered as an extended object of gyration 
radius $R_0$, related to the other particle parameters by $I=m R_0^2$. 

This system corresponds classically to a rigid body with spherical symmetry
where the orientation variables $\brho$ can describe for instance, the 
orientation of its principal axes of inertia in a suitable 
parametrization of the rotation group. This is a system of six degrees 
of freedom. Three represent the position of the center of charge 
${\bi r}$ and the other three $\brho$, represent the orientation of a 
Cartesian frame linked to that point ${\bi r}$. Since for this system 
there is no dependence on the acceleration, the centers of mass and 
charge will be represented by the same point. 

In the center of mass frame there is no current associated to this 
particle and therefore it has neither magnetic nor electric dipole 
structure. As seen in previous examples, all magnetic 
properties seem therefore to be related to the 
zitterbewegung part of spin and are absent in this rigid body-like model.

\section*{Relativistic particles}

\section{Relativistic point particle}
\index{relativistic point particle}\index{point particle!relativistic}
\label{sec:Point}

See the Appendix about the Poincar\'e group at the end of this chapter
for the group notation used throughout this section.

The kinematical space is the quotient 
structure $X={\cal P}/{\cal L}$, where ${\cal P}$ is the Poincar\'e 
group and the subgroup ${\cal L}$ is the Lorentz group. Then every point $x\in 
X$ is characterized by the variables $ x\equiv (t(\tau),{\bi 
r}(\tau))$, with domains $t\in\RR,\ {\bi r}\in \RR^3$ as the 
corresponding group parameters, in such a way that under the action of 
a group element $g\equiv(b,{\bi a},{\bi v},\bmu)$ of ${\cal P}$ they 
transform as:
 \begin{eqnarray}
t'(\tau)&=&\gamma t(\tau)+\gamma({\bi v}\cdot R(\bmu){\bi r}(\tau))/c^2+b,\label{eq:5.1a}\\ 
{\bi r}'(\tau)&=&R(\bmu){\bi r}(\tau)+\gamma{\bi v}t(\tau)+\frac{\gamma^2}{(1+\gamma)c^2}
({\bi v}\cdot R(\bmu){\bi r}(\tau)){\bi v}+{\bi a},\qquad\label{eq:5.1b}   
 \end{eqnarray} 
and are interpreted as the time and position of the system. 
If, as usual, we assume that the evolution parameter $\tau$ is invariant under the 
group, taking the $\tau$-derivatives of (\ref{eq:5.1a}) and 
(\ref{eq:5.1b}) we get
 \begin{eqnarray}
\dot t'(\tau)&=&\gamma\dot t(\tau)+\gamma({\bi v}\cdot 
R(\bmu)\dot{\bi r}(\tau))/c^2,\label{eq:tpunto}\\
\dot{\bi r}'(\tau)&=&R(\bmu)\dot{\bi r}(\tau)+\gamma{\bi v}\dot t(\tau)+
\frac{\gamma^2}{(1+\gamma)c^2}({\bi v}\cdot R(\bmu)\dot{\bi r}(\tau)){\bi v}.
 \label{eq:rpunto}
 \end{eqnarray}   

The homogeneity condition of the Lagrangian, in terms of the derivatives of the kinematical variables, 
reduces to three the 
number of degrees of freedom of the system. This leads to the general expression
 \begin{equation}
L=T\dot t+{\bi R}\cdot\dot{\bi r},
\label{eq:lagranpoin}
 \end{equation}
where $T=\partial L/\partial\dot t$ and $R_i=\partial L/\partial\dot r_i$, will be 
functions of $t$ and ${\bi r}$ and homogeneous functions of zero degree of $\dot t(\tau)$ and 
$\dot{\bi r}(\tau)$. Because the Lagrangian is invariant under ${\cal P}$, the 
functions $T$ and ${\bi R}$ transform under the group ${\cal P}$ in the form:
 \begin{eqnarray}
T'&=&\gamma T-\gamma ({\bi v}\cdot R(\bmu){\bi R}),\\ 
{\bi R}'&=&R(\bmu){\bi R}-\gamma{\bi v}T/c^2+
\frac{\gamma^2}{1+\gamma}({\bi v}\cdot R(\bmu){\bi R}){\bi v}/c^2. 
 \end{eqnarray} 
We thus see that $T$ and ${\bi R}$ are invariant under translations and therefore they 
must be functions independent of $t$ and ${\bi r}$.

The conjugate momenta of the independent degrees of freedom $q_i=r_i$ are $p_i=\partial L/\partial\dot r_i$, and 
consequently Noether's theorem (\ref{eq:constants}) leads to the following constants of the 
motion, that are calculated similarly as in the Galilei case except for the invariance
under pure Lorentz transformations. We have now no gauge function and the variations are
$\delta t={\bi r}\cdot{\delta{\bi v}}/c^2$, $M_i=r_i/c^2$ and $\delta{\bi r}=t\delta{\bi v}$,
$M_{ij}=t\delta_{ij}$ and thus we get:
 \begin{eqnarray}  
\hbox{\rm Energy}\quad H&=&-T,\label{eq:HP1}\\
\hbox{\rm linear momentum}\quad {\bi P}&=&{\bi R}={\bi p},\label{eq:pP1}\\
\hbox{\rm kinematical momentum}\quad {\bi K}&=&H{\bi r}/c^2-{\bi P}t,\label{eq:kP1}\\
\hbox{\rm angular momentum}\quad {\bi J}&=&{\bi r}\times{\bi P}.\label{eq:JP1}
 \end{eqnarray} 
The energy and the linear momentum transform as:
 \begin{eqnarray}
H'(\tau)&=&\gamma H(\tau)+\gamma ({\bi v}\cdot R(\bmu){\bi P}(\tau)),\label{eq:transH}\\
{\bi P}'(\tau)&=&R(\bmu){\bi P}(\tau)+\frac{\gamma {\bi v}}{c^2}H(\tau)+\frac{\gamma^2}{(1+\gamma)c^2}
({\bi v}\cdot R(\bmu){\bi P}(\tau)){\bi v}.\qquad\label{eq:transP}
 \end{eqnarray} 

They transform like the contravariant components of a four-vector $P^\mu\equiv(H/c,{\bi P})$.
The observables $c{\bi K}$ and ${\bi J}$ are the essential components of the antisymmetric
tensor $J^{\mu\nu}=-J^{\nu\mu}=x^\mu P^\nu-x^\nu P^\mu$, $cK_i=J^{i0}$ 
and $J_k=\epsilon_{kil}J^{il}/2$.

Taking the $\tau$ derivative of the kinematical momentum, $\dot{\bi K}=0$, 
we get ${\bi P}=H\dot{\bi r}/c^2\dot t=H{\bi 
u}/c^2$, where ${\bi u}={\dot{\bi r}}/{\dot{t}}$ is the velocity of the particle and the point ${\bi r}$ 
represents both the center of mass and center of charge position of the particle. 

The six conditions ${\bi P}=0$ and ${\bi K}=0$, imply ${\bi u}=0$ and 
${\bi r}=0$, so that the system is at rest and placed at the origin of 
the reference frame, similarly as in the nonrelativistic case. We 
again call this class of observers the center of mass observer.\index{center of mass!observer} 

From (\ref{eq:transH}) and (\ref{eq:transP}) we see that 
the magnitude $(H/c)^2-{\bi P}^2=m^2c^2$ is a Poincar\'e 
invariant and a constant of the motion. This defines the mass of the 
particle which we take as a positive number $m$. 
By using the expression of ${\bi P}=H{\bi u}/c^2$, we get
 \[
H=\pm mc^2(1-u^2/c^2)^{-1/2},
 \]
and the sign of $H$, which is another Poincar\'e invariant property, can be either positive or negative.
The velocity $u<c$, otherwise $H$ will be imaginary. If $u>c$ the
invariant $(H/c)^2-P^2<0$ and it is not possible to define the rest mass of the system.
By substitution of the found expressions for $T$ and ${\bi R}$ in (\ref{eq:lagranpoin}), 
there are two possible Lagrangians for a point particle of mass $m$, characterized by 
the sign of $H$
 \begin{equation} 
L=\mp mc\sqrt{c^2\dot t^2-\dot{\bi r}^2}.
 \label{eq:PointLag}
 \end{equation} 
Expansion of this Lagrangian to lowest order in $u/c$, in the case of positive $H$, we get
\[
L=-mc^2\dot t+\frac{m}{2}\,\frac{\dot{\bi r}^2}{\dot t},
\]
where the first term $-mc^2\dot t$ that can be withdrawn is just the equivalent to the Galilei
internal energy term $-H_0\dot t$ of (\ref{GaliL}). The Lagrangian with $H<0$ has as
nonrelativistic limit $-({m}/{2}){\dot{\bi r}^2}/{\dot t}$ which is not obtained in the Galilei
case.

The spin of this system, defined similarly as in the nonrelativistic
case,\index{spin}
 \begin{equation} 
{\bi S} \equiv {\bi J}-{\bi q}\times{\bi P}={\bi J}-\frac{c^2}{H}{\bi K}\times{\bi P}=0,
 \end{equation} 
vanishes, so that the relativistic point particle is also a spinless system.

\section{Relativistic spinning particles}
\index{relativistic particle}
\index{spinning particle!relativistic}
\label{sec:relspinningpar}

There are three maximal homogeneous spaces of ${\cal P}$, all of them at first
parameterized by the variables $(t,{\bi r},{\bi u},\brho)$, where the 
velocity variable ${\bi u}$ can be either $u<c$, $u=c$ or $u>c$. We shall call these
kinds of particles by the following names: The first one, since the motion of
the position of the charge ${\bi r}$
satisfies $u<c$, we call a {\bf Bradyon}, from the Greek term
$\beta\rho\alpha\delta\upsilon\varsigma\equiv$ slow. Bradyons are thus 
particles for which point ${\bi r}$ never reaches the speed of light.
The second class of particles $(u=c)$ will be called {\bf Luxons} because point ${\bi
r}$ is always moving at the speed of light for every observer, and finally those of the third group,
because $u>c$, are called {\bf Tachyons}, from the Greek 
$\tau\alpha\chi\upsilon\varsigma\equiv$ fast.

For the second class we use the Latin denomination Luxons in spite of
the Greek one of photons, because this class of particles will supply the
description not only of classical photons but also a classical model of
the electron. This class of models is very important and it has no
nonrelativistic limit. Therefore the models this manifold produce
have no nonrelativistic equivalent.

The first class corresponds to a kinematical space that is
the Poincar\'e group itself and produces models equivalent to the ones analyzed
in the non-relativistic case. To describe the classical electron and the photon
we shall consider next the case of luxons.

\section{Luxons}
 \index{luxons}
\index{relativistic particle!luxon}\label{sec:luxons}
Let us consider those mechanical systems whose kinematical space is 
the manifold $X$ generated by the variables $(t,{\bi r},{\bi 
u},\brho)$ with domains $t\in\RR,\ {\bi r}\in \RR^3,\ \brho\in \RR^3_c$ as in the 
previous case, and ${\bi u}\in \RR^3$ but now with $u=c$. Since $u=c$ we 
shall call this kind of particles {\bf Luxons}. This manifold is 
in fact a homogeneous space of the Poincar\'e group ${\cal P}$, and therefore,
according to our definition of elementary particle has to be considered as a 
possible candidate for describing the kinematical space of an elementary system. In fact, if we 
consider the point in this manifold
$x\equiv(0, 0,{\bi u},0)$, the little group that leaves $x$ invariant
is the one-parameter subgroup  ${\cal V}_u$ of 
pure Lorentz transformations in the direction of the vector ${\bi 
u}$. Then $X\sim{\cal P}/{\cal V}_u$, is a nine-dimensional homogeneous space. 

For this kind of systems the variables $t$, ${\bi r}$ transform according to (\ref{eq:5.1a}) and (\ref{eq:5.1b}), respectively
and the derivatives as in (\ref{eq:tpunto}) and (\ref{eq:rpunto}). For the velocity ${\bi u}$ the transformation
is obtained from (\ref{eq:boost}) and is
 \begin{equation}
{\bi u}'(\tau)=\frac{{R({\bmu}){\bi u}(\tau)+\gamma{\bi v}+\frac{\displaystyle{\gamma}^{2}}
{\displaystyle(1+\gamma)c^2}({\bi v}\cdot R({\bmu}){\bi u}(\tau)){\bi v}}}{{\gamma(1+{\bi v}\cdot R({\bmu})
\,{\bi u}(\tau)/c^2)}}.\label{eq:velo}
 \end{equation}
The general transformation of the orientation variables $\brho$ are obtained from (\ref{eq:rotation})
but now the functions ${\bi F}$ and $G$, which
involve some $\gamma(u)$ factors, become infinite and in the limit $u\to c$ they take the form
 \begin{equation} 
\brho'(\tau)=\frac{{\bmu+\brho(\tau)+\bmu\times\brho(\tau)+{\bi F}_c({\bi v},\bmu;{\bi u}(\tau),
\brho(\tau))}}{{1-\bmu\cdot\brho(\tau)+G_c({\bi v},\bmu;{\bi u}(\tau),\brho(\tau))}},
 \label{eq:5.2.1}
 \end{equation}
where the functions ${\bi F}_c$ and $G_c$ are given now by:
\begin{eqnarray}  
{\bi F}_c({\bi v},\bmu;{\bi u},\brho)&=&\frac{\gamma(v)}{(1+\gamma(v))c^2}\left[{\bi u}
\times{\bi v}+{\bi u}({\bi v}\cdot\bmu)+{\bi v}({\bi u}\cdot\brho)\right.\qquad\qquad\nonumber\\
&+&\,{\bi u}\times({\bi v}\times\bmu)+({\bi u}\times\brho)\times{\bi v}\ +({\bi u}\cdot\brho)({\bi v}\times\bmu)\nonumber\\
&+&\left.
({\bi u}\times\brho)({\bi v}\cdot\bmu)+({\bi u}\times\brho)\times({\bi v}\times\bmu)\right],
 \label{eq:F_c}
 \end{eqnarray}   
\begin{eqnarray}  
G_c({\bi v},\bmu;{\bi u},\brho)&=&\frac{\gamma(v)}{(1+\gamma(v))c^2}\left[{\bi u}\cdot{\bi v}+
{\bi u}\cdot({\bi v}\times\bmu)+{\bi v}\cdot({\bi u}\times\brho)\right.\qquad\nonumber\\
&-&\left.({\bi u}\cdot\brho)({\bi v}\cdot\bmu)+({\bi u}\times\brho)\cdot({\bi v}\times\bmu)\right].
 \label{eq:G_c}
 \end{eqnarray}   
Since $u'=u=c$, the absolute value of the velocity vector is conserved
and it means that ${\bi u}'$ can be obtained 
from ${\bi u}$ by an orthogonal transformation, so that the transformation equations of the
velocity under ${\cal P}$ can be expressed as:
 \begin{equation}
 {\bi u}'=R(\bphi){\bi u},
\label{eq:uRu}
 \end{equation}
where the kinematical rotation of parameter $\bphi$ is
 \begin{equation}
\bphi=\frac{{\bmu+{\bi F}_c({\bi v},\bmu;{\bi u}(\tau),0)}}{{1+G_c({\bi v},\bmu;{\bi u}(\tau), 0)}}.
 \label{eq:5.2.2}
 \end{equation} 
In this case there also exist among the kinematical variables the 
constraints ${\bi u}=\dot{\bi r}/\dot t$. 

Equation (\ref{eq:5.2.1}) also corresponds to 
\begin{equation}
R(\brho')=R(\bphi)R(\brho),
\label{eq:ropriro}
\end{equation} with 
the same $\bphi$ in both cases, as in (\ref{eq:5.2.2}). 

Since the variable $u(\tau)=c$, during the whole 
evolution, we can distinguish two different kinds of systems, because, 
by taking the derivative with respect to $\tau$ of this expression we 
get $\dot{\bi u}(\tau)\cdot{\bi u}(\tau)=0$, {\sl i.e.},  systems for which
$\dot{\bi u}=0$ or massless systems as we shall see, and systems where $\dot{\bi u}\neq 
0$ but always orthogonal to ${\bi u}$. These systems will correspond to 
massive particles whose charge internal motion occurs at the constant velocity 
$c$, although their center of mass moves with velocity below $c$.

\subsection{Massless particles. (The photon)}
 \index{photon}\index{relativistic particle!photon}\index{photon!classical} \index{classical!photon}
\label{sec:photon}

If $\dot{\bi u}=0$, ${\bi u}$ is constant and the system follows 
a straight trajectory with constant velocity, and therefore the 
kinematical variables reduce simply to $(t,{\bi r},\brho)$ with 
domains and physical meaning as usual as, time, position and orientation, respectively. 
The derivatives $\dot t$ and 
$\dot{\bi r}$ transform like (\ref{eq:tpunto}) and (\ref{eq:rpunto}) 
and instead of the variable $\dot{\brho}$ we shall consider the linear 
function $\bomega$ defined in (\ref{eq:4.1.5}) that transforms under ${\cal P}$:
 \begin{equation}  
\bomega'(\tau)=R(\bphi)\bomega(\tau),
 \label{eq:5.2.3}
 \end{equation}
where, again, $\bphi$ is given by (\ref{eq:5.2.2}). 

\begin{quotation}\footnotesize{
In fact, from (\ref{eq:ropriro}), since $\dot{\bi u}=0$,
taking the $\tau$-derivative, 
 \[ 
\dot R({\brho'})=R(\bphi)\dot R({\brho}),
\]
the antisymmetric matrix $\Omega=\dot R({\brho})R^{T}(\brho)$ has as essential
components the angular velocity $\bomega$, 
\begin{equation}
\Omega=\pmatrix{0&-{\omega}_z&{\omega}_y\cr
{\omega}_z&0&-{\omega}_x\cr
-{\omega}_y&{\omega}_x&0}.
\label{eq:omegia}\end{equation}
It transforms as
\[
\Omega'=\dot R({\brho'})R^{T}(\brho')=
R(\bphi)\dot R({\brho})R^{T}(\brho)R^{T}(\phi)
=R(\bphi)\Omega R^{T}(\phi),
 \] 
and this matrix transformation leads for its essential components to (\ref{eq:5.2.3}).}
\end{quotation}

For this system there are no constraints among the kinematical 
variables, and, since $\dot{\bi u}=0$, the general form of its 
Lagrangian is
 \begin{equation}
L=T\dot t+{\bi R}\cdot\dot{\bi r}+{\bi W}\cdot\bomega.
 \label{eq:5.2.4}
 \end{equation}  
Funtions $T=\partial L/\partial\dot t,\ R_i=\partial L/\partial\dot r^i,\ 
W_i=\partial L/\partial\omega^i$, will depend on the variables 
$(t,{\bi r},\brho)$ and are homogeneous functions of zero degree in terms 
of the derivatives of the kinematical variables $(\dot t,\dot{\bi r},\bomega)$. 
Since $\dot t\neq 0$ they will be expressed in terms of  ${\bi u}=\dot{\bi r}/\dot t$ and
${\bf\Omega}=\bomega/\dot t$, which are the true velocity and angular velocity of the 
particle respectively. 

Invariance of the Lagrangian under ${\cal P}$ leads to the following 
transformation form of these 
functions under the group ${\cal P}$: 
 \begin{equation}
T'=\gamma T-\gamma({\bi v}\cdot R(\bmu){\bi R}),
 \label{eq:5.2.5a}
 \end{equation}  
 \begin{equation}
{\bi R}'=R(\bmu){\bi R}-\gamma{\bi v}T/c^2+\frac{\gamma^2}{(1+\gamma)c^2}
({\bi v}\cdot R(\bmu){\bi R}){\bi v},
 \label{eq:5.2.5b}
 \end{equation}  
 \begin{equation}
{\bi W}'=R(\bphi){\bi W}.
 \label{eq:5.2.5c}
 \end{equation}  
They are translation invariant and therefore independent of 
$t$ and ${\bi r}$. They will be functions of only $(\brho,{\bi 
u},{\bf\Omega})$, with the constraint $u=c$. Invariance under 
rotations forbids the explicit dependence on $\brho$, so that the dependence of these
functions on $\brho$ and $\dot{\brho}$ variables is only through the angular velocity
$\bomega$.

Noether's theorem gives rise, as before, to the following constants of the motion:
 \begin{eqnarray}  
\hbox{Energy}\quad H&=&-T,\label{eq:5.2.6a}\\
\hbox{linear momentum}\quad {\bi P}&=&{\bi R},\label{eq:5.2.6b}\\
\hbox{kinematical momentum}\quad {\bi K}&=&H{\bi r}/c^2-{\bi P}\,t-{\bi W}\times{\bi u}/c^2,
\qquad\label{eq:5.2.6c}\\
\hbox{angular momentum}\quad {\bi J}&=&{\bi r}\times{\bi P}+{\bi W}.
\label{eq:5.2.6d}
 \end{eqnarray} 
In this case the system has no zitterbewegung because the Lagrangian does not depend 
on $\dot{\bi u}$ which vanishes. The particle, located at point ${\bi r}$, 
is moving in a straight trajectory at the speed of light
and therefore it is not possible to find an inertial rest frame observer. Although
we have no center of mass observer, we define
the spin by ${\bi S}={\bi J}-{\bi r}\times{\bi P}={\bi W}$.\index{spin}

If we take in (\ref{eq:5.2.6d}) the $\tau$-derivative we get $d{\bi 
S}/dt={\bi P}\times{\bi u}$. Since ${\bi P}$ and ${\bi u}$ are two 
non-vanishing constant vectors, then the spin has a constant time 
derivative. It represents a system with a continuously increasing 
angular momentum. This is not what we understand by an elementary 
particle except if this constant $d{\bi S}/dt=0$. Therefore for this
system the spin is a constant of the motion and ${\bi P}$ and ${\bi u}$
are collinear vectors.

Energy and linear momentum are in fact the components of a four-vector and with the spin 
they transform as
 \begin{equation}
H'=\gamma H+\gamma({\bi v}\cdot R(\bmu){\bi P}),
 \label{eq:5.2.5an}
 \end{equation}  
 \begin{equation}
{\bi P}'=R(\bmu){\bi P}+\gamma{\bi v}H/c^2+\frac{\gamma^2}{(1+\gamma)c^2}
({\bi v}\cdot R(\bmu){\bi P}){\bi v},
 \label{eq:5.2.5bn}
 \end{equation}  
 \begin{equation}
{\bi S}'=R(\bphi){\bi S}.
 \label{eq:5.2.5cn}
 \end{equation}  
The relation between ${\bi P}$ and ${\bi u}$ can be obtained from (\ref{eq:5.2.6c}), taking the 
$\tau$-derivative and the condition that the spin ${\bi W}$ is constant, 
$\dot{\bi K}=0=-H\dot{\bi r}/c^2+{\bi P}\dot t$, {\sl i.e.}, ${\bi P}=H{\bi 
u}/c^2$. If we take the scalar product of this expression with ${\bi u}$
we also get $H={\bi P}\cdot{\bi u}$.

Then, from (\ref{eq:5.2.5an}) and (\ref{eq:5.2.5bn}), an invariant and constant 
of the motion, which vanishes, is $(H/c)^2-{\bi P}^2$.  The mass of this system is zero.
It turns out that for this particle both $H$ and ${\bi P}$ are non-vanishing 
for every inertial observer. Otherwise, if one of them vanishes for a single observer
they vanish for all of them.
By (\ref{eq:5.2.5cn}), $S^2$ is another Poincar\'e invariant 
property of the system that is also a constant of the motion. 

The first part of the Lagrangian $T\dot t+{\bi R}\cdot\dot{\bi r}=-H\dot t+{\bi P}\cdot\dot{\bi r}$, 
which can be written as
$-(H-{\bi P}\cdot{\bi u})\dot t=0$, also vanishes. 
Then the Lagrangian is reduced to the third term ${\bi S}\cdot\bomega$. 

We see from (\ref{eq:uRu}) and (\ref{eq:5.2.5cn}) that the 
dimensionless magnitude $\epsilon={\bi S}\cdot{\bi u}/Sc$  is another 
invariant and constant of the motion, and we thus expect that the 
Lagrangian will be explicitly dependent on both constant parameters $S$ and $\epsilon$. 
Taking into account the transformation properties under ${\cal P}$ of ${\bi u}$, 
$\bomega$ and ${\bi S}$, given in (\ref{eq:uRu}), (\ref{eq:5.2.3}) and
(\ref{eq:5.2.5cn}) respectively, it turns out that the spin must necessarily 
be a vector function of ${\bi u}$ and $\bomega$. 

If the spin is not transversal, as it happens for real photons,
then ${\bi S}=\epsilon\, S{\bi u}/c$ where $\epsilon=\pm 1$, and thus 
the Lagrangian finally becomes:
 \begin{equation}
L=\left(\frac{\epsilon\,S}{c}\right)\,\frac{\dot{\bi r}\cdot\bomega}{\dot t}.
 \label{eq:5.2.7}
 \end{equation}  

From this Lagrangian we get that the energy is $H=-\partial 
L/\partial\dot t={\bi S}\cdot{\bf\Omega}$, where
${\bf\Omega}=\bomega/\dot{t}$ is the angular velocity of the particle. The linear momentum 
is ${\bi P}=\partial L/\partial\dot{\bi r}=\epsilon\, S\,{\bf\Omega}/c$, and, 
since ${\bi P}$ and ${\bi u}$ are parallel vectors, ${\bf\Omega}$ and 
${\bi u}$ must also be parallel, and if the energy is definite 
positive, then ${\bf\Omega}=\epsilon\Omega{\bi u}/c$.

This means that the energy $H=S\Omega$. For photons we know
that $S=\hbar$, and thus $H=\hbar\Omega=h\nu$. In this way the
frequency of a photon is the frequency of its rotational motion around 
the direction of its trajectory. We thus see that the spin and angular 
velocity for $H>0$ particles have the same direction, although they are not analytically 
related, because $S$ is invariant under ${\cal P}$ while $\Omega$ is not.

We say that the Lagrangian (\ref{eq:5.2.7}) represents a photon of spin 
$S$ and polarization $\epsilon$. A set of photons of this kind,
all with the same polarization, corresponds to circularly 
polarized light, as has been shown by direct measurement of the 
angular momentum carried by these photons.~\footnote{\hspace{0.1cm}R. A. Beth, {\sl 
Phys. Rev.} {\bf 50}, 115 (1936).} Left and right polarized photons 
correspond to $\epsilon=1$ and $\epsilon=-1$, respectively. Energy is 
related to the angular frequency $H=\hbar\Omega$, and linear momentum 
to the wave number ${\bi P}=\hbar{\bi k}$, that therefore  is 
related to the angular velocity vector by ${\bi k}=\epsilon{\bf\Omega}/c$. 
If it is possible to talk about the `wave-length' of a single photon this will be the
distance run by the particle during a complete turn.

\subsection{Massive particles. (The electron)} 
\index{electron}\index{relativistic particle!electron}\index{electron!classical} \index{classical!electron}
\label{sec:electron}

If we consider now the other possibility, $\dot{\bi u}\neq 0$ but 
orthogonal to ${\bi u}$, then variables $\dot{t}$ and $\dot{\bi r}$ 
transform as in the previous case (\ref{eq:tpunto}) and 
(\ref{eq:rpunto}), but for $\dot{\bi 
u}$ and $\bomega$ we have: 
 \begin{eqnarray}
\dot{\bi u}'&=&R(\bphi)\dot{\bi u}+\dot R(\bphi){\bi u},
 \label{eq:5.2.8a}\\
\bomega'&=&R(\bphi)\bomega+\bomega_\phi,
 \label{eq:5.2.8b}
 \end{eqnarray}  
where the rotation of parameter $\bphi$ is again given by (\ref{eq:5.2.2}) and vector $\bomega_\phi$ is:
 \begin{equation}
\bomega_\phi=\frac{{\gamma{R{\bi u}\times{\bi v}}-(\gamma-1){R({\bi u}
\times\dot{\bi u})}+2\gamma^2{({\bi v}\cdot R({\bi u}\times\dot{\bi u})){\bi v}}/
(1+\gamma)c^2}}{{\gamma(c^2+{\bi v}\cdot R{\bi u})}}.
 \label{eq:5.2.9}
 \end{equation}   
Expression (\ref{eq:5.2.8a}) is the $\tau$-derivative of (\ref{eq:uRu}) and 
can also be written in the form:
 \begin{equation}
\dot{\bi u}'=\frac{R(\bphi)\dot{\bi u}}{{\gamma(1+{\bi v}\cdot R(\bmu){\bi u}/c^2)}}.
 \label{eq:5.2.10}
 \end{equation}   
Expression (\ref{eq:5.2.8b}) comes from $R(\brho')=R(\bphi)R(\brho)$ and taking the $\tau$-derivative
of this expression $\dot{R}(\brho')=\dot{R}(\bphi)R(\brho)+R(\bphi)\dot{R}(\brho)$, because
parameter $\bphi$ depends on $\tau$ through the velocity ${\bi u}(\tau)$, and therefore
\[
\Omega'=\dot{R}(\brho'){R^T(\brho')}=R(\bphi)\Omega R^T(\bphi)+\dot{R}(\bphi)R^T(\bphi).
\]
$R(\bphi)\Omega R^T(\bphi)$ corresponds to $R(\bphi)\bomega$ and 
the antisymmetric matrix $\Omega_\phi=\dot{R}(\bphi)R^T(\bphi)$ has as 
essential components the $\bomega_\phi$
vector, {\sl i.e.}, equation (\ref{eq:5.2.9}). 

The homogeneity condition of the Lagrangian leads to the general form
 \begin{equation}
L=T\dot t+{\bi R}\cdot\dot{\bi r}+{\bi U}\cdot\dot{\bi u}+{\bi W}\cdot\bomega,
 \label{eq:5.2.11}
 \end{equation}  
where $T=\partial L/\partial\dot t,\ R_i=\partial L/\partial\dot r^i,\ 
U_i=\partial L/\partial\dot u^i$ and $W_i=\partial L/\partial\omega^i$, and 
Noether's theorem provides the following constants of the motion:
 \begin{eqnarray}
\hbox{Energy}\quad H&=&-T-(d{\bi U}/dt)\cdot{\bi u},\label{eq:5.2.12a}\\
\hbox{linear momentum}\quad {\bi P}&=&{\bi R}-(d{\bi U}/dt),\label{eq:5.2.12b}\\
\hbox{kinematical momentum}\quad {\bi K}&=&H{\bi r}/c^2-{\bi P}\,t-{\bi Z}\times{\bi u}/c^2,
\qquad\label{eq:5.2.12c}\\
\hbox{angular momentum}\quad{\bi J}&=&{\bi r}\times{\bi P}+{\bi Z}.\label{eq:5.2.12d}
 \end{eqnarray} 
In this case the function ${\bi Z}$ is defined as in the Galilei case, by
 \begin{equation}
{\bi Z}={\bi u}\times{\bi U}+{\bi W}.
 \label{eq:5.2.13}
 \end{equation}  

Expressions (\ref{eq:5.2.12a}, \ref{eq:5.2.12b}) imply that $H/c$ and ${\bi P}$ transform 
like the components of a four-vector, similarly as in 
(\ref{eq:transH}-\ref{eq:transP}), thus defining the invariant and constant of the 
motion $(H/c)^2-{\bi P}^2=m^2c^2$, in terms of the positive parameter $m$ which
is interpreted as the mass of the particle. 

Observable ${\bi Z}$ transforms as: 
 \begin{equation}
{\bi Z}'(\tau)=\gamma R(\bmu){\bi Z}(\tau)-\frac{\gamma^2}{(1+\gamma)c^2}({\bi v}\cdot 
R(\bmu){\bi Z}(\tau)){\bi v}+\frac{\gamma}{ c^2}({\bi v}\times 
R(\bmu)({\bi Z}(\tau)\times {\bi u})),
 \label{eq:5.2.14}
 \end{equation}   
an expression that corresponds to the transformation of an antisymmetric 
tensor $Z^{\mu\nu}$ with strict components $Z^{0i}=({\bi 
Z}\times{\bi u})^i/c$, and $Z^{ij}=\epsilon^{ijk}Z_k$.

By defining the relative position vector ${\bi k}={\bi Z}\times{\bi 
u}/H$, the kinematical momentum (\ref{eq:5.2.12c}) can be cast into the 
form
 \[
{\bi K}=H{\bi q}/c^2-{\bi P}\,t,
 \]
where ${\bi q}={\bi r}-{\bi k}$, represents the position of the center 
of mass of the particle.

The spin is defined as usual\index{spin}
 \begin{equation} 
{\bi S}={\bi J}-{\bi q}\times{\bi P}={\bi J}-\frac{c^2}{H}{\bi K}\times{\bi P},
 \label{eq:electspin}
 \end{equation} 
and is a constant of the motion. It takes the form
 \begin{equation} 
{\bi S}={\bi Z}+{\bi k}\times{\bi P}={\bi Z}+\frac{1}{H}({\bi Z}\times{\bi u})\times{\bi P}.
 \label{eq:electspin2}
 \end{equation} 

The helicity ${\bi S}\cdot{\bi P}={\bi Z}\cdot{\bi P}={\bi J}\cdot{\bi P}$, is also a 
constant of the motion. We can construct the constant Pauli-Lubanski
four-vector\index{Pauli-Lubanski four-vector}
 \begin{equation}  
w^\mu\equiv({\bi P}\cdot{\bi S},H{\bi S}/c), 
 \label{eq:PauliLub}
 \end{equation}  
with $-w^\mu w_\mu=m^2c^2S^2$, in terms of the invariant properties 
$m$ and $S$ of the particle.

If we take in (\ref{eq:5.2.12c}) the $\tau$-derivative and the 
scalar product with the velocity ${\bi u}$ we get the Poincar\'e 
invariant relation:
 \begin{equation}
H={\bi P}\cdot{\bi u}+\frac{1}{c^2}{\bi Z}\cdot\left(\frac{d{\bi u}}{ dt}\times{\bi u}\right).
 \label{eq:5.2.15}
 \end{equation}   

This will give rise to Dirac's Hamiltonian,\index{Dirac's Hamiltonian} $H=c{\bi P}\cdot\balpha+\beta 
mc^2$ when expressed in the quantum case, in terms of the $\balpha$ and $\beta$ Dirac
matrices. Since $c\balpha$ is usually interpreted as the local 
velocity operator ${\bi u}$ of the electron,~\footnote{\hspace{0.1cm}J.J. Sakurai, {\sl 
Advanced Quantum Mechanics}, Addison-Wesley Reading, MA (1967).} we have 
$H={\bi P}\cdot{\bi u}+\beta mc^2$ and this relation suggests the
identification
 \[
\beta=\frac{1}{mc^4}{\bi Z}\cdot\left(\frac{d{\bi u}}{dt}\times{\bi u}\right).
 \]
Here all magnitudes on the right-hand side are measured in the center 
of mass frame. We shall come back to this relation after quantization 
of this system.

The center of mass observer is defined by the conditions ${\bi 
P}={\bi K}=0$. For this observer ${\bi Z}={\bi S}$ is constant, $H=mc^2$ and 
thus from (\ref{eq:5.2.12c}) we get
 \begin{equation}
{\bi r}=\frac{1}{mc^2}\,{\bi S}\times{\bi u},
\label{eq:elecdina}
 \end{equation}
and the internal motion takes place in a plane 
orthogonal to the constant spin ${\bi S}$. 
The scalar product with ${\bi u}$ leads to ${\bi r}\cdot d{\bi r}/dt=0$, 
and thus the zitterbewegung radius is a constant. Taking the 
time derivative of both sides of (\ref{eq:elecdina}), we obtain $mc^2{\bi u}=({\bi S}\times d{\bi u}/dt)$, 
because the spin is constant in this frame, we get that ${\bi u}$ and ${\bi S}$ 
are orthogonal and therefore
 \begin{equation}
{\bi S}=m{\bi u}\times{\bi r}.
\label{eq:Smuxk}
 \end{equation}
Since $S$ and $u=c$ are constant, the motion is a circle of radius $R_0=S/mc$. 
For the electron we take $S=\hbar/2$, and the radius is 
$\hbar/2m_ec=1.93\times 10^{-13}$ m., half the Compton wave length of 
the electron. The frequency of this motion in the C.M. frame is 
$\nu=2m_ec^2/h=2.47\times 10^{20}$ s$^{-1}$, and 
$\omega=2\pi\nu=1.55\times 10^{21}$ rad s$^{-1}$. The ratio of this 
radius to the so-called classical radius 
$R_{cl}=e^2/8\pi\varepsilon_0m_ec^2=1.409\times 10^{-15}$ m, is 
precisely $R_{cl}/R_0=e^2/2\varepsilon_0hc=1/136.97=\alpha$, the fine 
structure constant. 

Motions of this sort, in which the particle is moving at the speed of light,
can be found in early literature, but the
distinction between the motion of center of charge and center of mass
is not sufficiently clarified.~\footnote{\hspace{0.1cm}M. Mathisson,
{\sl Acta Phys. Pol.} {\bf 6}, 163 (1937); {\bf 6}, 218 (1937)}${^,}$
\footnote{\hspace{0.1cm}M.H.L. Weyssenhof,
{\sl Acta Phys. Pol.} {\bf 9}, 46 (1947).
M.H.L. Weyssenhof and A. Raabe, 
{\sl Acta Phys. Pol.} {\bf 9}, 7 (1947); {\bf 9}, 19 (1947).}

Nevertheless,
in the model we are analyzing, the idea that the electron has a size of the order
of the zitterbewegung radius is a plausible macroscopic vision but is not
necessary to maintain any longer, because the only important point from the dynamical
point of view is the center of charge position, whose motion completely
determines the dynamics of the system. In this form, elementary particles, the kind of objects we are
describing, look like extended objects. Nevertheless, although some kind of related length can be defined,
they are dealt with as point particles with orientation because 
the physical attributes are all located at the single point ${\bi r}$.
The dynamics of equation (\ref{eq:elecdina}) can be represented in figure\ref{fig:elec} where we have separated
the two contributions to the total spin ${\bi Z}\equiv{\bi S}={\bi S}_u+{\bi S}_\omega$, related respectively to the orbital and rotational motion.
\cfigl{fig:elec}{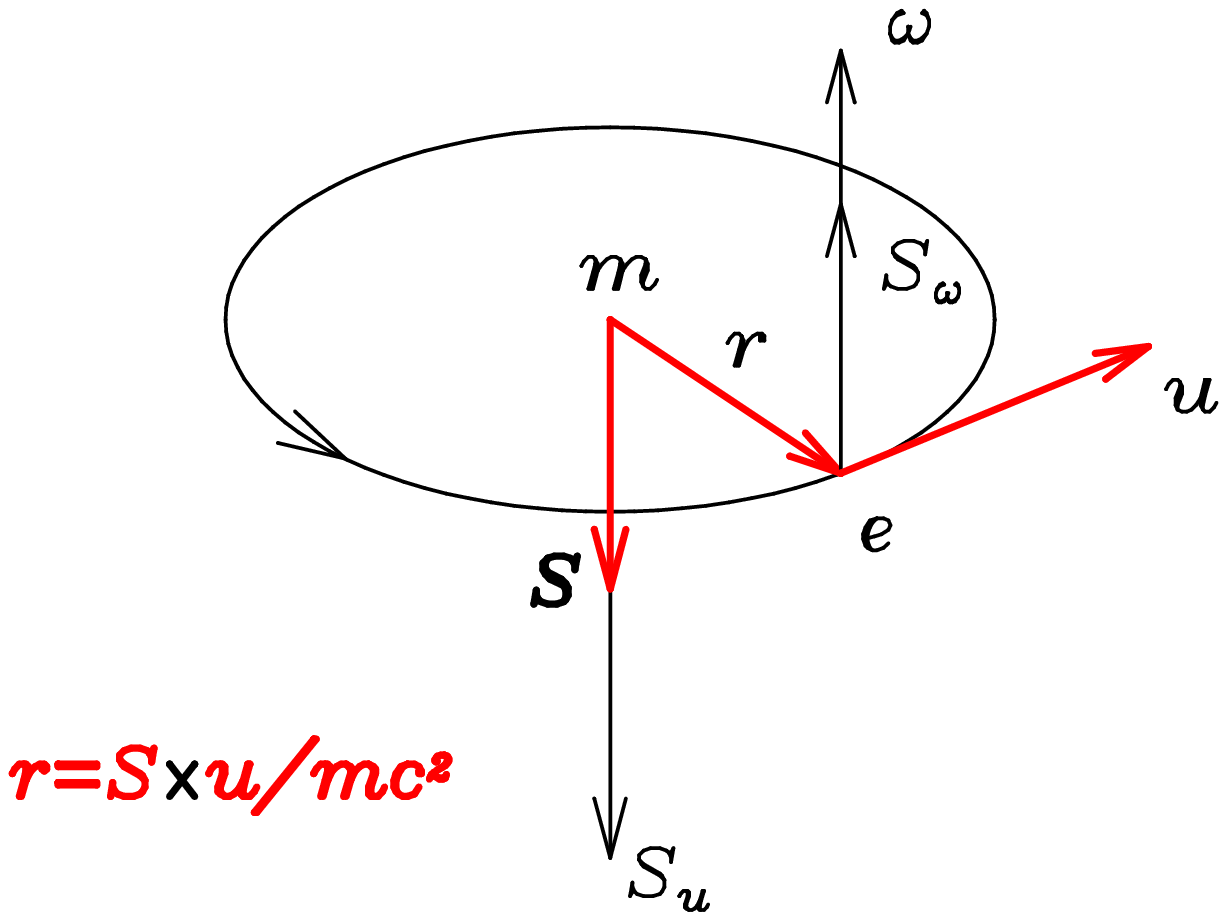}{Motion of the charge of the electron in the center of mass frame.}

The transformation equation for the function ${\bi Z}$, 
(\ref{eq:5.2.14}) can also be written as
 \begin{equation}
{\bi Z}'=\gamma(1+{\bi v}\cdot R(\bmu){\bi u}/c^2) R(\bphi){\bi Z},
 \label{eq:5.2.16}
 \end{equation}  
and therefore ${\bi Z}\cdot\dot{\bi u}={\bi Z}'\cdot\dot{\bi u}'$ and
${\bi Z}'\cdot{\bi u}'=\gamma(1+{\bi v}\cdot R(\bmu){\bi u}/c^2){\bi 
Z}\cdot{\bi u}$. Since it is orthogonal to ${\bi u}$ and $\dot{\bi 
u}$, for the center of mass observer, it is also orthogonal to ${\bi 
u}$ and $\dot{\bi u}$ for any other inertial observer. 

\begin{quotation}\footnotesize{
An alternative method of verifying this is to take the time derivative 
in (\ref{eq:5.2.12c}) and (\ref{eq:5.2.12d}), and thus
 \[ 
H{\bi u}-c^2{\bi P}-\frac{d{\bi Z}}{ dt}\times{\bi u}-{\bi Z}\times\frac{d{\bi u}}{ dt}=0, 
 \]
 \[ 
\frac{d{\bi Z}}{ dt}={\bi P}\times{\bi u},
 \] 
{\sl i.e.},
 \[ {\bi 
Z}\times\frac{d{\bi u}}{ dt}=(H-{\bi u}\cdot{\bi P}){\bi u}.
 \] 
and a final scalar product with ${\bi Z}$, leads to $(H-{\bi u}\cdot{\bi 
P}){\bi u}\cdot{\bi Z}=0$. The first factor does not vanish since the 
invariant $H^2/c^2-P^2=m^2c^2$ is positive definite and if $H={\bi 
u}\cdot{\bi P}$, then $({\bi u}\cdot{\bi P})^2/c^2-P^2$ with $u\le c$ 
is always negative, then ${\bi Z}\cdot{\bi u}=0$. If we take the time 
derivative of this last expression, with the condition that $d{\bi Z}/dt$ is orthogonal to 
${\bi u}$, we obtain ${\bi Z}\cdot\dot{\bi u}=0$. The observable ${\bi Z}$ has always 
the direction of the non-vanishing vector $\dot{\bi u}\times{\bi u}$ for
positive energy particles and the opposite direction for particles of negative energy.}
\end{quotation}

Equation (\ref{eq:5.2.15}) can be recast into the form
 \[
 \frac{H}{ c}c\dot t-{\bi P}\cdot\dot {\bi r}-\frac{1}{ c^2}{\bi Z}\cdot
(\dot {\bi u}\times {\bi u})=0,
 \] 
where the first two terms give rise to the invariant term $P_\mu\dot x^\mu=mc^2\dot t_{cm}$, 
and the third to the invariant relation
 \begin{equation}
{\bi Z}\cdot(\dot {\bi u}\times{\bi u})=mc^4\dot t_{cm}.
 \end{equation}
Here $t_{cm}$ is the time observable measured in the center of mass frame,
and the right-hand side, which is positive definite for particles,
implies that ${\bi Z}$ has precisely 
the direction of $\dot{\bi u}\times{\bi u}$. In the case of
antiparticles it has the opposite direction.

We see that the particle has mass and spin, and the center of charge moves 
in circles at the speed of light in a plane orthogonal to the spin, 
for the center of mass observer. All these features are independent of 
the particular Lagrangian of the type (\ref{eq:5.2.11}) we can 
consider. All that remains is to describe the evolution of the 
orientation and therefore its angular velocity. 
The analysis developed until now is compatible with many different
possibilities for the angular velocity. The behaviour of the angular velocity depends on the
particular model we work with.

To end this section and with the above model of the electron in mind, 
it is convenient to remember some of the features that Dirac~\footnote{\hspace{0.1cm}P.A.M. Dirac, 
{\sl The Principles of Quantum 
mechanics}, Oxford Univ. Press, 4th ed. Oxford (1967).} obtained for the motion of a free electron. Let 
point ${\bi r}$ be the position vector on which Dirac's spinor 
$\psi(t,{\bi r})$ is defined. When computing the velocity of point 
${\bi r}$, Dirac arrives at: 

a) The velocity ${\bi u}=i/\hbar[H,{\bi r}]=c\balpha$, is expressed in terms of 
$\balpha$ matrices and writes, {\sl 
`$\ldots$ a measurement of a component of the velocity of a free 
electron is certain to lead to the result $\pm c$'}. 

b) The linear momentum does not have the direction of this velocity 
${\bi u}$, but must be related to some average value of it: ${\ldots}$ 
{\sl `the $x_1$ component of the velocity, $c\alpha_1$, consists of 
two parts, a constant part $c^2p_1H^{-1}$, connected with the momentum 
by the classical relativistic formula, and an oscillatory part, whose 
frequency is at least $2mc^2/h$, ${\ldots}$'}. 

c) About the position ${\bi r}$: {\sl `The oscillatory part of $x_1$ is 
small, ${\ldots}$ , which is of order of magnitude $\hbar/mc$, 
${\ldots}$'}.

And when analyzing, in his original 1928 paper,~\footnote{\hspace{0.1cm}P.A.M. Dirac, 
{\sl Proc. Roy. Soc. Lon.} {\bf A117}, 610 (1928).} the 
interaction of the electron with an external electromagnetic field, 
after performing the square of Dirac's operator, he obtains two new 
interaction terms: 
 \begin{equation}
{e\hbar\over 2mc}{\bf\Sigma}\cdot{\bi B}+{ie\hbar\over 2mc}\balpha\cdot{\bi E},
 \label{eq:D8}
 \end{equation}
where the electron spin is written as ${\bi S}=\hbar{\bf\Sigma}/2$ and 
 \[ 
{\bf\Sigma}=\pmatrix{\bsigma&0\cr 0&\bsigma\cr},
 \] 
in terms of $\sigma$-Pauli matrices and ${\bi E}$ 
and ${\bi B}$ are the external electric and magnetic fields, 
respectively. He says, {\sl `The electron will therefore behave as 
though it has a magnetic moment $(e\hbar/2mc)\,{\bf\Sigma}$ and an 
electric moment $(ie\hbar/2mc)\,\balpha$. The magnetic moment
\index{magnetic moment}\index{moment!magnetic dipole}
is just that assumed in the spinning electron model' }({\rm Pauli model}). `{\sl The 
electric moment, being a pure imaginary, we should not expect to 
appear in the model.'} \index{moment!electric dipole}

However, if we look at our classical model, we see that for 
the center of mass observer, there is a 
non-vanishing electric and magnetic dipole moment \index{magnetic moment}\index{electric dipole} 
 \begin{equation} 
{\bi d}=e{\bi k}={e\over mc^2}{\bi S}\times{\bi u},\quad\bmu=
{e\over2}{\bi k}\times\frac{d{\bi k}}{dt}=-{e\over2m}{\bi Y},
 \label{eq:D9}
 \end{equation} 
where ${\bi S}$ is the total spin and ${\bi Y}=-m{\bi k}\times d{\bi k}/dt$
is the zitterbewegung part of spin.
The time average value of ${\bi d}$ is zero, and the average value of 
$\bmu$ is the constant vector $\bmu$. 

This classical model gives rise to the same kinematical prediction as 
the nonrelativistic model described in Sec.\ref{sec:galispinmag}. 
If the charge of the particle is negative, the current of 
Fig.\ref{fig:elec} produces a magnetic moment that necessarily has 
the same direction as the spin. If the electron spin and magnetic 
moments are antiparallel, then we need another contribution to the 
total spin, different from the zitterbewegung. All real experiments to 
determine very accurately the gyromagnetic ratio are based on the 
determination of precession frequencies, but these precession 
frequencies are independent of the spin orientation. However, the 
difficulty to separate electrons in a Stern-Gerlach type experiment, 
suggests to perform polarization experiments in order to determine in 
a direct way whether spin and magnetic moment for elementary particles are either parallel or 
antiparallel. We have suggested a couple of plausible experiments to determine the relative
orientation between the spin and magnetic moment of free electrons and also for electrons in the outer
shell of atoms\footnote{M. Rivas, {\sl Are the electron spin parallel or antiparallel vectors?,
ArXiv:physics/0112057}.}

Another consequence of the classical model is that it enhances the 
role of the so-called
 \index{minimal coupling}
minimal coupling interaction $j_\mu A^\mu$. 
The magnetic properties of the electron are produced by the current of 
its internal motion and not by some possible distribution of magnetic 
dipoles, so that the only possible interaction of a point charge at ${\bi r}$
with the external electromagnetic field is that of the current $j^\mu$, associated
to the motion of point ${\bi r}$, with the external potentials.

\section{The dynamical equation of the spinning electron}
We have seen that for relativistic particles with $u=c$ and ${\bi u}$ and $\dot{\bi u}$
orthogonal vectors, the position vector ${\bi r}$ moves in circles according to the dynamical equation
(\ref{eq:elecdina}) in the center of mass frame, as depicted in figure \ref{fig:elec}. But this solution
is independent of the particular Lagrangian we choose as an invariant function of the kinematical variables
and their derivatives, which accomplish with this orthogonality ${\bi u}\cdot\dot{\bi u}=0$, requirement.
We are going to analyze this dynamical equation for any arbitrary inertial observer.

Let us consider the trajectory ${\bi r}(t)$, $t\in[t_1,t_2]$ followed by a point of a system for an 
arbitrary inertial observer $O$. Any other inertial observer $O'$ 
is related to the previous one by a transformation
of a kinematical group such that their relative space-time measurements of any space-time event are given by
\[
t'=T(t,{\bi r}; g_1,\ldots,g_r),\quad {\bi r}'={\bi R}(t,{\bi r}; g_1,\ldots,g_r),
\]
where the functions $T$ and ${\bi R}$ define the action of the kinematical group $G$, 
of parameters $(g_1,\ldots,g_r)$, on space-time. Then the description of the trajectory of that point 
for observer $O'$ is obtained from
\[
t'(t)=T(t,{\bi r}(t); g_1,\ldots,g_r),\quad {\bi r}'(t)={\bi R}(t,{\bi r}(t); g_1,\ldots,g_r),\quad \forall t\in[t_1,t_2].
\]
If we eliminate $t$ as a function of $t'$ from the first equation and substitute into the second 
we shall get
\begin{equation}
{\bi r}'(t')={\bi r}'(t'; g_1,\ldots,g_r).
 \label{eq:rdet}
 \end{equation}
Since observer $O'$ is arbitrary, equation (\ref{eq:rdet}) represents the complete set of 
trajectories of the point for all inertial observers. 
Elimination of the $r$ group parameters among the function ${\bi r}'(t')$
and their time derivatives will give us the differential equation satisfied by the trajectory of the point. 
This differential equation is invariant by construction because it is independent
of the group parameters and therefore independent of the inertial observer.
If $G$ is either the Galilei or Poincar\'e group
it is a ten-parameter group so that we have to work out in general up to the fourth derivative 
to obtain sufficient equations to eliminate the ten group parameters. 
Therefore the order of the differential equation is dictated by the
number of parameters and the structure of the kinematical group.

\subsection{The relativistic spinning electron}

Let us assume the above electron model. Since the charge is moving at the speed
of light for the center of mass observer $O^*$ it is moving 
at this speed for every other inertial observer $O$. 
Now, the relationship of space-time measurements between the center of mass observer and 
any arbitrary inertial observer is given by:
 \begin{eqnarray*}
t(t^*;g)&=&
\gamma\left(t^*+{\bi v}\cdot R(\balpha){\bi r}^*(t^*)\right)+b,\\
{\bi r}(t^*;g)&=&
R(\balpha){\bi r}^*(t^*)+\gamma{\bi v}t^*+\frac{\gamma^2}{1+\gamma}
\left({\bi v}\cdot R(\balpha){\bi r}^*(t^*)\right){\bi v}+{\bi a}.
 \end{eqnarray*}
With the shorthand notation for the following expressions:
\[
{\bi K}(t^*)=R(\balpha){\bi r}^*(t^*),
\quad {\bi V}(t^*)=R(\balpha)\frac{d{\bi r}^*(t^*)}{dt^*}=
\frac{d{\bi K}}{dt^*},\quad
\frac{d{\bi V}}{dt^*}=-{\bi K},
\]
\[
B(t^*)={\bi v}\cdot{\bi K},\quad
A(t^*)={\bi v}\cdot{\bi V}=\frac{dB}{dt^*},
\quad\frac{dA}{dt^*}=-B 
\]
we obtain
\begin{eqnarray}
{\bi r}^{(1)}&=&\frac{1}{\gamma(1+A)}\left({\bi V}+\frac{\gamma}{1+\gamma}
(1+\gamma+\gamma A){\bi v}\right),\label{eq:r1}\\
{\bi r}^{(2)}&=&\frac{1}{\gamma^2(1+A)^3}\left(-(1+A){\bi K}+B{\bi V}
+\frac{\gamma}{1+\gamma}\,B{\bi v}\right),\label{eq:r2}
\end{eqnarray}
\[
{\bi r}^{(3)}=\frac{1}{\gamma^3(1+A)^5}\left(-3B(1+A){\bi K}-(1+A-3B^2){\bi V}+\right.\]
\begin{equation}
\qquad\left.\frac{\gamma}{1+\gamma}\,(A(1+A)+3B^2){\bi v}\right)\label{eq:r3}
\end{equation}
\[
{\bi r}^{(4)}=\frac{1}{\gamma^4(1+A)^7}\left((1+A)(1-2A-3A^2-15B^2){\bi K}-\right.\]
\[
\qquad B(7+4A-3A^2-15B^2){\bi V}-\]
\begin{equation}
\qquad\left.\frac{\gamma}{1+\gamma}\,(1-8A-9A^2-15B^2)B{\bi v}\right).\label{eq:r4}
\end{equation}
From this we get
 \begin{eqnarray}
\left({\bi r}^{(1)}\cdot{\bi r}^{(1)}\right)^2&=&1,\quad\left({\bi r}^{(1)}\cdot{\bi r}^{(2)}\right)=0,
\label{eq:inv1}\\
 \left({\bi r}^{(2)}\cdot{\bi r}^{(2)}\right)&=&-\left({\bi r}^{(1)}\cdot{\bi r}^{(3)}\right)=\frac{1}{\gamma^4(1+A)^4},
\label{eq:inv2}\\
\left({\bi r}^{(2)}\cdot{\bi r}^{(3)}\right)&=&-\frac{1}{3}\left({\bi r}^{(1)}\cdot{\bi r}^{(4)}\right)=\frac{2B}{\gamma^5(1+A)^6},
\label{eq:inv3}\\
\left({\bi r}^{(3)}\cdot{\bi r}^{(3)}\right)&=&\frac{1}{\gamma^6(1+A)^8}\left(1-A^2+3B^2\right),
\label{eq:inv4}\\
\left({\bi r}^{(2)}\cdot{\bi r}^{(4)}\right)&=&\frac{1}{\gamma^6(1+A)^8}\left(-1+2A+3A^2+9B^2\right),
\label{eq:inv5}\\
\left({\bi r}^{(3)}\cdot{\bi r}^{(4)}\right)&=&\frac{1}{\gamma^7(1+A)^{10}}\left(1+A+3B^2\right)4B.
\label{eq:inv6}
 \end{eqnarray}
From equations (\ref{eq:inv2})-(\ref{eq:inv4}) we can express the magnitudes $A$, $B$ and $\gamma$ in terms of these
scalar products between the different time derivatives $({\bi r}^{(i)}\cdot{\bi r}^{(j)})$. The constraint
that the velocity is 1 implies that all these and further scalar products for higher derivatives can be expressed
in terms of only three of them.
If the three equations (\ref{eq:r1})-(\ref{eq:r3})
are solved in terms of the unknowns ${\bi v}$, ${\bi V}$ and ${\bi K}$ and substituded into (\ref{eq:r4}),
we obtain the differential equation satisfied by the charge position, for any arbitrary inertial observer
\[
{\bi r}^{(4)}-\frac{3({\bi r}^{(2)}\cdot{\bi r}^{(3)})}{({\bi r}^{(2)}\cdot{\bi r}^{(2)})}\,{\bi r}^{(3)}+\]
\begin{equation}
\qquad\left(\frac{2({\bi r}^{(3)}\cdot{\bi r}^{(3)})}{({\bi r}^{(2)}\cdot{\bi r}^{(2)})}-
\frac{3({\bi r}^{(2)}\cdot{\bi r}^{(3)})^2}{4({\bi r}^{(2)}\cdot{\bi r}^{(2)})^2}-({\bi r}^{(2)}\cdot{\bi r}^{(2)})^{1/2}\right){\bi r}^{(2)}=0.\label{eq:elecbuena}
\end{equation}
It is a fourth order ordinary differential equation which contains as solutions 
motions at the speed of light. In fact, if $({\bi r}^{(1)}\cdot{\bi r}^{(1)})=1$, then by derivation we have $({\bi r}^{(1)}\cdot{\bi r}^{(2)})=0$
and the next derivative leads to  $({\bi r}^{(2)}\cdot{\bi r}^{(2)})+({\bi r}^{(1)}\cdot{\bi r}^{(3)})=0$. If we take this into account
and make the scalar product of (\ref{eq:elecbuena}) with ${\bi r}^{(1)}$, we get $({\bi r}^{(1)}\cdot{\bi r}^{(4)})+3({\bi r}^{(2)}\cdot{\bi r}^{(3)})=0$,
which is another relationship between the derivatives as a consequence of $|{\bi r}^{(1)}|=1$.
It corresponds to a helical motion since the term in the first derivative ${\bi r}^{(1)}$ is lacking.

\subsection{The center of mass}

The center of mass position is defined by
\begin{equation}
{\bi q}={\bi r}+\frac{2({\bi r}^{(2)}\cdot{\bi r}^{(2)})\,{\bi r}^{(2)}}{({\bi r}^{(2)}\cdot{\bi r}^{(2)})^{3/2}+({\bi r}^{(3)}\cdot{\bi r}^{(3)})-
\frac{\displaystyle{3({\bi r}^{(2)}\cdot{\bi r}^{(3)})^2}}{\displaystyle{4({\bi r}^{(2)}\cdot{\bi r}^{(2)})}}}.
 \label{eq:cmq}
 \end{equation}
We can check that both ${\bi q}$ and ${\bi q}^{(1)}$ vanish for the center of mass observer. 
Then, the fourth order dynamical equation for the position of the 
charge can also be rewritten here
as a system of two second order differential equations for the positions ${\bi q}$ and ${\bi r}$
\begin{equation}
{\bi q}^{(2)}=0,\quad {\bi r}^{(2)}=
\frac{1-{\bi q}^{(1)}\cdot{\bi r}^{(1)}}{({\bi q}-{\bi r})^2}\left({\bi q}-{\bi r}\right),
\label{eq:q2r2}\end{equation}
a free motion for the center of mass and a kind of central motion for the 
charge around the center of mass.

For the non-relativistic electron we get in the low velocity case ${\bi q}^{(1)}\to0$
and $|{\bi q}-{\bi r}|=1$, the equations of the Galilei case 
\begin{equation}
{\bi q}^{(2)}=0,\quad {\bi r}^{(2)}=
{\bi q}-{\bi r}.
\label{eq:q321}\end{equation}
a free motion for the center of mass and a harmonic motion
around ${\bi q}$ for the position of the charge. 

\subsection{Interaction with some external field}

The free equation for the center of mass motion ${\bi q}^{(2)}=0$, represents the conservation
of the linear momentum $d{\bi P}/dt=0$. But the linear momentum is 
written in terms of the center of mass velocity as
${\bi P}=m\gamma(q^{(1)}){\bi q}^{(1)}$, so that the free dynamical 
equations (\ref{eq:q2r2}) in the presence
of an external field should be replaced by
\begin{equation}
{\bi P}^{(1)}={\bi F},\quad {\bi r}^{(2)}=
\frac{1-{\bi q}^{(1)}\cdot{\bi r}^{(1)}}{({\bi q}-{\bi r})^2}\left({\bi q}-{\bi r}\right),
\label{eq:Pr2}\end{equation}
where ${\bi F}$ is the external force and the second equation is left unchanged because we consider, 
even with interaction, the same definition of the center of mass position. 
\[
\frac{d{\bi P}}{dt}=m\gamma(q^{(1)}){\bi q}^{(2)}+m\gamma(q^{(1)})^3({\bi q}^{(1)}\cdot{\bi q}^{(2)}){\bi q}^{(1)}
\]
we get
\[
m\gamma(q^{(1)})^3({\bi q}^{(1)}\cdot{\bi q}^{(2)})={\bi F}\cdot{\bi q}^{(1)}
\]
and by leaving the highest derivative ${\bi q}^{(2)}$ on the left hand side we finally get the differential
equations which describe the evolution of a relativistic spinning electron in the presence of an 
external electromagnetic field:
\begin{eqnarray}
m{\bi q}^{(2)}&=&\frac{e}{\gamma(q^{(1)})}\left[{\bi E}+{\bi r}^{(1)}\times{\bi B}-{\bi q}^{(1)}\left(\left[{\bi E}
+{\bi r}^{(1)}\times{\bi B}\right]\cdot{\bi q}^{(1)}\right)\right],\label{eq:q2}\\
{\bi r}^{(2)}&=&\frac{1-{\bi q}^{(1)}\cdot{\bi r}^{(1)}}{({\bi q}-{\bi r})^2}\left({\bi q}-{\bi r}\right).\label{eq:r2}
 \end{eqnarray}

\section{Appendix: Galilei group}
 \index{Galilei group}\index{kinematical group!Galilei group} \index{group!Galilei group}
\label{sec:galigroup}

The Galilei group is a group of space-time transformations characterized by ten 
parameters $g\equiv(b,{\bi a},{\bi v},{\balpha})$. The action of $g$ on a 
space-time point $x\equiv(t,{\bi r})$ is given by $x'=gx$, and is considered in the form
\[
x'=\exp(bH)\exp({\bi a}\cdot{\bi P})\exp({\bi v}\cdot{\bi K})\exp(\balpha\cdot{\bi J})\,x
\]
as the action of a rotation followed by a pure Galilei transformation and finally a space 
and time translation. In this way all parameters that define each one-parameter subgroup
are normal, because the exponential mapping works. Explicitly
 \begin{eqnarray} 
t'&=&t+b,\label{tgali}\\ 
{\bi r}'&=&R({\balpha}){\bi r}+{\bi v}t+{\bi a},
\label{rgali}
 \end{eqnarray} 
and the composition law of the group $g''=g'g$ is: 
\begin{eqnarray} 
b''&=&b'+b,\label{eq:ttr1}\\
{\bi a}''&=&R({\balpha'}){\bi a}+{\bi v}'b+{\bi a}',\\
{\bi v}''&=&R({\balpha'}){\bi v}+{\bi v}',\\
R({\balpha''})&=&R({\balpha'})R({\balpha}).
\label{eq:n(3.10.2)} 
 \end{eqnarray} 

For rotations we shall alternatively use two different 
parametrizations. One is the normal or canonical parametrization in 
terms of a three vector $\balpha=\alpha{\bi n}$, where ${\bi n}$ is a 
unit vector along the rotation axis, and $\alpha\in[0,\pi]$ is the 
clockwise rotation angle in radians, when looking along ${\bi n}$. 
Another, in terms of a three vector $\bmu={\bi n}\tan(\alpha/2)$, 
which is more suitable to represent algebraically the composition of 
rotations.

The rotation matrix $R(\balpha)=\exp(\balpha\cdot{\bi J})$ is expressed in terms of the normal
parameters $\alpha_i$ and in terms of the antisymmetric matrix generators $J_i$ which have the
usual matrix representation
\[
J_1=\pmatrix{0&0&0\cr 0&0&-1\cr 0&1&0\cr},\;\;
J_2=\pmatrix{0&0&1\cr 0&0&0\cr -1&0&0\cr},\;\;
J_3=\pmatrix{0&-1&0\cr 1&0&0\cr 0&0&0\cr},
\]
and satisfy the commutation relations $[J_i,J_k]=\epsilon_{ikl}J_l$, such that if we 
write the normal parameters $\balpha=\alpha{\bi n}$ in terms of the rotation angle $\alpha$
and the unit vector ${\bi n}$ along the rotation axis, it is written as
\begin{equation}
R(\balpha)_{ij}=\delta_{ij}\cos\alpha+n_in_j(1-\cos\alpha)-\epsilon_{ijk}n_k\sin\alpha,
\quad i,j,k=1,2,3.
\label{eq:rotalfa}
\end{equation}
In the parametrization $\bmu={\bi n}\tan(\alpha/2)$, the rotation matrix is
\begin{equation}
R(\bmu)_{ij}=\frac{1}{1+\mu^2}\left((1-\mu^2)\delta_{ij}+2\mu_i\mu_j-2\epsilon_{ijk}\mu_k\right),
\quad i,j,k=1,2,3.
\label{eq:rotmu}
\end{equation}
In terms of these variables, $R(\bmu'')=R(\bmu')R(\bmu)$ is equivalent to
\begin{equation}
{\bmu''}=\frac{{\bmu'}+{\bmu}+{\bmu'}\times{\bmu}}{1-{\bmu'}\cdot{\bmu}}.
\label{eq:composirot}
\end{equation}
This can be seen in a simple manner by using the homomorphism between the rotation
group and the group $SU(2)$. The matrix generators of $SU(2)$ are ${\bi J}=-i\bsigma/2$ 
in terms of Pauli matrices $\bsigma$. In the normal parametrization the rotation matrix 
$\exp({\balpha\cdot{\bi J}})=\exp(-i\balpha\cdot\bsigma/2)$ is written in the form\index{group!$SU(2)$ group}\index{group!rotation group}\index{group!$SO(3)$ group}
\[
R(\balpha)=\cos(\alpha/2)\ID-i({\bi n}\cdot\bsigma)\sin(\alpha/2).
\]
By defining $\bmu={\bi n}\tan(\alpha/2)$, this rotation matrix is expressed as
\begin{equation}
R(\bmu)=\frac{1}{\sqrt{1+\mu^2}}\left(\ID-i\bmu\cdot\bsigma\right),
\end{equation}
where $\ID$ is the $2\times2$ unit matrix and in this form we can get the composition law 
(\ref{eq:composirot}).~\footnote{\hspace{0.1cm}D. Hestenes, {\sl Space-time algebra}, Gordon and Breach, NY (1966).}

\begin{quotation}\footnotesize{
If the rotation is of value $\pi$, then eqs. (\ref{eq:rotalfa}) or
(\ref{eq:rotmu}) lead to 
\[
R({\bi n},\pi)_{ij}=-\delta_{ij}+2n_in_j.
\]
Even if the two rotations $R(\bmu)$ and $R(\bmu')$ involved in (\ref{eq:composirot}) are of value $\pi$, although
$\tan(\pi/2)=\infty$, this expression is defined and gives:
\[
{\bi n}''\tan(\alpha''/2)=\frac{{\bi n}\times{\bi n}'}{{\bi n}\cdot{\bi n}'}.
\]
The absolute value of this relation leads to
$\tan(\alpha''/2)=\tan\theta$, {\sl i.e.}, $\alpha''=2\theta$, where $\theta$ is
the angle between the two unit vectors ${\bi n}$ and ${\bi n}'$. We obtain the known
result that every rotation of value $\alpha$ around an axis ${\bi n}$ can be obtained as the
composition of two rotations of value $\pi$ around two axes orthogonal to ${\bi n}$
and separated by an angle $\alpha/2$.}
\end{quotation}

For the orientation variables we shall use throughout the book the early Greek
variables $\balpha,\bbeta,\ldots$ whenever we consider the normal
paramet\-rization, while for the $\tan(\alpha/2)$ parametrization we will
express rotations in terms of the intermediate Greek variables $\bmu,\bnu,\brho,\ldots\;$. In this last notation,
transformation equations (\ref{eq:ttr1}-\ref{eq:n(3.10.2)}) should be replaced by 
\begin{eqnarray} 
b''&=&b'+b,\label{eq:ttr1n}\\
{\bi a}''&=&R({\bmu'}){\bi a}+{\bi v}'b+{\bi a}',\\
{\bi v}''&=&R({\bmu'}){\bi v}+{\bi v}',\\
\bmu''&=&\frac{{\bmu'}+{\bmu}+{\bmu'}\times{\bmu}}{1-{\bmu'}\cdot{\bmu}}.
\label{eq:n(3.10.2n)} 
 \end{eqnarray} 

The neutral element of the Galilei group is $(0,{\bf 0},{\bf 0},{\bf 0})$ and the inverse of every 
element is
 \[ 
(b,{\bi a},{\bi v},{\balpha})^{-1}=(-b,-R(-{\balpha})({\bi a}-b{\bi v}),-R(-{\balpha}){\bi v},-{\balpha}).
 \] 
The generators of the group in the realization (\ref{tgali}, \ref{rgali}) are
the differential operators
 \begin{equation} 
H=\partial/\partial t,\quad P_i=\partial/\partial r_i,\quad 
K_i=t\partial/\partial r_i,\quad J_k=\varepsilon_{kli}r_l\partial/\partial r_i
 \end{equation} 
and the commutation rules of the Galilei Lie algebra are
 \begin{equation} 
[{\bi J},{\bi J}]=-{\bi J}, \quad [{\bi J},{\bi P}]=-{\bi P}, \quad [{\bi J},{\bi K}]=-
{\bi K}, \quad [{\bi J},H]=0,
\label{eq:commgali1}
 \end{equation} 
 \begin{equation} 
[H,{\bi P}]=0, \; [H,{\bi K}]={\bi P}, \; [{\bi P},{\bi P}]=0, \;
[{\bi K},{\bi K}]=0, \; [{\bi K},{\bi P}]=0.
\label{eq:commgali2}
  \end{equation} 
All throughout this book, except when explicitly stated, we shall use the following
shorthand notation for commutators of scalar and 3-vector operators, that as usual,
are represented by bold face characters:
\begin{eqnarray*}
{[{\bi A},{\bi B}]}&=& {\bi C},\quad\Longrightarrow \quad [A_i,B_j]=\epsilon_{ijk}C_k,\\ 
{[{\bi A},{\bi B}]}&=& C,\quad\Longrightarrow \quad [A_i,B_j]=\delta_{ij}C,\\ 
{[{\bi A}, B]}&=& {\bi C},\quad\Longrightarrow \quad [A_i,B]=C_i,\\ 
{[B,{\bi A}]}&=& {\bi C},\quad\Longrightarrow \quad [B,A_i]=C_i,
\end{eqnarray*}
where $\delta_{ij}=\delta_{ji}$ is Kronecker's delta and $\epsilon_{ijk}$ is the completely
antisymmetric symbol, so that Latin indexes match on both sides of commutators.

The group action (\ref{tgali})-(\ref{rgali}) represents the relationship between 
the coordinates $(t,{\bi r})$ of a space-time event as measured by the 
inertial observer $O$ and the corresponding coordinates $(t',{\bi 
r}')$ of the same space-time event as measured by another inertial observer 
$O'$. The ten group parameters have the following meaning. If we 
consider the event $(0,{\bi 0})$ measured by $O$, for instance the 
flashing of a light beam from its origin at time $t=0$, it takes the values 
$(b,{\bi a})$ in $O'$, where $b$ is the time parameter that represents the 
time translation and ${\bi a}$ is the space translation. The parameter 
${\bi v}$ of dimensions of velocity represents the velocity of the 
origin of the Cartesian frame of $O$ as measured by $O'$, and finally  
the parameters $\balpha$, or $R({\balpha})$, represent the orientation of the 
Cartesian frame of $O$  as measured by $O'$. In a certain sense the 
ten parameters $(b,{\bi a},{\bi v},{\balpha})$ with dimensions 
respectively of time, position, velocity and orientation describe
the relative motion of the Cartesian frame of $O$ by $O'$.

The Galilei group has non-trivial exponents given by~\footnote{\hspace{0.1cm}V. Bargmann, {\sl Ann. Math.} {\bf 5}, 1 
(1954).}
  \index{Galilei group!exponent}
 \begin{equation} 
\xi(g,g')=m\left(\frac{1}{2}{\bi v}^2b'+{\bi v}\cdot R(\balpha){\bi a}'\right).
\label{exponent}
 \end{equation} 
They are characterized by the non-vanishing parameter $m$.

The central extension of the Galilei group~\footnote{\hspace{0.1cm}J.M. Levy-Leblond, 
{\sl Galilei Group and 
Galilean Invariance}, in E.M. Loebl, {\sl Group Theory and its 
applications}, Acad. Press, NY (1971), vol. 2, p.~221. } 
is an 11-parameter 
 \index{Galilei group!central extension}
group with an additional generator $I$ which commutes with the other ten,
\begin{equation} 
[I,{H}]=[I,{\bi P}]=[I,{\bi K}]=[I,{\bi J}]=0,
\end{equation} 
and the remaining 
commutation relations are the same as above 
(\ref{eq:commgali1}, \ref{eq:commgali2}), except the last one which appears as
 \begin{equation} 
[K_i,P_j]=-m\delta_{ij}I,\quad \hbox{\rm or}\quad [{\bi K},{\bi P}]=-mI,
\label{commkp}
 \end{equation}  
using our shorthand notation, in terms of a non-vanishing parameter $m$. If we define the 
following polynomial operators on the group algebra
 \begin{equation}  
{\bi W}=I{\bi J}-\frac{1}{m}{\bi K}\times{\bi P},\quad U=IH-\frac{1}{2m}{\bi P}^2,
 \end{equation}  
$U$ commutes with all generators of the extended Galilei group and ${\bi W}$ 
satisfies the commutation relations:
 \index{Galilei!extended group}
 \[ 
[{\bi W},{\bi W}]=-I{\bi W},\quad [{\bi J},{\bi W}]=-{\bi W},\quad [{\bi W},{\bi P}]=[{\bi 
W},{\bi K}]=[{\bi W},H]=0,
 \] 
so that ${\bi W}^2$ also commutes with all generators. It turns out that
the extended Galilei group has three functionally independent Casimir operators which, in those 
representations in which the operator $I$ becomes the unit operator, 
for instance in irreducible representations, 
are interpreted as the mass, $M=mI$, the internal energy $H_0=H-
P^2/2m$, and the absolute value of spin
 \index{Casimir operator} \index{Galilei group!Casimir operators}\index{Casimir operators!extended Galilei group}
 \index{internal energy}
 \index{spin}
 \begin{equation}  
S^2=\left({\bi J}-\frac{1}{m}{\bi K}\times{\bi P}\right)^2. \label{spin2}
 \end{equation}  
The spin operator ${\bi S}$ in those representations in which $I=\ID$, 
satisfy the commutation relations:
 \[ 
[{\bi S},{\bi S}]=-{\bi S},\quad [{\bi J},{\bi S}]=-{\bi S},\quad [{\bi S},{\bi P}]=
[{\bi S},H]=[{\bi S},{\bi K}]=0,
 \] 
{\sl i.e.}, it is an angular momentum operator, transforms 
like a vector under rotations and is invariant under space and time 
translations and under Galilei boosts, respectively. It reduces to the 
total angular momentum operator ${\bi J}$ in those frames in which ${\bi P}={\bi K}=0$.
 \index{Galilei!boost}

\section{Appendix: Poincar\'e group}
 \index{Poincar\'e group}\index{kinematical group!Poincar\'e group} \index{group!Poincar\'e group}
\label{sec:Poingroup}

The Poincar\'e group is the group of transformations of 
Minkowski's space-time that leave invariant
the separation between any two close space-time events $ds^2=\eta_{\mu\nu}dx^\mu dx^\nu$. We shall consider
the contravariant components $x^\mu\equiv(ct,{\bi r})$, and $x'=gx$ is expressed as
${x'}^{\mu}={\Lambda^{\mu}}_\nu\,x^\nu+a^{\mu}$, in terms of a constant matrix $\Lambda$ and
a constant translation four-vector $a^\mu\equiv(cb,{\bi a})$. We take for the covariant components
of Minkowski's metric tensor $\eta_{\mu\nu}\equiv$ diag$(1,-1,-1,-1)$. 
Then $d{x'}^{\mu}={\Lambda^{\mu}}_\nu dx^\nu$ and $ds^2=\eta_{\mu\nu}d{x'}^{\mu}d{x'}^{\nu}=
\eta_{\sigma\rho}dx^\sigma dx^\rho$ implies for the matrix $\Lambda$
\begin{equation}
\eta_{\mu\nu}{\Lambda^{\mu}}_\sigma{\Lambda^{\nu}}_\rho=\eta_{\sigma\rho}.
\label{eq:LambdaLor}
\end{equation}
Relations (\ref{eq:LambdaLor}) represent ten conditions among the 16 components of the
matrix $\Lambda$, so that each matrix depends on six essential parameters, which
can be chosen in many ways. Throughout this book we shall take three of them as 
the components of the relative velocity ${\bi v}$ between inertial observers and the remaining three 
as the orientation $\balpha$ of their Cartesian frames, expressed in a suitable
parametrization of the rotation group.

Therefore, every element of the Poincar\'e group ${\cal P}$ will 
be represented, as in the previous 
case of the Galilei group, by the ten parameters $g \equiv (b,{\bi a},{\bi v},{\balpha})$ and 
the group action on a space-time point $x\equiv(t,{\bi r})$ will be 
interpreted in the same way, {\sl i.e.}, $x'=gx$: 
 \begin{equation} 
x'=\exp(bH)\exp({\bi a}\cdot{\bi P})\exp({\bbeta}\cdot{\bi K})\exp({\balpha}\cdot{\bi J}) x,
\label{eq:groupPoinx}
 \end{equation} 
as the action of a rotation followed by a boost or pure Lorentz transformation 
and finally a space and time translation. 
It is explicitly given on the space-time variables by
 \begin{eqnarray}  
t'&=&\gamma t+\gamma({\bi v}\cdot R({\bmu}){\bi r})/c^2+b,\label{eq:3.11.1}\\ 
{\bi r}'&=&R({\bmu}){\bi r}+\gamma{\bi v}t+\gamma^2({\bi v}\cdot R({\bmu})
{\bi r}){\bi v}/(1+\gamma)c^2+{\bi a}. \label{eq:3.11.2}
 \end{eqnarray}  
Parameter $\bbeta$ in (\ref{eq:groupPoinx}) is the normal parameter for the pure Lorentz 
transformations, that in terms of the relative velocity among 
observers ${\bi v}$ is expressed as ${\bbeta}/\beta\tanh\beta={\bi 
v}/c$ as we shall see below. The dimensions and domains of the parameters $b$, 
${\bi a}$ and ${\bmu}$ are the same as those of the Galilei group, and 
the parameter ${\bi v}\in\RR^3$, with the upper bound $v<c$, has also 
dimensions of velocity. The physical meaning of these ten parameters, 
that relate any two inertial observers, is the same as in the Galilei 
case. The parameter ${\bi v}$ is the velocity of observer $O$, as 
measured by $O'$, and $R(\bmu)$ represents the orientation of 
Cartesian frame $O$ relative to $O'$, once $O'$ is boosted with velocity 
${\bi v}$. The factor $\gamma(v)=(1-v^2/c^2)^{-1/2}$.

The composition law of the group is obtained from $x''=\Lambda' x'+a'=\Lambda'(\Lambda x+a)+a'$ 
that by identification with $x''=\Lambda''x+a''$ reduces to $\Lambda''=\Lambda'\Lambda$ and $a''=\Lambda' a+a'$, 
{\sl i.e.}, the composition law of the Lorentz transformations, that we will find in the next Section 
\ref{sec:Lorgroup}, and a Poincar\'e transformation $(\Lambda',a')$ of 
the four-vector $a^\mu$. 
In this parametrization $g''=g'g$, is:~\footnote{\hspace{0.1cm}M.Rivas, M.Valle and 
J.M.Aguirregabiria, {\sl Eur. J. Phys.} {\bf 6}, 128 (1986).}  
 \begin{eqnarray}  
b^{''}&=&\gamma'b+\gamma'({\bi v}'\cdot R({\bmu'}){\bi a})/c^2+b',\label{eq:poinb}\\
{\bi a}^{''}&=&R({\bmu'}){\bi a}+\gamma'{\bi v}'b+\frac{{\gamma}^{'2}}
{(1+\gamma')c^2}({\bi v}'\cdot R({\bmu'}){\bi a}){\bi v}'+{\bi a}',\label{eq:poina}\\
{\bi v}^{''}&=&\frac{{R({\bmu'}){\bi v}+\gamma'{\bi v}'+\frac{\displaystyle{\gamma}^{'2}}
{\displaystyle(1+\gamma')c^2}({\bi v}'\cdot R({\bmu'}){\bi v}){\bi v}'}}{{\gamma'(1+{\bi v}'\cdot R({\bmu'})
\,{\bi v}/c^2)}},\label{eq:boost}\\
{\bmu^{''}}& = &\frac{{{\bmu'}+{\bmu}+{\bmu'}\times{\bmu}+{\bi F}
({\bi v}',{\bmu'},{\bi v},{\bmu})}}{{1-{\bmu'}\cdot{\bmu}+G({\bi v}',{\bmu'},{\bi v},{\bmu})}},
 \label{eq:rotation} 
 \end{eqnarray} 
where ${\bi F}({\bi v}',{\bmu'},{\bi v},{\bmu})$ and $G({\bi v}',{\bmu'},{\bi v},{\bmu})$ 
are the real analytic functions: 
 \begin{eqnarray}
{\bi F}({\bi v}',{\bmu'},{\bi v},{\bmu})&=&\frac{\gamma\gamma'}{{(1+\gamma)(1+\gamma')c^2}}
{\left[ {\bi v}\times {\bi v}'+{\bi v} ({\bi v}'\cdot{\bmu'})+{\bi v}'({\bi v}\cdot{\bmu}) \right.}\quad\nonumber\\
&+&{\bi v}\times({\bi v}'\times{\bmu}')+({\bi v}\times{\bmu})\times {\bi v}'+({\bi v}\cdot{\bmu})({\bi v}'\times{\bmu'})
\nonumber\\
&+&\left.({\bi v}\times{\bmu})({\bi v}'\cdot{\bmu'})+({\bi v}\times{\bmu})\times({\bi v}'\times{\bmu'})
 \right],\qquad
 \label{eq:3.11.3a}
 \end{eqnarray} 
 \begin{eqnarray}
G({\bi v}',{\bmu'},{\bi v},{\bmu})&=&\frac{\gamma\gamma'}{(1+\gamma)(1+\gamma')c^2}{\left[
{\bi v}\cdot {\bi v}'+{\bi v}\cdot({\bi v}'\times{\bmu'})+{\bi v}'\cdot({\bi v}\times{\bmu})
\right.}\nonumber\\
&-&\left.({\bi v}\cdot{\bmu})({\bi v}'\cdot{\bmu'})+({\bi v}\times{\bmu})\cdot
({\bi v}'\times{\bmu'}) \right]. 
 \label{eq:3.11.3b}
 \end{eqnarray} 

The unit element of the group is $(0,{\bf 0},{\bf 0},{\bf 0})$ and the inverse of any arbitrary 
element $(b,{\bi a},{\bi v},{\bmu})$ is 
 \[
(-\gamma b+\gamma {\bi v}\cdot{\bi a}/c^2,-R(-{\bmu})({\bi a}-\gamma {\bi 
v}b+\frac{{\gamma}^2}{{(1+\gamma)c^2}}({\bi v}\cdot{\bi a}){\bi v}),-R(-{\bmu}){\bi v},-{\bmu}).
 \]

The group generators in the realization (\ref{eq:3.11.1}, 
\ref{eq:3.11.2}), and in terms of the 
normal parameters $(b,{\bi a},\bbeta,\balpha)$, are
 \[
H=\partial/\partial t,\; P_i=\partial/\partial r_i,\;
K_i=ct\partial/\partial r_i+(r_i/c)\partial/\partial t,\;
J_k={\varepsilon_{kl}}^ir_l\partial/\partial r_i.
 \]
Thus, ${\bi K}$ and ${\bi J}$ are dimensionless and the commutation 
relations become
 \begin{equation}
[{\bi J},{\bi J}]=-{\bi J}, \; [{\bi J},{\bi P}]=-{\bi P}, \; [{\bi J},{\bi K}]=-
{\bi K}, \; [{\bi J},H]=0,\;[H,{\bi P}]=0,
 \label{eq:commpoin1}
 \end{equation}
 \begin{equation}
[H,{\bi K}]=c{\bi P}, \; [{\bi P},{\bi P}]=0, \; [{\bi K},
{\bi K}]={\bi J}, \; [{\bi K},{\bi P}]=-H/c.
 \label{eq:commpoin2}
 \end{equation}

If, as usual, we call $x^0=ct$, $P_0=H/c$, and $K_i=J_{0i}=-
J_{i0}$ and $J_k=\frac{1}{2}\epsilon_{klr}J_{lr}$, $x_\mu=\eta_{\mu\nu}x^\nu$, 
$\mu=0,1,2,3$ and $\partial_\nu\equiv\partial/\partial x^\nu$, then,                  
 \[
P_\mu=\partial_\mu,\quad J_{\mu\nu}=-J_{\nu\mu}=x_\mu\partial_\nu-x_\nu\partial_\mu.
 \] 
In covariant notation the commutation relations appear: 
 \begin{eqnarray}  
[P_\mu,P_\nu]&=&0,\nonumber\cr
[J_{\mu\nu},P_\sigma]&=&-\eta_{\mu\sigma}P_\nu+ \eta_{\nu\sigma}P_\mu,\nonumber\cr
[J_{\mu\nu},J_{\rho\sigma}]&=&-\eta_{\mu\rho}J_{\nu\sigma}-\eta_{\nu\sigma}J_{\mu\rho}+
\eta_{\nu\rho}J_{\mu\sigma}+\eta_{\mu\sigma}J_{\nu\rho}.\nonumber
 \end{eqnarray}  

The Poincar\'e group has two functionally independent Casimir 
invariants. One is interpreted as the squared mass of the system, 
\index{Casimir!invariant}
 \begin{equation} 
P^\mu P_\mu=(H/c)^2-{\bi P}^2=m^2c^2,
 \end{equation} 
and the other is the square of the Pauli-Lubanski four-vector $W^\mu$. 
The Pauli-Lubanski four-vector
\index{Pauli-Lubanski four-vector} is defined as
 \begin{equation} 
W^\mu=\frac{1}{2}\varepsilon^{\mu\nu\sigma\lambda}\,P_\nu J_{\sigma\lambda}\equiv({\bi P}\cdot{\bi J},H{\bi J}/c-{\bi K}\times{\bi P})\equiv
({\bi P}\cdot{\bi S},H{\bi S}/c),
 \label{eq:cas1}
 \end{equation} 
which is by construction orthogonal to $P_\mu$, {\sl i.e.}, $W^\mu P_\mu=0$.

It is related to the spin of the system ${\bi S}$, defined 
through the relation 
 \begin{equation}
H{\bi S}/c=H{\bi J}/c-{\bi K}\times{\bi P},
\label{eq:spinPauliLub}
 \end{equation}
so that its time component $W^0={\bi P}\cdot{\bi S}={\bi P}\cdot{\bi J}$
is the helicity of the particle, and the spatial part is the vector (\ref{eq:spinPauliLub}).

The other Casimir operator is thus\index{Casimir operators!Poincar\'e group}
 \begin{equation} 
W^\mu W_\mu=({\bi P}\cdot{\bi J})^2-(H{\bi J}/c-{\bi K}\times{\bi P})^2=-m^2c^2S^2,
 \label{eq:cas2}
 \end{equation} 
where it depends on $S^2$, the absolute value squared of the spin.
We see in the relativistic case that the two parameters $m$ and $S$
characterize the two Casimir invariants and therefore they are the
intrinsic properties of the elementary particle the formalism provides.
In the quantum case, since the representation must be irreducible
$S^2=s(s+1)\hbar^2$, for any $s=0,1/2,1,\ldots$, depending on the value
of the quantized spin of the particle, but in the classical case $S^2$ can take any continuous value.

These $W^\mu$ operators satisfy the commutation relations:
 \begin{equation} 
[W^\mu,W^\nu]=\epsilon^{\mu\nu\sigma\rho}W_\sigma P_\rho,
 \end{equation} 
where we take $\epsilon^{0123}=+1$, and
 \begin{equation} 
[P^\mu,W^\nu]=0,\qquad [M_{\mu\nu},W_\sigma]=-\eta_{\mu\sigma}W_\nu+ \eta_{\nu\sigma}W_\mu.
 \end{equation}   
The Poincar\'e group has no non-trivial exponents, so that gauge functions when
restricted to homogeneous spaces of ${\cal P}$ vanish. 

\subsection{Lorentz group}
 \index{Lorentz group}\index{group!Lorentz group}
\label{sec:Lorgroup}

The Lorentz group ${\cal L}$ is the subgroup of transformations of the form 
 \index{Lorentz group}
$(0,{\bi 0},{\bi  v},{\bmu})$, and every Lorentz transformation 
$\Lambda({\bi v},{\bmu})$ will be interpreted as $\Lambda({\bi 
v},{\bmu})=L({\bi v})R({\bmu})$, as mentioned before where $L({\bi v})$ is a boost or
pure Lorentz transformation and $R({\bmu})$ a spatial rotation. 
Expressions (\ref{eq:boost}, \ref{eq:rotation}) come from $\Lambda({\bi 
v}'',{\bmu''})=\Lambda({\bi v}',{\bmu}')\Lambda({\bi v},{\bmu})$. 
Expression (\ref{eq:boost}) is the relativistic composition of velocities
since 
 \begin{eqnarray*} 
L({\bi v}'')R({\bmu''})&=&L({\bi v}')R({\bmu'})L({\bi 
v})R({\bmu})
\\
&=&L({\bi v}')R({\bmu'})L({\bi v})R(-{\bmu'})R({\bmu'})R({\bmu}),
 \end{eqnarray*} 
but the conjugate of the boost $R({\bmu'})L({\bi v})R(-{\bmu'})=L(R({\bmu'}){\bi v})$ is another boost
and thus
 \[ 
L({\bi v}'')R({\bmu''})=L({\bi v}')L(R({\bmu'}){\bi v})R({\bmu'})R({\bmu}).
 \]
The product $L({\bi v}')L(R({\bmu'}){\bi v})=L({\bi v}'')R({\bi w})$ where
${\bi v}''$ is the relativistic composition of the velocities ${\bi 
v}'$ and $R({\bmu'}){\bi v}$, and $R({\bi w})$ is the Thomas-Wigner rotation 
 \index{Thomas-Wigner rotation}
associated to the boosts $L({\bi v}')$ and $L(R({\bmu'}){\bi v})$. 

Therefore, expression (\ref{eq:boost}) is equivalent to 
\begin{equation} 
L({\bi v}'')=L({\bi v}')L(R({\bmu'}){\bi v})R(-{\bi w}), 
\end{equation}
and (\ref{eq:rotation}) is
\begin{equation}
R({\bmu''})=R({\bi w})R({\bmu'})R({\bmu})\equiv R({\bphi})R({\bmu}).
\end{equation}
The Thomas-Wigner rotation matrix $R({\bi w})$ is: 
 \[ 
R({\bi w})_{ij}=\delta_{ij}+\frac{1}{1+\gamma''}\left(\frac{{\gamma'}^2}{ c^2}\left(\frac{1-\gamma}{1+\gamma'}\right)v'_iv'_j+
\frac{\gamma^2}{c^2}\left(\frac{1-\gamma'}{1+\gamma}\right)R'_{ik}v_kR'_{jl}v_l\; \right. 
 \]
 \[ 
\left.+\frac{\gamma'\gamma}{ 
c^2} (v'_iR'_{jk}v_k-v'_jR'_{ik}v_k)+
\frac{2{\gamma'}^2\gamma^2(v'_kR'_{kl}v_l)}{(1+\gamma')(1+\gamma)c^2}
v'_iR'_{jk}v_k \right),
 \] 
and the factor 
 \[ 
\gamma''=\gamma'\gamma\left(1+\frac{{\bi v}'\cdot R(\bmu){\bi v}}{ c^2}\right).
 \] 
Matrix $R({\bi w})$ is written in terms of the vector parameter ${\bi w}$, which is a
function of ${\bi v}'$, $\bmu'$ and ${\bi v}$, given by
\begin{equation}
{\bi w}=\frac{{\bi F}({\bi v}',{\bf 0},R(\bmu'){\bi v},{\bf 0})}
{1+G({\bi v}',{\bf 0},R(\bmu'){\bi v},{\bf 0})},
\label{eq:WigThomw}
\end{equation}
and the parameter $\bphi$, such that $R(\bphi)=R({\bi w})R({\bmu'})$ is
\begin{equation}
\bphi=\frac{\bmu'+{\bi F}({\bi v}',\bmu',{\bi v},{\bf 0})}{1+G({\bi v}',\bmu',{\bi v},{\bf 0})}.
\label{eq:WigThom}
\end{equation}
If any one of the two velocities ${\bi v}$ or ${\bi v}'$ vanishes, $R({\bi w})_{ij}=\delta_{ij}$. 

The composition law is obtained by the homomorphism between the \index{group!$SL(2,\CC)$ group}
Lorentz group ${\cal L}$ and the group $SL(2,\CC)$ of $2\times2$ complex matrices of determinant $+1$. 
The Lie algebra of this group has as generators ${\bi J}=-i\bsigma/2$ and ${\bi K}=\bsigma/2$, where
$\sigma_i$ are Pauli spin matrices. A rotation of 
angle $\alpha$ around a rotation
axis given by the unit vector ${\bi n}$ is 
given by the $2\times2$ unitary matrix $\exp(\balpha\cdot{\bi J})$,
 \begin{equation}  
R(\balpha)=\cos(\alpha/2)\sigma_0- i{\bi n}\cdot\bsigma\sin(\alpha/2).
 \end{equation} 
In terms of the vector $\bmu=\tan(\alpha/2){\bi n}$,
 \begin{equation} 
R(\bmu)=\frac{1}{\sqrt{1+\mu^2}}\big(\sigma_0-i\bmu\cdot\bsigma\big),
 \label{eq:3.13.1}
 \end{equation} 
where $\sigma_0$ is the $2\times2$ unit matrix.
 A pure Lorentz transformation of normal parameters
$\beta_i$ is represented by the hermitian matrix $\exp(\bbeta\cdot{\bi K})$. 
This matrix is:
 \begin{equation} 
L(\bbeta)=\cosh(\beta/2)\sigma_0+\frac{\bsigma\cdot\bbeta}{\beta}\sinh(\beta/2).
 \end{equation}  
In terms of the relative velocity parameters, taking into account the functions 
$\cosh\beta=\gamma(v)$, $\sinh\beta=\gamma v/c$ and the trigonometric relations 
$\cosh(\beta/2)=\sqrt{(\cosh\beta+1)/2}$ 
and $\tanh(\beta/2)=\sinh\beta/(1+\cosh\beta)$, the matrix can be written as
 \begin{equation} 
L({\bi v})=\sqrt{\frac{1+\gamma}{2}}\left(\sigma_0+\frac{\gamma}{1+\gamma} 
\frac{\bsigma\cdot{\bi v}}{ c}\right).
 \label{eq:3.13.2}
 \end{equation} 

Then, every element of $SL(2,\CC)$ is parametrized by the six real 
numbers $({\bi v},\bmu)$, and interpreted as
 \begin{equation} 
A({\bi v},\bmu)=L({\bi v})R(\bmu).
 \label{eq:3.13.3}
 \end{equation} 

We thus see that every $2\times2$ matrix $A\in SL(2,\CC)$ can be 
written in terms of a complex four-vector $a^\mu$ and the four Pauli 
matrices $\sigma_\mu$. As $A=a^\mu\sigma_\mu$, and det$A=1$ leads to 
$a^\mu a_\mu=1$ or $(a^0)^2-{\bi a}^2=1$. 
The general form of (\ref{eq:3.13.3}) is
 \begin{equation} 
A({\bi v},\bmu)=\sqrt{\frac{1+\gamma}{2(1+\mu^2)}}\left[\sigma_0\left(1-
i\frac{\bmu\cdot{\bi u}}{1+\gamma}\right)+\bsigma\cdot\left(\frac{{\bi 
u}+{\bi u}\times\bmu}{1+\gamma}-i\bmu\right)\right],
 \label{eq:3.13.4}
 \end{equation} 
here the dimensionless vector ${\bi u}=\gamma(v){\bi v}/c$.

Conversely, since ${\rm Tr}\;(\sigma_\mu\sigma_\nu)=2\delta_{\mu\nu}$, we obtain
 $a^\mu=(1/2){\rm Tr}\;(A\sigma_\mu)$. If we 
express (\ref{eq:3.13.4}) in the form $A({\bi v},\bmu)=a^\mu\sigma_\mu$ 
we can determine $\bmu$ and ${\bi v}$, and thus ${\bi u}$, from the components of the complex four-vector
 $a^\mu$ as:
 \begin{eqnarray} 
\bmu&=&-\frac{{\rm Im}\,({\bi a})}{ {\rm Re}\,(a^0)}, \label{eq:bmu}\\
{\bi u}&=&2\left[{\rm Re}\,(a^0) {\rm Re}\,({\bi a})+{\rm Im}\,(a^0) {\rm Im}\,({\bi 
a})+{\rm Re}\,({\bi a})\times{\rm Im}\,({\bi a})\right],\qquad
\label{eq:bu}
 \end{eqnarray} 
where ${\rm Re}\,(a^\mu)$ and ${\rm Im}\,(a^\mu)$ are the real and imaginary parts
of the corresponding components of the four-vector $a^\mu$. When ${ {\rm Re}\,(a^0)}=0$
expression (\ref{eq:bmu}) is defined and represents a rotation of value $\pi$ 
along the axis in the direction of vector ${{\rm Im}\;({\bi a})}$.

If we represent every Lorentz transformation in terms of a rotation and a boost, {\sl i.e.},
in the reverse order, $\Lambda({\bi v},\bmu)=R(\bmu)L({\bi v})$, then the general
expression of $A$ is the same as (\ref{eq:3.13.4}) with a change of sign in 
the cross product term ${\bi u}\times\bmu$. Therefore, the decomposition
is also unique, the rotation $R(\bmu)$ is the same as before but the Lorentz boost is
given in terms of the variables $a^\mu$ by
\[
{\bi u}=2\left[{\rm Re}\,(a^0) {\rm Re}\,({\bi a})+{\rm Im}\,(a^0) {\rm Im}\,({\bi 
a})+{\rm Im}\,({\bi a})\times{\rm Re}\,({\bi a})\right].
\]
Note the difference in the third term which is reversed when compared with (\ref{eq:bu}).

In the four-dimensional representation of the Lorentz group on Minkowski space-time,
a boost is expressed as $L(\bbeta)=\exp(\bbeta\cdot{\bi K})$ in terms of the 
dimensionless normal parameters
$\beta_i$ and the $4\times4$ boost generators $K_i$ given by
\[
K_1=\pmatrix{0&1&0&0\cr 1&0&0&0\cr 0&0&0&0\cr 0&0&0&0\cr},\,
K_2=\pmatrix{0&0&1&0\cr 0&0&0&0\cr 1&0&0&0\cr 0&0&0&0\cr},\,
K_3=\pmatrix{0&0&0&1\cr 0&0&0&0\cr 0&0&0&0\cr 1&0&0&0\cr}.
\]
If we call $B=\bbeta\cdot{\bi K}\equiv\sum_i\beta_iK_i$, we have
\[
B=\pmatrix{0&\beta_1&\beta_2&\beta_3\cr \beta_1&0&0&0\cr \beta_2&0&0&0\cr \beta_3&0&0&0\cr},\quad 
B^2=\pmatrix{\beta^2&0&0&0\cr 0&\beta_1\beta_1&\beta_1\beta_2&\beta_1\beta_3\cr 0&\beta_2\beta_1&\beta_2\beta_2&\beta_2\beta_3\cr 
0&\beta_3\beta_1&\beta_3\beta_2&\beta_3\beta_3\cr},
\]
with $\beta^2=\beta_1^2+\beta_2^2+\beta_3^2$ and $B^3=\beta^2B$, and so on for the remaining powers of $B$,
so that the final expression for $L(\bbeta)=\exp(\bbeta\cdot{\bi K})$ is
 \[
\pmatrix{C&(\beta_1/\beta)S&(\beta_2/\beta)S&(\beta_3/\beta)S\cr 
(\beta_1/\beta)S&1+\frac{\displaystyle{\beta_1\beta_1}}{\displaystyle{\beta^2}}(C-
1)&\frac{\displaystyle{\beta_1\beta_2}}{\displaystyle{\beta^2}}(C-
1)&\frac{\displaystyle{\beta_1\beta_3}}{\displaystyle{\beta^2}}(C-
1)\cr
(\beta_2/\beta)S&\frac{\displaystyle{\beta_2\beta_1}}{\displaystyle{\beta^2}}(C-
1)&1+\frac{\displaystyle{\beta_2\beta_2}}{\displaystyle{\beta^2}}(C-
1)&\frac{\displaystyle{\beta_2\beta_3}}{\displaystyle{\beta^2}}(C-
1)\cr
(\beta_3/\beta)S&\frac{\displaystyle{\beta_3\beta_1}}{\displaystyle{\beta^2}}(C-
1)&\frac{\displaystyle{\beta_3\beta_2}}{\displaystyle{\beta^2}}(C-
1)&1+\frac{\displaystyle{\beta_3\beta_3}}{\displaystyle{\beta^2}}(C-
1)\cr}
 \]  
where $S=\sinh\beta$ and $C=\cosh\beta$. 
What is the physical interpretation of $\beta_i$? Let us assume that observers $O$ and $O'$
relate their space-time measurements $x$ and $x'$ by ${x'}^{\mu}={L(\bbeta)^{\mu}}_\nu x^\nu$. Observer
$O$ sends at time $t$ and at a later time $t+dt$ two light signals from a source placed
at the origin of its Cartesian frame. These two signals when measured by $O'$ take place at
points ${\bi r'}$ and ${\bi r'}+d{\bi r'}$ and at instants $t'$ and $t'+dt'$, respectively. 
Then they are related by
 \[
c dt'={L^0}_0c dt,\qquad  d{x'}^{i}={L^{i}}_0c dt
 \] 
because $dx^i=0$. The quotient $d{x'}^{i}/dt'$ is just the velocity of the light source $v^i$, 
{\sl i.e.}, of the origin of the $O$ frame as measured by observer $O'$, 
and then this velocity $v^i=c{L^i}_0/{L^0}_0=c(\beta_i/\beta)S/C$, such that the relation between
the normal parameters and the relative velocity between observers is
 \[
\frac{\bi v}{c}=\frac{\bbeta}{\beta}\,\tanh\beta
 \] 
and therefore $\tanh\beta=v/c$. Function $\cosh\beta\equiv\gamma(v)=(1-
v^2/c^2)^{-1/2}$ and when the transformation is expressed in terms
of the relative velocity it takes the form of the symmetric matrix:
 \begin{equation}
L({\bi v})=\pmatrix{\gamma&\gamma{v_x/ c}& \gamma{v_y/ c}& 
\gamma{v_z/c}\cr 
\gamma{v_x/c}&1+{\displaystyle v_x^2\over 
\displaystyle c^2}{\displaystyle\gamma^2\over\displaystyle\gamma+1}&{\displaystyle v_xv_y\over\displaystyle 
c^2}{\displaystyle\gamma^2\over\displaystyle\gamma+1}&{\displaystyle v_xv_z\over\displaystyle c^2}{\displaystyle\gamma^2\over\displaystyle\gamma+1}\cr 
\gamma{v_y/ c}&{\displaystyle v_yv_x\over\displaystyle c^2}{\displaystyle\gamma^2\over\displaystyle\gamma+1}& 1+{\displaystyle v_y^2\over\displaystyle 
\displaystyle c^2}{\displaystyle\gamma^2\over\displaystyle 
\gamma+1}&{\displaystyle v_yv_z\over\displaystyle 
c^2}{\displaystyle\gamma^2\over\displaystyle\gamma+1}\cr 
\gamma{v_z/ c}&{\displaystyle v_zv_x\over 
\displaystyle c^2}{\displaystyle\gamma^2\over\displaystyle \gamma+1}&{\displaystyle v_zv_y\over 
\displaystyle c^2}{\displaystyle\gamma^2\over\displaystyle \gamma+1}&1+{\displaystyle v_z^2\over 
\displaystyle c^2}{\displaystyle\gamma^2\over\displaystyle \gamma+1}\cr}.
 \label{eq:3.12.1}
 \end{equation} 
The inverse transformation $L^{-1}({\bi v})=L(-{\bi v})$\index{Lorentz!boost}. 
The orthogonal $4\times4$ rotation matrix takes the block form
\begin{equation}
R(\bmu)=\pmatrix{1&{\bf 0}\cr {\bf 0}&\widetilde{R}(\bmu)},
\label{eq:1Rmu}
\end{equation}
\index{Lorentz transformation}where $\widetilde{R}(\bmu)$ is the $3\times3$ orthogonal matrix (\ref{eq:rotalfa}).
When a Lorentz transformation is expressed in the form $\Lambda({\bi v},\bmu)=L({\bi v})R(\bmu)$,
then by construction the first column of $\Lambda({\bi v},\bmu)$ is just the first
column of (\ref{eq:3.12.1}) where the velocity parameters ${\bi v}$ are defined. 
Therefore, given the general Lorentz transformation $\Lambda({\bi v},\bmu)$, from its first
column we determine the parameters ${\bi v}$ and thus the complete $L({\bi v})$ can be worked out.
The rotation involved can be easily calculated as $L(-{\bi v})\Lambda({\bi v},\bmu)=R(\bmu)$. If expressed
in the reverse order $\Lambda({\bi v},\bmu)=R(\bmu)L({\bi v})$, then it is the first row
of $\Lambda$ that coincides with the first row of (\ref{eq:3.12.1}). It turns out
that, given any general Lorentz transformation $\Lambda({\bi v},\bmu)$,
then $\Lambda({\bi v},\bmu)=L({\bi v})R(\bmu)=R(\bmu)L({\bi v}')$
with the same rotation in both sides as derived in (\ref{eq:bmu}) 
and $L({\bi v}')= R(-\bmu)L({\bi v})R(\bmu)=L(R(-\bmu){\bi v})$, 
{\sl i.e}, the velocity ${\bi v}'=R(-\bmu){\bi v}$. In any case, the decomposition of a general Lorentz 
transformation as a product of a rotation and a boost is a unique one, in terms of the
same rotation $R(\bmu)$ and a boost to be determined, depending on the order in which we take
these two operations.
\index{Lorentz transformation!decomposition}

Matrix $\Lambda$ can be considered as a tetrad ({\sl i.e.}, a set of
four orthonormal four-vectors, one time-like and the other three
space-like) attached by observer $O'$ to the origin of observer $O$. In
fact, if the matrix is considered in the form $\Lambda({\bi v},\bmu)=L({\bi
v})R(\bmu)$, then the first column of $\Lambda$ is the four-velocity of the
origin of the $O$ Cartesian frame and the other three columns are just the
three unit vectors of the $O$ reference frame, rotated with rotation $R(\bmu)$
and afterwards boosted with $L({\bi v})$.

\chapter{Quantization of the models}
\label{ch:quantization}

Quantization of generalized Lagrangian systems will suggest that wave 
functions for elementary particles must be squared integrable functions 
defined on the kinematical 
space.

We shall use Feynman's quantization method to show the structure of the wave function and the way
it transforms under the kinematical or symmetry group of the theory. Once the
Hilbert space structure of the state space is determined, this leads to a specific
representation of the generators of the group as self-adjoint 
operators and the remaining analysis
is done within the usual quantum mechanical context, {\sl i.e.}, by choosing the 
complete commuting set of operators to properly determine a set of orthogonal basis
vectors of the Hilbert space. Special emphasis is devoted to the analysis of the 
different angular momentum operators the formalism supplies. They have a similar structure 
to the classical ones, and this will help us to properly obtain the identification of the
spin observable.

The structure of the spin operator depends on the kind of translation invariant
kinematical variables we use to describe the particle, and the way these 
variables transform under the rotation group. Since in the Galilei and Poincar\'e 
case, as we have seen previously,
these variables are the velocity ${\bi u}$ and orientation 
$\balpha$ and they transform in the same way
under rotations in both approaches, then
the structure of the spin operator is exactly the same in both relativistic and
nonrelativistic formalisms. 

As we have seen in the classical description the position of the charge of the particle and 
its center of mass are different points, and spin is related to the 
rotation and internal motion (zitterbewegung) of the charge around the 
center of mass of the particle. The magnetic properties of the particle are
connected only with the motion of the charge and therefore to the zitterbewegung part
of spin. It is this double spin structure that gives rise to the concept of gyromagnetic
ratio when expressing the magnetic moment in terms of the total spin. 
If the Lagrangian shows no dependence 
on the acceleration, the spin is only of rotational nature, and the 
position and center of mass position define the same point. Spin 1/2 
particles arise if the corresponding classical model rotates but no half integer 
spins are obtained for systems with spin of orbital nature related only to the 
zitterbewegung. On the manifold spanned by non-compact variables ${\bi u}$
no half-integer spins can be found, because the spin operator has the form of
an orbital angular momentum and eigenvectors are but spherical harmonics.

Dirac's equation will be obtained when quantizing the classical relativistic 
spinning particles whose center of charge is circling around its 
center of mass at the speed $c$. In that case, the internal orientation of the 
electron completely characterizes its Dirac algebra.

\section{Feynman's quantization of Lagrangian systems} 
\index{Feynman's quantization}\index{quantization!Feynman method}
\label{sec:quantize}

Let us consider a generalized Lagrangian system as described in 
previous chapters and whose evolution is considered on the kinematical 
space between points $x_1$ and $x_2$. 

For quantizing these generalized Lagrangian systems we shall follow 
Feynman's path integral method.~\footnote{\hspace{0.1cm}R.P. Feynman and A.R. Hibbs, {\sl Quantum Mechanics 
and Path Integrals}, MacGraw Hill, NY (1965), p. 36.}
The Uncertainty 
Principle is introduced in Feynman's approach by the condition that if 
no measurement is performed to determine the trajectory followed by 
the system from $x_1$ to $x_2$, then all paths $x(\tau)$ are allowed with the 
same probability. Therefore a probability definition $P[x(\tau)]$, must be given for every path.

But instead of defining the probability associated to each possible path $P[x(\tau)]$, this is calculated
in terms of a {\bf probability amplitude}, $\phi[x(\tau)]$ for that path such that 
$P[x(\tau)]=|\phi[x(\tau)]|^2$, where $0\le P\le 1$. But in general $\phi$ does not need to be
a positive real number; in fact it is a complex number. Thus, to every possible trajectory followed by 
the system, $x(\tau)$ in $X$ space, Feynman associates a complex 
number $\phi[x(\tau)]$ called the probability amplitude of this 
alternative, given by
 \begin{equation} 
\phi[x(\tau)]=N\exp\left\{{i\over\hbar}\int_{\tau_1}^{\tau_2} L(x(\tau),\dot x(\tau)) 
d\tau\right\}=N\exp\left\{{i\over\hbar}\,A_{[x]}(x_1,x_2) \right\},
 \label{eq:8}
 \end{equation}
where $N$ is a path independent normalization 
factor, and where the phase of this complex number in units of $\hbar$ is the classical 
action of the system $A_{[x]}(x_1,x_2)$ along the path $x(\tau)$. 
Once we perform the integration along the path, this 
probability amplitude becomes clearly a function of the initial and final 
points in $X$ space, $x_1$ and $x_2$, respectively. 

In this Feynman statistical procedure, the probability amplitude of 
the occurrence of any alternative of a set of independent alternatives
is the sum of the corresponding probability amplitudes of the different
independent events. The probability of the whole process is the square of the absolute 
value of the total probability amplitude. This produces the effect that the probability
of the whole process can be less than the probability of any single alternative of the set. 
This is what Feynman calls {\bf interfering statistics}.
\index{interfering statistics}\index{Feynman!interfering statistics}

Then, the total probability amplitude that the system arrives at point $x_2$ 
coming from $x_1$, {\sl i.e.}, Feynman's kernel $K(x_1,x_2)$, is obtained as 
the sum or integration over all paths, of terms of the form of Eq. 
(\ref{eq:8}). Feynman's kernel $K(x_1,x_2)$, will be in general a 
function, or more precisely a distribution, on the $X\times X$ manifold. If 
information concerning the initial point is lost, and the final point 
is left arbitrary, say $x$, the kernel reduces to the probability 
amplitude for finding the system at point $x$, {\sl i.e.}, 
the usual interpretation of the quantum mechanical wave function 
$\Phi(x)$. By the above discussion we see that wave functions must be 
complex functions of the kinematical variables. 

We thus see that Feynman's quantization method enhances the role of the kinematical
variables to describe the quantum state of an arbitrary system, in spite of the independent
degrees of freedom. We consider that this is one of the reasons why the kinematical variables
have to play a leading role also in the classical approach.

We are used to consider in quantum mechanics, instead of a single function $\Phi(x)$,
multicomponent wave functions, {\sl i.e}, a set of linearly independent 
functions $\psi_i(t,{\bi r})$
defined on space-time and labeled with a discrete subindex that runs over a finite range, 
such that it can be considered as a vector valued function in a finite dimensional complex vector
space. In general this finite space carries some irreducible representation of the rotation
group and each component $\psi_i$ represents a definite spin state of the system. 
Nevertheless, our wave function $\Phi(x)$ depends on more variables than space-time variables.
Once we define later the complete commuting set of observables to obtain, in terms of their
simultaneous eigenvectors, an orthonormal basis for the Hilbert space of states, we shall find
that $\Phi(x)$ can be separated in two parts. One part $\phi(t,{\bi r})$ depending on
space-time variables and another part $\chi$
that depends on the remaining translation invariant kinematical variables, that in our case
will reduce to the velocity ${\bi u}$ and orientation $\balpha$. It is this possible separation  
of our wave function that will produce the emergence of the different components of the usual
formalism.

\subsection{Transformation of the wave function}

To see how the wave function transforms between inertial observers, 
and therefore to obtain its transformation equations under the kinematical groups,
let us consider that $O$ and $O'$ are two inertial observers related by means of a 
transformation $g\in G$, such that the kinematical variables transform 
as: 
 \begin{equation}
{x'}^{i}=f^i(x,g).
 \end{equation}  

If observer $O$ considers that the system follows the path $\bar x(\tau)$, then it follows for 
$O'$ the path $\bar x'(\tau)=f(\bar x(\tau),g)$ and because the action 
along classical paths transforms according to Eq. (\ref{eq:2.5.3}), 
the probability amplitude for observer $O'$ is just
 \[
\phi'[\bar x'(\tau)]=N\exp\left\{\frac{i}{\hbar}\int_{\tau_1}^{\tau_2} L(\bar x'(\tau),\dot{\bar x'}(\tau)) 
d\tau\right\}\qquad\qquad
 \]
 \[=N\exp\left\{\frac{i}{\hbar}\int_{\tau_1}^{\tau_2} L(\bar x(\tau),\dot{\bar x}(\tau)) 
d\tau\right\}\exp\left\{\frac{i}{\hbar}\int_{\tau_1}^{\tau_2} \frac{d\alpha(g;\bar x(\tau))}{d\tau}d\tau\right\},
 \]
{\sl i.e.},
\[
\phi'[\bar x'(\tau)]=\phi[\bar x(\tau)]\,\exp\left\{\frac{i}{\hbar}\left(\alpha(g;x_2)-\alpha(g;x_1)\right)\right\},
\]
where the last phase factor is independent of the integration path. If we add all 
probability amplitudes of this form, it turns out that Feynman's kernel transforms as:
 \begin{equation}
K'(x'_1,x'_2)=K(x_1,x_2)\exp\left\{\frac{i}{\hbar}\left(\alpha(g;x_2)-\alpha(g;x_1)\right)\right\}.
 \end{equation}
If information concerning the initial point $x_1$ is lost, the wave 
function transforms as the part related to the variables $x_2$, up to 
an arbitrary function on $G$,
 \begin{equation} 
\Phi'(x'(x))=\Phi'(gx)=\Phi(x)\exp\left\{\frac{i}{\hbar}\left(\alpha(g;x)+\theta(g)\right)\right\},
\label{eq:transfi}
 \end{equation}
or in terms of unprimed $x$ variables
 \begin{equation}
\Phi'(x)=\Phi(g^{-1}x)\exp\left\{{i\over\hbar}\left(\alpha(g;g^{-1}x)+\theta(g)\right)\right\},
 \label{eq:9}
 \end{equation}
where $\theta(g)$ is some function defined on $G$ but independent of $x$. 

Since our system is somewhere in $X$ space, the probability of finding the system
anywhere is 1. Then we have to define the way of adding 
probabilities at different points $x\in X$.
If we define a measure on $X$, $\mu(x)$, such that $d\mu(x)$ is the volume element in $X$ space
and $|\Phi(x)|^2d\mu(x)$ is interpreted as the probability of finding the system inside 
the volume element $d\mu(x)$ around point $x$, the probability of 
finding it anywhere in $X$ must be unity, so that
 \begin{equation} 
\int_X |\Phi(x)|^2 d\mu(x)=1.
 \end{equation}  
Since from (\ref{eq:9})
 \begin{equation} 
|\Phi'(x')|^2=|\Phi(x)|^2,
 \label{eq:10}
 \end{equation}
it is sufficient for the conservation of probability to assume 
that the measure to be defined $\mu(x)$ is
group invariant. In that case, equation (\ref{eq:10}) implies also that inertial 
observers measure locally the same probability. This will have strong consequences
about the possibility of invariance of the formalism under arbitrary changes of 
phase of the wave function. But the phase can be changed in a different manner
at different points $x$. We can use this fact to further impose the local 
gauge invariance of the theory. It must be remarked that this arbitrary change of phase $\beta(x)$
is not only a phase on space-time, but rather on the whole kinematical space of the system
and this enlarges the possibilities of analyzing different transformation groups that 
can be more general than the original kinematical groups, because they act on a larger manifold.

Consequently, the 
Hilbert space ${\cal H}$ whose unit rays represent the pure states of 
the system is the space of squared-integrable functions $\LL^2(X,\mu)$ 
defined on the kinematical space $X$, $\mu(x)$ being an invariant 
measure such that the scalar product on ${\cal H}$ is defined as
 \begin{equation} 
<\Phi|\Psi>=\int_X \Phi^*(x)\Psi(x) d\mu(x),
 \label{eq:11}
 \end{equation}  
$\Phi^*(x)$ being the complex conjugate function of $\Phi(x)$. There is an
arbitrariness in the election of the invariant measure $\mu(x)$ but this will
be guided by physical arguments. Nevertheless, the invariance condition will
restrict the possible measures to be used.

\subsection{Representation of Observables}

Wigner's theorem,~\footnote{\hspace{0.1cm}E.P. Wigner, {\sl Group theory and its 
application to the quantum mechanics of atomic spectra}, Acad. Press, NY
(1959).}$^,$\footnote{\hspace{0.1cm}V. Bargmann, {\sl J. Math. Phys}. {\bf 5}, 862 (1964).} 
implies that to every symmetry $g\in G$ of a continuous 
group, there exists a one to one mapping of unit rays into unit rays 
that is induced on ${\cal H}$ by a unitary operator $U(g)$ defined up 
to a phase that maps a wave function defined on $x$ into an arbitrary 
wave function of the image unit ray in $x'$. 
The Relativity Principle is a strong symmetry of physical systems that defines the equivalence
between the set of inertial observers whose space-time measurements are related
by means of a transformation of a kinematical group G.
Now, if we interpret $\Phi(x)$ 
as the wave function that describes the state of the system for the 
observer $O$ and $\Phi'(x)$ for $O'$, then we have
 \begin{equation} 
U(g)\Phi(x)=\Phi'(x)=\Phi(g^{-1}x)\exp\left\{{i\over\hbar}\alpha(g;g^{-1}x)+\theta(g)\right\}.
 \end{equation}  

Since the $\theta(g)$ function gives rise to a constant phase we can 
neglect it and then take as the definition of the unitary 
representation of the group $G$ on Hilbert space ${\cal H}$
 \begin{equation} 
\Phi'(x)=U(g)\Phi(x)=\Phi(g^{-1}x)\exp\left\{{i\over\hbar}\alpha(g;g^{-1}x)\right\}.
 \label{eq:12}
 \end{equation}  

Gauge functions satisfy (\ref{eq:2.6.2}), and therefore the phase term can be replaced by 
 \begin{equation}
\alpha(g;g^{-1}x)=-\alpha(g^{-1};x)+\alpha(0;x)+\xi(g,g^{-1})=
-\alpha(g^{-1};x)+\zeta(g),
 \end{equation}
because gauge functions can always be 
chosen such that $\alpha(0;x)=0$ and the group function $\zeta(g)=\xi(g,g^{-1})$ 
giving rise also to a constant phase, can be suppressed. 
We thus define the transformation of the wave function by
 \begin{equation}
\Phi'(x)=U(g)\Phi(x)=\Phi(g^{-1}x)\exp\left\{-{i\over\hbar}\alpha(g^{-1};x)\right\}.
 \label{eq:123}
 \end{equation}
If the unitary operator is represented in terms of the corresponding 
self-adjoint generators of the Lie algebra in the form
 \begin{equation} 
U(g)=\exp\left\{-\frac{i}{\hbar}\,g^\sigma X_\sigma \right\},
 \end{equation}
then, for an infinitesimal transformation of parameters $\delta g^\sigma$ its inverse transformation
has infinitesimal parameters $-\delta g^\sigma$, we obtain at first order in $\delta g^\sigma$
\[
U(\delta g)\Phi(x)=\left(\ID-\frac{i}{\hbar}\delta g^\sigma X_\sigma \right)\Phi(x)=\Phi(x)-\frac{i}{\hbar}\delta g^\sigma X_\sigma\,\Phi(x),
\]
while
\[
\Phi(\delta g^{-1}x)\equiv\Phi(f(x,\delta g^{-1}))=\Phi(x)-\delta g^{\sigma} u^i_\sigma (x)\frac{\partial\Phi(x)}{\partial x^i},
\]and
\[
\exp\left\{-{i\over\hbar}\alpha(\delta g^{-1};x)\right\}=1-{i\over\hbar}\alpha(\delta g^{-1};x).
\]
But because $\alpha(0;x)=0$,
\[
\alpha(\delta g^{-1};x)=\frac{\partial\alpha(g;x)}{\partial g^\sigma}\bigg|_{g=0}\,(-\delta g^\sigma),
\]
and the substitution of the above terms in (\ref{eq:123}) and further identification of the first order terms 
in $\delta g^\sigma$ imply that the 
self-adjoint operators $X_\sigma$ when acting on the wave functions 
have the differential representation
 \begin{equation} 
X_\sigma={\hbar\over i}\,u^j_\sigma(x)\,{\partial\over\partial x^j}-v_\sigma(x),
 \label{eq:13}
 \end{equation}
where
 \begin{equation}
u^j_\sigma(x)={\partial f^j(x,g)\over\partial g^\sigma}\bigg|_{g=0},
\qquad v_\sigma(x)= {\partial\alpha(g;x)\over\partial g^\sigma}\bigg|_{g=0}.
\label{eq:uvalfa}
 \end{equation}  

If we restrict ourselves to transformations of the enlarged 
configuration space $(t,q_i)$ that can be extended to the whole 
kinematical space $x\equiv(t,q_i,\ldots,q_i^{(k-1)})$, then, using the same 
notation as in (\ref{eq:timepar})-(\ref{eq:qspar}), if the infinitesimal transformation is of the form
\[
t'=t+M_\sigma\delta g^\sigma,\; q'_i=q_i+M_{i\sigma}\delta g^\sigma,\; \ldots,
{q'}^{(k-1)}_i=q^{(k-1)}_i+M^{(k-1)}_{i\sigma}\delta g^\sigma,
\]
these generators take the form
 \begin{equation} 
X_\sigma=\frac{\hbar}{i}\left(M_\sigma\frac{\partial}{\partial t}+
M_{i\sigma}\frac{\partial}{\partial q_i}+\ldots+M^{(k-1)}_{i\sigma}
\frac{\partial}{\partial q_i^{(k-1)}}\right)-v_\sigma(x).
 \label{eq:generators}
 \end{equation}
When compared with the Noether constants of the motion 
(\ref{eq:constant}) written in the form
 \begin{equation} 
-N_\sigma=-H\,M_\sigma+ p_{(s+1)}^iM_{i\sigma}^{(s)}-B_\sigma(x),
\label{eq:constantes} 
 \end{equation} 
we see a certain kind of {\bf `correspondence 
recipe'}.\index{correspondence recipe} When 
restricted to kinematical groups, the functions $B_{\sigma}(x)$ of 
(\ref{eq:constant}), are obtained from the Lagrangian gauge functions 
$\alpha(g;x)$, by (\ref{eq:1B}), which is exactly the same derivation as the functions 
$v_\sigma(x)$ above in (\ref{eq:uvalfa}).\index{group generators}\index{group generators!self adjoint}
Now, by identifying the different classical observables and generalized momenta
that appear here in (\ref{eq:constantes}) with the corresponding differential operators of
(\ref{eq:generators}) that multiply the corresponding $M^{(s)}_{i\sigma}$ function, we get:
the generalized Hamiltonian $H=p^i_{(s)}q_i^{(s)}-L$, which is multiplied in (\ref{eq:constantes}) by the function $M_\sigma$, is identified with the operator
$i\hbar{\partial}/{\partial}t$ which is also in front of the function $M_\sigma$ in (\ref{eq:generators}), and 
similarly, the generalized momentum 
$p^i_{(s+1)}$, the factor that multiplies the function $M^{(s)}_{i\sigma}$, 
with the differential operator $-i\hbar\partial/\partial q^{(s)}_i$, for $s=0,\ldots,k-1$, because 
the functions $v_\sigma(x)=B_\sigma(x)$, are the same.

Remember that $p^i_{(s+1)}$ and $q_i^{(s)}$ are
canonical conjugate variables. Then, each generalized momentum is replaced by $(\hbar/i)$ times
the differential operator that differentiates with respect to its conjugate generalized 
coordinate and the generalized Hamiltonian by $i\hbar\partial/\partial t$.

The Heisenberg representation is that representation in which the time dependence
has been withdrawn from the wave function by means of a time dependent unitary transformation. 
Then the wave function in this representation depends on the kinematical
variables with the time excluded, {\sl i.e.}, it depends only on the generalized coordinates $q^{(r)}_i$.
Therefore, when acting on the wave function in the Heisenberg representation $\psi(q_i,q^{(1)}_i,\ldots,q^{(k-1)}_i)$,
the observables $q^{(r)}_i$ and $p^j_{(s)}$ satisfy the canonical commutation relations
\[
[q^{(r)}_i,p^j_{(s+1)}]=i\hbar\delta^j_i\delta^r_s.
\]

If functions $v_\sigma(x)$ in (\ref{eq:13}) vanish, the $X_\sigma$ generators 
satisfy the commutation relations of the group $G$. But if some $v_\sigma(x)\neq0$
the $X_\sigma$ generators do not satisfy in general the commutation 
relations of the initial group $G$ where they come from, 
but rather the commutation relations of a 
central extension of $G$. The group representation is not a true 
representation but a projective representation of $G$ as shown by Bargmann.~\footnote{\hspace{0.1cm}V. Bargmann, {\sl Ann. Math.} {\bf 59}, 1 (1954). } 

In fact, from (\ref{eq:12}) we get 
 \[
U(g_1)\Phi(x)=\Phi(g_1^{-1}x)\exp\{{i\over\hbar}\alpha(g_1;g_1^{-1}x)\},
 \]
acting now on the left with $U(g_2)$,
 \[
U(g_2)U(g_1)\Phi(x)=U(g_2)\Phi(g_1^{-1}x)\exp\{{i\over\hbar}
\alpha(g_1;g_1^{-1}x)\}
 \]
 \begin{equation}
=\Phi((g_2g_1)^{-1}x)\exp\{{i\over\hbar}\alpha(g_2;g_2^{-1}x)\}
\exp\{{i\over\hbar}\alpha(g_1;(g_2g_1)^{-1}x)\},
\label{eq:gaug19}
 \end{equation}
while acting on $\Phi(x)$ with $U(g_2g_1)$,
 \begin{equation}
U(g_2g_1)\Phi(x)=\Phi((g_2g_1)^{-1}x)\exp\{{i\over\hbar}
\alpha(g_2g_1;(g_2g_1)^{-1}x)\}.
\label{eq:gaug20} 
\end{equation} 

If we define $(g_2g_1)^{-1}x=g_1^{-1}g_2^{-1}x=z$, then $g_1z=g_2^{-1}x$ and because gauge 
functions satisfy (\ref{eq:2.6.2}), we write
 \begin{equation}
\alpha(g_2;g_1z)+\alpha(g_1;z)=\alpha(g_2g_1;z)+\xi(g_2,g_1),
\label{eq:gaug21} 
 \end{equation} 
and by comparing (\ref{eq:gaug19}) with (\ref{eq:gaug20}), taking into account (\ref{eq:gaug21}), 
we obtain
 \begin{equation}
U(g_2)U(g_1)\Phi(x)=U(g_2g_1)\Phi(x)\exp\{{i\over\hbar}\xi(g_2;g_1)\}.
 \end{equation}
Since $\Phi(x)$ is arbitrary, 
we have a projective unitary representation of the group $G$ characterized by the non-trivial exponent
$\xi(g,g')$. 

For both Galilei and Poincar\'e particles the kinematical space is the 
ten-dimen\-sional 
manifold spanned by the variables $(t,{\bi r},{\bi u},\balpha)$, 
$t$ being the time, ${\bi r}$ the charge position, ${\bi u}$ the velocity and 
$\balpha$ the orientation of the particle. Thus in the quantum formalism the 
wave function of the most general elementary particle is a squared-integrable function 
$\Phi(t, {\bi r},{\bi u},\balpha)$ of these kinematical variables. For point 
particles, the kinematical space is just the four-dimensional space-time, so 
that wave functions are only functions of time and position, but spinning 
particles will have to depend on the additional variables like velocity and 
orientation. The spin structure will thus be related to these additional variables.

\subsection{Nonrelativistic spinning particles. Bosons}
\index{quantization!bosons}\index{bosons}
\label{sec:Nonrelbosons}

Now let us apply the formalism to the most interesting case of spinning particles. 
Let us consider next Galilei particles 
with (anti)orbital spin. This corresponds for example to systems for which $X={\cal 
G}/SO(3)$ and thus the kinematical variables are time, position and velocity. 
A particular classical example is given in Chapter 2, Section \ref{sec:galispin} 
by the free Lagrangian 
 \begin{equation}
L={m\over2}\left({d{\bi r}\over dt}\right)^2-{m\over2\omega^2}
\left({d{\bi u}\over dt}\right)^2,
 \end{equation}  
with ${\bi u}=d{\bi r}/dt$. 
For the free particle, the center of mass ${\bi q}={\bi r}-{\bi k}$ has a straight 
motion while the relative position vector ${\bi k}$ follows an elliptic 
trajectory with frequency 
$\omega$ around its center of mass, being the spin related to this 
internal motion. It is expressed as ${\bi S}=-m{\bi k}\times d{\bi k}/dt$.

The kinematical variables transform under ${\cal G}$ in the form
 \begin{eqnarray}
t'(\tau)&=&t(\tau)+b,\label{eq:t4}\\ 
{\bi r}'(\tau)&=&R(\balpha){\bi r}(\tau)+{\bi v}t(\tau)+{\bi a},\label{eq:r4}\\
{\bi u}'(\tau)&=&R(\balpha){\bi u}(\tau)+{\bi v}.\label{eq:u4}
 \end{eqnarray}
The wave functions are functions on $X$ and thus functions of the 
variables $(t,{\bi r},{\bi u})$. On this kinematical space the gauge 
function is the same as in (\ref{eq:4.1.9}),
where $m$ defines again the mass of the 
system.  Taking into 
account as in the previous example the correspondence recipe
for the Hamiltonian $H\to i\hbar\partial/\partial t$, the first generalized 
momentum ${\bi p}_1\equiv{\bi P}\to-
i\hbar\nabla$ and the other generalized momentum 
${\bi p}_2\equiv{\bi U}\to-i\hbar\nabla_u$, the generators of the projective representation 
are given by
 \begin{equation} 
H=i\hbar\frac{\partial}{\partial t},\quad {\bi P}=
\frac{\hbar}{i}\nabla,\quad {\bi K}=m{\bi r}-
t\,\frac{\hbar}{i}\nabla-\frac{\hbar}{i}\,\nabla_u,
 \end{equation} 
 \begin{equation}
{\bi J}={\bi r}\times\frac{\hbar}{i}\,\nabla+{\bi u}\times\frac{\hbar}
{i}\nabla_u={\bi L}+{\bi Z},
 \end{equation}
where $\nabla$ is the gradient operator with respect to
${\bi q}_1\equiv{\bi r}$ variables and $\nabla_u$ the gradient 
operator with respect to the ${\bi q}_2\equiv{\bi u}$ variables. 
It is important to stress that this representation of the generators is independent 
of the particular Lagrangian that 
describes the system. It depends only on the kinematical variables 
$(t,{\bi r},{\bi u})$
and the usual Galilei gauge function.

If we define ${\bi q}={\bi r}-{\bi k}=({\bi K}+{\bi P}t)/m$, 
it satisfies the commutation relations with ${\bi P}$, 
\[
[ q_i, P_j]=i\hbar\,\delta_{ij}, 
\]
which are the canonical commutation relations between the linear 
momentum and position for a point particle
and therefore these canonical commutation relations between the total linear momentum
and the center of mass position for a spinning particle
are already contained in the commutation relations of the extended Lie algebra of the kinematical group.
Therefore the quantum mechanical operator
\begin{equation}
{\bi q}={\bi r}-\frac{\hbar}{im}\nabla_u,
\label{eq:centerofmQ}
\end{equation}
can be interpreted as the center of mass position operator\index{center of mass!position}.
Discussion of other possibilities for the center of mass position operator 
can be found in the book by the author.

In this representation, one Casimir operator is the internal energy $H-{\bi 
P}^2/2m$. We see that the spin operator is defined as usual
 \[
{\bi S}={\bi J}-\frac{1}{m}{\bi K}\times{\bi P}=
{\bi u}\times{\bi U}+{\bi k}\times{\bi P}={\bi u}\times\frac{\hbar}
{i}\nabla_u+\frac{\hbar}
{im}\nabla_u\times\frac{\hbar}{i}\nabla;
 \]
written in terms of two non-commuting terms, it satisfies
 \[ 
[{\bi S},{\bi S}]=i\hbar{\bi S},\quad [{\bi J},{\bi S}]=i\hbar{\bi S},
\quad [{\bi S},{\bi P}]=[{\bi S},H]=[{\bi S},{\bi K}]=0, 
 \] 
{\sl i.e.}, it is an angular momentum operator, transforms like a vector under 
rotations and is invariant under space and time translations and under 
Galilei boosts, respectively. The second part of the spin operator is of order
$\hbar^2$ so that it produces a very small correction to the first ${\bi Z}$ part.

Operators ${\bi Z}$ satisfy the commutation relations
 \[
[{\bi Z},{\bi Z}]=i\hbar{\bi Z},\quad [{\bi J},{\bi Z}]=
i\hbar{\bi Z},\quad [{\bi Z},{\bi P}]=[{\bi Z},H]=0,
 \]
 \[
[{\bi Z},{\bi K}]=-i\hbar{\bi U}=-\hbar^2\nabla_u,
 \] 
{\sl i.e.}, ${\bi Z}$ is an angular momentum operator, transforms like a vector under 
rotations and is invariant under space and time translations but not under 
Galilei boosts. It is usually considered as the quantum mechanical spin operator.

We see however, that the angular 
momentum operator ${\bi J}$ is split into two commuting terms ${\bi r}\times{\bi 
P}$ and ${\bi Z}$. They both commute with $H$, but the 
first one is not invariant under space translations. The ${\bi Z}$ 
operators are angular momentum operators that only differentiate the 
wave function with respect to the velocity variables, and consequently 
commute with $H$ and ${\bi P}$, and although it is not the true
Galilei invariant spin operator, we can find simultaneous 
eigenstates of the three commuting operators $H-{\bi P}^2/2m$, $Z^2$ 
and $Z_3$. Because the ${\bi Z}$ operators only affect the wave function in 
its dependence on ${\bi u}$ variables, we can choose functions with the 
variables separated in the form $\Phi(t,{\bi r},{\bi 
u})=\sum_i\psi_i(t,{\bi r})\chi_i({\bi u})$ so that
 \begin{equation}
(H-{\bi P}^2/2m)\psi_i(t,{\bi r})=E\psi_i(t,{\bi r}),
\label{eq:Hp2}
 \end{equation}
 \begin{equation}
Z^2\chi_i({\bi u})=z(z+1)\hbar^2\chi_i({\bi u}),
 \end{equation}
 \begin{equation}
Z_3\chi_i({\bi u})=m_z\hbar\chi_i({\bi u}).
 \end{equation}
The space-time dependent 
wave function $\psi_i(t,{\bi r})$, satisfies 
Sch\-roedinger's equation and is uncoupled with the spin part $\chi({\bi u})$. 

Due to the structure of $Z^2$ in terms of the 
${\bi u}$ variables, which is that of an orbital angular momentum, the 
spin part of the wave function is of the form
 \begin{equation}
\chi({\bi u})=f(u)Y_z^{m_z}(\theta,\phi),
 \end{equation}
$f(u)$ being an arbitrary 
function of the modulus of ${\bi u}$ and $Y_z^{m_z}(\theta,\phi)$ the 
spherical harmonics on the direction of ${\bi u}$. 

For the center of mass observer, ${\bi S}={\bi Z}$ and both angular 
momentum operators are the same. But for an arbitrary observer, ${\bi 
Z}$ operators do not commute with the boosts generators so that its 
absolute value is not Galilei invariant, while ${\bi S}$ is. But the 
splitting of the wave function into a multiple-component function that 
reflects its spin structure is an intrinsic property that can be done 
in any frame.

It turns out that if for an arbitrary observer ${\bi Z}$ is not the 
spin of the system, ${\bi r}\times{\bi P}$ is not the conserved 
orbital angular momentum, because ${\bi r}$ does not represent the 
position of the center of mass of the particle. 

When there is an interaction with an external electromagnetic field, equation (\ref{eq:Hp2})
is satisfied for the mechanical parts $H_m=H-e\phi$ and ${\bi P}_m={\bi P}-e{\bi A}$ and we thus obtain 
the usual equation
 \begin{equation}
\left(H-e\phi-\frac{({\bi P}-e{\bi A})^2}{2m}\right)\psi_i(t,{\bi r})=E\psi_i(t,{\bi r}).
\label{eq:Hp2bis}
 \end{equation}

This formalism, when the classical spin is of orbital nature, does not lead 
to half integer spin values, and therefore, from the quantum 
mechanical point of view these particles can be used only as models for representing
bosons. 

\subsection{Nonrelativistic spinning particles. Fermions}
\index{quantization!fermions}\index{fermions}
\label{sec:Nonrelferm}

Other examples of nonrelativistic spinning particles are those which have 
orientation and thus angular velocity. For instance, if $X={\cal 
G}/\RR_v^3$, $\RR_v^3$ being the subgroup $\{\RR^3,+\}$ of pure Galilei transformations, 
then the kinematical space is spanned by the variables $(t,{\bi 
r},\balpha)$. This corresponds for instance to the Lagrangian system 
described by
 \begin{equation}
L={m\over2}\left({d{\bi r}\over dt}\right)^2+{I\over2}\,\bomega^2.
 \end{equation}  

The particle travels freely at constant velocity while it rotates with 
constant angular velocity $\bomega$. The classical spin is just ${\bi 
S}=I\bomega$, and the center of charge and center of mass represent 
the same point. 

To describe orientation we can think of the three orthogonal unit 
vectors ${\bi e}_i$, $i=1,2,3$ linked to the body, similarly as in a 
rigid rotator. If initially they are taken parallel to the spatial 
Cartesian axis of the laboratory inertial frame, then their nine components considered by columns 
define an orthogonal rotation matrix $R_{ij}(\balpha)$ that describes 
the triad evolution with the initial condition 
$R_{ij}(t=0)=\delta_{ij}$. 

Now, kinematical variables $t$, ${\bi r}$ and $\brho$ transform under ${\cal G}$ in the form
 \begin{eqnarray} 
t'(\tau)&=&t(\tau)+b,\label{eq:t5}\\
{\bi r}'(\tau)&=&R(\balpha){\bi r}(\tau)+{\bi v}t(\tau)+{\bi a},\label{eq:r5}\\
\brho'(\tau)&=&\frac{\bmu+\brho(\tau)+\bmu\times\brho(\tau)}{1-\bmu\cdot\brho(\tau)}.\label{eq:composrot}
 \end{eqnarray} 
On the corresponding Hilbert space, the Galilei generators are given by:
 \begin{equation}
H=i\hbar{\partial\over\partial t},\quad {\bi P}={\hbar\over i}\nabla,
\quad {\bi K}=m{\bi r}-t\,\frac{\hbar}{i}\,\nabla,
 \label{eq:16}
 \end{equation} 
 \begin{equation}
{\bi J}={\hbar\over i}\,{\bi r}\times\nabla+ 
\frac{\hbar}{2i}\left\{\nabla_\rho+\brho\times\nabla_\rho+\brho 
(\brho\cdot\nabla_\rho)\right\}={\bi  L}+{\bi W},
 \label{eq:17}
 \end{equation} 
$\nabla_\rho$ being the gradient operator with respect to the 
$\brho$ variables and in the 
$\brho$ parameterization of the rotation group. 

The ${\bi W}$ part comes from the general group analysis. 
The group generators in this parametrization $X_i$ will be obtained 
from (\ref{eq:composrot}) and according to (\ref{eq:auxilifunc}) 
and (\ref{eq:generat}). They are obtained as
\[
X_i=\left.\left(\frac{\partial{\rho'}^k}{\partial\mu^i}\right)\right|_{\mu=0}
\;\frac{\partial}{\partial\rho^k},
\]
that can be written in vector notation as
 \[
{\bi X}=\nabla_\rho+\brho\times\nabla_\rho+\brho(\brho\cdot\nabla_\rho)
 \]
They satisfy the commutation relations
 \[
[X_i,X_k]=-2\epsilon_{ikl}X_l
 \]
and therefore operators $W_k={\displaystyle \hbar\over\displaystyle 2i}X_k$, 
or in vector notation
 \begin{equation}
{\bi W}=\frac
{\hbar}{2i}\left\{\nabla_\rho+\brho\times\nabla_\rho+\brho(\brho\cdot\nabla_\rho)\right\},
\label{eq:spinoper}
\end{equation}
will satisfy the angular momentum commutation relations
\begin{equation}
[{\bi W},{\bi W}]=i\hbar {\bi W}.
\label{eq:spincomm}
\end{equation}
In this way since ${\bi L}$ and ${\bi W}$ commute among each other, we also get $[{\bi J},{\bi J}]=i\hbar {\bi J}$.

In this example the center of mass and center of charge are the same 
point, ${\bi L}={\bi r}\times{\bi P}$ is the orbital angular momentum associated to the center of mass motion
and ${\bi W}\equiv{\bi S}$ is the spin operator. The spin operator commutes with 
$H$, ${\bi P}$ and ${\bi K}$ and the wave function can be separated as 
$\Phi(t,{\bi r},\brho)=\sum_i\psi_i(t,{\bi r})\chi_i(\brho)$ leading 
to the equations
 \begin{equation}
(H-{\bi P}^2/2m)\psi_i(t,{\bi r})=E\psi_i(t,{\bi r}),
 \end{equation}
 \begin{equation} 
S^2\chi_i(\brho)=s(s+1)\hbar^2\chi_i(\brho),
 \label{eq:18}
 \end{equation} 
 \begin{equation} 
S_3\chi_i(\brho)=m_s\hbar\chi_i(\brho).
 \label{eq:19}
 \end{equation}  

Bopp and Haag~\footnote{\hspace{0.1cm}F. Bopp and R. Haag, {\sl Z. Naturforschg.} 
{\bi 5a}, 644 (1950).} succeeded in finding $s=1/2$ solutions for the 
system of equations (\ref{eq:18}) and (\ref{eq:19}). They are called 
Wigner's functions.~\footnote{\hspace{0.1cm}L.C. Biedenharn and J.D. Louck, {\sl 
Angular Momentum in Quantum Physics. Theory and Application},
Cambridge U. P., Cambridge, England (1989). } Solutions of (\ref{eq:18}) for 
arbitrary spin $s$ are but a linear combination of the matrix elements 
of a $(2s+1)\times(2s+1)$ irreducible matrix representation of the 
rotation group as can be derived from the Peter-Weyl theorem on finite 
representations of compact groups.~\footnote{\hspace{0.1cm}A.R. Edmonds, {\sl Angular Momentum in 
Quantum Mechanics}, Princeton U. P., Princeton NJ (1957).}$^,$\footnote{\hspace{0.1cm}N. Ja. Vilenkin, {\sl Fonctions sp\'{e}ciales 
et Th\'{e}orie de la repr\'{e}sentation des groups}, Dunod, Paris (1969).}$^,$\footnote{\hspace{0.1cm}A.O. Barut 
and R. Raczka, {\sl Theory of group representations and applications}, PWN, Warszawa (1980).} We shall deal with the 
$s=1/2$ functions in the Appendix Section \ref{sec:genspinors}, where explicit expressions 
and a short introduction to the Peter-Weyl theorem, will be given.          

To describe fermions, the classical particles must necessarily 
have compact orientation variables as kinematical variables, otherwise no spin $1/2$
values can be obtained when the classical spin is related only to the zitterbewegung.

\section{Appendix: Spinors}
\index{spinors}
\label{sec:genspinors}

In this section of mathematical content we shall review the main properties of spinors,
in particular those connected with the possible representation of the wave function
to describe spin 1/2 particles. We shall describe the 
representations in terms of eigenfunctions of the different commuting spin
operators. But it must be remarked that in addition to the spin operators in
the laboratory frame we also have spin operators in the body frame, because our general
spinning particle has orientation, and therefore, a local Cartesian
frame linked to its motion. This produces the result that for a spin $1/2$ particle the wave function
necessarily is a four-component object.

The general wave function is a function of 
the ten kinematical variables, $\Phi(t,{\bi r},{\bi u},\brho)$, 
and the spin part of the system related to the translation invariant kinematical 
variables ${\bi u}$ and $\brho$ is
\begin{equation}
{\bi S}={\bi u}\times{\bi U}+{\bi W}={\bi Y}+{\bi W},
\label{eq:spinSZW}
\end{equation}
where ${\bi Y}$ and ${\bi W}$ are given by
\begin{equation}
{\bi Y}={\bi u}\times\frac{\hbar}{i}\nabla_u,\quad {\bi W}=\frac{\hbar}{2i}\left\{\nabla_\rho+\brho\times\nabla_\rho+
\brho(\brho\cdot\nabla_\rho)\right\},
\label{eq:spinZW}
\end{equation}
in the $\tan(\alpha/2)$ representation of the rotation group, as has been deduced
in previous sections. $\nabla_u$ and $\nabla_\rho$ are respectively
the gradient operators with respect to ${\bi u}$ and $\brho$ variables. These operators always commute
with the $H=i\hbar\partial/\partial t$ and ${\bi P}=-i\hbar\nabla$ operators, and therefore they are
translation invariant. This feature allows the separation of the general wave function in terms of space-time variables
and velocity-orientation variables to describe the translation invariant properties of the system.

The above spin operators satisfy the commutation relations
\begin{equation}
[{\bi Y},{\bi Y}]=i\hbar{\bi Y},\quad [{\bi W},{\bi W}]=i\hbar{\bi W},\quad [{\bi Y},{\bi W}]=0,
\label{eq:ZWWZ}
\end{equation}
and thus
\[
[{\bi S},{\bi S}]=i\hbar{\bi S}.
\]

Because we are describing the orientation of the particle by attaching to it 
a system of three unit vectors ${\bi e}_i$, whose orientation in space is described by 
variables $\brho$ or $\balpha$, then,
if at initial instant $\tau=0$ we choose the body axes coincident with the laboratory axes, the components
of the unit vectors ${\bi e}_i$ at any time are\index{orientation}
 \begin{equation}
({{\bi e}_i})_j=R_{ji}(\balpha)=\delta_{ji}\cos\alpha+n_jn_i(1-\cos\alpha)-
\epsilon_{jik}n_k\sin\alpha,
 \label{eq:14}
 \end{equation}
in the normal parametrization and also in the $\brho$ parametrization by
 \begin{equation}
({{\bi e}_i})_j=R_{ji}(\brho)={1\over1+\rho^2}\big((1-\rho^2)\delta_{ji}+
2\rho_j\rho_i-2\epsilon_{jik}\rho_k\big),
 \label{eq:15}
 \end{equation}
where the Cartesian components of the rotation axis unit vector ${\bi n}$ are:
 \begin{equation} 
n_1=\sin\theta\cos\phi,\qquad n_2=\sin\theta\sin\phi,\qquad n_3=\cos\theta,
\label{eq:ncompon}
 \end{equation}
where $\theta$ is the polar angle and $\phi$ the usual azimuth angle. Explicitly:
 \begin{eqnarray*}
{e_1}_1&=&\cos\alpha+\sin^2\theta\cos^2\phi(1-\cos\alpha),\\
{e_1}_2&=&\cos\theta\sin\alpha+\sin^2\theta\sin\phi \cos\phi(1-\cos\alpha),\\
{e_1}_3&=&-\sin\theta\sin\phi\sin\alpha+\sin\theta\cos\theta\cos\phi(1-\cos\alpha),
\\
\\
{e_2}_1&=&-\cos\theta\sin\alpha+\sin^2\theta\sin\phi \cos\phi(1-\cos\alpha),\\
{e_2}_2&=&\cos\alpha+\sin^2\theta\sin^2\phi(1-\cos\alpha),\\
{e_2}_3&=&\sin\theta\cos\phi\sin\alpha+\sin\theta\cos\theta\sin\phi(1-\cos\alpha),
\\
\\
{e_3}_1&=&\sin\theta\sin\phi\sin\alpha+\sin\theta\cos\theta\cos\phi(1-\cos\alpha),\\
{e_3}_2&=&-\sin\theta\cos\phi\sin\alpha+\sin\theta\cos\theta\sin\phi(1-\cos\alpha),\\
{e_3}_3&=&\cos\alpha+\cos^2\theta(1-\cos\alpha),
 \end{eqnarray*}  
in the $\balpha=\alpha{\bi n}$, or normal parametrization of the rotation group. 
In the $\brho=\tan(\alpha/2){\bi n}$ parametrization the body frame is
 \begin{eqnarray*}
{e_1}_1&=&(1+\rho_1^2-\rho_2^2-\rho_3^2)/(1+\rho^2),\\
{e_1}_2&=&(2\rho_1\rho_2+2\rho_3)/(1+\rho^2),\\
{e_1}_3&=&(2\rho_1\rho_3-2\rho_2)/(1+\rho^2),\\
\\
{e_2}_1&=&(2\rho_2\rho_1-2\rho_3)/(1+\rho^2),\\
{e_2}_2&=&(1-\rho_1^2+\rho_2^2-\rho_3^2)/(1+\rho^2),\\
{e_2}_3&=&(2\rho_2\rho_3+2\rho_1)/(1+\rho^2),\\
\\
{e_3}_1&=&(2\rho_1\rho_3+2\rho_2)/(1+\rho^2),\\
{e_3}_2&=&(2\rho_3\rho_2-2\rho_1)/(1+\rho^2),\\
{e_3}_3&=&(1-\rho_1^2-\rho_2^2+\rho_3^2)/(1+\rho^2),
 \end{eqnarray*}  
where $\rho^2\equiv\rho_1^2+\rho_2^2+\rho_3^2=\tan^2(\alpha/2)$.

In addition to the different components of
the spin operators $S_i$, $Y_i$ and $W_i$ in the laboratory frame, we also have another set of
spin operators. They are the spin projections on the body axes ${\bi e}_i$, {\sl i.e.}, the operators
$R_i={\bi e}_i\cdot{\bi S}$, $M_i={\bi e}_i\cdot{\bi Y}$ and $T_i={\bi e}_i\cdot{\bi W}$, respectively.
In particular, spin operators $T_i$, collecting terms from (\ref{eq:15}) and (\ref{eq:spinZW}),
take the expression
\begin{eqnarray*}
T_i=\sum_{k=1}^{k=3} (e_i)_k W_k&=&\frac{\hbar}{2i(1+\rho^2)}\sum_{k=1}^{k=3}\left((1-\rho^2)
\delta_{ik}+2\rho_i\rho_k-2\epsilon_{kij}\rho_j\right)\\
&\,&\times\left(\frac{\partial}{\partial\rho_k}+\epsilon_{klr}\rho_l\frac{\partial}{\partial\rho_r}+\rho_k(\brho\cdot\nabla_\rho)\right),
\end{eqnarray*}
and after some tedious manipulations we reach the final result, written
in vector notation as
\begin{equation}
{\bi T}=\frac{\hbar}{2i}\left\{\nabla_\rho-\brho\times\nabla_\rho+
\brho(\brho\cdot\nabla_\rho)\right\}.
\label{eq:spinT}
\end{equation}
We see, by inspection, that this result can also
be obtained from the expression of ${\bi W}$ in (\ref{eq:spinZW}), just by replacing
$\brho$ by $-\brho$, followed by a global change of sign. This is because we describe the orientation
of the particle by vector $\brho$ in the laboratory frame from the active viewpoint, {\sl i.e.}, with the laboratory reference 
frame fixed. However, its orientation with respect to the body frame is described
by the motion of the laboratory frame, whose orientation for the body is $-\brho$, 
and the global change of sign comes from the 
change from the active point of view to the passive one. 
This is the difference in the spin description in one frame or another.

It satisfies the following commutation relations
\[
[{\bi T},{\bi T}]=-i\hbar{\bi T},\quad [{\bi T},{\bi W}]=0.
\]
and in general all spin projections on the body frame $R_i$, $M_i$ and $T_i$, 
commute with all the spin projections on the laboratory frame $S_i$, $Y_i$ and $W_i$.
This is in agreement with the quantum mechanical uncertainty principle, because spin components with respect
to different frames are compatible observables.

To find eigenstates of the spin operator we have to solve equations of the form:
\[
S^2\chi({\bi u},\brho)=s(s+1)\hbar^2\chi({\bi u},\brho),\quad
S_3\chi({\bi u},\brho)=m\hbar\chi({\bi u},\brho).
\]
But we also have the orientation of the particle, and therefore the spin projections 
on the body axes. These projections commute with $S^2$ and $S_3$,
and it is possible to choose another commuting spin 
operator, like the $T_3$ operator, and therefore our 
wave function can be taken also as an eigenvector of $T_3$,
\[
T_3\chi({\bi u},\brho)=n\hbar\chi({\bi u},\brho),
\]
so that the complete commuting set of operators that describe the spin structure must 
also include spin projections on the body axes.

The spin squared operator is
 \begin{equation}
S^2={\bi Y}^2+{\bi W}^2+2{\bi Y}\cdot{\bi W},
\label{eq:24n}
 \end{equation}
and we see from (\ref{eq:ZWWZ}) that is expressed as the sum of three commuting 
terms and its eigenvectors can be obtained as the simultaneous 
eigenvectors of the three commuting operators on the right-hand side 
of (\ref{eq:24n}). Operators ${\bi Y}$ and ${\bi W}$ produce derivatives of the wave function
with respect to ${\bi u}$ and $\brho$ variables, separately.
Thus, each $\chi({\bi u},\brho)$ can again be 
separated as
 \begin{equation}
\chi({\bi u},\brho)=\sum_jU_{j}({\bi u})\,V_{j}(\brho),
\label{eq:UVpartwave}
 \end{equation}  
where the sum runs over a finite range, and where $U_{j}({\bi u})$ will be eigenfunctions of ${\bi Y}^2$
and $V_{j}(\brho)$ of ${\bi W}^2$, respectively.

Functions $U_{j}({\bi u})$ are multiples of spherical harmonics defined on the 
orientation of the velocity vector ${\bi u}$, because the ${\bi Y}$ 
operator has the structure of an orbital angular momentum in terms of 
the ${\bi u}$ variables, and thus its eigenvalues are integer numbers. The global factor
left out is an arbitrary function depending on the absolute value of the velocity $u$.

It turns out that to find the most general spinor is necessary to seek also solutions of the
$V_{j}(\brho)$ part, depending on the orientation variables. This goal will be achieved
in the next section, where we consider the action of the rotation group on itself as 
a transformation group.

\subsection{Spinor representation on SU(2)}
\index{spinors!on $SU(2)$}
\label{sec:spinors}

We shall describe now in detail the orientation part of the general wave function, $V(\brho)$.
If there is no contribution to spin from the zitterbewegung part ${\bi Y}$,
the spin operator (\ref{eq:spinSZW}) reduces to the ${\bi W}$ operator given in (\ref{eq:spinZW}).
To solve the corresponding eigenvalue equations we shall first represent the
spin operators in spherical coordinates.

If we represent vector $\brho=\tan(\alpha/2){\bi n}=r{\bi n}$ in spherical coordinates $(r,\theta,\phi)$, with 
$r=|\brho|=\tan(\alpha/2)$ and $\theta$ and $\phi$ the usual polar and azimuth
angles, respectively, then unit vector ${\bi n}$ has the Cartesian
components given in (\ref{eq:ncompon}). If from now on we take $\hbar=1$,
the spin operators (\ref{eq:spinZW}) are represented by the differential operators
 \[
W_1={1\over 2i}\left[(1+r^2)\sin\theta\,\cos\phi\,{\partial\over\partial r}+ 
\left({1\over
r}\cos\theta\,\cos\phi-\sin\phi\right){\partial\over\partial\theta}
-\left({\sin\phi\over r\sin\theta}+{\cos\theta\, 
\cos\phi\over\sin\theta}\right)\,{\partial\over\partial\phi}\right],
 \]
 \[
W_2=\frac{1}{2i}\left[(1+r^2)\sin\theta\,\sin\phi\,\frac{\partial}{\partial r}+ \left(\frac{1}{r}\cos\theta\,\sin\phi+\cos\phi\right)\frac{\partial}{\partial\theta} 
-\left({\cos\theta\,\sin\phi\over\sin\theta}-{\cos\phi\over r\sin\theta}\right) 
{\partial\over\partial\phi}\right],
 \]
 \[
W_3={1\over 2i}\left[(1+r^2)\cos\theta\,{\partial\over\partial r}-
{\sin\theta\over r}{\partial\over\partial\theta}+ 
{\partial\over\partial\phi}\right].
 \] 
The Casimir operator of the rotation group $W^2$ is:
 \[
W^2=-\frac{1+r^2}{4}\left[(1+r^2)\frac{\partial^2}{\partial r^2}+\frac{2(1+r^2)}{r}\frac{\partial}{\partial r}
+\frac{1}{r^2}\left\{\frac{\partial^2}{\partial\theta^2}+ 
\frac{\cos\theta}{\sin\theta}\frac{\partial}{\partial\theta} 
+{1\over\sin^2\theta}\frac{\partial^2}{\partial\phi^2}\right\}\right].
 \]

The up and down spin operators defined as usual by $W_{\pm}=W_1\pm iW_2$ are 
 \[
W_+=\frac{e^{i\phi}}{2i}\left[(1+r^2)\sin\theta\,\frac{\partial}{\partial r}+ 
\left(\frac{\cos\theta +ir}{r}\right)\,\frac{\partial}{\partial\theta} -
\left(\frac{r\cos\theta -i}{r\sin\theta}\right)
\frac{\partial}{\partial\phi}\right],
 \] 
 \[
W_-=\frac{e^{-i\phi}}{2i}\left[(1+r^2)\sin\theta\,\frac{\partial}{\partial r}+ 
\left(\frac{\cos\theta -ir}{r}\right)\,\frac{\partial}{\partial\theta} -
\left(\frac{r\cos\theta +i}{r\sin\theta}\right)
\frac{\partial}{\partial\phi}\right].
 \]
They satisfy the commutation relations
\[
[W_3,W_+]=W_+,\quad [W_3,W_-]=-W_-,\quad [W_+,W_-]=2W_3.
\]
We can check that $(W_i)^*=-W_i$ and $W_+=-(W_-)^*$, where $^*$ means to take the complex conjugate
of the corresponding operator.

If $F_s^m(r,\theta,\phi)$ is an eigenfunction of $W^2$ and $W_3$, it satisfies
the differential equations:
 \[
W^2F_s^m(r,\theta,\phi)=s(s+1)F_s^m(r,\theta,\phi),\quad
W_3F_s^m(r,\theta,\phi)=mF_s^m(r,\theta,\phi).
 \]
To find solutions of the above system we know that we can proceed in the following way.
Let us compute first the eigenfunctions of the form $F_s^s$. Then operator $W_+$
annihilates this state $W_+F_s^s=0$ and by acting on this function with operator
$W_-$ we can obtain the remaining eigenstates $F_s^m$ of the same irreducible representation
characterized by parameter $s$ and for $-s\le m\le s$. Then our task will
be to obtain first the $F_s^s$ functions.

Now, let us consider eigenfunctions $F_s^s$ that can be written in
separate variables as $F_s^s(r,\theta,\phi)=A(r)B(\theta)C(\phi)$. Then
 \[
W_3 A(r)B(\theta)C(\phi)=s A(r)B(\theta)C(\phi)
 \]
gives rise to
 \[
(1+r^2)\cos\theta A'BC-{\sin\theta\over r}AB'C+ABC'=2is ABC
 \]
where $A'$ is the derivative of $A$ and so on, and by dividing both sides by $ABC$ we have
 \[
(1+r^2)\cos\theta\frac{A'(r)}{A(r)}-\frac{\sin\theta}{r}\frac{B'(\theta)}
{B(\theta)}+\frac{C'(\phi)}{C(\phi)}=2is.
 \]
Now, the third term on the left-hand side must be a constant, because the remaining
terms are functions independent of $\phi$. Therefore, this term is written as $C'(\phi)/C(\phi)=ik$ and thus 
$C(\phi)=e^{ik\phi}$ up to an arbitrary constant factor. Since
$C(\phi+2\pi)=C(\phi)$ this implies that the constant $k$ must be an
integer. The other two functions satisfy
 \begin{equation}
r(1+r^2)\cos\theta A'B-\sin\theta AB'+ir(k-2s)AB=0.
\label{eq:ABecua}
 \end{equation}
If there exist solutions with real functions $A$ and $B$, then necessarily $k=2s$
so that the eigenvalue $s$ can be any integer or half integer, and
equation (\ref{eq:ABecua}) can be separated in the form:
 \begin{equation}
r(1+r^2)\frac{A'(r)}{A(r)}=\frac{\sin\theta}{\cos\theta}\,\frac{B'(\theta)}{B(\theta)}
=p=\hbox{\rm constant},
 \label{eq:condic}
 \end{equation}
where, up to constant factors, the general solution is
 \[
A(r)=\left({r^2\over 1+r^2}\right)^{p/2},\quad
B(\theta)=(\sin\theta)^p.
 \]

By acting on this solution $F_s^s\equiv A(r)B(\theta)C(\phi)$, with $W_+$,
since $W_+ F_s^s=0$, it gives:
 \[
r(1+r^2)\sin^2\theta A'B+(\sin\theta\cos\theta+ir\sin\theta)AB'-
2s(ir\cos\theta+1)AB=0.
 \]
By dividing all terms by $AB$, taking into account (\ref{eq:condic}), we get the condition
$(p-2s)(1+ir\cos\theta)=0$. Then there exist real solutions in separate variables
whenever $p=2s=k$. They are given, up to a constant factor, by
 \begin{equation}
F_s^s(r,\theta,\phi)=\left(\frac{r^2}{1+r^2}\right)^s(\sin\theta)^{2s}e^{i2s\phi}.
\label{eq:ssspinor}
 \end{equation}

For $s=1/2$ and after the action of $W_-$ we obtain the two orthogonal spinors
 \[
\Psi_{1/2}^{1/2}=\frac{r}{\sqrt{1+r^2}}\sin\theta\; e^{i\phi},\qquad
W_-\Psi_{1/2}^{1/2}=\Psi_{1/2}^{-1/2}=\frac{r\cos\theta+i}{\sqrt{1+r^2}},
 \]
that produce a two-dimensional representation of the rotation group. We can
similarly check that $W_-\Psi_{1/2}^{-1/2}=0$.

By inspection of the structure of $W_{\pm}$ operators,
if we take the complex conjugate of expression $W_+F_s^s=0$ we get
$-W_-({F_s^s})^*=0$ and therefore
$({F_s^s})^*\sim G_s^{-s}$ so that taking the complex conjugate spinors
of the above representation we obtain another pair of orthogonal $s=1/2$
spinors,
 \[
\widetilde \Psi_{1/2}^{1/2}=\frac{r\cos\theta-i}{\sqrt{1+r^2}},\qquad 
\widetilde \Psi_{1/2}^{-1/2}=\frac{r}{\sqrt{1+r^2}}\sin\theta\;
e^{-i\phi}.
 \]

The remaining representations for higher spins can thus be obtained by
the same method, or by taking tensor products of the above two-dimensional
representations. For instance, for $s=1$ we can obtain the following three
orthogonal representations. From (\ref{eq:ssspinor}) with $s=1$ and acting with the $W_-$ operator we get
\begin{eqnarray*}
\Psi_1^1=(\Psi_{1/2}^{1/2})^2 &=& \frac{r^2}{{1+r^2}}\;\sin^2\theta\;e^{i2\phi},\\
\Psi_1^0=(\Psi_{1/2}^{1/2})(\Psi_{1/2}^{-1/2})&=&\frac{r}{1+r^2}\;\sin\theta\,(i+r\cos\theta) \;e^{i\phi},\\
\Psi_1^{-1}=(\Psi_{1/2}^{-1/2})^2&=&\frac{(i+r\cos\theta)^2}{1+r^2},
 \end{eqnarray*}
that can also be obtained as the tensor product $\Psi\otimes\Psi$. 

If we work in the normal or canonical representation of the rotation
group, where the parameters are $\balpha=\alpha{\bi n}$, this amounts to replacing
the variable $r=\tan(\alpha/2)$ in terms of parameter $\alpha$ and expressing the 
differential operator $\partial/\partial r$ in terms of $\partial/\partial\alpha$, 
and then the spin operators are given by
  \[
W_1=\frac{1}{2i}\left[2\sin\theta\,\cos\phi\,{\partial\over\partial\alpha}+ 
\left({\cos\theta\,\cos\phi\over\tan(\alpha/2)}-\sin\phi\right){\partial\over\partial\theta}
-\left({\sin\phi\over\tan(\alpha/2)\sin\theta}+{\cos\theta\, 
\cos\phi\over\sin\theta}\right)\,{\partial\over\partial\phi}\right],
  \]
  \[
W_2=\frac{1}{2i}\left[2\sin\theta\,\sin\phi\,{\partial\over\partial\alpha}+ 
\left({\cos\theta\,\sin\phi\over\tan(\alpha/2)}+\cos\phi\right){\partial\over\partial\theta}
-\left(\frac{\cos\theta\,\sin\phi}{\sin\theta}-\frac{\cos\phi}
{\tan(\alpha/2)\sin\theta}\right){\partial\over\partial\phi}\right],
  \]
 \[
W_3=\frac{1}{2i}\left[2\cos\theta\,{\partial\over\partial\alpha}-
{\sin\theta\over\tan(\alpha/2)}{\partial\over\partial\theta}+ 
{\partial\over\partial\phi}\right],
 \]
  \[
W^2=-\left[{\partial^2\over\partial\alpha^2}+{1\over\tan(\alpha/2)}
{\partial\over\partial\alpha}
+{1\over 4\sin^2(\alpha/2)}\left\{{\partial^2\over\partial\theta^2}+ 
{\cos\theta\over\sin\theta}{\partial\over\partial\theta} 
+{1\over\sin^2\theta}{\partial^2\over\partial\phi^2}\right\}\right],
  \]
  \[
W_+=\frac{e^{i\phi}}{2i}\left[2\sin\theta\frac{\partial}{\partial\alpha}+ 
\left(\frac{\cos\theta}{\tan(\alpha/2)}
+i\right)\frac{\partial}{\partial\theta}-\left(\frac{\cos\theta\tan(\alpha/2)-
i}{\tan(\alpha/2)\sin\theta}\right)\frac{\partial}{\partial\phi}\right],
 \]
 \[
W_-=\frac{e^{-i\phi}}{2i}\left[2\sin\theta\frac{\partial}{\partial\alpha}+ 
\left(\frac{\cos\theta}{\tan(\alpha/2)}
-i\right)\frac{\partial}{\partial\theta}-\left(\frac{\cos\theta\tan(\alpha/2)+i}{\tan(\alpha/2)\sin\theta}\right)\frac{\partial}{\partial\phi} \right]
 \]
and the orthogonal spinors of the two two-dimensional representations can
be written as
 \begin{equation}
\Psi_{1/2}^{1/2}=i\sin\frac{\alpha}{2}\sin\theta\; e^{i\phi},\qquad
\Psi_{1/2}^{-1/2}=\cos\frac{\alpha}{2}-i\sin\frac{\alpha}{2}\cos\theta \label{eq:rot11}
 \end{equation}
and
 \begin{equation}
\widetilde\Psi_{1/2}^{1/2}=\cos\frac{\alpha}{2}+i\sin\frac{\alpha}{2}\cos\theta,\qquad
\widetilde\Psi_{1/2}^{-1/2}=-i\sin\frac{\alpha}{2}\sin\theta\;e^{-i\phi}.\label{eq:rot22}
 \end{equation}

We have mentioned that the different spinors are orthogonal. 
To endow the group manifold with a Hilbert space structure it is necessary
to define a hermitian, definite positive, scalar product. The Jacobian matrix
of variables $\brho'$ in terms of variables $\brho$ given in (\ref{eq:composrot}), 
has the determinant
\[
\det 
\left(\frac{\partial{\rho'}^i}{\partial\rho^j}\right)=\frac{(1+\mu^2)^2}{(1-
\bmu\cdot\brho)^4},
\]
and thus the transformation of the volume element
\[
d^3\rho'=\frac{(1+\mu^2)^2}{(1-
\bmu\cdot\brho)^4}\,d^3\rho.
\]
We also get from (\ref{eq:composrot}) that
\[
1+{\rho'}^2=\frac{(1+\mu^2)}{(1-
\bmu\cdot\brho)^2}\,(1+\rho^2)
\]
and then the measure
\[
\frac{d^3\rho'}{(1+{\rho'}^2)^2}=\left(\frac{(1-\bmu\cdot\brho)^2}{(1+\mu^2)(1+\rho^2)}\right)^2
\frac{(1+\mu^2)^2}{(1-\bmu\cdot\brho)^4}d^3\rho
=\frac{d^3\rho}{(1+\rho^2)^2}
\]
is in fact an invariant measure.

In spherical coordinates it is written as
\[
\frac{r^2\sin\theta}{(1+r^2)^2}\;dr d\theta d\phi
\]
and in the normal representation is
\[
\sin^2(\alpha/2)\sin\theta d\alpha d\theta d\phi.
\]

Since the rotation group is a double-connected group, the above measure must be defined
on a simply connected manifold, {\sl i.e.}, on the universal covering group of $SO(3)$, 
which is $SU(2)$. The $SU(2)$ group manifold in the
normal representation is given by the three-dimensional sphere of radius $2\pi$ and where points on the surface
of this sphere represent a unique $SU(2)$ element, namely the $2\times2$ unitary matrix $-\ID$. 
The normalized invariant measure becomes
 \begin{equation}
d\mu_N(\alpha,\theta,\phi)\equiv\frac{1}{4\pi^2}\,\sin^2(\alpha/2)\sin\theta\, d\alpha\,d\theta\, d\phi.
\label{eq:normmeas}
 \end{equation}  

Therefore, the hermitian scalar product will be defined as
 \begin{equation}
<f|g>=\frac{1}{4\pi^2}\int_0^{2\pi}d\alpha\int_0^{\pi} d\theta\int_0^{2\pi} 
d\phi\;f^*(\alpha,\theta,\phi)g(\alpha,\theta,\phi)\sin^2(\alpha/2)\sin\theta,
\label{eq:scalrpro}
 \end{equation}  
where $f^*$ is the complex conjugate function of $f$. 

All the previous computed spinors are orthogonal vectors with respect to the group invariant
measure (\ref{eq:normmeas}). In particular,
the normalized $s=1/2$ spinors are those given in (\ref{eq:rot11})-(\ref{eq:rot22}), 
multiplied by $\sqrt{2}$.

The spin projection operators on the body axis ${\bi e}_i$ linked to the particle, are given
in (\ref{eq:spinT}) in the $\brho$ parametrization, and we have seen
that they differ from the spin operators ${\bi W}$ only in the change of $\brho\to-\brho$,
and a global change of sign. 
In the normal parametrization this corresponds to the change $\alpha\to -\alpha$, followed
by a global change of sign.

It can be checked as mentioned before, that
 \begin{equation}
[T_i,T_k]=-i\epsilon_{ikl}\, T_l,
\label{eq:Tspincomm}
 \end{equation} 
 \begin{equation}
[W_i,T_k]=0.
\label{eq:TSspincomm}
 \end{equation} 

Since $W^2=T^2$ we can find simultaneous 
eigenvectors of the operators
$W^2$, $W_3$ and $T_3$, which will be denoted by
$D_{mn}^{(s)}(\balpha)$ in such a way that
 \begin{eqnarray*}
W^2 D_{mn}^{(s)}(\balpha)&=&s(s+1)D_{mn}^{(s)}(\balpha),\\
W_3 D_{mn}^{(s)}(\balpha)&=&m D_{mn}^{(s)}(\balpha),\\
T_3 D_{mn}^{(s)}(\balpha)&=&n D_{mn}^{(s)}(\balpha).
 \end{eqnarray*} 

Since $W_3(\alpha)D_{mn}^{(s)}(\alpha)=mD_{mn}^{(s)}(\alpha)$, by producing the change
$\alpha\to-\alpha$ we get $W_3(-\alpha)D_{mn}^{(s)}(-\alpha)=mD_{mn}^{(s)}(-\alpha)$
and the subsequent global change of sign
it reduces to
\[
-W_3(-\alpha)D_{mn}^{(s)}(-\alpha)=T_3(\alpha)D_{mn}^{(s)}(-\alpha)=-mD_{mn}^{(s)}(-\alpha),
\]
so that the above spinors (\ref{eq:rot11})-(\ref{eq:rot22}) are also eigenvectors of $T_3$.

With this notation, the four normalized spinors,
denoted by the corresponding eigenvalues $|s,m,n>$, are
 \begin{eqnarray}
\Phi_1&=& |1/2,1/2,1/2>\qquad=\sqrt{2} (\cos(\alpha/2)+
i\cos\theta\sin(\alpha/2)), \label{eq:Fi1}\\
\Phi_2&=& |1/2,-1/2,1/2>\quad\;=i\sqrt{2} \sin(\alpha/2)\,\sin\theta 
e^{-i\phi},\label{eq:Fi2}\\
\Phi_3&=& |1/2,1/2,-1/2>\quad\;=i\sqrt{2} \sin(\alpha/2)\,\sin\theta 
e^{i\phi}.\label{eq:Fi3}\\
\Phi_4&=& |1/2,-1/2,-1/2>\;\;=\sqrt{2} 
(\cos(\alpha/2)-i\cos\theta\sin(\alpha/2)),\label{eq:Fi4}\\
\end{eqnarray}
They form an orthonormal set with respect to the normalized invariant 
measure (\ref{eq:normmeas}) and with the scalar product defined in (\ref{eq:scalrpro}).
We can check that the lowering operators $W_-\Phi_1=\Phi_2$, $W_-\Phi_2=0$, $W_-\Phi_3=\Phi_4$, $W_-\Phi_4=0$, and simmilarly
$T_-\Phi_1=0$, $T_-\Phi_3=\Phi_1$, $T_-\Phi_2=0$, and $T_-\Phi_4=\Phi_2$, and the corresponding up relations when acting with
the rising operators $W_+$ and $T_+$, respectively. Remark that because the opposite sign in the commutation
relations of the $T_i$ operators, here the $T_{\pm}$ operate in the reverse direction.

The important feature is that if the system has spin $1/2$, although
the $s=1/2$ irreducible representations of the rotation group are
two-dimensional, to describe the spin part of the wave function we need
a function defined in the above four-dimensional complex Hilbert space,
because to describe orientation we attach some local frame to the particle, and therefore in
addition to the spin values in the laboratory frame we also have as
additional observables the spin projections in the body axes, which can 
be included within the set of commuting operators.

\subsection{Matrix representation of internal observables}

The matrix representation of any observable $A$ that acts on the orientation variables 
or in this internal
four-dimensional space spanned by these spin 1/2 wave functions $\Phi_i$, is obtained 
as $A_{ij}=<\Phi_i|A\Phi_j>$, $i,j=1,2,3,4$. Once these four normalized basis 
vectors are fixed, when acting on the subspace they span, the 
differential operators $W_i$ and $T_i$ have the $4\times4$ block
matrix representation 
 \begin{equation}
{\bi S}\equiv{\bi W}={\hbar\over2}\pmatrix{\bsigma&0\cr 0&\bsigma\cr},
 \label{eq:q31Z}
 \end{equation}
 \begin{equation}
 T_1={\hbar\over2}\pmatrix{0&\ID\cr \ID&0\cr},\quad 
 T_2={\hbar\over2}\pmatrix{0&i\ID\cr -i\ID&0\cr},\quad 
 T_3={\hbar\over2}\pmatrix{\ID&0\cr 0&-\ID\cr},
 \label{eq:q32T}
 \end{equation}  
where $\bsigma$ are the three Pauli matrices and $\ID$ represents the 
$2\times2$ unit matrix. We have included Planck's constant into the angular momentum operators.

If we similarly compute the matrix elements of the nine components of 
the unit vectors $({\bi e}_i)_j$, $i,j=1,2,3$ 
we obtain the nine traceless hermitian matrices
 \begin{equation} 
{\bi e}_1={1\over3}\pmatrix{0&\bsigma\cr \bsigma&0\cr},\;
{\bi e}_2={1\over3}\pmatrix{0&i\bsigma\cr -i\bsigma&0\cr},\;
{\bi e}_3={1\over3}\pmatrix{\bsigma&0\cr 0&-\bsigma\cr}.
 \label{eq:q33T}
 \end{equation}  
We can check that the $T_i={\bi S}\cdot{\bi e}_i={\bi e}_i\cdot{\bi S}$.
We see that the different components of the unit vectors ${\bi e}_i$,
in general do not commute. The eigenvalues of every component ${e_i}_j$, in this matrix representation
of definite spin, are $\pm1/3$. However, the matrix representation of the square of any component
is $({e_i}_j)^2=\ID/3$, so that the magnitude squared of each 
vector ${\bi e}_i^2=\sum_j({e_i}_j)^2=\ID$ 
when acting on these wave functions. 
The eigenvalues of the squared operator $({e_i}_j)^2$ are not the
squared eigenvalues of ${e_i}_j$.
This is because the function ${e_i}_j\Phi_k$ does not belong 
in general to the same space spanned by the
$\Phi_k$, $k=1,\ldots,4$ although this space is invariant space for operators $W_i$ and $T_j$.
In fact, each function ${e_i}_j\Phi_k$ is a linear combination of a spin $1/2$ and a spin $3/2$
wave function. 

We do not understand why any component of a classical unit vector ${e_i}_j$ of a Cartessian frame, 
can have as eigenvalues $\pm1/3$ in the quantum case and its square $({e_j}_j)^2=\ID/3$ instead
of $\ID/9$.

\subsection{Peter-Weyl theorem for compact groups}
\index{Peter-Weyl theorem}
\label{sec:Pet-Weyl}

The above spinors can also be obtained by making use of an important
theorem for representations of compact groups, known as the Peter-Weyl
theorem,~\footnote{\hspace{0.1cm}N. Ja. Vilenkin, {\sl Fonctions sp\'eciales et 
Th\'eorie de la repr\'esentation des groupes}, Dunod, Paris (1969), p. 39. \\A.O. 
Barut and R. Raczka, {\sl Theory of group representations and applications},
PWN-Polish Scientific Publishers, Warszawa (1980), p. 174.\\ F. Peter and H. Weyl, {\sl 
Math. Ann.} {\bf 7}, 735 (1927).} which is stated without proof that can be read in 
any of the mentioned references.

\vglue 0.5cm

\begin{quotation}
\noindent\normalsize
{\bf Theorem.- } Let $D^{(s)}(g)$ be a complete system of non-equivalent, unitary,
irreducible representations of a compact group $G$, labeled by the parameter $s$.
Let $d_s$ be the dimension of each representation and $D^{(s)}_{ij}(g)$, $1\le 
i,j\le d_s$ the corresponding matrix elements. Then, the functions 
\[
\sqrt{d_s}\,D^{(s)}_{ij}(g),\quad 1\le i,j\le d_s
 \]
form a complete orthonormal system on $G$, with respect to some normalized invariant 
measure $\mu_N(g)$ defined on this group, {\sl i.e.},
\begin{equation}
\int_G \;\sqrt{d_s}\,D^{(s)*}_{ij}(g)\;\sqrt{d_r}\,D^{(r)}_{kl}(g)\; d\mu_N(g)=\delta^{sr}\delta_{ik}\delta_{jl}. 
\end{equation}
\end{quotation}
\vglue 0.5cm

\noindent
That the set is complete means that every square integrable function defined on $G$, $f(g)$, 
admits a series expansion, convergent in norm, in terms of the above orthogonal functions
$D^{(s)}_{ij}(g)$, in the form
\[
f(g)=\sum_{s,i,j}\;a^{(s)}_{ij}\,\sqrt{d_s}\,D^{(s)}_{ij}(g),
\]
where the coefficients, in general complex numbers $a^{(s)}_{ij}$, are obtained by
 \[
a^{(s)}_{ij}=\int_G \sqrt{d_s}\,D^{(s)*}_{ij}(g)\, f(g) d\mu_N(g).
 \]
In our case $SU(2)$, as a group manifold, is the simply connected three-dimensional sphere
of radius $2\pi$, with the normalized measure as seen before (\ref{eq:normmeas}),
 \[
d\mu_N(\alpha,\theta,\phi)={1\over4\pi^2}\sin\theta\sin(\alpha/2)^2\, d\alpha d\theta d\phi.
 \]  

In the normal parametrization, the two-dimensional representation of $SU(2)$ corresponds to the 
eigenvalue $s=1/2$ of $S^2$ and the matrix representation is given by
\[
D^{(1/2)}(\balpha)=\cos(\alpha/2)\ID-i\sin(\alpha/2){({\bi u}\cdot\bsigma)},
\]
{\sl i.e.},
 \[
D^{(1/2)}(\balpha)=\pmatrix{\cos(\alpha/2)-i\cos\theta\sin(\alpha/2)&-
i\sin\theta\sin(\alpha/2)\,e^{-i\phi}\cr -
i\sin\theta\sin(\alpha/2)\,e^{i\phi}&\cos(\alpha/2)+i\cos\theta 
\sin(\alpha/2)\cr}.
 \] 

If we compare these four matrix components with the four orthogonal spinors given in
(\ref{eq:Fi1})-(\ref{eq:Fi4}) we see that
\begin{equation}
D^{(1/2)}(\balpha)=\frac{1}{\sqrt{2}}\pmatrix{\Phi_4&-\Phi_2\cr
-\Phi_3&\Phi_1\cr}
\label{eq:matDa}
\end{equation}
In the three-dimensional representation of $SO(3)$, considered as a representation of SU(2)
 \[
D^{(1)}_{ij}(\balpha)=\delta_{ij}\cos\alpha+u_iu_j(1-\cos\alpha)+\epsilon_{ikj}u_k\sin\alpha\equiv {e_j}_i
 \] 
we get another set of nine orthogonal functions. Multiplied by $\sqrt{3}$ they form another orthonormal set
orthogonal to the previous four spinors. It is a good exercise to check this orthogonality among
these functions.

\subsection{General spinors}
\index{general spinors}\index{spinors}
\label{sec:generalspinors}

In the case that the zitterbewegung content of the spin is not vanishing
we can also obtain spin 1/2 wave-functions as the irreducible representations
contained in the tensor product of integer and half-integer spin states coming from the $U({\bi u})$
and $V(\brho)$ part of the general wave function (\ref{eq:UVpartwave}).

The total spin operator of the system is of the form
\[
{\bi S}={\bi u}\times{\bi U}+{\bi W}={\bi Y}+{\bi W},
\]
where ${\bi Y}=-i\hbar\nabla_u$ and ${\bi W}$ is given in (\ref{eq:spinZW}).
Spin projections on the body axes, {\sl i.e.}, operators $T_i={\bi e}_i\cdot{\bi W}$, are
described in (\ref{eq:spinT}). They satisfy the commutation relations
\[
[{\bi Y},{\bi Y}]=i{\bi Y},\quad [{\bi W},{\bi W}]=i{\bi W},\quad [{\bi T},{\bi T}]=i{\bi T},
\]
\[
[{\bi Y},{\bi W}]=0,\quad [{\bi Y},{\bi T}]=0,\quad [{\bi W},{\bi T}]=0.
\]
These commutation relations are invariant under the change $\brho$ by $-\brho$ in the definition of
the operators ${\bi W}$ and ${\bi T}$, because they are changed into each other. 
The expression of the body frame unit vectors ${\bi e}_i$
is given in (\ref{eq:14}) and (\ref{eq:15}).

We can see that these unit vector components and spin operators $W_i$ and $T_j$
satisfy the following properties:

{\bf 1)$\;$} ${e_i}_j(-\alpha,\theta,\phi)=-{e_j}_i(\alpha,\theta,\phi)$.

{\bf 2)$\;$} ${\bi e}_i\cdot{\bi W}\equiv\sum_j {e_i}_jW_j=T_i$.

{\bf 3)$\;$} $\sum_j {\bi e}_j T_j={\bi W}$.

{\bf 4)$\;$} For all $i,j$, the action $W_i {e_j}_i=0$, with no addition on index $i$.

{\bf 5)$\;$}  For all $i,j$, the action $T_i {e_i}_j=0$, with no addition on index $i$.

{\bf 6)$\;$}  For all $i,j,k$, with $i\neq j$, we have that $W_i {e_k}_j+W_j {e_k}_i=0$, and in the
case that $i=j$, it leads to property 4.

{\bf 7)$\;$}  For all $i,j,k$, with $i\neq j$, we have that $T_i {e_j}_k+T_j {e_i}_k=0$, and similarly
as before in the case $i=j$  it leads to property 4.

This implies that ${\bi e}_i\cdot{\bi W}={\bi W}\cdot{\bi e}_i=T_i$, because of property 4, since when 
acting on an arbitray function $f$, 
\[
({\bi W}\cdot{\bi e}_i) f\equiv \sum_j W_j ({e_i}_j f)=f \sum_j W_j ({e_i}_j)+\sum_j {e_i}_j W_j (f)=T_i(f),
\]
because $\sum_j W_j{e_i}_j=0$.

In the same way $\sum_j {\bi e}_j T_j\equiv\sum_j T_j{\bi e}_j ={\bi W}$.

Now we fix the value of spin. Particles of different values of spin can
be described. Let us consider systems that take the lowest admissible spin values.
For spin 1/2 particles, if we take first for simplicity 
eigenfunctions $V(\brho)$ of $W^2$ with eigenvalue 1/2, 
and then since the total spin has to be 1/2, the orbital ${\bi Y}$ 
part can only contribute with spherical harmonics of value $y=0$ and 
$y=1$. 

If there is no zitterbewegung spin, $y=0$, and
Wigner's functions can be 
taken as simultaneous eigenfunctions of the three commuting $W^2$, 
$W_3$, and $T_3$ operators, and the normalized 
eigenvectors $|w,w_3,t_3>$ are explicitly given by the functions (\ref{eq:Fi1}-\ref{eq:Fi4}).

If we have a zitterbewegung spin of value $y=1$, then the $U({\bi u})$ part contributes with 
the spherical harmonics
\begin{eqnarray}
Y_1^1(\tilde\theta,\tilde\phi)&\equiv&|1,1>=-\sin(\tilde\theta)e^{i\tilde\phi}\sqrt{\frac{3}{8\pi}},
\label{eq:esferic1}\\
Y_1^0(\tilde\theta,\tilde\phi)&\equiv&|1,0>=\cos(\tilde\theta)\sqrt{\frac{3}{4\pi}},
\label{eq:esferic2}\\
Y_1^{-1}(\tilde\theta,\tilde\phi)&\equiv&|1,-1>=\sin(\tilde\theta)e^{-i\tilde\phi}\sqrt{\frac{3}{8\pi}},
\label{eq:esferic3}
\end{eqnarray}
normalized with respect to the measure
\[
\int_{0}^\pi \int_{0}^{2\pi} \sin(\tilde\theta) d\tilde\theta d\tilde\phi,
\]
which are the indicated eigenfunctions $|y,y_3>$
of ${\bi Y}^2$ and $Y_3$, and where the variables $\tilde\theta$ and $\tilde\phi$
determine the orientation of the velocity ${\bi u}$.

The tensor product representation of the rotation group constructed
from the two irreducible representations ${\bf 1}$ associated to the 
spherical harmonics (\ref{eq:esferic1})-(\ref{eq:esferic3}) 
and ${\bf 1/2}$ given in (\ref{eq:Fi1})-(\ref{eq:Fi4}) is split
into the direct sum
${\bf 1}\otimes{\bf 1/2}={\bf 3/2}\oplus{\bf 1/2}$. 

The following functions of five variables $\tilde\theta$,
$\tilde\phi$, $\alpha$, $\theta$ and $\phi$, where variables $\tilde\theta$ and
$\tilde\phi$ correspond to the ones of the spherical harmonics $Y_l^m$, 
and the remaining $\alpha$, $\theta$ and $\phi$, 
to the previous spinors $\Phi_i$, are normalized spin 1/2 
functions $|s,s_3,t_3>$ that are eigenvectors of total spin $S^2$, and $S_3$ and $T_3$
operators
\begin{eqnarray}
\Psi_1 &\equiv&|1/2,1/2,1/2>\qquad =\frac{1}{\sqrt{3}}\left(Y_1^0 \Phi_1-\sqrt{2}Y_1^1 \Phi_2 \right),
\label{eq:psi1}
\\
\Psi_2 &\equiv&|1/2,-1/2,1/2>\quad\; =\frac{1}{\sqrt{3}}\left(-Y_1^0 \Phi_2+\sqrt{2}Y_1^{-1} \Phi_1 \right),
\label{eq:psi2}
\\
\Psi_3 &\equiv&|1/2,1/2,-1/2>\quad\; =\frac{1}{\sqrt{3}}\left(Y_1^0 \Phi_3-\sqrt{2}Y_1^1 \Phi_4 \right), 
\label{eq:psi3}
\\
\Psi_4 &\equiv&|1/2,-1/2,-1/2>\;\;=\frac{1}{\sqrt{3}}\left(-Y_1^0 \Phi_4+\sqrt{2}Y_1^{-1} \Phi_3 \right),
\label{eq:psi4}\quad
\end{eqnarray}
such that $\Psi_2=S_-\Psi_1$ and similarly $\Psi_4=S_-\Psi_3$, and also that
$\Psi_3=T_-\Psi_1$, and $\Psi_4=T_-\Psi_2$. 
They are no longer eigenfunctions of the $W_3$ operator, although they span
an invariant vector space for $S^2$, $S_3$ and $T_3$
operators. In the above basis (\ref{eq:psi1})-(\ref{eq:psi4}) formed by orthonormal vectors $\Psi_i$, the matrix
representation of the spin is
 \begin{equation}
{\bi S}={\bi Y}+{\bi W}={\hbar\over2}\pmatrix{\bsigma&0\cr 0&\bsigma\cr},
 \label{eq:qs31}
 \end{equation}
while the matrix representation of the ${\bi Y}$ and ${\bi W}$ part is
 \begin{equation}
{\bi Y}={2\hbar\over3}\pmatrix{\bsigma&0\cr 0&\bsigma\cr},\quad
{\bi W}={-\hbar\over6}\pmatrix{\bsigma&0\cr 0&\bsigma\cr},
 \label{eq:qlz31}
 \end{equation}
which do not satisfy commutation relations of angular momentum operators because
the vector space spanned by the above basis is not an invariant space for these operators ${\bi Y}$
and ${\bi W}$.

The spin projection of the ${\bi W}$ part on the body axis,
{\sl i.e.}, the ${\bi T}$ operator, takes the same form as before (\ref{eq:q32T})
 \begin{equation}
 T_1={\hbar\over2}\pmatrix{0&\ID\cr \ID&0\cr},\quad 
 T_2={\hbar\over2}\pmatrix{0&i\ID\cr -i\ID&0\cr},\quad 
 T_3={\hbar\over2}\pmatrix{\ID&0\cr 0&-\ID\cr},
 \label{eq:qt32}
 \end{equation}
because $\Psi_1$ and $\Psi_2$ functions are eigenfunctions of $T_3$ with eigenvalue
$1/2$, while $\Psi_3$ and $\Psi_4$ are of eigenvalue $-1/2$, and thus the spinors $\Psi_i$
span an invariant space for $S_i$ and $T_j$ operators. In fact the basis is formed
by simultaneous eigenfunctions of total spin $S^2$, $S_3$ and $T_3$, and the ket representation
is the same as in the case of the $\Phi_i$ given in (\ref{eq:Fi1})-(\ref{eq:Fi4}).

The expression in this basis of the components of the unit vectors ${\bi
e}_i$ are represented by
 \begin{equation} 
{\bi e}_1=-{1\over9}\pmatrix{0&\bsigma\cr \bsigma&0\cr},\;
{\bi e}_2=-{1\over9}\pmatrix{0&i\bsigma\cr-i\bsigma&0\cr},\;
{\bi e}_3=-{1\over9}\pmatrix{\bsigma&0\cr 0&-\bsigma\cr}.
 \label{eq:qs33}
 \end{equation}

\chapter{Dirac equation and analysis of Dirac algebra}

\section{Quantization of the $u=c$ model}
For Luxons we have the nine-dimensional 
homogeneous space of the Poincar\'e group, spanned by the ten variables 
$(t,{\bi r},{\bi u},\balpha)$ similarly as before, but 
now ${\bi u}$ is restricted to $u=c$. For this system, since ${\bi 
u}\cdot\dot{\bi u}=0$ and $\dot{\bi u}\neq0$, we are describing particles with a circular 
internal orbital motion at the constant speed $c$.

In the center of mass frame, (see Fig.\ref{fig:electronu}) the 
center of charge describes a circle of radius $R_0=S/mc$ at the 
constant speed $c$, the spin being 
orthogonal to the charge trajectory plane 
and a constant of the motion in this frame. 
Let us consider the quantization of this $u=c$ model
whose dynamical equation is given by (\ref{eq:elecdina}).

If we analyse this particle in the centre of mass frame it becomes a system of three degrees of freedom.
These are the $x$ and $y$ coordinates of the point charge on the plane and the phase $\alpha$
of the rotation of the body axis with angular velocity $\omega$. But this phase is the same as the phase of the
orbital motion, as we shall see later, and because this motion is a circle of constant radius only one degree of freedom is left,
for instance the $x$ coordinate.
In the centre of mass frame the system is equivalent to a one-dimensional harmonic oscillator 
of angular frequency $\omega=mc^2/S$ in its ground state. 

Identification of the ground energy 
of the one-dimensional harmonic oscillator $\hbar\omega/2$ with the
rest energy of the system in the center of mass frame $+mc^2$, for $H>0$ particles, implies that
the classical constant parameter $S=\hbar/2$. All Lagrangian systems defined with this kinematical space,
irrespective of the particular Lagrangian we choose,
have this behaviour and represent spin $1/2$ particles when quantized.

\section{Dirac's equation}
\index{Dirac's equation}
\label{sec:Diracequ}

The kinematical variables of this system transform under ${\cal P}$ according to 
 \begin{eqnarray}
t'(\tau)&=&\gamma t(\tau)+\gamma({\bi v}\cdot R(\bmu){\bi r}(\tau))/c^2+b,
\label{eq:et}\\ 
{\bi r}'(\tau)&=&R(\bmu){\bi r}(\tau)+\gamma{\bi v}t(\tau)+\frac{\gamma^2}
{(1+\gamma)c^2}({\bi v}\cdot R(\bmu){\bi r}(\tau)){\bi v}+{\bi a},\qquad\qquad{}
\label{eq:er}\\
{\bi u}'(\tau)&=&\frac{{R(\bmu){\bi u}(\tau)+\gamma{\bi v}+({\bi v}\cdot 
R(\bmu){\bi u}(\tau)){\bi v}\gamma^2/(1+\gamma)c^2}}{{\gamma(1+{\bi v}
\cdot R(\bmu){\bi u}(\tau)/c^2)}},\label{eq:eu}\\ 
\brho'(\tau)&=&\frac{{\bmu+\brho(\tau)+\bmu\times\brho(\tau)+
{\bi F}_c({\bi v},\bmu;{\bi u}(\tau),\brho(\tau))}}{{1-\bmu\cdot\brho(\tau)
+G_c({\bi v},\bmu;{\bi u}(\tau),\brho(\tau))}},
 \label{eq:ero}
 \end{eqnarray}

\cfigl{fig:electronu}{electronu.eps}{Motion of the charge in the C.M. frame.}

\noindent where the functions ${\bi F}_c$ and $G_c$ are given in (\ref{eq:F_c}) 
and (\ref{eq:G_c}), respectively.
When quantized, the wave function of the system is a function 
$\Phi(t,{\bi r},{\bi u},\brho)$ of these kinematical variables. For 
the Poincar\'e group all exponents and thus all gauge functions on 
homogeneous spaces are equivalent to zero, and the Lagrangians for 
free particles can thus be taken strictly invariant. Projective 
representations reduce to true representations so that the ten 
generators on the Hilbert space, taking into account 
(\ref{eq:et})-(\ref{eq:ero}) and (\ref{eq:uvalfa}) are given by:
 \begin{equation} 
H=i\hbar\frac{\partial}{\partial t},\quad {\bi P}=\frac{\hbar}{i}\nabla,
\quad {\bi K}={\bi r}\,\frac{i\hbar}{c^2}\,\frac{\partial}{\partial t}-t\,
\frac{\hbar}{i}\nabla-\frac{1}{c^2}{\bi S}\times{\bi u},
 \end{equation}
 \begin{equation}
{\bi J}={\bi r}\times\frac{\hbar}{i}\,\nabla +{\bi S},
 \end{equation}
where as we shall see, the angular momentum operator ${\bi S}$ represents Dirac's spin operator
and is given by the differential operator
 \begin{equation}
{\bi S}={\bi u}\times\frac{\hbar}{i}\,\nabla_u+\frac{\hbar}{2i}\,
\left\{\nabla_\rho+\brho\times\nabla_\rho+\brho(\brho\cdot\nabla_\rho) 
\right\}={\bi u}\times{\bi U}+{\bi W}={\bi S}_u+{\bi S}_\omega,
 \end{equation}
and where the differential 
operators $\nabla_u$ and $\nabla_\rho$ are the corresponding gradient 
operators with respect to the ${\bi u}$ and $\brho$ variables as in 
the Galilei case. The operator ${\bi S}$ is not a constant of the motion
even for the free particle, and although it is not the angular momentum of the system with respect
to its center of mass we keep this notation because it is the equivalent to Dirac's spin operator.
Of course, it reduces to the true spin ${\bi S}$ in the center of mass frame.

To obtain the complete commuting set of observables we start with the 
Casimir invariant operator, or Klein-Gordon operator
 \begin{equation}
H^2-c^2{\bi P}^2=m^2c^4.
 \label{eq:20}
 \end{equation}
In the above representation, $H$ and ${\bi P}$ only differentiate the wave 
function with respect to time $t$ and position ${\bi r}$, respectively. Since the 
spin operator ${\bi S}$ operates only on the velocity and orientation variables, it 
commutes with the Klein-Gordon operator (\ref{eq:20}). Thus, we can 
find simultaneous eigenfunctions of (\ref{eq:20}), $S^2$, and $S_3$. 
This allows us to try solutions in separate variables so that the wave 
function can be written
 \begin{equation}
\Phi(t,{\bi r},{\bi u},\brho)=\sum_i \psi_i(t,{\bi r})
\chi_i({\bi u},\brho),
 \end{equation}
where $\psi_i(t,{\bi r})$ are the space-time components and the $\chi_i({\bi u},\brho)$ 
represent the internal spin structure. Consequently
 \begin{equation}
(H^2-c^2{\bi P}^2-m^2c^4)\,\psi_i(t,{\bi r})=0,
 \label{eq:21}
 \end{equation}
{\sl i.e.}, space-time components satisfy the Klein-Gordon equation, while the internal structure 
part satisfies
 \begin{equation}
S^2\chi_i({\bi u},\brho)=s(s+1)\hbar^2\chi_i({\bi u},\brho),
 \label{eq:22}
 \end{equation}
 \begin{equation}
S_3\chi_i({\bi u},\brho)=m_s\hbar\chi_i({\bi u},\brho).
 \label{eq:23}
 \end{equation}  
Eigenfunctions of the above type have been found in Section \ref{sec:genspinors}, in
particular we are interested in solutions that give rise to spin $1/2$ particles.
These solutions, which are also eigenvectors of the spin projection on the body axis $T_3$,
become a four-component wave function.

For spin 1/2 particles, if we take first for simplicity 
eigenfunctions $\chi(\brho)$ of $S^2$ with eigenvalue 1/2, 
then since the total spin has to be 1/2, the orbital zitterbewegung part ${\bi Y}={\bi u}\times{\bi U}$ 
can only contribute with spherical harmonics of value $y=0$ and 
$y=1$. This means that we can find at least two different kinds of elementary particles of spin $1/2$,
one characterized by the singlet $y=0$ (lepton?) and another by $y=1$ (quark?) in three possible states according to the
component $y_3$. If we call to the spin part $Y$ the {\bf colour}, we can have colourless and coloured 
systems of spin $1/2$. The three different colours $y_3$ are unobservable because the $\Psi_i$
states (\ref{eq:psi1}-\ref{eq:psi4}) are eigenstates of $S_3$ and $T_3$ but not eigenstates of $Y_3$. 
Nevertheless this interpretation of this spin part as representing the colour, 
as in the standard model, is still unclear and will be discussed in more detail in the forthcoming
workshop Spin05 next week.

For $y=0$, the spin 1/2 functions $\chi_i(\brho)$ are linear combinations 
of the four $\Phi_i$ functions (\ref{eq:Fi1})-(\ref{eq:Fi4})
and in the case $y=1$ they are linear combinations of the four $\Psi_i$
of (\ref{eq:psi1})-(\ref{eq:psi4}), such that the factor function in front
of the spherical harmonics is 1 because for this model $u=c$ is a constant.
It turns out that the Hilbert space that describes the internal structure of 
this particle is isomorphic to the four-dimensional Hilbert space 
$\CC^4$. 

If we have two arbitrary directions in space characterized by the unit 
vectors ${\bi u}$ and ${\bi v}$ respectively, and $S_{\bi u}$ and 
$S_{\bi v}$ are the corresponding angular momentum projections $S_{\bi u}={\bi 
u}\cdot{\bi S}$ and $S_{\bi v}={\bi v}\cdot{\bi S}$, then $S_{-{\bi 
u}}=-S_{\bi u}$, and $[S_{\bi u},S_{\bi v}]=i\hbar S_{{\bi 
u}\times{\bi v}}$. In the case of the opposite sign commutation relations of operators $T_i$, 
we have for instance for the spin projections $[T_1,T_2]=-i\hbar T_3$, thus suggesting that 
${\bi e}_1\times{\bi e}_2=-{\bi e}_3$, and any cyclic permutation $1\to2\to3$, 
and thus ${\bi e}_i$ vectors linked to the body, not only have as eigenvalues $\pm1/3$, 
but also behave in the quantum case as a left-handed system. In 
this case ${\bi e}_i$ vectors are not arbitrary vectors in space, but 
rather vectors linked to the rotating body and thus they are not 
compatible observables, so that any measurement to 
determine, say the components of ${\bi e}_i$, will produce some interaction with the body 
that will mask the measurement of the others. We shall use 
this interpretation of a left-handed system for particles later, when we analyse the chirality
in section \ref{sec:chirality}. For antiparticles it will behave as a right handed one. 

Operators $S_i$ and $T_i$ have the 
matrix representation obtained before which is just
 \begin{equation}
{\bi S}\equiv{\bi W}={\hbar\over2}\pmatrix{\bsigma&0\cr 0&\bsigma\cr},
 \label{eq:espinS}
 \end{equation}
 \begin{equation}
 T_1={\hbar\over2}\pmatrix{0&\ID\cr \ID&0\cr},\quad 
 T_2={\hbar\over2}\pmatrix{0&i\ID\cr -i\ID&0\cr},\quad 
 T_3={\hbar\over2}\pmatrix{\ID&0\cr 0&-\ID\cr},
 \label{eq:spinTm}
 \end{equation}  
where we represent by $\bsigma$ the three Pauli matrices and $\ID$ is the 
$2\times2$ unit matrix. 

Similarly, the matrix elements of the nine components of 
the unit vectors $({\bi e}_i)_j$, $i,j=1,2,3$ give rise to the two alternative sets
of representations depending on whether the zitterbewegung contribution is $y=0$ or $y=1$. In the first case
we get
 \begin{equation} 
{\bi e}_1={1\over3}\pmatrix{0&\bsigma\cr \bsigma&0\cr},\;
{\bi e}_2={1\over3}\pmatrix{0&i\bsigma\cr -i\bsigma&0\cr},\;
{\bi e}_3={1\over3}\pmatrix{\bsigma&0\cr 0&-\bsigma\cr},
 \label{eq:q33}
 \end{equation}  
while in the $y=1$ case the representation is
 \begin{equation} 
{\bi e}_1=-{1\over9}\pmatrix{0&\bsigma\cr \bsigma&0\cr},\;
{\bi e}_2=-{1\over9}\pmatrix{0&i\bsigma\cr-i\bsigma&0\cr},\;
{\bi e}_3=-{1\over9}\pmatrix{\bsigma&0\cr 0&-\bsigma\cr}.
 \label{eq:qeijz=1}
 \end{equation}
It must be remarked that the different components of the observables 
${\bi e}_i$ are not compatible in general, because they are represented 
by non-commuting operators.

We finally write the wave function for spin 1/2 particles in the 
following form for $y=0$
 \begin{equation}
\Phi_{(0)}(t,{\bi r},{\bi u},\balpha)=\sum_{i=1}^{i=4}\psi_i(t,{\bi r})\Phi_i(\alpha,\theta,\phi),
 \end{equation}  
independent of the ${\bi u}$ variables, and
in the case $y=1$ by
 \begin{equation}
\Phi_{(1)}(t,{\bi r},{\bi u},\balpha)=\sum_{i=1}^{i=4}\psi_i(t,{\bi r})\Psi_i(\widetilde\theta,\widetilde\phi;\alpha,\theta,\phi).
 \end{equation}  
where $\widetilde\theta$ and $\widetilde\phi$ represent the direction of vector ${\bi u}$.
Then, once the $\Phi_i$ or $\Psi_j$ functions that describe the internal structure 
are identified with the four orthogonal unit vectors of the internal 
Hilbert space $\CC^4$, the wave function becomes a four-component 
space-time wave function, and the six spin components $S_i$ and $T_j$ 
and the nine vector components $({{\bi e}_i})_j$, together the 
$4\times4$ unit matrix, completely exhaust the 16 linearly independent 
$4\times4$ hermitian matrices. They form a vector basis of Dirac's algebra, 
such that any other translation invariant internal observable that describes internal 
structure, for instance internal velocity and acceleration, angular velocity, etc.,
must necessarily be expressed as a real linear combination of the mentioned 
16 hermitian matrices. We shall see in Sec.~\ref{sec:Diracalgebra} that the internal orientation completely 
characterizes its internal structure. 

The spin operator ${\bi S}={\bi u}\times{\bi U}+{\bi W}$ which, as seen in (\ref{eq:qs31})
and (\ref{eq:espinS}), coincides with the usual matrix representation of Dirac's spin operator.

If we consider the expression of the kinematical momentum for $u=c$ particles 
\[
{\bi K}=\frac{H}{c^2}{\bi r}-t{\bi P}-\frac{1}{c^2}{\bi S}\times{\bi u}
\]
and we take the time derivative of this expression followed by the 
scalar product with ${\bi u}$, it leads to the Poincar\'e invariant operator 
(Dirac's operator):
 \begin{equation}
H-{\bi P}\cdot{\bi u}-{1\over c^2}\left({d{\bi u}\over dt}
\times{\bi u}\right)\cdot{\bi S}=0.
 \label{eq:q34}
 \end{equation} 

When Dirac's operator acts on a general wave function $\Phi_{(0)}$ or $\Phi_{(1)}$, we know 
that $H$ and ${\bi P}$ have the differential representation given by 
(\ref{eq:16}) and the spin the differential representation 
(\ref{eq:spinZW}), or the equivalent matrix representation (\ref{eq:espinS}), 
but we do not know how to represent the action of the velocity ${\bi 
u}$ and the $(d{\bi u}/dt)\times{\bi u}$ observable. However, we know 
that for this particle ${\bi u}$ and $d{\bi u}/dt$ are orthogonal 
vectors and together with vector ${\bi u}\times d{\bi u}/dt$ they form an 
orthogonal right-handed system, and in the center of mass frame the 
particle describes a circle of radius $R_0=\hbar/2mc$
in the plane spanned by ${\bi u}$ and $d{\bi u}/dt$.

\cfigl{fig:PauliDirac}{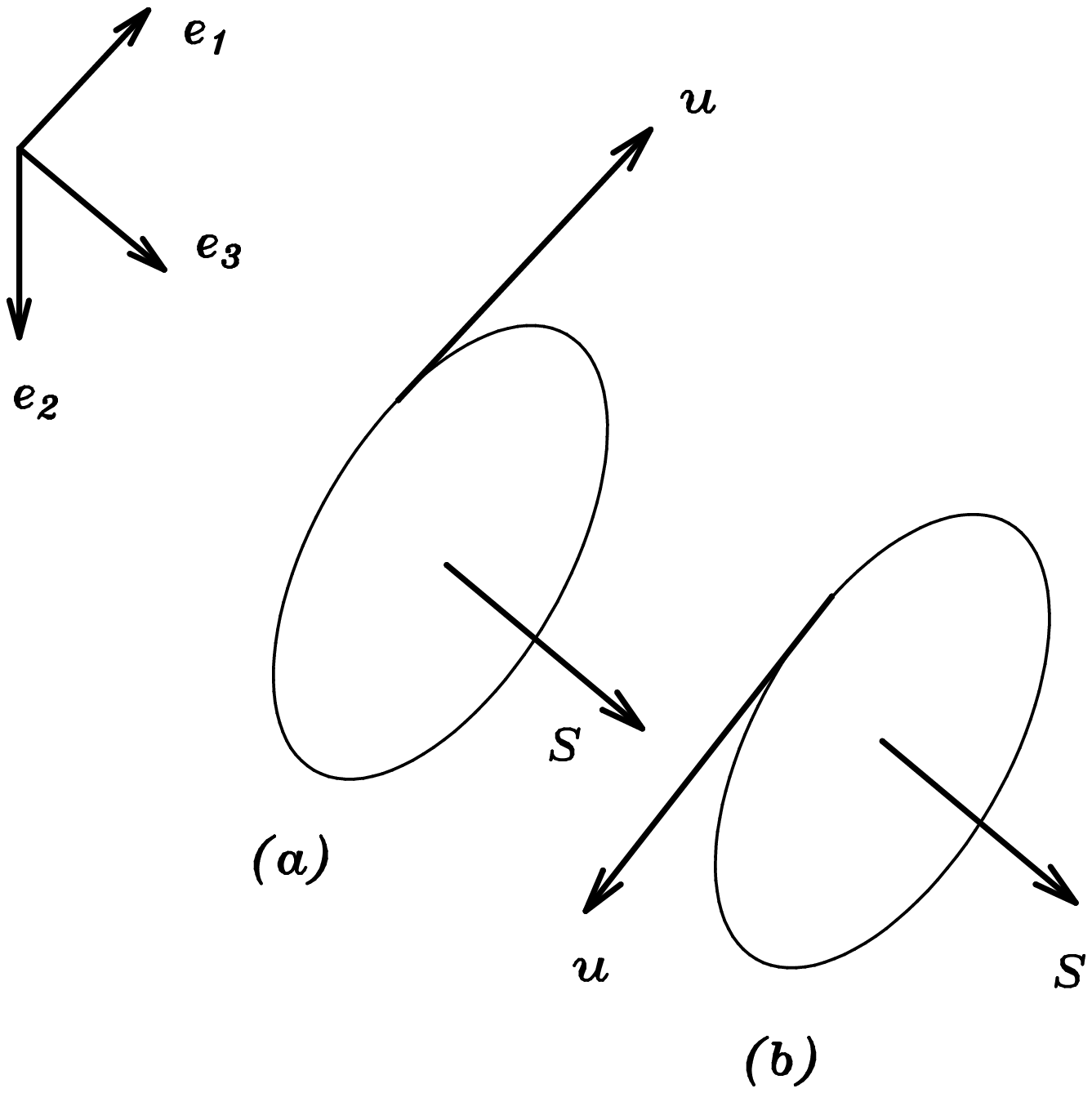}{Representation of the local body frame and the different observables for 
the (a) $H>0$ solution and (b) $H<0$ solution. 
This orientation produces Dirac equation in the Pauli-Dirac representation}

Let us consider first the case $y=0$.
Since ${\bi u}$ and $d{\bi u}/dt$ are translation invariant observables they 
will be elements of Dirac's algebra, and it turns out that we can
relate these three vectors with the left-handed orthogonal system 
formed by vectors ${\bi e}_1$, ${\bi e}_2$ and ${\bi e}_3$ with representation
(\ref{eq:q33}). Then,
as shown in part  $(a)$ of Figure \ref{fig:PauliDirac} for the $H>0$ system, we have ${\bi u}=a{\bi e}_1$ 
and $d{\bi u}/dt\times {\bi u}=b{\bi e}_3$, where $a$ and $b$ are constant
positive real numbers. Then the third term in Dirac's operator is $
(b/c^2){\bi e}_3\cdot{\bi S}=(b/c^2) T_3$, and (\ref{eq:q34}) 
operator becomes
 \begin{equation}
H-a{\bi P}\cdot{\bi e}_1-\frac{b}{c^2} T_3=0.
 \label{eq:q35}
 \end{equation}  
If we make the identification with the $H<0$ solution
of part $(b)$ of Figure \ref{fig:PauliDirac}, the relation of the above
observables is opposite to the previous one but now with the coefficients $-a$
and $-b$, respectively, {\sl i.e.}, we get
 \begin{equation}
H+a{\bi P}\cdot{\bi e}_1+\frac{b}{c^2} T_3=0,
 \label{eq:q36}
 \end{equation}  
which clearly corresponds to the change $H\to -H$ in equation (\ref{eq:q35}).

Multiplying (\ref{eq:q36}) by (\ref{eq:q35}) we obtain an expression whihch is satisfied by both
particle and antiparticle
 \begin{equation} 
H^2-{a^2\over 9} {\bi P}^2\ID-{b^2\hbar^2\over4c^4}\ID=0,
 \end{equation}
and which is an algebraic relation between $H^2$ and $P^2$.
By identification of this expression with the 
Klein-Gordon operator (\ref{eq:20}), which also contains both $H>0$ and $H<0$ solutions,
leads to $a=3c$ and $b=2mc^4/\hbar=c^3/R_0$ and by substitution in (\ref{eq:q35})
we obtain Dirac's equation:
 \begin{equation}
H-c{\bi P}\cdot\balpha-\beta mc^2=0,
 \end{equation}
where Dirac's matrices $\balpha$ and $\beta$ are represented by 
 \begin{equation}
\balpha=\pmatrix{0&\bsigma\cr \bsigma&0\cr},\quad \beta=\pmatrix{\ID&0\cr 0&-\ID\cr},
 \end{equation}
and thus Dirac's gamma matrices are\index{Dirac's gamma matrices}\index{gamma matrices}
 \begin{equation}
\gamma^0\equiv\beta=\pmatrix{\ID&0\cr 0&-\ID\cr},\quad 
\bgamma\equiv\gamma^0\balpha=\pmatrix{0&\bsigma\cr -\bsigma&0\cr},
 \end{equation}
{\sl i.e.}, Pauli-Dirac representation,\index{Pauli-Dirac representation}
where $3{\bi e}_1$ plays the role of a unit vector in the direction of the velocity. Substitution
into (\ref{eq:q36}) corresponds to the equivalent representation with the change $\gamma^\mu\to-\gamma^\mu$.

This representation is compatible with the acceleration $d{\bi u}/dt$ lying 
along the vector ${\bi e}_2$. In fact, in the center of mass 
frame and in the Heisenberg representation, Dirac's Hamiltonian reduces to
$H=\beta mc^2$, and the time derivative of any observable $A$ is 
obtained as
 \begin{equation} 
\frac{dA}{dt}=\frac{i}{\hbar}[H,A]+\frac{\partial A}{\partial t},
 \end{equation}
such that for the velocity operator ${\bi u}=c\balpha$,
 \begin{equation} 
{d{\bi u}\over dt}={i\over\hbar}[mc^2\beta,c\balpha]= {2mc^3\over\hbar} 
\pmatrix{0&i\bsigma\cr-i\bsigma&0\cr}={c^2\over R_0} 3 {\bi e}_2,
 \end{equation}
$c^2/R_0$ being the constant modulus of the acceleration in this frame, 
and where $3{\bi e}_2$ plays the role of a unit vector along that direction.

The time derivative of this Cartesian system is 
 \begin{eqnarray}
{d{\bi e}_1\over dt}&=&{i\over\hbar}[\beta mc^2,{\bi e}_1]=
{c\over R_0}{\bi e}_2,\label{eq:e1}\\
{d{\bi e}_2\over dt}&=&{i\over\hbar}[\beta mc^2,{\bi e}_2]=-
{c\over R_0}{\bi e}_1,\\ {d{\bi e}_3\over dt}&=&{i\over\hbar}
[\beta mc^2,{\bi e}_3]=0,\label{eq:e3}
 \end{eqnarray}
since ${\bi e}_3$ is orthogonal to the 
trajectory plane and does not change, and where $c/R_0=\omega$ is the 
angular velocity of the internal orbital motion. This time evolution of the 
observables ${\bi e}_i$ is the correct one if assumed to be a rotating left-handed 
system of vectors as shown in Figure \ref{fig:PauliDirac}-$(a)$. It is for this reason
that we considered at the beginning of this chapter that the body frame rotates
with the same angular velocity as the orbital motion of the charge.

To be consistent with the above consideration as $3{\bi e}_i$ as unit vectors, this means that the spin
in the center of mass frame should be along $3{\bi e}_3$. This is the case for the upper components
while for the lower components (which in this representation correspond to $H<0$ states) the orientation
is the opposite. This means that for particles the corresponding set of axis forms a left handed system
while for antiparticles they behave as a right handed system, showing a clear chirality difference 
between particles and antiparticles.

In general
 \[
\frac{d{\bi S}}{dt}=\frac{i}{\hbar}[H,{\bi S}]=
\frac{i}{\hbar}[c{\bi P}\cdot\balpha+\beta mc^2,{\bi S}]=
c{\bi P}\times{\balpha}\equiv{\bi P}\times{\bi u},
 \]
is not a constant of the motion, but for the 
center of mass observer, this spin operator ${\bi u}\times{\bi U}+{\bi W}$ reduces to the equivalent of the classical 
spin of the particle ${\bi S}$ and is constant in this frame:
 \begin{equation}
\frac{d{\bi S}}{dt}=\frac{i}{\hbar}[\beta mc^2,{\bi S}]=0.
 \end{equation} 

Only the $T_3$ spin component on the body axis remains constant while the 
other two $T_1$ and $T_2$ change because of the rotation of the 
corresponding axis, 
 \begin{eqnarray}
{d T_1\over dt}&=&{i\over\hbar}[\beta mc^2, T_1]={c\over R_0} T_2,\\
{d T_2\over dt}&=&{i\over\hbar}[\beta mc^2, T_2]=-{c\over R_0} T_1,\\
{d T_3\over dt}&=&{i\over\hbar}[\beta mc^2, T_3]=0.
 \end{eqnarray} 

When analyzed from the point of view of an arbitrary observer, the classical 
motion is a helix and the acceleration is not of 
constant modulus $c^2/R_0$, and the spin operator ${\bi S}$ is no longer a constant of the 
motion, because it is the total angular momentum ${\bi J}={\bi r}\times{\bi 
P}+{\bi S}$ that is conserved. 

Identification of the internal variables with different real linear 
combinations of the ${\bi e}_i$ matrices lead to different equivalent 
representations of Dirac's matrices, and thus to different
expressions of Dirac's equation.

\cfigl{fig:Weyl}{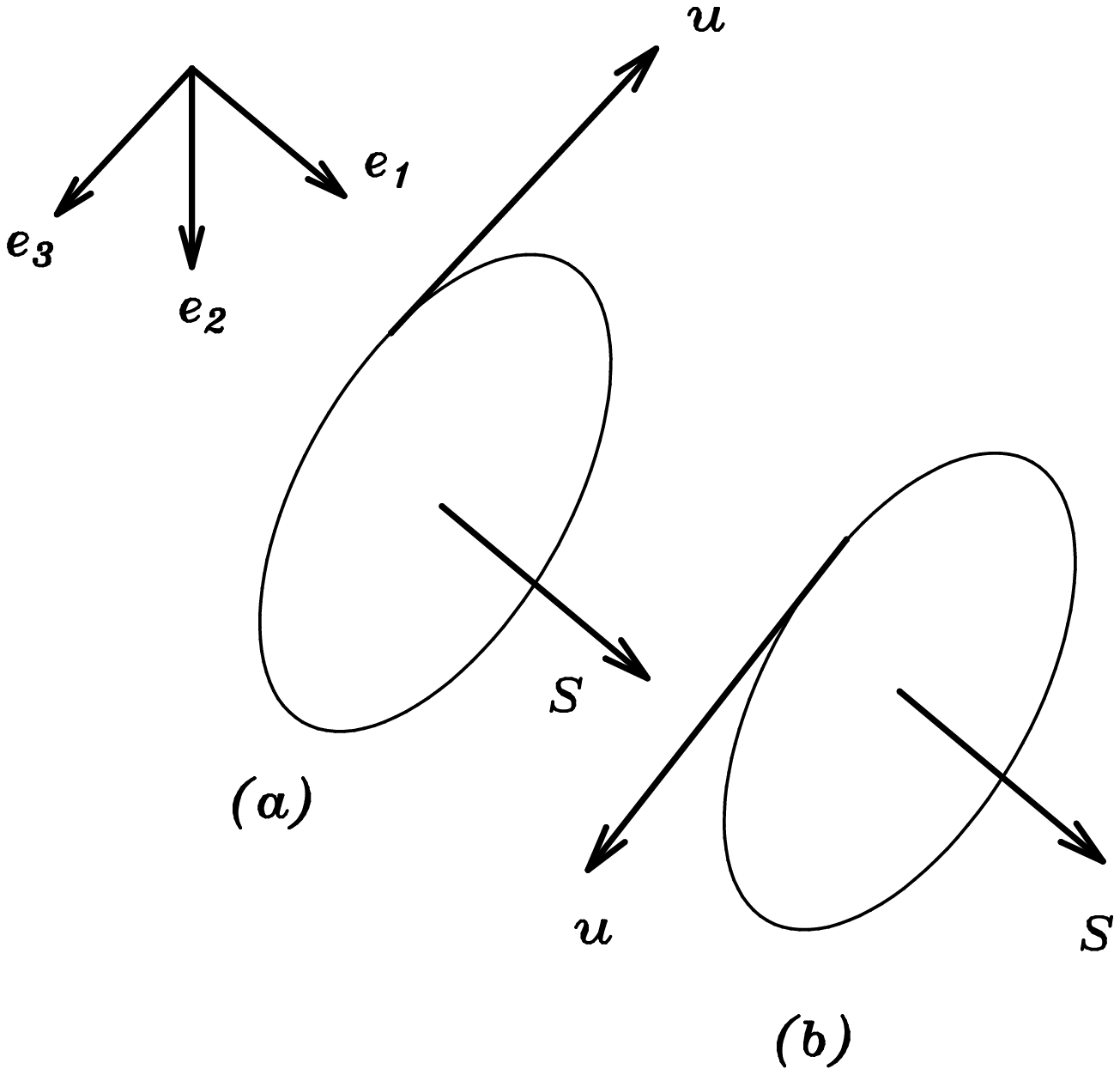}{Orientation in the Weyl representation.}

For instance if we make the identification suggested by Figure~\ref{fig:Weyl}, 
${\bi u}=-a{\bi e}_3$ and the observable 
$d{\bi u}/dt\times {\bi u}=b{\bi e}_1$ with positive constants $a$ and $b$, 
we obtain by the same method
 \begin{equation}
\beta=\pmatrix{0&\ID\cr \ID&0\cr},\qquad 
\balpha=\pmatrix{-\bsigma&0\cr 0&\bsigma\cr},
 \end{equation}
and thus gamma matrices \index{gamma matrices}
 \begin{equation}
\gamma^0\equiv\beta=\pmatrix{0&\ID\cr \ID&0\cr},\qquad 
\bgamma\equiv\gamma^0\balpha=\pmatrix{0&\bsigma\cr -\bsigma&0\cr},
 \end{equation}  
{\sl i.e.}, Weyl's representation. \index{Weyl's representation}

When we compare both representations, we see that Weyl's 
representation is obtained from Pauli-Dirac representation if we 
rotate the body frame $\pi/2$ around ${\bi e}_2$ axis. Then the 
corresponding rotation operator 
 \[
R(\pi/2,{\bi e}_2)=\exp(\frac{i}{\hbar}\frac{\pi}{2}
{\bi e}_2\cdot{\bi S})=\exp(\frac{i}{\hbar}\frac{\pi}{2}T_2)=
\frac{1}{\sqrt{2}}\pmatrix{\ID&-\ID\cr \ID&\ID\cr}.
 \]

We can check that $R\,\gamma^\mu_{PD}\,R^{\dagger}=\gamma^{\mu}_{W}$, where 
$\gamma^{\mu}_{PD}$ and $\gamma^{\mu}_{W}$ are gamma matrices in the 
Pauli-Dirac and Weyl representation, respectively.

We can similarly obtain Dirac's equation in the case of zitterbewegung $y=1$,
by using the set of matrices (\ref{eq:qeijz=1}) instead of (\ref{eq:q33}), because 
they are multiples of each other and only some intermediate constant 
factor will change.

\subsection{PCT Invariance}

\cfigl{fig:inverP}{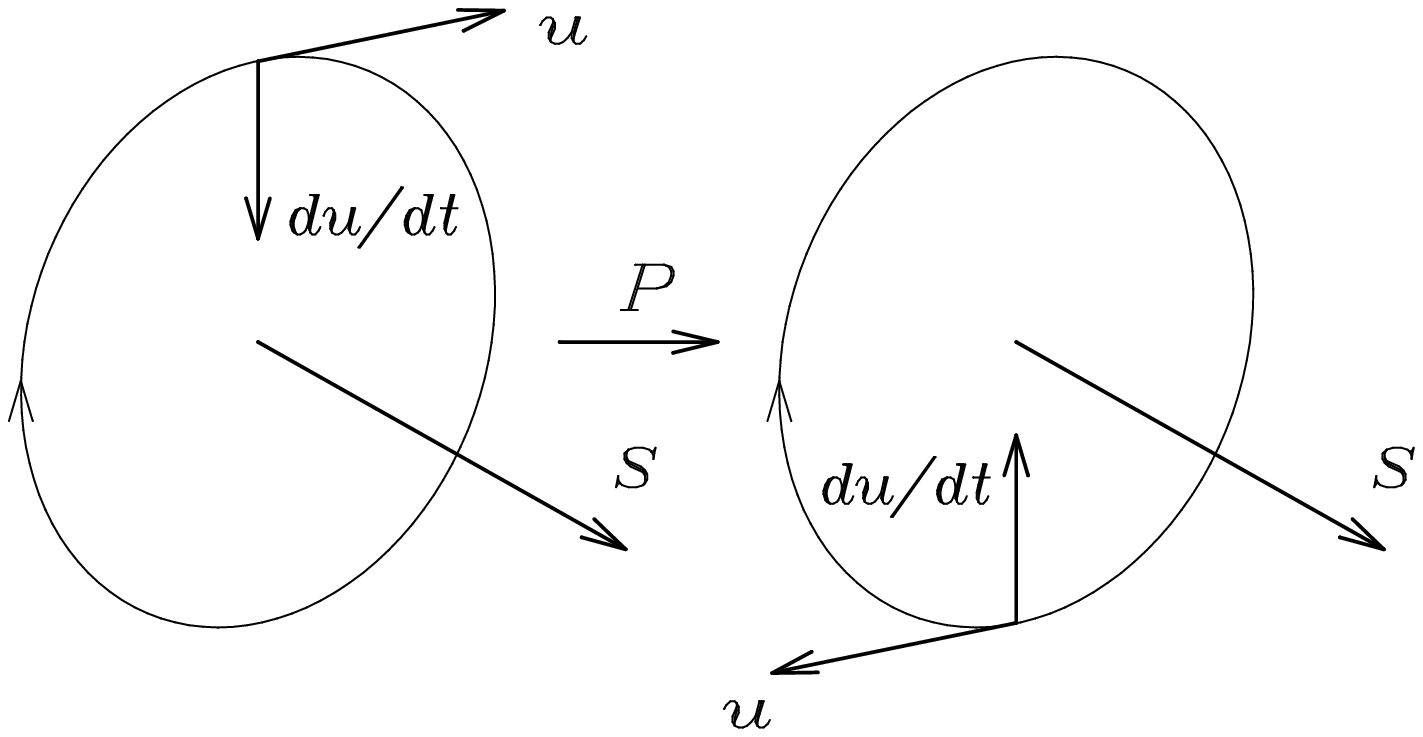}{Space reversal of the electron in the center of mass frame
is equivalent to a rotation
of value $\pi$ along $\protect{\bi S}$.}

In Figure \ref{fig:inverP} we represent the parity reversal of the 
description of the electron as given by this model of luxon which is circling around the
center of mass at the velocity $c$
and that under ${P}$ and in the center of mass frame it changes according to
\[
P:\{{\bi r}\to-{\bi r},{\bi u}\to-{\bi u},d{\bi u}/dt\to-d{\bi u}/dt,{\bi S}\to{\bi S},H\to H\}.
\]
In the Pauli-Dirac representation as we see in Figure \ref{fig:PauliDirac}, 
this amounts to a rotation of value $\pi$ around
axis ${\bi e}_3$ and thus
\[
P\equiv R(\pi,{\bi e}_3)=\exp(i\pi{\bi e}_3\cdot{\bi S}/\hbar)=\exp(i\pi T_3/\hbar)=i\gamma_0,
\]
which is one of the possible representations of the parity operator $\pm\gamma_0$ or $\pm i\gamma_0$.
In Weyl's representation this is a rotation of value $\pi$ around ${\bi e}_1$ which gives again
$P\equiv i\gamma_0$.

\cfigl{fig:inverT}{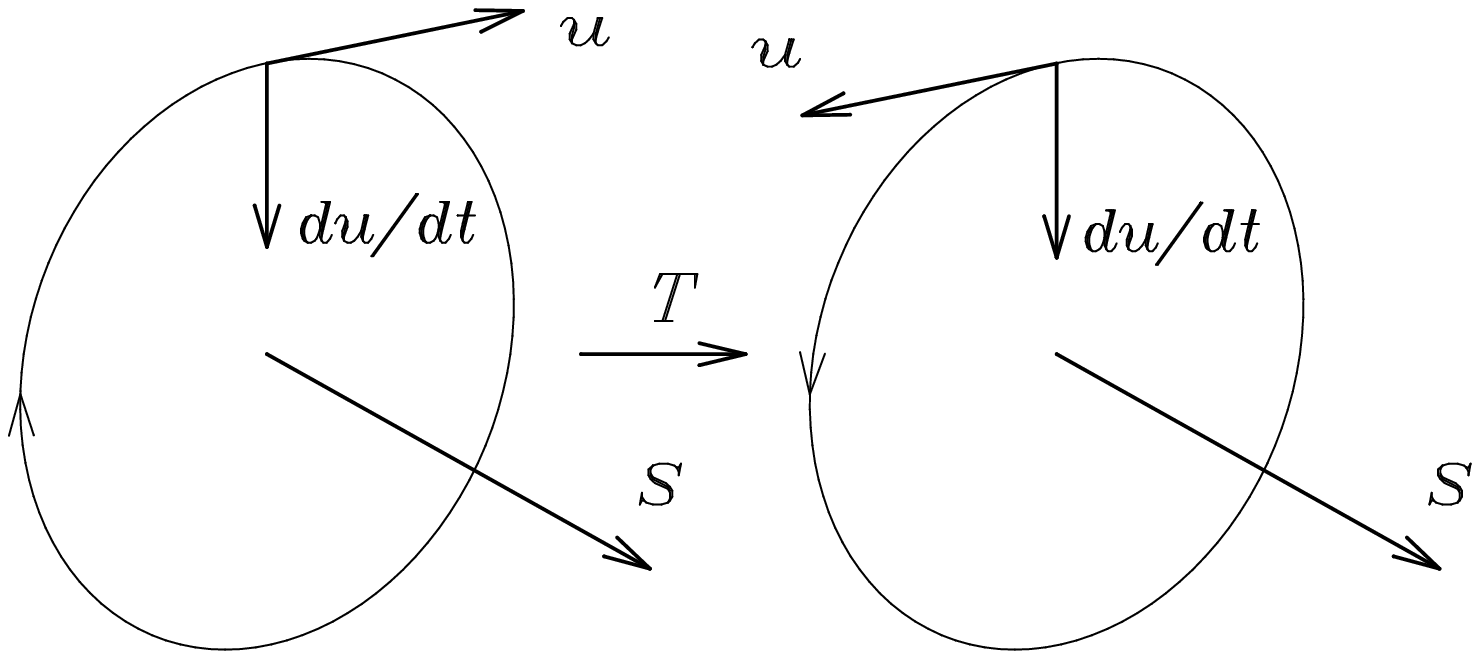}{Time reversal of the electron produces a particle of negative energy.}

In Figure \ref{fig:inverT} we represent its time reversal 
also in the center of mass frame
\[
T:\{{\bi r}\to{\bi r},{\bi u}\to-{\bi u},d{\bi u}/dt\to d{\bi u}/dt,{\bi S}\to{\bi S}, H\to -H\},
\]
but this corresponds to a particle of of $H<0$ such 
that the relative orientation
of spin, velocity and position, given by equation (\ref{eq:Smuxk}) agrees 
with the motion depicted in this figure.

A Dirac particle is a mechanical system whose intrinsic attributes are mass $m>0$ and spin $\hbar/2$.
We also see that the sign of $H$ is also Poincar\'e invariant and it is also an intrinsic property
which establishes two different systems of the same value of $m$ and $S$. The system with $H>0$ is called
the particle and the other with $H<0$ the antiparticle. The value of the mass attribute 
is introduced by hand. To characterize its interaction with an external electromagnetic field, 
we also introduce by hand another intrinsic property
the electric charge $e$, located at the point ${\bi r}$. 
This implies that in addition to the mechanical
properties $m$ and $S$ the system has as electromagnetic properties the electric charge $e$ and because of the
charge location separated from its center of mass and its motion at the speed of light, 
an electric dipole moment ${\bi d}$ and a magnetic moment $\bmu$, respectively.
The electric charge can also have either a positive or negative sign. 
\[
\pmatrix{S\cr m\cr H\cr {e}\cr \bmu\cr {\bi d}}\quad P\;\Rightarrow\quad
\pmatrix{S\cr m\cr H\cr {e}\cr \bmu\cr -{\bi d}}\quad T\;\Rightarrow\quad
\pmatrix{S\cr m\cr -H\cr e\cr -\bmu\cr -{\bi d}}\quad C\;\Rightarrow\quad
\pmatrix{S\cr m\cr -H\cr -{e}\cr \bmu\cr {\bi d}}
\]
The $PCT$ transformation transforms particle into antiparticle and conversely, while keeping
invariant the mechanical attributes $m$ and $S$ and the electromagnetic attributes $\bmu$
and ${\bi d}$. The $PCT$ invariance of the system
establishes a relationship between the sign of $H$ and the sign of $e$, although an indeterminacy
exists in the election of the sign of the charge of the particle. The product $eH$ is $PCT$ invariant.

This implies that particle and antiparticle have a magnetic moment and an oscillating electric dipole
in a plane orthogonal to the spin. Once the spin direction is fixed, the magnetic moment of both
have the same relative orientation with the spin, either parallel or antiparallel, according to the election
of the sign of the electric charge. The electric dipole moment oscillates leftwards for particles
and rightwards for antiparticles which shows a difference between them which is called {\it chirality}.
If as usual we call the electron to the system of negative electric charge {\it the particle}, 
the above $PCT$ transformation
transforms the system $(a)$ of figure \ref{fig:muyd} into the system $(b)$. If what we call the particle
is of positive electric charge, then the spin and magnetic moment are opposite to each other
for both particle and antiparticle.

\cfigl{fig:muyd}{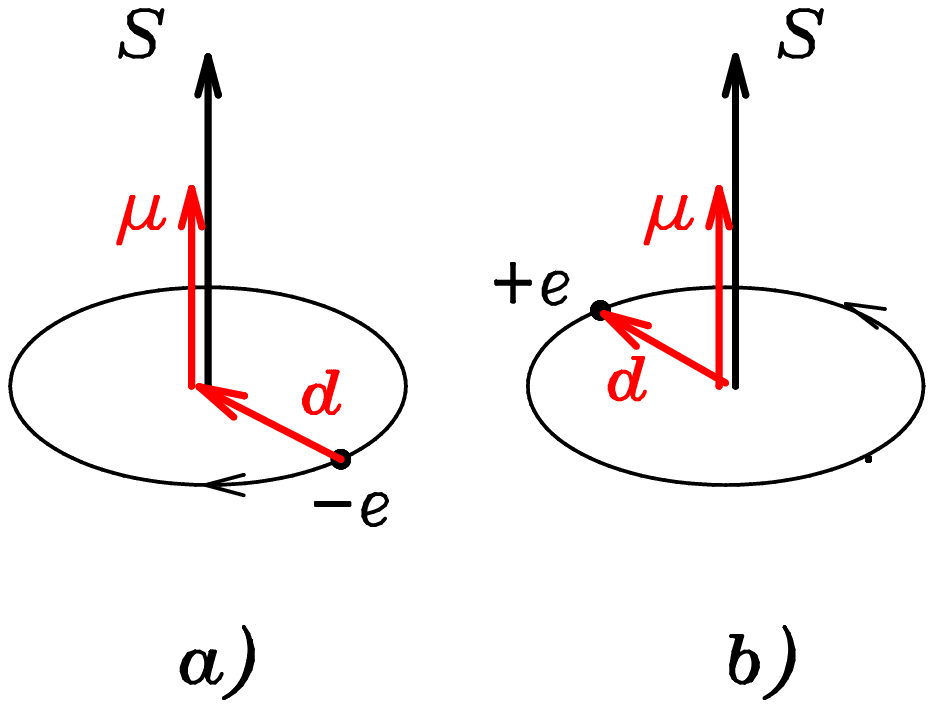}{Electromagnetic attributes ${\bmu}$ and ${\bi d}$ for $(a)$ a negatively charged particle
and its $PCT$ transformed $(b)$, and their relative orientation with the spin, in the center of mass frame.
The electric dipole of the particle oscillates leftwards and rightwards for the antiparticle.}

However, to our knowledge no explicit direct measurement of the relative orientation 
between spin and magnetic moment of the free electron, can be found in the literature although very high precision experiments
are performed to obtain the absolute value of $g$, the gyromagnetic ratio. 

A plausible indirect experiment \footnote{M.Rivas, {\it Are the electron spin and magnetic moment parallel or antiparallel vectors?}, LANL ArXiv:physics/0112057.}
has been proposed to measure the relative orientation between spin and magnetic moment for 
one outer electron atoms like Rb or Cs.

Rb$^{87}$ atoms have one electron at the level $5s$. Its nucleus has
spin $3/2$ and the ground state of the atom has a total spin 1, and therefore the outer electron
has its spin in the opposite direction to the spin of the nucleus.
The magnetic moment of the atom is basically the magnetic moment of this outer electron because
the inner shells are full and the magnetic moment of the nucleus is relatively smaller.

Ultracold Rb$^{87}$ atoms in an external magnetic field will be oriented with their magnetic moments pointing along
the field direction. If in this direction we send a beam of circularly polarized photons of sufficient energy
$\sim 6.8$GHz to produce the corresponding hyperfine transition to flip the electron spin in the opposite direction
and thus leaving the atom in a spin 2 state, only those photons with the spin opposite to the spin
of the outer electron will be absorbed. Measuring the spin orientation of the circularly polarized beam
will give us the spin orientation of the electron thus showing its relationship with
the magnetic moment orientation. 
Now the task is to check also the relative orientation for positrons.

\subsection{Chirality}
\label{sec:chirality}

The classical model which satisfies Dirac's equation when quantized gives rise to two possible
physical systems of $H>0$ and $H<0$. The $H>0$ is usually called the particle. According 
to the previous analysis the internal motion of the charge takes place on a plane 
orthogonal to the spin
direction and in a leftward sense when we fix as positive the spin direction. 
For the antiparticle the motion
is rightwards. For particles, the local orientable frame of unit vectors ${\bi e}_i$ behaves
as a left handed system rotating with an angular velocity in the opposite direction to the spin, 
while for antiparticles it can be considered as a right handed one.

\cfigl{fig:antipar}{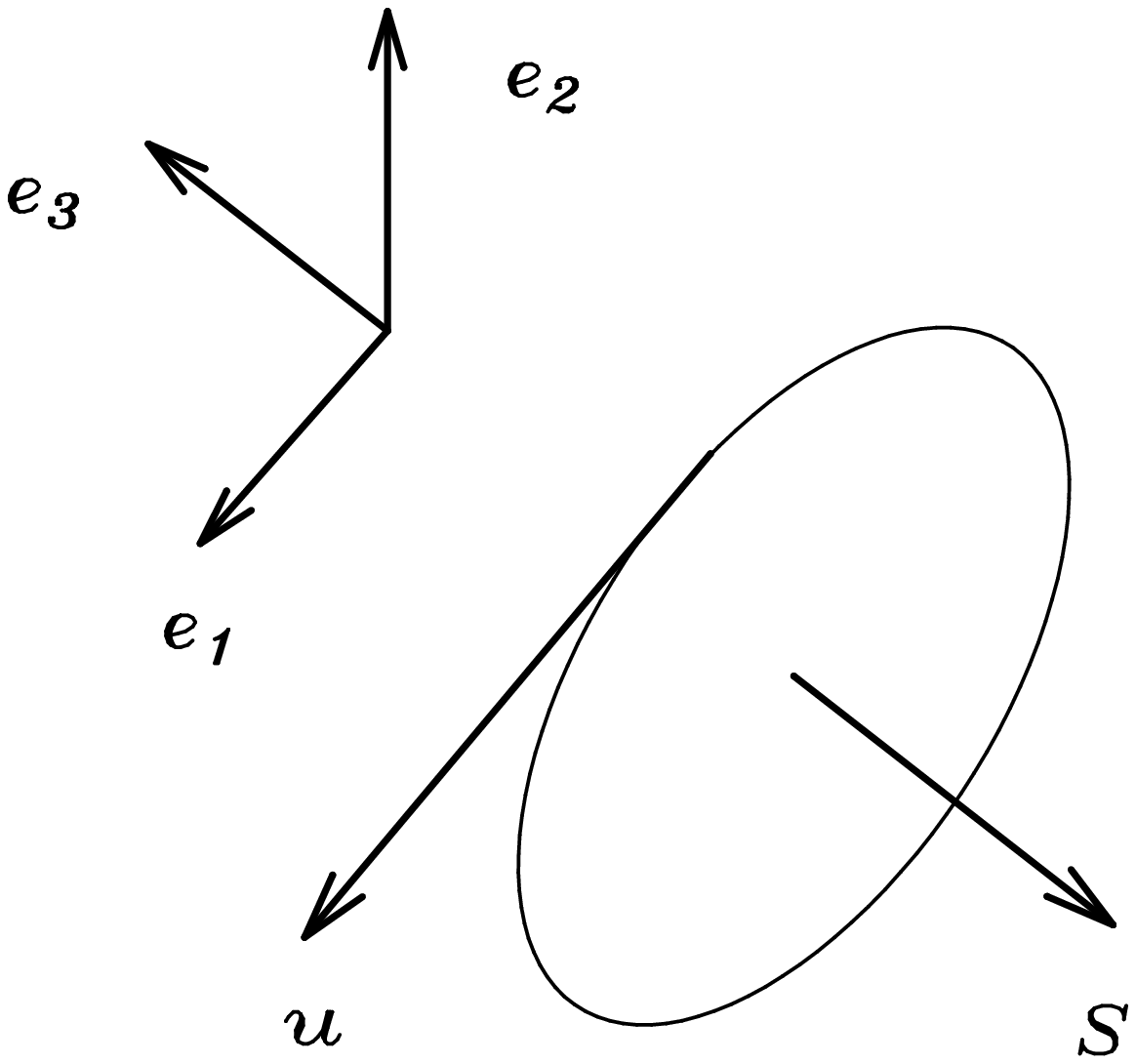}{Relative orientation of the body axis for the antiparticle that
leads to Pauli-Dirac representation.
It behaves as a rotating right handed Cartessian frame around the spin direction.}

If we should have started the analysis by considering first the antiparticle, then in order to get
the same Pauli-Dirac representation as before we have to consider the body axis as the ones depicted
in figure \ref{fig:antipar}, {\it i.e.}, in the opposite direction to the ones we chose before and this 
leads by the same arguments that the $\gamma^\mu$ matrices have to replaced by the $-\gamma^\mu$, so that the 
Hamiltonian in the center of mass frame is $-\beta mc^2$. In this way the motion of the body
frame, instead of (\ref{eq:e1}-\ref{eq:e3}) is
 \begin{eqnarray}
{d{\bi e}_1\over dt}&=&{i\over\hbar}[-\beta mc^2,{\bi e}_1]=
-\omega{\bi e}_2,\label{eq:ed1}\\
{d{\bi e}_2\over dt}&=&{i\over\hbar}[-\beta mc^2,{\bi e}_2]=
\omega{\bi e}_1,\\ {d{\bi e}_3\over dt}&=&{i\over\hbar}
[-\beta mc^2,{\bi e}_3]=0,\label{eq:ed3}
 \end{eqnarray}
with $\omega={c/R_0}$, which clearly corresponds to a rotating right handed system with an angular velocity
around the spin direction.

Matter is left and antimatter is right in this kind of models as far as the charge motion
and the rotation of the local body frame are concerned, so that particles and antiparticles
show a clear chirality. 

Although the local motion of the charge, which takes place in a region of order of Compton's wavelength,
is probably physically unobservable, this motion corresponds nevertheless 
to the oscillation of the instantaneous electric dipole moment,
which oscillates at very high frequency, but its sense of motion, once the spin direction is fixed, 
reflects this difference between particle and antiparticle. This electric dipole motion is independent
of whether the particle is positively or negatively charged. 

Finally, when we compare the spin operator and the vector ${\bi e}_3$ we see
\[
{\bi S}=\frac{\hbar}{2}\pmatrix{\bsigma&0\cr 0&\bsigma\cr},\quad {\bi e}_3={1\over3}\pmatrix{\bsigma&0\cr 0&-\bsigma\cr}.
\]
that the two upper components of the Dirac spinor correspond to positive energy solutions and therefore
the upper components of these operators are related by ${\bi S}\sim{\bi e}_3$, while the lower components correspond to negative energy solutions
and for this components these operators behave as ${\bi S}\sim -{\bi e}_3$, a vector relationship which
is clearly depicted in the figures \ref{fig:PauliDirac} and \ref{fig:antipar} respectively.

\section{Dirac's algebra}
\index{Dirac's algebra}
\label{sec:Diracalgebra}

The three spatial spin components $S_i$, the three spin projections 
on the body frame $T_j$ and the nine components of the body frame $({\bi e}_i)_j$, 
$i,j=1,2,3$, whose matrix representations are given in the $y=0$ case
in (\ref{eq:q33}) or in (\ref{eq:qeijz=1}) in the $y=1$ case, together 
with the $4\times4$ unit matrix $\ID$, form a set of 16 linearly independent 
hermitian matrices. They are a linear basis of Dirac's algebra, and satisfy 
the following commutation relations:
 \begin{equation}
[S_i,S_j]=i\hbar\epsilon_{ijk}S_k, \qquad [T_i,T_j]=-i\hbar\epsilon_{ijk}T_k,
 \qquad [S_i,T_j]=0,
 \label{eq:qA1}
 \end{equation}  
 \begin{equation}
[S_i,({\bi e}_j)_k]=i\hbar\epsilon_{ikr}({\bi e}_j)_r,\qquad 
[T_i,({\bi e}_j)_k]=-i\hbar\epsilon_{ijr}({\bi e}_r)_k,
 \label{eq:qA2}
 \end{equation}
and the scaled $3{\bi e}_i$ vectors in the $y=0$ case
 \begin{equation}
[(3{\bi e}_i)_k,(3{\bi e}_j)_l]=\frac{4i}{\hbar}\left(\delta_{ij} 
\epsilon_{klr}S_r-\delta_{kl}\epsilon_{ijr}T_r\right),
 \label{eq:qA3}
 \end{equation}
showing that the ${\bi e}_i$ operators transform like vectors under 
rotations but they are not commuting observables. In the case $y=1$, the scaled $-9{\bi e}_i$, 
satisfy the same relations.

If we fix the pair of indexes $i$, and $j$, then the set of four 
operators $S^2$, $S_i$, $T_j$ and $({\bi e}_j)_i$ form a complete 
commuting set. In fact, the wave 
functions $\Phi_i$, $i=1,\ldots,4$, given before (\ref{eq:Fi1})-(\ref{eq:Fi4}), are simultaneous 
eigenfunctions of $S^2$, $S_3$, $T_3$ and $({\bi e}_3)_3$ with 
eigenvalues $s=1/2$ and for $s_3$, $t_3$, and $e_{33}$ are the following 
ones:
 \begin{equation} 
\Phi_1=\,|1/2,1/2,1/3>,\qquad \Phi_2=\,|-1/2,1/2,-1/3>,
 \end{equation}  
 \begin{equation}
\Phi_3=\,|1/2,-1/2,-1/3>,\qquad \Phi_4=\,|-1/2,-1/2,1/3>,
 \end{equation}  
and similarly for the $\Psi_j$ spinors of (\ref{eq:psi1})-(\ref{eq:psi4})
 \begin{equation} 
\Psi_1=\,|1/2,1/2,-1/9>,\qquad \Psi_2=\,|-1/2,1/2,1/9>,
 \end{equation}  
 \begin{equation}
\Psi_3=\,|1/2,-1/2,1/9>,\qquad \Psi_4=\,|-1/2,-1/2,-1/9>.
 \end{equation}  

The basic observables satisfy the following anticommutation relations:
 \begin{equation}
\{S_i,S_j\}=\{T_i,T_j\}=\frac{\hbar^2}{2}\, \delta_{ij}\ID,
 \label{eq:qA4}
 \end{equation} 
 \begin{equation}
\{S_i,T_j\}={\hbar^2\over2}\,(3{\bi e}_j)_i,
 \label{eq:qA5}
 \end{equation} 
 \begin{equation}
\{S_i,(3{\bi e}_j)_k\}=2\,\delta_{ik} T_j,\qquad \{T_i,(3{\bi e}_j)_k\} 
=2\,\delta_{ij} S_k,
 \label{eq:qA6}
 \end{equation} 
 \begin{equation}
\{({\bi e}_i)_j,({\bi e}_k)_l\}={2\over9}\,\delta_{ik} 
\delta_{jl}\ID+{2\over3}\epsilon_{ikr}\epsilon_{jls}({\bi e}_r)_s.
 \label{eq:qA7}
 \end{equation} 

If we define the dimensionless normalized matrices:
 \begin{equation}
a_{ij}=3({\bi e}_i)_j,\;(\hbox{\rm or}\,a_{ij}=-9({\bi e}_i)_j),\qquad s_i={2\over\hbar}S_i,\qquad t_i={2\over\hbar}T_i,
 \label{eq:qA8}
 \end{equation}  
together with the $4\times4$ unit matrix $\ID$, they form a set of 16 matrices 
$\Gamma_\lambda$, $\lambda=1,\ldots,16$ that are hermitian, unitary, 
linearly independent and of unit determinant. They are the orthonormal basis of the corresponding 
Dirac's Clifford algebra.

The set of 64 unitary matrices of determinant $+1$, $\pm\Gamma_\lambda$, 
$\pm i\Gamma_\lambda$, $\lambda=1,\ldots,16$ form a finite subgroup of 
$SU(4)$. Its composition law can be obtained from:
 \begin{eqnarray} 
a_{ij}\,a_{kl}&=&\,\delta_{ik}\delta_{jl}\ID+i\delta_{ik}\epsilon_{jlr}\,s_r-
i\delta_{jl}\epsilon_{ikr}\,t_r+\epsilon_{ikr}\epsilon_{jls}\,a_{rs},\label{eq:qA9}\\  
a_{ij}\,s_k&=&\,i\epsilon_{jkl}\,a_{il}+\delta_{jk}\,t_i,\label{eq:qA10}\\  
a_{ij}\,t_k&=&\,-i\epsilon_{ikl}\,a_{lj}+\delta_{ik}\,s_j,\label{eq:qA11}\\  
s_i\,a_{jk}&=&\,i\epsilon_{ikl}\,a_{jl}+ \delta_{ik} t_j,\label{eq:qA12}\\  
s_i\,s_j&=&i\epsilon_{ijk}\,s_k +\delta_{ij}\ID,\label{eq:qA13}\\  
s_i\,t_j&=&t_j\,s_i=\,a_{ji},\label{eq:qA14}\\  
t_i\,a_{jk}&=&\,-i\epsilon_{ijl}\,a_{lk}+ \delta_{ij} s_k,\label{eq:qA15}\\  
t_i\,t_j&=&\,-i\epsilon_{ijk}\,t_k +\delta_{ij}\ID, 
 \label{eq:qA16}
 \end{eqnarray}
and similarly we can use these expressions to derive the commutation and 
anticommutation relations (\ref{eq:qA1}-\ref{eq:qA7}).

Dirac's algebra is generated by the four Dirac gamma matrices $\gamma^\mu$, 
$\mu=0,1,2,3$ that satisfy the anticommutation relations
 \begin{equation}
\{\gamma^\mu,\gamma^\nu\}=2\eta^{\mu\nu}\ID,
 \label{eq:qA17}
 \end{equation} 
$\eta^{\mu\nu}$ being Minkowski's metric tensor.

Similarly it can be generated by the following four observables, for 
instance: $S_1$, $S_2$, $T_1$ and $T_2$. In fact by (\ref{eq:qA13})
and (\ref{eq:qA16}) we 
obtain $S_3$ and $T_3$ respectively and by (\ref{eq:qA14}), the remaining elements.

Classically, the internal orientation of an electron is characterized by the 
knowledge of the components of the body frame $({\bi e}_i)_j$, $i,j=1,2,3$ 
that altogether constitute an orthogonal matrix. To completely characterize 
in a unique way this orthogonal matrix we need at least four of these 
components. In the quantum version, the knowledge of four $({\bi e}_i)_j$ 
matrices and by making use of (\ref{eq:qA9})-(\ref{eq:qA16}), allows us to recover the remaining 
elements of the complete Dirac algebra. It is in this sense that {\sl 
internal orientation} of the electron completely characterizes its internal 
structure. Dirac's algebra of translation invariant observables of the electron
can be generated by the orientation operators.

\chapter{Some spin features}

\section{Gyromagnetic ratio}
\index{gyromagnetic ratio |(}
\label{sec:gyromagnetic}  

The $g=2$ gyromagnetic ratio of the electron was considered for years 
a success of Dirac's electron theory.~\footnote{\hspace{0.1cm}P.A.M. Dirac, {\sl 
Proc. Roy. Soc. London} {\bf A117}, 610 (1928).} Later, Levy-Leblond~\footnote{\hspace{0.1cm}J.M. 
Levy-Leblond, {\sl Comm. Math. Phys.} {\bf 6}, 286 (1967).} 
obtained similarly $g=2$ but from a $s=1/2$ 
nonrelativistic wave equation. Proca~\footnote{\hspace{0.1cm} A. Proca, {\sl Compt. Rend.} 
{\bf 202}, 1420 (1936); 
{\sl Journ. Phys. Radium}, {\bf 49}, 245 (1988).} found $g=1$ for 
spin 1 particles  and this led Belinfante~\footnote{\hspace{0.1cm}F.J. Belinfante, {\sl Phys. Rev.}
 {\bf 92}, 997 (1953).} to 
conjecture that the gyromagnetic ratio for elementary systems is 
$g=1/s$, irrespective of the value $s$ of its spin. He showed this to 
be true for quantum systems of spin $3/2$, and a few years later the 
conjecture was analyzed and checked  by Moldauer and Case~\footnote{\hspace{0.1cm} P.A. Moldauer and 
K.M. Case, {\sl Phys. Rev.} {\bf 102}, 279 (1956).} to be right for any half-integer 
spin, and by Tumanov~\footnote{\hspace{0.1cm}V.S. Tumanov, 
{\sl Sov. Phys. JETP}, {\bf 19}, 1182 (1964).} for the value $s=2$. 
In all these cases a minimal 
electromagnetic coupling was assumed. 

Weinberg~\footnote{\hspace{0.1cm}S. Weinberg, in {\sl Lectures on Elementary 
Particles and Quantum Field Theory}, edited by S. Deser, M. Grisaru 
and H. Pendleton, MIT press, Cambridge, MA (1970), p. 283. } 
made the prediction $g=2$ for the intermediate 
bosons of the weak interactions when analyzing the interaction of $W$ 
bosons with the electromagnetic field by requiring a good high-energy 
behavior of the scattering amplitude. The discovery of the charged 
$W^\pm$ spin 1 bosons with $g=2$, contradictory to Belinfante's conjecture, 
corroborated Weinberg's prediction and raised the question as to 
whether $g=2$ for any elementary particle of arbitrary spin. 

Jackiw~\footnote{\hspace{0.1cm}R. Jackiw, {\sl Phys. Rev.} D {\bf 57}, 2635 (1998).} has given
another dynamical argument confirming that the gyromagnetic ratio of 
spin-1 fields is $g=2$, provided a 
nonelectromagnetic gauge invariance is accepted. He also gives some {\sl ad 
hoc} argument for $s=2$ fields, consistent with the $g=2$ prescription. 

Ferrara {\sl et al.}~\footnote{\hspace{0.1cm}S. Ferrara, M. Porrati and V.L. Telegdi, {\sl Phys. Rev.}
D {\bf 46}, 3529 (1992).} in a Lagrangian approach for massive 
bosonic and fermionic strings, by the requirement of a smooth 
fixed-charge $M\to0$ limit, get $g=2$ as the most natural value for 
particles of arbitrary spin. However the only known particles which 
fulfill this condition are leptons and charged $W^\pm$ bosons, {\sl i.e.}, 
charged fermions and bosons of the lowest admissible values of spin. No 
other higher spin charged elementary particles have been found.

The aim of this section, instead of using dynamical arguments as in the 
previous attempts, is to give 
a kinematical description of the gyromagnetic ratio of elementary 
particles~\footnote{\hspace{0.1cm}M. Rivas, J.M.Aguirregabiria and A. Hern\'andez, {\sl
Phys. Lett.} {\bf A 257}, 21 (1999).} which is based upon the double content of their spin 
operator structure.

The general structure of the quantum mechanical angular momentum operator in either relativistic or
nonrelativistic approach is 
\begin{equation}
{\bi J}={\bi r}\times\frac{\hbar}{i}\nabla+{\bi S}={\bi r}\times{\bi 
P}+{\bi S},
\end{equation}
where the spin operator is
 \begin{equation} 
{\bi S}={\bi u}\times\frac{\hbar}{i}\nabla_u+{\bi W},
 \label{eq:spinq} 
 \end{equation} 
and $\nabla_u$ is the gradient operator with respect 
to the velocity variables and ${\bi W}$ is a linear 
differential operator that operates only on the orientation variables 
$\balpha$ and therefore commutes with the other. For instance, in the $\brho={\bi n}\tan(\alpha/2)$ 
parametrization ${\bi 
W}$ is written as
\begin{equation}
{\bi W}=\frac{\hbar}{2i}\left[\nabla_\rho+{\brho}\times\nabla_\rho+ 
{\brho}({\brho}\cdot\nabla_\rho)\right].
\label{eq:Da}
\end{equation}

The first part in (\ref{eq:spinq}), related to the zitterbewegung spin, has 
integer eigenvalues because it has the form of an orbital 
angular momentum in terms of the ${\bi u}$ variables. Half-integer 
eigenvalues come only from the operator (\ref{eq:Da}). This operator ${\bi W}$ takes into 
account the change of orientation, {\sl i.e.}, the rotation of the 
particle.

We have seen in either relativistic or non-relativistic examples that if 
the only spin content of the particle 
${\bi S}$ is related to the zitterbewegung part ${\bi Z}={\bi u}\times{\bi U}$, then the relationship between the magnetic moment 
and zitterbewegung spin is given by
 \begin{equation} 
\bmu=\frac{e}{2}\;{\bi k}\times\frac{d{\bi k}}{dt}=-\frac{e}{2m}{\bi Z},
\label{eq:mu31}
 \end{equation}
{\sl i.e.}, with a normal up to a sign gyromagnetic ratio $g=1$. 
If the electron has a gyromagnetic ratio 
$g=2$, this implies necessarily that another part of the spin is 
coming from the angular velocity of the body, but producing no contribution to the
magnetic moment.

Therefore for the electron, both parts ${\bi W}$ and ${\bi Z}$ contribute to the total
spin. But the ${\bi W}$ part is related to the angular 
variables that describe orientation
and does not contribute to the separation ${\bi k}$ between the center 
of charge and the center of mass. It turns out that the magnetic 
moment of a general particle is still related to the motion of the 
charge by the expression (\ref{eq:mu31}), {\sl i.e.}, in terms of the 
${\bi Z}$ part but not to the total spin ${\bi S}$. It is precisely 
when we try to express the magnetic moment in terms of the total spin that 
the concept of gyromagnetic ratio arises.

Now, let us assume that both ${\bi Z}$ and ${\bi W}$ terms 
contribute to the total spin ${\bi S}$ with their lowest admissible 
values. 

For Dirac's particles, the classical zitterbewegung 
is a circular motion at the speed of light of radius $R=S/mc$ and 
angular frequency $\omega=mc^2/S$, in a plane orthogonal to the total 
spin. The total spin ${\bi S}$ and the ${\bi Z}$ part, are both orthogonal to 
this plane and can be either parallel or antiparallel. 
Let us define the gyromagnetic ratio by $Z=gS$. 
For the lowest admissible values of the quantized spins 
$z=1$ and $w=1/2$ in the opposite direction this gives rise to a 
total $s=1/2$ perpendicular to the zitterbewegung plane and then $g=2$.

For $s=1$ particles the lowest possible values compatible with the 
above relative orientations are $z=2$ and $w=1$ in the opposite 
direction, thus obtaining again $g=2$. The possibility $z=1$ and 
$w=0$ is forbidden in the relativistic case because necessarily 
$w\neq0$ to describe vector bosons with a multicomponent wave-function.

\section{Instantaneous electric dipole}
\index{electric dipole}\index{electric dipole!instantaneous}\index{dipole!electric}
\label{sec:elecdipole}  

The internal motion of the charge of the electron in the center of mass frame
is a circle at the speed of light. 
The position of the charge in this frame is related 
to the total spin by eq. (\ref{eq:elecdina}), {\sl i.e.},
\begin{equation}
{\bi k}=\frac{1}{mc^2}\,{\bi S}\times{\bi u},
\label{eq:elecdina2}
\end{equation}
where ${\bi S}$ is the total constant spin and ${\bi u}=d{\bi k}/dt$, with $u=c$ is the velocity of the charge. 
In addition to this motion there is a rotation of a local frame linked to the particle
that gives rise to some angular velocity, but this rotation has no effect on the electric dipole structure.
(See Fig. \ref{fig:motion} where the angular velocity and the local frame are not depicted).

\cfigl{fig:motion}{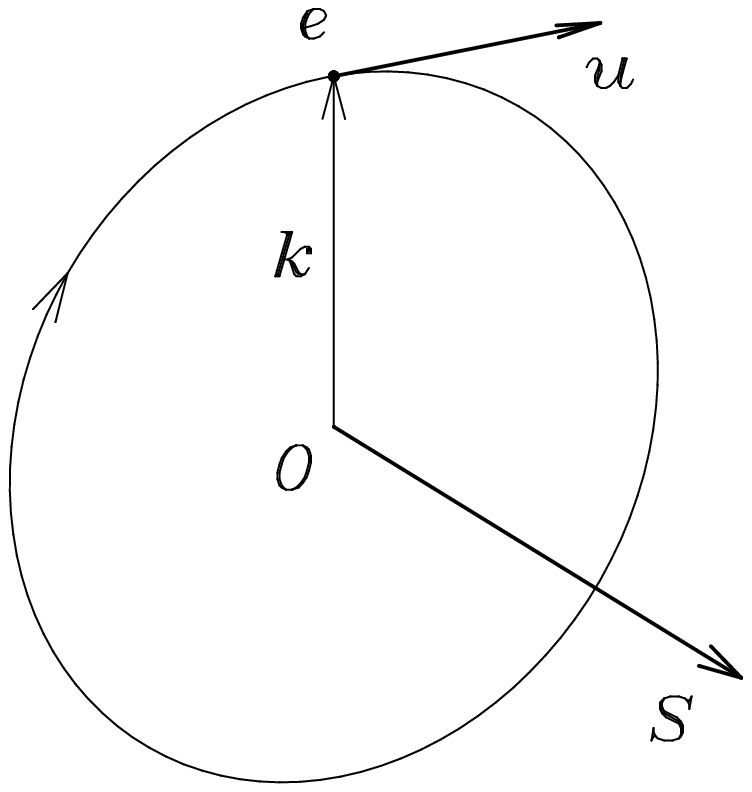}{Electron charge motion in the C.M. frame.} 

Now, from the point of view of the center of mass observer, the particle behaves as though it has a magnetic moment
related to the particle current by the usual classical expression
\[
\bmu=\frac{1}{2}\int {\bi k}\times{\bi j}\; d^3{\bi r}=\frac{e}{2}\;{\bi k}\times\frac{d{\bi k}}{dt},
\]
where $e$ is the charge and ${\bi j}({\bi r}-{\bi k})=e\, d{\bi k}/dt\;\delta^3({\bi r}-{\bi k})$ is the 
particle current density. 
The orbital term ${\bi k}\times d{\bi k}/dt$ is related to the zitterbewegung part of spin
that quantizes with integer values and which for spin $1/2$ and spin $1$ charged particles 
is twice the total spin ${\bi S}$, giving rise to a pure kinematical 
interpretation of the gyromagnetic ratio $g=2$ for this model as seen in the previous section.

But also in the center of mass frame the particle has an oscillating instantaneous electric dipole moment
${\bi d}=e{\bi k}$, that is thus related to the total spin by
\begin{equation}
{\bi d}=\frac{e}{mc^2}\,{\bi S}\times{\bi u}.
\label{eq:dipole}
\end{equation}

This instantaneous electric dipole, which fulfills the usual definition of
the momentum of the point charge $e$ with respect to the origin of the
reference frame, is translation invariant because it is expressed in
terms of a relative position vector ${\bi k}$. It can never be
interpreted as some kind of fluctuation of a spherical symmetry of a charge distribution.
Even in this kind of model, it is not necessary to talk about charge
distributions, because all particle attributes are defined at single
points.

In his original 1928 article,~\footnote{\hspace{0.1cm}P.A.M. Dirac, 
{\sl Proc. Roy. Soc. London}, {\bf A117}, 610 (1928).} Dirac obtains that the Hamiltonian for the electron has, in addition
to the Hamiltonian of a free point particle of mass $m$, two new terms that in the presence 
of an external electromagnetic field are\index{Dirac's Hamiltonian!electric and magnetic dipole terms}
\begin{equation}
\frac{e\hbar}{2m}{\bf\Sigma}\cdot{\bi B}+\frac{ie\hbar}{2mc}{\balpha}\cdot{\bi E}=-\bmu\cdot{\bi B}-{\bi d}\cdot{\bi E},
\label{eq:Hamilt}
\end{equation}
where 
\[
{\bf\Sigma}=\pmatrix{\bsigma&0\cr 0&\bsigma\cr},\quad {\rm and}\quad \balpha=\gamma_0\bgamma,
\]
{\sl i.e.}, ${\bf\Sigma}$ is expressed in terms of $\bsigma$ Pauli-matrices and $\balpha$ is Dirac's
velocity operator when written in terms of Dirac's gamma matrices. 

We shall show that the quantum counterpart of expression (\ref{eq:dipole}) is in fact
the electric dipole term of Dirac's Hamiltonian (\ref{eq:Hamilt}). 
The remaining part of this section is to consider the representation
of the `cross' product in (\ref{eq:dipole}) in terms of the matrix (or geometric) product of the elements
of Dirac's algebra that represent the quantum version of the above observables, so that a short explanation
to properly interpret these observables as elements of a Clifford algebra is given in what follows.

Both, velocity operator ${\bi u}=c\balpha$ and spin operator ${\bi S}$ are bivectors in Dirac's
algebra, considered as elements of the Geometric or Clifford algebra of space-time 
in the sense of Hestenes.~\footnote{\hspace{0.1cm}D. Hestenes, {\sl Space-Time algebra}, Gordon and Breach, NY (1966);
D. Hestenes and G. Sobczyk, {\sl Clifford Algebra to Geometric Calculus}, D. Reidel Pub. Co. Dordrecht, (1984).}

In fact, Dirac's alpha matrices are written as a product of two gamma matrices $\alpha_i=\gamma_0\gamma_i$
and also the spin components $S_j=(i\hbar/2)\, \gamma_k\gamma_l$, $j,k,l$ cyclic $1,2,3$, and where the four gamma matrices,
$\gamma_\mu$, $\mu=0,1,2,3$ are interpreted as the four basic vectors of Minkowski's space-time 
that generate Dirac's Clifford algebra. They satisfy $\gamma_\mu\cdot\gamma_\nu=\eta_{\mu\nu}$, {\sl i.e.}, $\gamma_0^2=1$
and $\gamma_i^2=-1$, where the dot means the inner product in Dirac's Clifford algebra.
We thus see that velocity and spin belong to the even subalgebra of Dirac's 
algebra and therefore they also belong to Pauli algebra
or geometric algebra of three-dimensional space. Under spatial
inversions $\gamma_0\to\gamma_0$ and $\gamma_i\to-\gamma_i$, 
the velocity operator changes its sign and it is thus a spatial vector, 
while the spin is invariant under this transformation as it corresponds to a spatial bivector
or pseudovector.

\cfigl{fig:vectors}{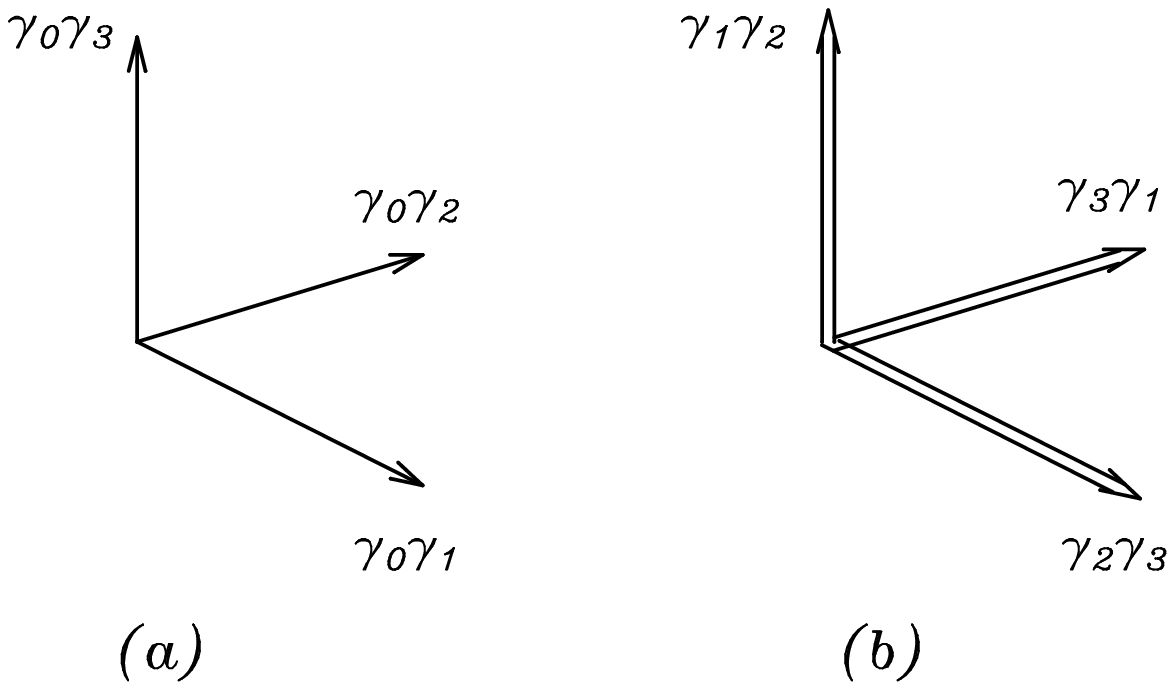}{A basis for vectors $(a)$ and bivectors (pseudovectors) $(b)$ of Pauli algebra.} 

The relationship between the cross product and the outer and inner product of two vectors ${\bf a}$ 
and ${\bf b}$ in Pauli algebra is, 
\begin{equation}
{\bi a}\times{\bi b}=-i{\bi a}\wedge{\bi b}={\bi b}\cdot (i{\bi a}),
\label{eq:geomal1}
\end{equation}
where $\wedge$ represents the symbol for the outer product in geometric algebra, 
the imaginary unit $i$ represents the unit three-vector or pseudoscalar 
and $i{\bi a}$ is the dual bivector of vector ${\bi a}$.

The inner product of a vector ${\bi b}$ and a bivector $A$ is expressed in terms of the geometric product in the form
\begin{equation}
{\bi b}\cdot{A}=\frac{1}{2}({\bi b}A-A{\bi b})
\label{eq:geomal2}
\end{equation}
where in Dirac's or Pauli algebra the geometric product ${\bi b}A$ is just
the ordinary multiplication of matrices. 

If we choose a basis of vectors and pseudovectors as in Fig.~\ref{fig:vectors}, where the double-lined
objects of part $(b)$ represent the dual vectors of the corresponding spatial bivectors, 
and express in these bases the observables of Fig. \ref{fig:motion}, then the spatial velocity vector 
${\bi u}=c\gamma_0\gamma_2$
and the pseudovector ${\bi S}=(\hbar/2)\gamma_2\gamma_3$ and therefore, using (\ref{eq:geomal1}) and (\ref{eq:geomal2})
we get
\[
{\bi S}\times{\bi u}={\bi u}\cdot (i{\bi S})=\frac{ic\hbar}{2}\left(\frac{1}{2}\left(\gamma_0\gamma_2\gamma_2\gamma_3
-\gamma_2\gamma_3\gamma_0\gamma_2\right)\right)=\frac{-ic\hbar}{2}\gamma_0\gamma_3.
\]
Now vector ${\bi k}=R\gamma_0\gamma_3$ with $R=\hbar/2mc$, 
and substituting in (\ref{eq:dipole}) we get the desired result.

\section{Classical Tunneling}
\label{sec:tunnel}
\index{Tunnel effect}\index{Tunnel effect!classical crossing}

As a consequence of the zitterbewegung and therefore
of the separation between the center of mass and center of charge, we shall see that 
spinning particles can have a non-vanishing crossing of potential barriers.

Let us consider a spinning particle with spin of (anti)orbital type, 
as described in Section \ref{sec:galispin}, under the influence of a potential 
barrier. The Langrangian of this system is given by:
 \begin{equation}
L=\frac{m}{2}\frac{\dot{\bi r}^2}{\dot t}-
\frac{m}{2\omega^2}\frac{\dot{\bi u}^2}{\dot t}-eV({\bi r})\dot t.
 \end{equation}
Sharp walls correspond classically to infinite forces so that 
we shall consider potentials that give rise to finite forces like 
those of the shape depicted in Fig.~\ref{fig:triangl}, where $V_0$ 
represents the top of the potential. 

\cfigl{fig:triangl}{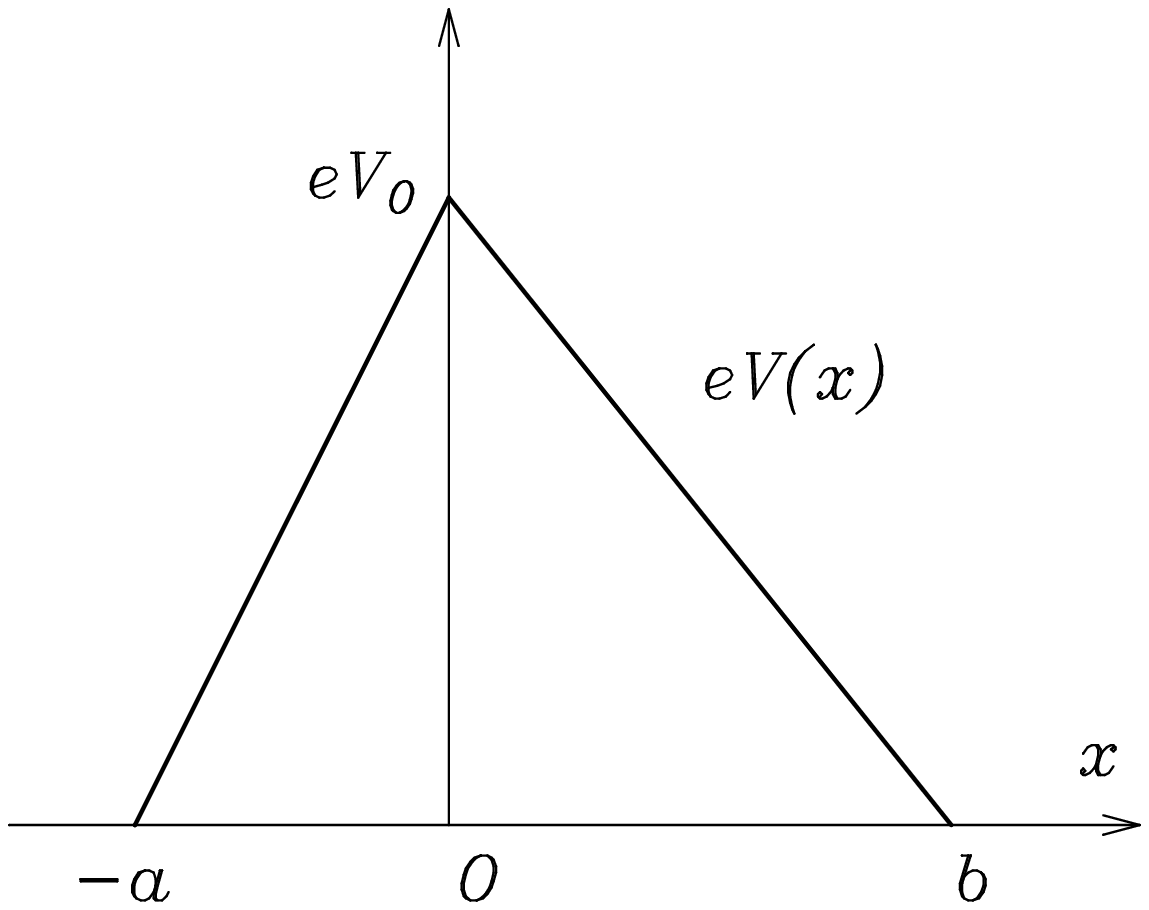}{Triangular potential barrier.}

Then the external force $F(x)$, 
is constant and directed leftwards in the region $x\in (-a,0)$ and 
rightwards for $x\in (0,b)$, vanishing outside these regions. 

Potentials of this kind can be found for instance in the simple experiment
depicted in Figure \ref{fig:canonelec} in which an electron beam, accelerated
with some acceleration potential $V_a$, is sent into the uniform field region
of potential $V_0$ contained between the grids or plates $A$, $C$ and $B$. 

\cfigl{fig:canonelec}{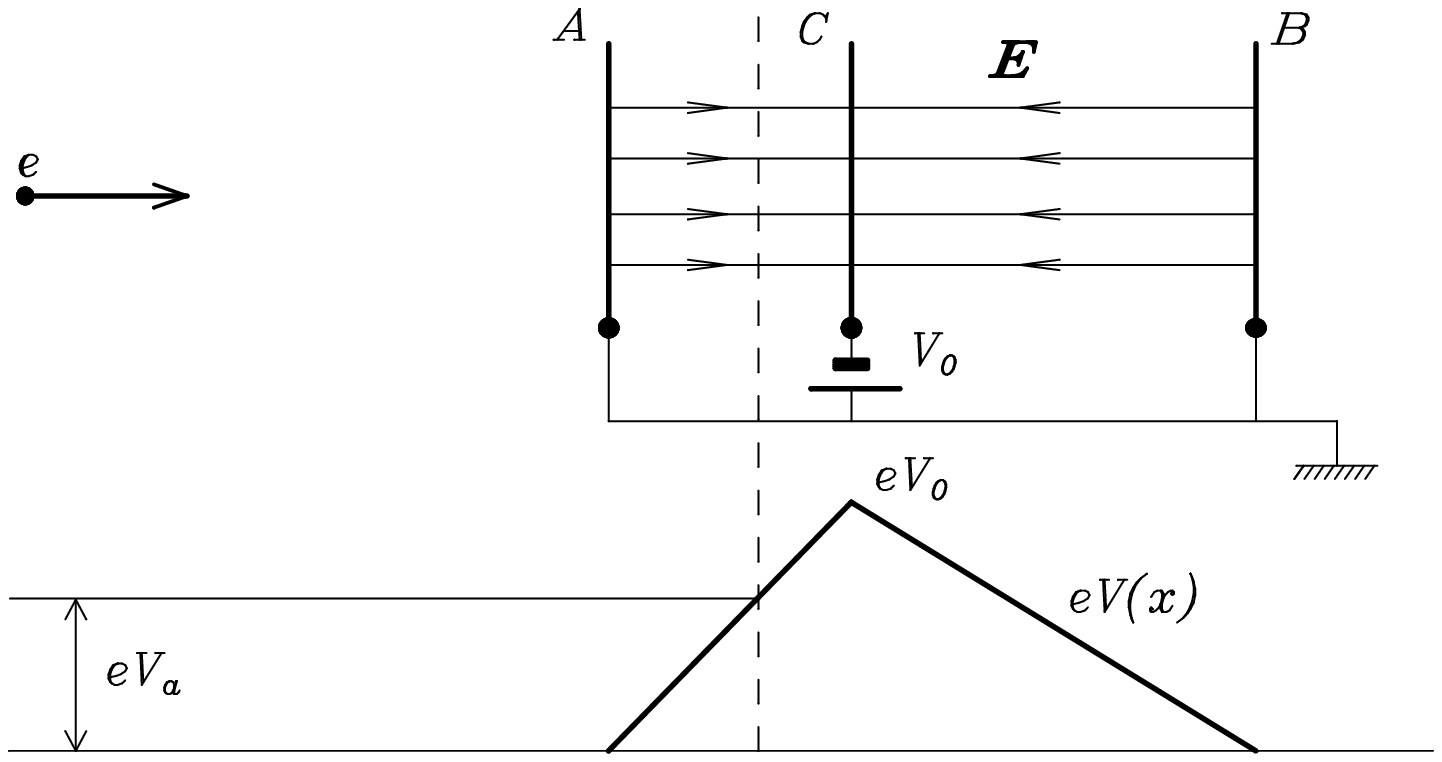}{Electron beam into a potential barrier. A classical 
spinless electron never crosses the dotted line. A spinning particle of the same energy might cross the barrier.}

In Figure \ref{fig:canonelec} from a strict classical viewpoint a spinless electron stops at the dotted line
and is rejected backwards. But a classical spinning electron can cross the barrier provided its kinetic
energy is above some minimum value, although below the top of the potential. This minimum value
depends on the separation between plates.

Let us assume for simplicity that the spin is pointing up or down in 
the $z$ direction such that the point charge motion takes place in the 
$XOY$ plane. Let $q_x$, $q_y$ and $q_z=0$, be the coordinates of the 
center of mass and $x$, $y$ and $z=0$, the position of the charge. 

The dynamical equations are
 \begin{equation} 
\frac{d^2q_x}{ dt^2}=\frac{1}{ m}F(x),\quad\frac{d^2q_y}{ dt^2}=0,
 \label{eq:qxqy}
 \end{equation} 
 \begin{equation} 
\frac{d^2x}{ dt^2}+\omega^2(x-q_x)=0,\quad\frac{d^2y}{ dt^2}+\omega^2(y-
q_y)=0,
 \end{equation} 
where
 \[
F(x)=\left\{\begin{array}{ll}-{eV_0/a},&\mbox{for $x\in(-a,0)$},\\
{eV_0/b},&\mbox{for $x\in(0,b)$},\\
0, &\mbox{otherwise}.
\end{array}\right.
 \]

Equations (\ref{eq:qxqy}) are nonlinear and we have not been 
able to obtain an analytical solution in closed form. We shall try to 
find a numerical solution. To make the corresponding numerical 
analysis we shall define different dimensionless variables. Let $R$ be 
the average separation between the center of charge and center of 
mass. In the case of circular internal motion, it is just the radius 
$R_0$ of the zitterbewegung. Then we define the new dimensionless 
position variables: 
 \[
\hat q_x=q_x/R,\quad \hat q_y=q_y/R,\quad \hat x=x/R,\quad \hat y=y/R,
\quad \hat a=a/R,\quad \hat b=b/R. 
 \]

The new dimensionless time variable $\alpha=\omega t$ is just the 
phase of the internal motion, such that the dynamical equations become 
 \[
\frac{d^2\hat q_x}{ d\alpha^2}=A(\hat x),\quad\frac{d^2\hat q_y}{d\alpha^2}=0,
 \]
 \[
\frac{d^2\hat x}{ d\alpha^2}+\hat x-\hat q_x=0,\quad\frac{d^2\hat y}{d\alpha^2}+\hat y-\hat q_y=0,
 \]
where $A(\hat x)$ is given by
 \[
A(\hat x)=\left\{\begin{array}{ll}
-{eV_0/\hat am\omega^2R^2},&\mbox{for $\hat x\in(-\hat a,0)$},\\
{eV_0/\hat bm\omega^2R^2},&\mbox{for $\hat x\in(0,\hat b)$},\\0, &\mbox{otherwise}.
\end{array}\right.
 \]

In the case of the relativistic electron, the internal velocity of the 
charge is $\omega R=c$, so that 
the parameter $e/mc^2=1.9569\times10^{-6}$V$^{-1}$, and for 
potentials of order of 1 volt we can take the dimensionless 
parameter $eV_0/m\omega^2R^2=1.9569\times10^{-6}$. 

If we choose as initial conditions for the center of mass motion 
 \[
\hat q_y(0)=0,\quad d\hat q_y(0)/d\alpha=0,
 \] 
then the center of mass is moving along the $OX$ axis. The above 
system reduces to the analysis of the one-dimensional motion where the 
only variables are $\hat q_x$ and $\hat x$. Let us call from now on 
these variables $q$ and $x$ respectively and remove all hats from the 
dimensionless variables. Then the dynamical equations to be solved 
numerically are just 
 \begin{equation} 
\frac{d^2 q}{ d\alpha^2}=A(x),\quad \frac{d^2 x}{d\alpha^2}+x-q=0,
 \end{equation}  
where $A(x)$ is given by 
 \begin{equation}
A(x)=\left\{\begin{array}{ll}
-1.9569\times 10^{-6}\, a^{-1}V_0 ,&\mbox{for $x\in(-a,0)$},\\
1.9569\times 10^{-6}\, b^{-1}V_0,&\mbox{for $x\in(0,b)$},\\
0,&\mbox{otherwise}.\end{array}\right.
 \end{equation} 

Numerical integration has been performed by means of the computer 
package {\sl Dynamics Solver}.~\footnote{\hspace{0.1cm}J.M. Aguirregabiria, {\sl Dynamics Solver}, computer program for solving
different kinds of dynamical systems, which is available from his author 
through the web site {\tt <http://tp.lc.ehu.es/jma.html>} 
at the server of the Theoretical Physics dept. of The University of the Basque Country,
Bilbao (Spain).} The quality of the numerical 
results is tested by using the different integration schemes this 
program allows, ranging from the very stable embedded Runge-Kutta code 
of eight order due to Dormand and Prince to very fast extrapolation 
routines. All codes have adaptive step size control and we check that 
smaller tolerances do not change the results. 

\cfigl{fig:Pot1-1}{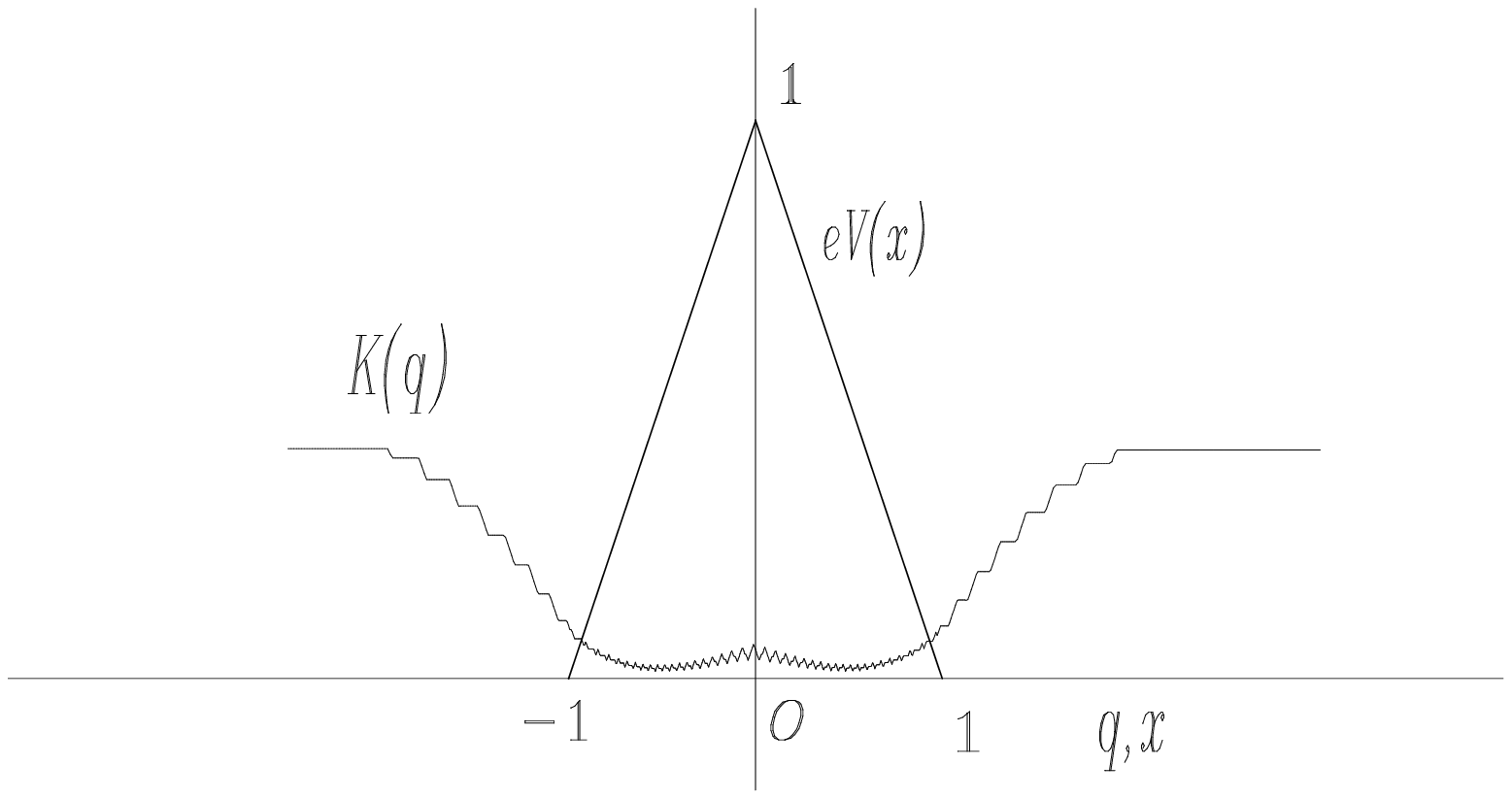}{Kinetic Energy during the crossing for the values $a=b=1$.}

With $a=b=1$, and in energy units such that the top of the barrier is 
1, if we take an initial kinetic energy $K$ below this threshold, 
$K=m\dot q(0)^2/2eV_0=0.41$ we obtain for the center of mass motion 
the graphic depicted in Fig.~\ref{fig:Pot1-1}, where is shown the 
variation of the kinetic energy of the particle $K(q)$, with the 
center of mass position during the crossing of the barrier. There is 
always crossing with a kinetic energy above this value. In 
Fig.~\ref{fig:Pot1-10}, the same graphical evolution with $a=1$ and $b=10$ 
and $K=0.9055$ for a potential of $10^3$ Volts in which the different 
stages in the evolution are evident. Below the initial 
values for the kinetic energy of $0.4$ and $0.9$ respectively, the 
particle does not cross these potential barriers and it is rejected 
backwards. 

If in both examples the parameter $a$ is ranged from 1 to 0.05, thus 
making the left slope sharper, there is no appreciable change in the 
crossing energy, so that with $a=1$ held fixed we can compute the 
minimum crossing kinetic energies for different $b$ values, $K_c(b)$. 

\cfigl{fig:Pot1-10}{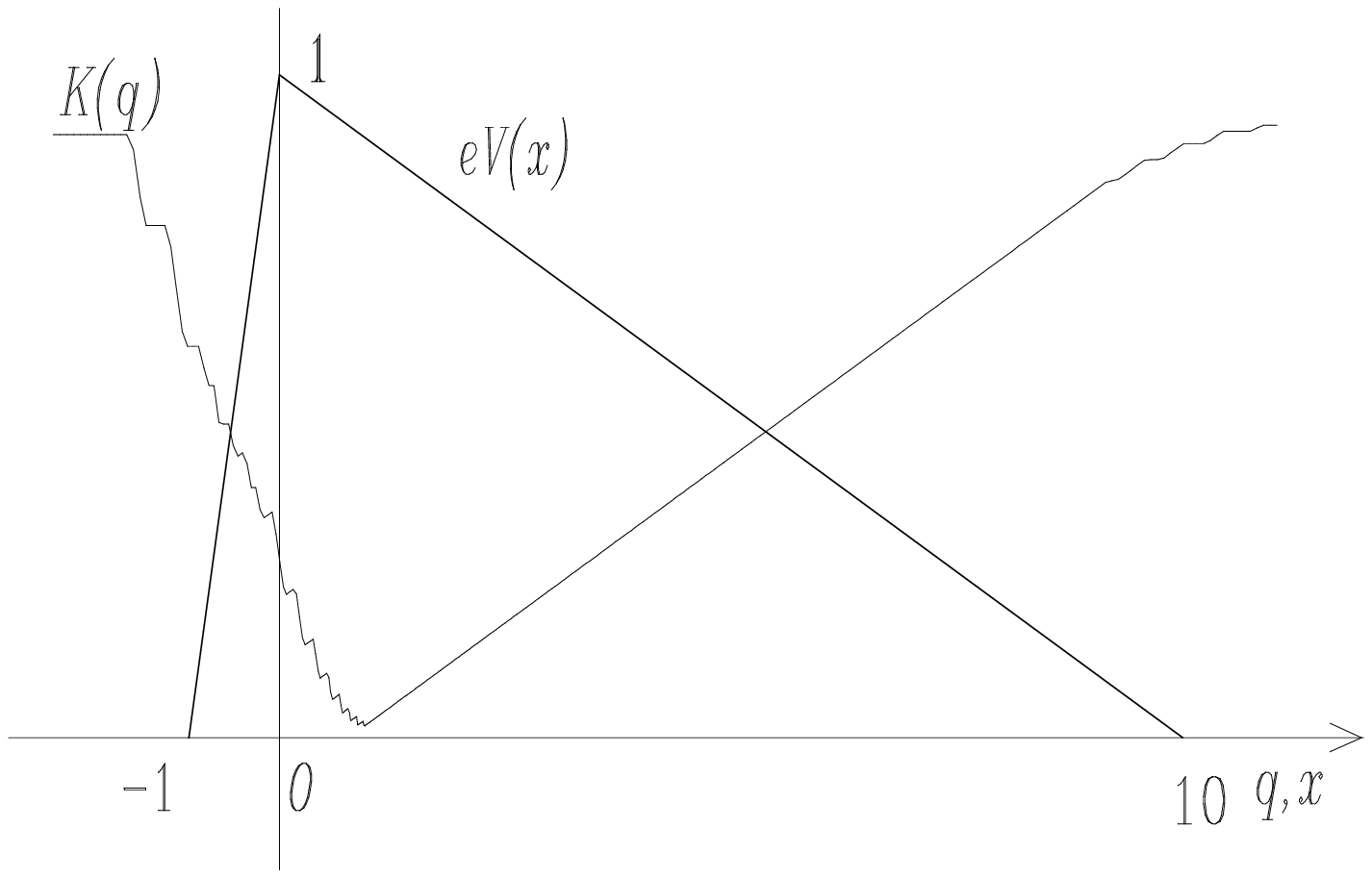}{Kinetic Energy during the crossing for the values $a=1$, $b=10$.}

To compare this model with the quantum tunnel effect\index{tunnel effect!quantum}, 
let us quantize the system. In the quantization of generalized Lagrangians developed in the
Chapter \ref{ch:quantization}, the wave function for this system is a 
squared-integrable function $\psi(t,{\bi r},{\bi u})$, of the seven 
kinematical variables and the generators of 
the Galilei group have the form: 
 \begin{equation}
H=i\hbar\frac{\partial}{\partial t},\; {\bi P}=-i\hbar\nabla,\;
{\bi K}=m{\bi r}-t{\bi P}+i\hbar\nabla_u,\; {\bi J}={\bi r}\times{\bi P}+{\bi Z},
 \label{operadores}
 \end{equation}
where $\nabla_u$ is the gradient operator with respect to the ${\bi u}$ 
variables. These generators satisfy the commutation relations of the extended
Galilei group,~\footnote{\hspace{0.1cm}J.M. Levy-Leblond, {\sl Galilei Group and 
Galilean Invariance}, in E.M. Loebl, {\sl Group Theory and its 
applications}, Acad. Press, NY (1971), vol. 2, p.~221. } and the spin operator is given by 
${\bi Z}=-i\hbar{\bi u}\times\nabla_u$.

One Casimir operator of this extended Galilei group 
is the Galilei invariant internal energy of the system ${\cal E}$, which in the 
presence of an external electromagnetic field and with the minimal 
coupling prescription is written as,
 \begin{equation}
{\cal E}=H-eV-\frac{1}{2m}({\bi P}-e{\bi A})^2,
 \label{eq:casimir}
 \end{equation}
where $V$ and ${\bi A}$ are the external scalar and vector potentials, 
respectively.

In our system ${\bi A}=0$, and $V$ is only a function of the $x$ 
variable. It turns out that because of the structure of the above 
operators we can find simultaneous eigenfunctions of the following 
observables: the Casimir operator (\ref{eq:casimir}), $H$, $P_y$, $P_z$, 
$Z^2$ and $Z_z$. The particle moves along the $OX$ axis, with the spin 
pointing in the $OZ$ direction, and we look for solutions which are 
eigenfunctions of the above operators in the form:
 \begin{equation} 
\left(H-eV(x)-\frac{1}{2m}{\bi P}^2\right)\psi={\cal E}\psi,\;
H\psi=E\psi,\; P_y\psi=0,\; P_z\psi=0,
 \end{equation} 
 \begin{equation} Z^2\psi=s(s+1)\hbar^2\psi,\quad Z_z\psi=\pm s\hbar\psi,
 \end{equation}
so that $\psi$ is independent of $y$ and 
$z$, and its time dependence is of the form $\exp(-iEt/\hbar)$. Since 
the spin operators produce derivatives only with respect to the velocity 
variables, we can look for solutions with the variables separated in 
the form: 
 \[ 
\psi(t,x,{\bi u})=e^{-iEt/\hbar}\phi(x)\chi({\bi u}), 
 \] 
and thus 
 \begin{equation}
\left(\frac{\hbar^2}{2m}\frac{d^2}{dx^2}+E-eV(x)-{\cal E}\right)\phi(x)=0,
 \label{eq:Schroed} 
 \end{equation} 
 \begin{equation}
Z^2\chi({\bi u})=s(s+1)\hbar^2\chi({\bi u}),\quad 
Z_z\chi({\bi u})=\pm s\hbar\chi({\bi u}),
 \label{eq:espin}
 \end{equation}
where the spatial part $\phi(x)$, is uncoupled with 
the spin part $\chi({\bi u})$, and $E-eV(x)-{\cal E}$ represents the 
kinetic energy of the system. The spatial part satisfies the one-dimensional 
Schroedinger equation, and the spin part is independent of 
the interaction, so that the probability of quantum tunneling is 
contained in the spatial part and does not depend on the particular 
value of the spin. If the particle is initially on the left-hand side 
of the barrier, with an initial kinetic energy $E_0=E-{\cal E}$, then
we can determine the quantum probability for crossing for $a=1$ and 
different values of the potential width $b$.

The one-dimensional quantum mechanical problem of the spatial part for 
the same one-dimensional 
potential depicted in Fig.~\ref{fig:triangl} is:~\footnote{\hspace{0.1cm}L. 
Landau and E. Lifchitz, {\sl M\'ecanique quantique}, Mir Moscow (1988), 3rd. edition.}
 \begin{equation} 
\phi(x)=\left\{\begin{array}{ll}
e^{ikx}+Re^{-ikx},&x\le -a,\\
C_1{\rm Ai}(D(1-G+\frac{\displaystyle{x}}{\displaystyle{a}})+C_2{\rm Bi}(D(1-G+\frac{\displaystyle{x}}{\displaystyle{a}}),&-a\le x\le 0,\\
C_3{\rm Ai}(L(1-G-\frac{\displaystyle{x}}{\displaystyle{b}}))+C_4{\rm Bi}(L(1-G-\frac{\displaystyle{x}}{\displaystyle{b}})),&0\le x\le b,\\
Te^{ikx}, &x\ge b,\\
\end{array}\right.
 \end{equation} 
where $x$ is the same dimensionless position variable as before, and the 
constants
 \begin{equation}  
k=\sqrt{\frac{E}{2mc^2}},\; D=\sqrt[3]{\frac{eV_0a^2}{2mc^2}},\;
L=\sqrt[3]{\frac{eV_0b^2}{2mc^2}},\; G=\frac{E}{eV_0}.
 \end{equation}  
Functions ${\rm Ai}(x)$ and ${\rm Bi}(x)$ are the Airy functions of $x$.
The six integration constants $R$, $T$, and $C_i, i=1,2,3,4$, can be 
obtained by assuming continuity of the functions and their first order 
derivatives at the separation points of the different regions. The 
coefficient $|R|^2$ represents the probability of the particle to be 
reflected by the potential and  $|T|^2$ its probability of crossing.

\cfigl{fig:Probab}{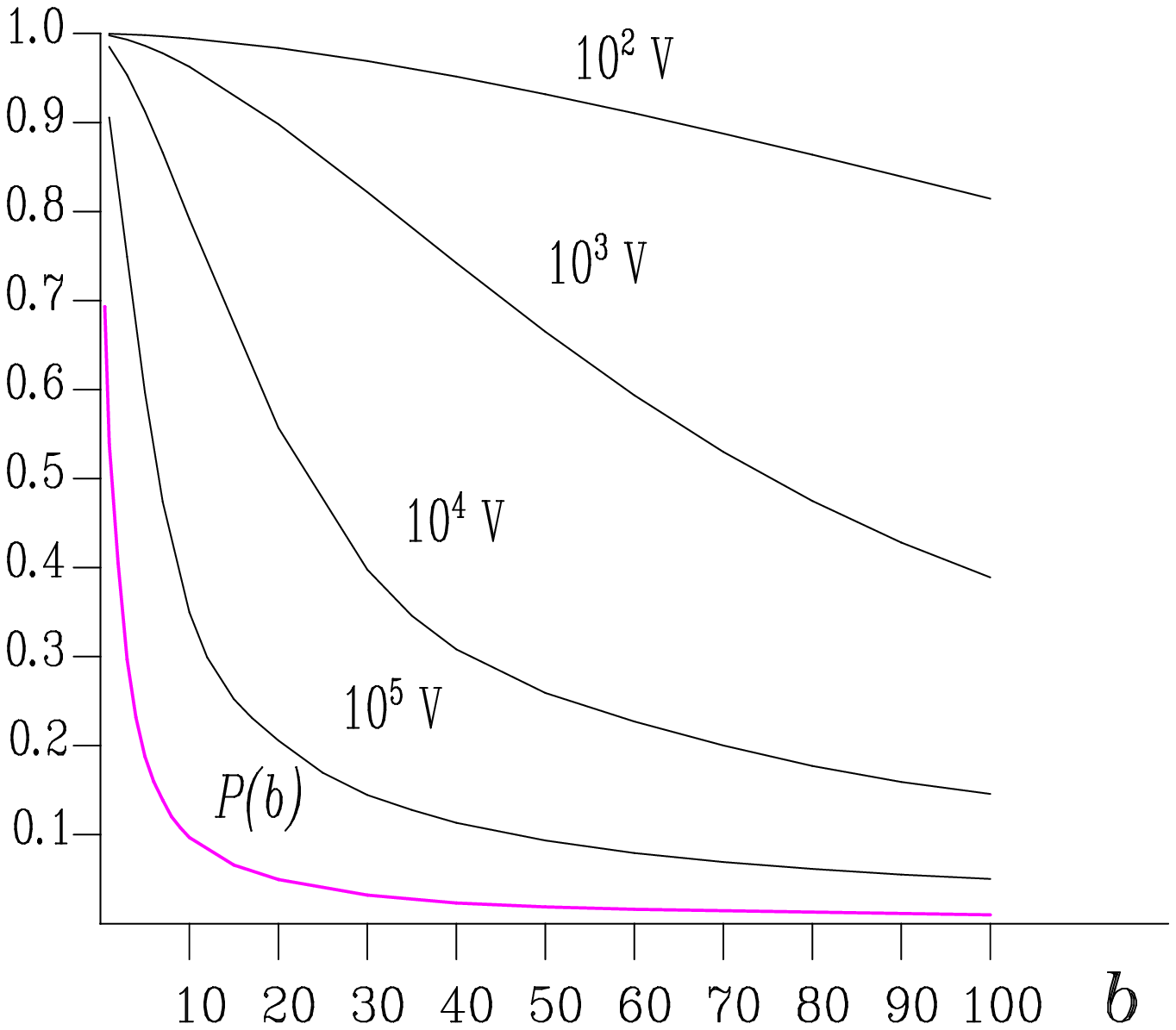}{Classical and Quantum Probability of crossing for different potentials.}

Computing the $T$ amplitude for $a=1$ and different values of the 
potential width $b$, and for energies below the top of the barrier 
$eV_0$, we show in Fig.~\ref{fig:Probab}, the average probability for 
quantum tunneling for four different potentials $V_0$ of 
$10^2$, $10^3$, $10^4$ and $10^5$ Volts. This average probability has been 
computed by assuming that on the left of the barrier there is a 
uniform distribution of particles of energies below $eV_0$. 

If we consider for the classical spinning particle the same uniform 
distribution of particles, then, the function $P(b)=1-K_c(b)$, where 
$K_c(b)$ is the minimum dimensionless kinetic energy for crossing computed before, 
represents the ratio of the particles that with kinetic energy below 
the top of the potential cross the barrier because of the spin 
contribution. 

This function $P(b)$, is also depicted in 
Fig.~\ref{fig:Probab}. We see that for the different potentials shown in 
that figure the classical average probability of crossing is smaller 
than the quantum one, but for stronger potentials this classical 
probability, coming from the spin contribution, becomes relatively important. 

Because the tunnel effect is a function of $\hbar$ and the spin of elementary 
particles is also of order of $\hbar$ it is very difficult to separate 
from the outcome of a real experiment involving elementary particles, 
which part is due to a pure quantum effect and which is the contribution 
to crossing coming from the spin structure. From (\ref{eq:Schroed}) and (\ref{eq:espin}) 
it is clear that the quantum 
probability of tunneling is independent of the spin. 

To test experimentally 
this contribution, it will be necessary to perform separate 
experiments with particles of the same mass and charge but with 
different values of the spin. Thus, the difference in the outcome will 
be related to the spin contribution. This can be accomplished for 
instance, by using ions of the type $A^{++}$ that could be either in a 
singlet, $(s=0)$ state or in a triplet $(s=1)$ state. 

But if there exists a contribution to crossing not included in the usual quantum mechanical analysis
we have to modify the quantum mechanical equations.
To be consistent with the above analysis the Schroedinger-Pauli equation
should be modified to include the additional electric dipole term. A term of the 
form $-eER\cos\omega t$, where $E$ is the external electric field and $R$ the radius of
the zitterbewegung, should be considered to solve the corresponding quantum wave function.
This term is of the order of the separation $R$ between the center of mass and center of
charge, which is responsible for the classical crossing. This additional electric dipole term
is already included in Dirac's equation but is suppressed when taking the low velocity limit, 
as it corresponds to this low energy example. Nevertheless, although this is a low energy
process and the time average value of the electric dipole vanishes, there are very high field gradients. 

We see that the separation between the center of mass and center of 
charge that gives rise to the spin structure of this particle model 
justifies that this system can cross a potential barrier even if its 
kinetic energy is below the top of the potential. 

\subsection{Spin polarized tunneling}
\index{Spin polarized tunneling}
\index{tunneling!Spin polarized}
\label{sec:spinpoltunnel}

I like to point out the following ideas to discuss whether 
they can be useful in connection with the interpretation of 
the magnetoresistance of polycrystaline films. This is known in 
the literature as the {\bf spin polarized tunneling}.~\footnote{\hspace{0.1cm}V.N. Dobrovolsky, D.I. Sheka and B.V. 
Chernyachuk, Surface Science {\bf 397}, 333 (1998);
P.~Raychaudhuri, T.K. Nath, A.K. Nigam and R.~Pinto, cond-mat/9805258, preprint.}

The main feature of the ``classical'' spin polarized tunneling we have seen
in the previous section is not a matter of whether tunneling is classical or not, 
because this is a nonsense question. Matter at this scale is interpreted under 
quantum mechanical rules. But if we use a model of a classical 
spinning particle that, when polarized orthogonal to the direction of 
motion, produces a crossing that is not predicted by the Schroedinger-Pauli 
equation, it means that this quantum mechanical equation is lacking some term. 
The coupling term $-\bmu\cdot{\bi B}$, between the magnetic moment and magnetic field that 
gives rise to the Pauli equation, is inherited from Dirac's electron theory. 
But Dirac's equation also predicts another term $-{\bi d}\cdot{\bi E}$, of the coupling of an 
instantaneous electric dipole with the electric field. It is this oscillating electric 
dipole term that we believe is lacking in quantum mechanical wave equations. 
In general, the average value of this term in an electric field of smooth 
variation is zero. But in high intensity 
fields or in intergranular areas in which the effective potentials are low, 
but their gradients could be very high, this average value should not be negligible. 

The conduction of electrons in synterized materials is completely different 
than the conduction on normal conductors. The material is not a continuous crystal.
It is formed by small grains that are bound together by the action of some external
pressure. 
If we can depict roughly the electric current flow, 
this is done by the jumping of electrons from grain to grain, through 
a tunneling process in which there is some estimated effective potential 
barrier confined in the gap between grains. 
Therefore these materials show in general a huge resistivity when compared with 
true conductors. 

The form of this potential is unknown. The simplest one is to assume a wall of thickness $d$,
the average separation between grains, and height $h$. But it can also be estimated as one of the
potentials of the former example.
What we have shown previously is that for every potential barrier, there is 
always a minimum energy, below the top of the potential, that electrons 
above that energy cross with probability 1 when polarized orthogonal to the motion, even 
within a classical interpretation. But this effect is not predicted 
by ``normal'' quantum mechanics because tunneling is spin independent.

Now, let us assume that we are able to estimate some average
effective potential barrier in the intergranular zone of this polycristaline material. If the 
corresponding minimum crossing energy of this barrier for polarized 
electrons is below the Fermi level, then, when we introduce a magnetic 
field in the direction of the film and the magnetic domains in the grains become 
polarized, all electrons above that minimum energy of crossing will flow from grain 
to grain as in a good conductor, with a classical probability 1. 
That's all. Here the difficulty is to estimate properly this potential 
barrier and therefore the corresponding classical crossing energy.

It can be argued that the presence of the magnetic field to polarize electrons produces a change in the
energy of particles. Nevertheless, even for a magnetic field of the order of 1 Tesla
and in a potential barrier of 1 Volt, the magnetic term $-\bmu\cdot{\bi B}$ contributes
with an energy of order of $\pm5.7\times10^{-5}$eV, which does not modify the quantum 
probability of crossing.

\section{Formation of bound pairs of electrons}
\label{sec:2elec}

If we have the relativistic and nonrelativistic differential equations satisfied 
by the charge
of the spinning electrons
we can analyse as an example, 
the interaction among them by assuming a Coulomb interaction between their charges.
In this way we have a system of differential equations of the form (\ref{eq:q2}, \ref{eq:r2}) 
in the relativistic case
or in the form (\ref{eq:n151}, \ref{eq:n152}) in the nonrelativistic, for each particle.
For instance, the external field acting on charge $e_1$ is replaced by the Coulomb 
field created by the other charge $e_2$
at the position of $e_1$, and simmilarly for the other particle.
The integration is performed numerically by means of the numerical integration program {\it Dynamics Solver}
\footnote{J.M. Aguirregabiria, {\sl Dynamics Solver}, computer program for solving
different kinds of dynamical systems, which is available from his author 
through the web site {\tt <http://tp.lc.ehu.es/jma.html>} 
at the server of the Theoretical Physics Dept. of The University of the Basque Country,
Bilbao, Spain.}.

\cfigl{fig2}{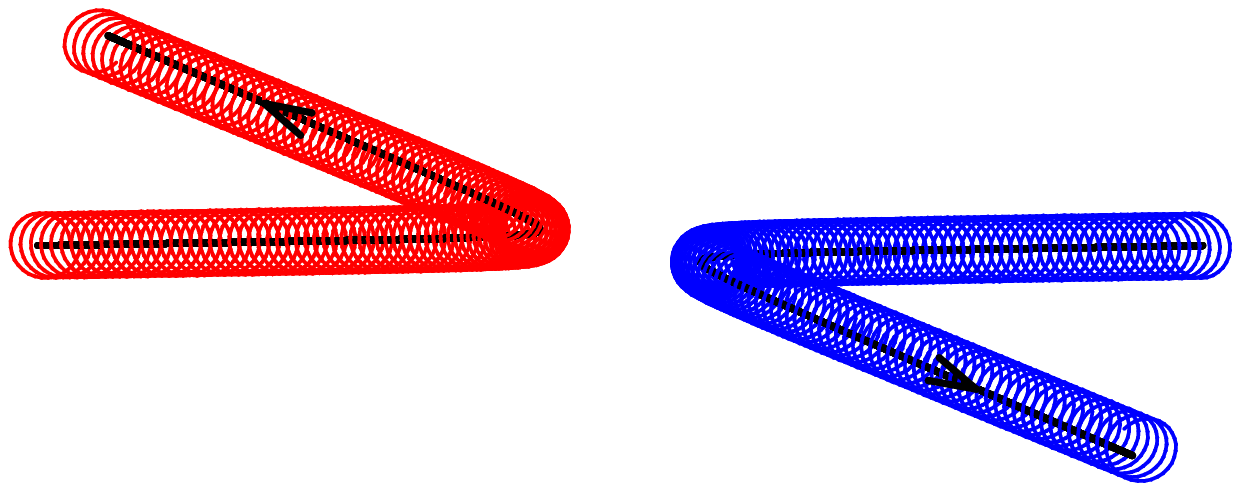}{Scattering of two spinning electrons with the spins parallel, in their center of mass frame.
It is also depicted the scattering of two spinless electrons with the same energy and linear momentum.}

In Figure \ref{fig2} we represent the scattering of two spinning electrons 
analysed in their center of mass frame. We send the particles with their spins parallel and 
with a nonvanishing impact parameter. In addition to the curly motion of their charges we can also 
depict the trajectories of their center of masses. If we compare 
this motion with the Coulomb interaction
of two spinless electrons coming from the same initial position and with the same 
velocity as the center of mass 
of the spinning electrons we obtain the solid trajectory marked with an arrow. 
Basically this corresponds to the
trajectory of the center of mass of each spinning particle provided the two particles 
do not approach each other below Compton's wave length. 
This can be understood because the average position of the center of charge
of each particle aproximately coincides with its center of mass and as far as they do not approach
each other too much the average Coulomb force is the same. The difference comes out when we consider
a very deep interaction or very close initial positions.

In Figure \ref{fig3} we represent the initial positions for a pair of particles
with the spins parallel. The initial separation $a$ of their center of masses is 
a distance below Compton's wave length. We also consider that initially the center of mass of each
particle is moving with a velocity $v$ as depicted.
That spins are parallel is reflected
by the fact that the internal motions of the charges, represented by the oriented circles that surround
the corresponding center of mass, have the same orientation.
It must be remarked that the charge motion around its center of mass can be characterised
by a phase. The phases of each particle are chosen 
opposite to each other.
We also represent the repulsive Coulomb force $F$ computed in terms of the separation 
of the charges. 
This interacting force $F$ has also been attached to the corresponding center of mass,
so that the net force acting on point $m_2$ is directed towards point $m_1$, and conversely. 
We thus see there that a repulsive 
force between the charges represents an atractive force between their center of masses when located
at such a short distance.

\cfigl{fig3}{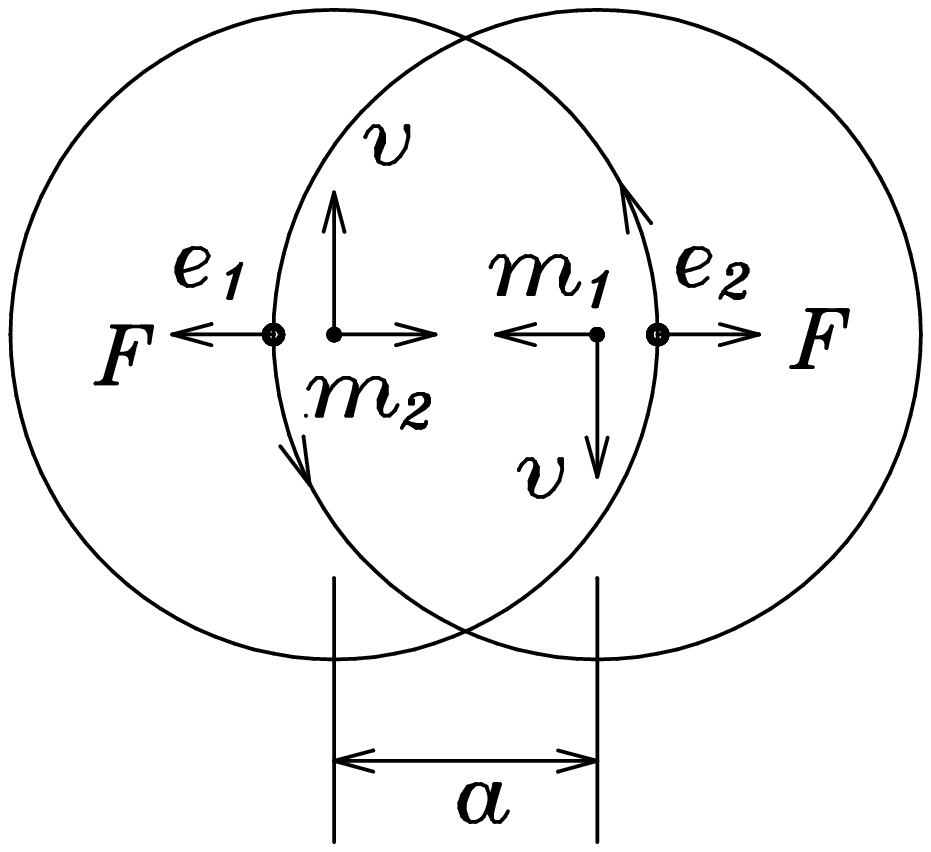}{Initial position and velocity of the center of mass and charges for a bound motion 
of a two-electron system with parallel spins. The circles would correspond to the 
trajectories of the charges if considered free. The interacting Coulomb force $F$ is computed
in terms of the separation distance between the charges.}

In Figure \ref{fig4} we depict the evolution of the charges and masses 
of this two-electron system for $a=0.4\lambda_C$ and $v=0.004c$ during a short time interval. 
Figure \ref{fig5} represents only the motions of the center of masses
of both particles for a longer time. 
It shows that the center of mass of each particle 
remains in a bound region.

\cfigl{fig4}{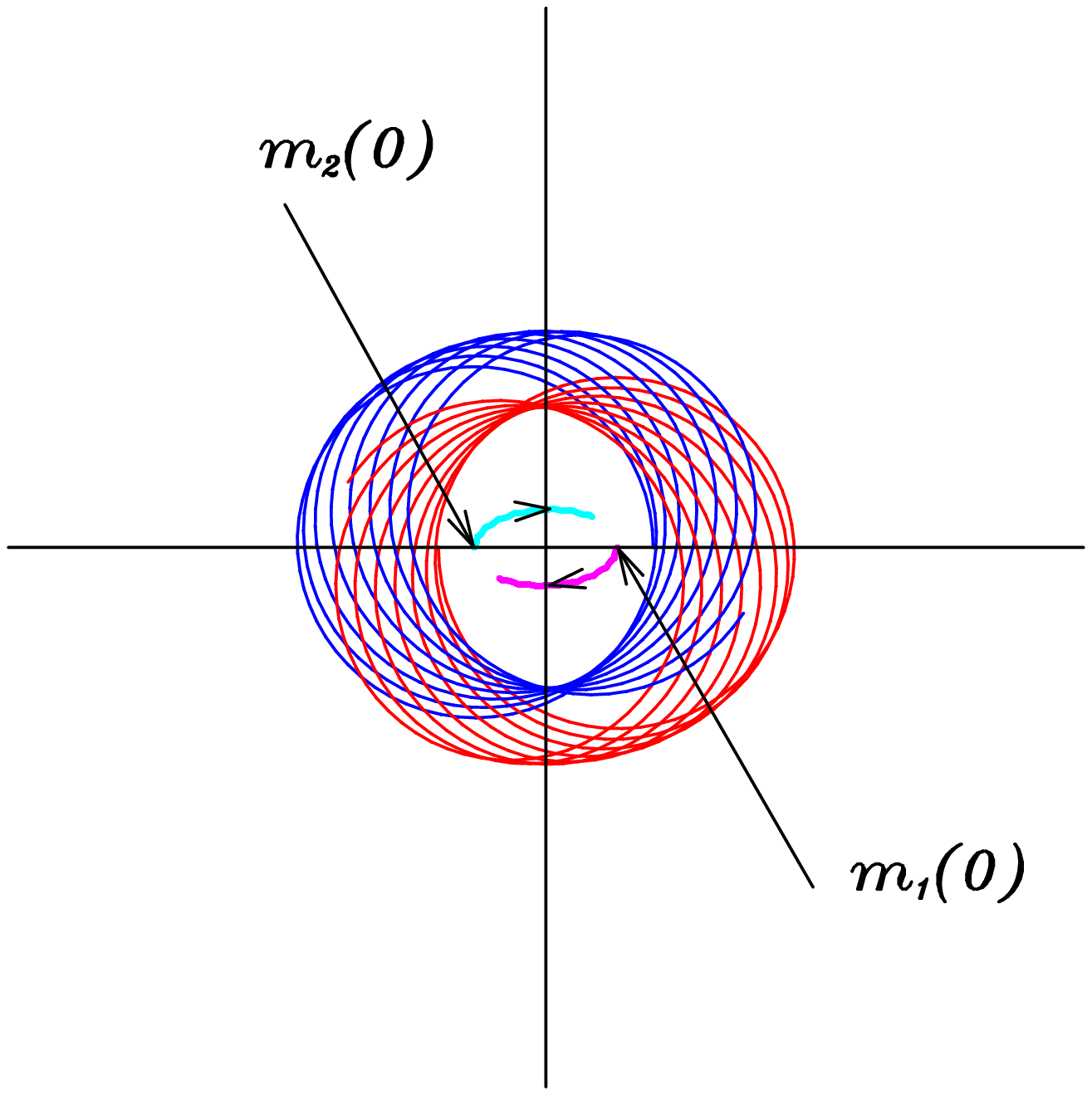}{Bound motion of two electrons with parallel spins during a short period of time}

The evolution of the charges is not shown in this figure because it blurs the picture
but it can be inferred from the previous figure.
We have found bound motions at least for the range $0\le a\le 0.8\lambda_C$ and velocity
$0\le v\le 0.01c$. We can also obtain similar bound motions if the initial velocity $v$ 
has a component along the $OX$ axis. 
The bound motion is also obtained for different initial charge positions
as the ones depicted in Figure \ref{fig3}. This range for the relative phase 
depends on $a$ and $v$ but in general the bound motion is more likely if the 
initial phases of the charges are opposite to each other.

We thus see that if the separation 
between the center of mass and center of charge of a particle ({\it zitterbewegung})
is responsible for its spin structure then this atractive 
effect and also a spin polarised tunneling effect can be easily interpreted.

\cfigl{fig5}{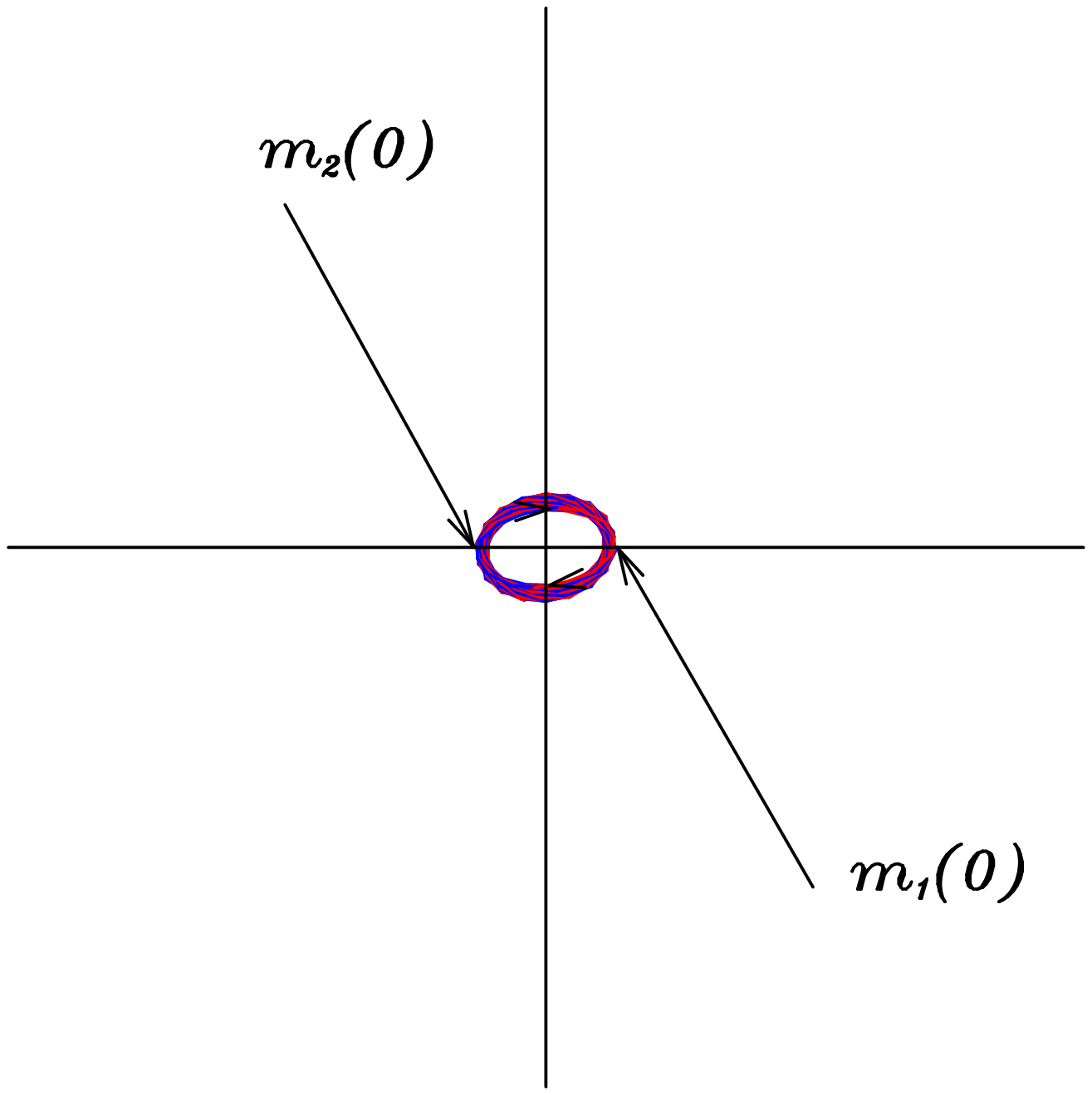}{Evolution of the center of mass of both particles for a larger time}

A bound motion for classical spinless electrons is not possible. 
We can conclude that one of the salient features of this example
is the existence from the classical 
viewpoint of bound states for spinning electron-electron interaction. 
It is the spin structure which contributes to the prediction of new physical phenomena.
If two electrons have their center of masses separated by 
a distance greater than Compton's wave length
they always repel each other as in the spinless case.
But if two electrons have their center of masses separated by a distance less 
than Compton's wave length
they can form from the classical viewpoint bound states provided some 
initial conditions on their relative 
initial spin orientation, position of the charges and 
center of mass velocity are fulfilled. 

The example analysed gives just a classical prediction, not a quantum one, 
associated to a model that satisfies
Dirac's equation when quantised. The possible 
quantum mechanical bound states if they exist, must be obtained from the 
corresponding analysis of two interacting quantum Dirac particles, bearing in mind 
that the classical bound states are not forbiden from the classical viewpoint.
Bound states for a hydrogen atom can exist from the classical viewpoint for any
negative energy and arbitrary angular momentum. 
It is the quantum analysis of the atom that gives the correct
answer to the allowed bound states. 

\newpage

{\Large\bf References}

These are the main references of the author and colaborators about this formalism. References to other
works are included in the main text as footnotes.

\vspace{1cm}
{\Large\bf Book}

\noindent M. Rivas,\\ {\it Kinematical theory of spinning particles},\\ 
Fundamental Theories of Physics Series, vol 116, Kluwer, Dordrecht (2001).

\vspace{1cm}
{\Large\bf Articles}

\noindent M. Rivas,\\ {\it Classical Particle Systems: I. 
Galilei free particles}, J. Phys. {\bf A 18}, 1971 (1985).\\
{\it Classical Relativistic Spinning Particles}, 
J. Math. Phys. {\bf 30}, 318 (1989).\\
{\it Quantization of generalized 
spinning particles. New derivation of Dirac's equation},\\ 
J. Math. Phys. {\bf 35}, 3380 (1994). \\
{\it Is there a classical
spin contribution to the tunnel effect?}, Phys. Lett. {\bf A 248}, 279 (1998).\\
{\it The dynamical equation of the spinning electron}, J. Phys. A, {\bf 36}, 4703, (2003),\\ and also LANL ArXiv:physics/0112005.\\
{\it Are the electron spin and magnetic moment parallel or antiparallel vectors?},\\ LANL ArXiv:physics/0112057.\\
M. Rivas, J.M. Aguirregabiria and A. Hern\'andez,\\ {\it A pure kinematical explanation of 
the gyromagnetic ratio $g=2$ of leptons and charged bosons}, 
Phys. Lett. {\bf A 257}, 21 (1999).

\end{document}